\documentclass{dissertation} 
\usepackage{booktabs}
\usepackage{siunitx} 
\usepackage[final]{pdfpages} 
\usepackage{gensymb} 
\usepackage{afterpage} 
\usepackage{adjustbox} 
\usepackage{colortbl} 
\usepackage{rotating} 
\usepackage{microtype} 
\usepackage{textcomp} 
\usepackage{textgreek}
\usepackage{float}

\usepackage{bbm,tensor,amsthm} 
\usepackage{comment}
\usepackage{subcaption}
\usepackage[shortlabels]{enumitem} 
\usepackage{shorttoc}

\usepackage{doi} 
\usepackage[natbibapa]{apacite} 

\addto{\captionsenglish}{} 

\newcommand{\pb}[2]{\ensuremath{\lf\{#1,#2 \rt\}}}
\newcommand{\mean}[1]{\ensuremath{\lf\langle #1 \rt\rangle }}

\newcommand{\deby}[2]{\ensuremath{\frac{\rm d #1}{\rm d #2}}}
\newcommand{\diby}[2]{\ensuremath{\frac{\partial #1}{\partial #2}}}
\newcommand{\ddiby}[2]{\ensuremath{\frac{\delta #1}{\delta #2}}}

\newcommand{\st}[1]{#1} 

\newcommand{\ud}[2]{\ensuremath{\indices{^{\st{#1}}_{\st{#2}}}}}
\newcommand{\du}[2]{\ensuremath{\indices{_{\st{#1}}^{\st{#2}}}}}

\newcommand{\T}[2]{\ensuremath{T\indices{_{(\st{#1})}^{\st{#2}}_\alpha}}}
\newcommand{\dT}[2]{\ensuremath{\dot{T}\indices{_{(\st{#1})}^{\st{#2}}_\alpha}}}

\def\lf {\ensuremath{\left}}  
\def\rt {\ensuremath{\right}}
\def\de {{\rm d}}  

\def\Lie { \ensuremath{\mathfrak {L}} }
\def\Sec {Section~}
\def\chap {Chapter~}
\def\fig {Figure~}
\def\eqn {Equation~}
\def\Xinv { \ensuremath{X_\text{inv}} }

\def\gauge {gauge }
\def\Ham { \ensuremath{\mathcal {H}} }
\def\Janus {\ensuremath{\mathcal {J}}}

\newif\ifchapcomp 
\newif\ifdumb 
\def\pdots{}

\begin{document}

\chapcompfalse

\title[How to count what counts]{ Gauge symmetry and the arrow of time }
\author{Sean}{Gryb}

\frontmatter

\begin{titlepage}

\thispagestyle{empty}
{\flushleft\Huge\titlestyle\titlefont\bfseries\makeatletter
	\@title \\

	\vspace{1em}

\begin{minipage}{0.65\textwidth}
	\begin{flushleft}
		\Large \@subtitle
	\end{flushleft}
\end{minipage}
\makeatother}

\vfill

\begin{flushright}
{\Large\titlestyle\titlefont\bfseries\makeatletter
	\@firstname\ \@lastname
\makeatother}
\end{flushright}

\cleardoublepage

\newpage \thispagestyle{empty}

{\fontsize{11pt}{11pt}\selectfont 

\vspace{1em}

\includegraphics[width=7.38cm]{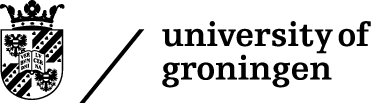}

\vspace{4em}

\begin{center}
	{\Huge\titlefont\bfseries 
		{\makeatletter
			\@title \\
		\makeatother}
	}

	\vspace{1em}

	{\Large\titlefont
		{\makeatletter
			\@subtitle \\
		\makeatother}
	}	

	\vspace{4em}

	{ \large\bfseries PhD Thesis\\ }

	\vspace{4em}
	
	to obtain the degree of PhD at the\\
	University of Groningen\\
	on the authority of the\\
	Rector Magnificus Prof. J.M.A. Scherpen\\
	and in accordance with\\
	the decision by the  College of Deans.\\

	\vspace{1em}

	This thesis was defended in public on\\

	\vspace{1em}

	Thursday 11 September 2025 at 12.45 hours\\

	\vspace{4em}

	by\\

	\vspace{4em}

	{\Large\bfseries
		{\makeatletter
			\@firstname\ \@lastname\\
		\makeatother}
	}

	\vspace{1em}

	born on 22 December 1981\\
	in Ottawa, Canada

\end{center}

\newpage\thispagestyle{empty}
{\flushleft

	{\bfseries Supervisors\\}

	Prof. J.W. Romeijn\\
	Prof. D. Roest\\

	\vspace{2em}

	{\bfseries Co-supervisor\\}
	
	Dr. S. Friederich\\

	\vspace{2em}

{\bfseries Assessment committee\\}

	Prof. R. Loll\\
	Prof. J. Uffink\\
	Prof. B.W. Roberts\\
}
} 

\cleardoublepage
\end{titlepage}

\dedication{For Nienke, Tyler and Kyra, who count more than anything.}

\shorttoc{Short Contents}{0}
\tableofcontents

\chapter*{Abstract}
\addcontentsline{toc}{chapter}{Abstract}

In this thesis, I achieve two main goals. The first is to give a universal definition of gauge symmetry that contains a minimal set of dynamic and epistemic ingredients. The second is to use this definition to motivate a new explanation of the Arrow of Time that evades the standard objections plaguing existing proposals.

My definition of gauge symmetry is inspired by a proposal due to Dirac that identifies gauge structure with representational structure that is underdetermined by the dynamical laws. To arrive at my definition, I generalise Dirac's analysis in two ways. First, I make use of an explicit account of representation to cleanly separate puzzles that result from well-known problems of model-building from those that result directly from gauge symmetry. Then, I extend Dirac's formalism to apply beyond standard Hamiltonian systems. When applied to well-studied examples of symmetry, my definition is consistent with expectations. But when applied to a particular kind of scaling symmetry called \emph{dynamical similarity}, my definition implies that dynamical similarity should be treated as a gauge symmetry of cosmology. While this result is found to be consistent with certain standard practices in cosmology, broader implications for empirical and conceptual problems in cosmology arise.

One such implication regards the explanation of the Arrow of Time. When implementing dynamical similarity as a gauge symmetry using my extension of Dirac's method, the resulting theory can contain dynamically stable \emph{attractor} states and dynamically privileged states called \emph{Janus points} even though the underlying dynamics is time-reversal invariant. When such structures exist, I say that a \emph{Janus--Attractor scenario} has been realised and show how such a scenario can generally be used to define an Arrow of Time pointing from the Janus point to an attractor for an observer near that attractor.

I then consider two empirical problems that, I argue, underpin the Arrow of Time in our Universe: the \emph{smoothness problem}, which involves explaining the relative smoothness of the early Universe, and the \emph{red-shift problem}, which involves explaining the wildly out-of-equilibrium behaviour of the red-shift manifest in the large monotonic values of the Hubble parameter in the past. Two corresponding models are then constructed: a self-gravitating Newtonian $N$-body model and a Friedmann--Lema\^itre--Robertson--Walker cosmology. I show that, under certain reasonable assumptions, gauge-fixing dynamical similarity in the $N$-body model solves the smoothness problem and doing the same in the cosmological model solves red-shift problem. These solutions arise as realisations of a Janus--Attractor scenario and suggest a general mechanism for a more comprehensive explanation of the many aspects of the AoT in our Universe.

\chapter*{Samenvatting}
\addcontentsline{toc}{chapter}{Samenvatting}

{\selectlanguage{dutch}
	
In dit proefschrift realiseer ik twee hoofddoelstellingen. De eerste is het geven van een universele definitie van ijksymmetrie die een minimale set dynamische en epistemische elementen bevat. De tweede is het gebruiken van deze definitie om een nieuwe verklaring voor de pijl van de tijd te motiveren die de gebruikelijke bezwaren tegen bestaande voorstellen ontwijkt.

Mijn definitie van ijksymmetrie is geïnspireerd op een voorstel van Dirac, waarin ijkstructuur wordt geïdentificeerd met representatiestructuur die niet volledig wordt bepaald door de dynamische wetten. Om tot mijn definitie te komen, generaliseer ik Diracs analyse op twee manieren. Ten eerste maak ik gebruik van een expliciete theorie van representatie om verwarring tussen bekende problemen van modelvorming en de gevolgen van ijksymmetrie te vermijden. Vervolgens breid ik Diracs formele aanpak uit zodat deze ook buiten standaard Hamiltoniaanse systemen toepasbaar is. Toegepast op bekende voorbeelden van symmetrie, stemt mijn definitie overeen met de verwachtingen. Maar toegepast op een specifiek type schalingssymmetrie, genaamd dynamische gelijkenis, impliceert mijn definitie dat deze symmetrie als een ijksymmetrie van de kosmologie moet worden beschouwd. Deze conclusie blijkt in overeenstemming te zijn met bepaalde gangbare praktijken in de kosmologie, maar leidt ook tot bredere empirische en conceptuele implicaties.

Een daarvan betreft de verklaring van de pijl van de tijd. Wanneer dynamische gelijkenis wordt geïmplementeerd als een ijksymmetrie volgens mijn uitbreiding van Diracs methode, kan de resulterende theorie dynamisch stabiele \emph{aantrekkingspunten} en dynamisch bevoorrechte toestanden bevatten, zogenoemde \emph{Januspunten}, ondanks dat de onderliggende dynamica tijdomkeerinvariant is. In gevallen waarin zulke structuren bestaan, spreek ik van een gerealiseerd \emph{Janus--Aantrekkingspunt (JA)-scenario} en toon ik aan hoe zo'n scenario algemeen kan worden gebruikt om een pijl van de tijd te definiëren die loopt van het Januspunt naar een aantrekkingspunt voor een waarnemer nabij dat aantrekkingspunt.

Vervolgens behandel ik twee empirische problemen die, zo als ik betoog, ten grondslag liggen aan de pijl van de tijd in ons universum: het gladheidsprobleem, dat vraagt om een verklaring voor de relatief gladde begintoestand van het vroege heelal, en het roodverschuivingsprobleem, dat vraagt om een verklaring voor het extreme niet-evenwichtsgedrag van de roodverschuiving, zichtbaar in de grote monotoon dalende waarden van de Hubbelparameter in het verleden. Ik construeer twee corresponderende modellen: een zelfgraviterend Newtoniaans $N$-deeltjesmodel en een FLRW-kosmologie. Ik laat zien dat, onder bepaalde redelijke aannames, het ijk-vastleggen van dynamische gelijkenis in het $N$-deeltjesmodel het gladheidsprobleem oplost, en als we hetzelfde doen in het kosmologisch model dat dat het roodverschuivingsprobleem oplost. Deze oplossingen vormen realisaties van een JA-scenario en suggereren een algemeen mechanisme voor een meer omvattende verklaring van de vele aspecten van de pijl van de tijd in ons universum.

}

\chapter*{Foreword}
\addcontentsline{toc}{chapter}{Foreword}
\setheader{Foreword}

\subsection*{Acknowledgements}

I am heavily indebted to many people without whom this dissertation would not have been possible. First, I would like to acknowledge the incredible support, advice and guidance I received from Simon Friederich and Jan--Willem Romeijn. Their mentorship and expertise was essential in helping me to develop the ideas of this thesis and their support came at a vital crossroad in my life. They helped me find my voice within the philosophy of physics. My long-time collaborator and friend Karim Th\'ebault was also essential during this process. Thanks for your patience and insight during my endless brainstorming (i.e., rants) and for providing helpful mentorship at just the right time. Another important person that was essential to this work is David Sloan. He introduced me to concept of attractors in contact systems. His kindness over the years have been invaluable. A source of continued inspiration for my has been Julian Barbour, who introduced me to the concept of a Janus point and who has always pushed me to express physics in terms of the minimal, necessary ingredients.

Thanks to Diederik Roest for support, the Young Academy Groningen for funding the PhD scholarship that supported this work, and the Philosophy Faculty and the University College of the University of Groningen for hosting me.

The ideas in this thesis take their roots in \emph{shape dynamics}, an attempt to describe the Universe in a scale-free way. This work therefore depends on the genius of all the ``shapers'' I have had the pleasure of frequently interacting with over the years. In addition to Julian Barbour and Dave Sloan, who are mentioned above, these include Henrique Gomes, Tim Koslowski, Flavio Mercati, Pedro Naranjo and Vasudev Shyam. Other people who have contributed to the programme and who have greatly influenced me are Edward Anderson, Steffen Geilen, Hans Westman, Derek Wise, and Tom Zlosnik. Alessandro Bravetti helpfully commented on the sections on contact geometry. Lee Smolin, though perhaps not an official shaper, has contributed greatly to the programme and has been incredibly formative in my development as a physicist and philosopher.

The philosophy of physics community has been a constant source of support, inspiration and expertise. Jeremy Butterfield, in particular, has provided invaluable mentorship at crucial moments in my career and has helped shape the ideas of this thesis. Others who have provided relevant support and insights through innumerable discussions include Harvey Brown, Guido Bacciagaluppi, Clara Bradley, Erik Curiel, Sam Fletcher, Stephan Hartmann, James Ladyman, Dennis Lehmkuhl, Renate Loll, Fred Muller, Patricia Palacios, Oliver Pooley, James Read, Bryan Roberts, Dominic Ryder, Jos Uffink, Manus Visser, and David Wallace among many others I have unintentionally neglected.

Finally, I'd like to thank all the friends and family who have supported me over the years. In particular, I must express my endless gratitude to my wife Nienke, whose undying support has kept me afloat, and to my beautiful children Tyler and Kyra, who provide endless inspiration and joy.

\subsection*{Author declaration}

Some material from this thesis has been adapted from published work or work in preparation as outlined below. For all collaborations listed below, authorship was alphabetical (i.e., there are no first authors) but I was the primary author of the excerpts that have been adapted for this thesis.
\begin{itemize}
    \item Some passages from \Sec\ref{sub:dyn of sim} about the nature of dynamical similarity have been adapted from \cite{Gryb:2021qix}.
    \item Excepts from \Sec\ref{sec:lagrangian_variational_principle} to \ref{sec:the_dirac_algorithm} have been adapted from \cite{gryb_thebault_book}.
    \item Parts of \Sec\ref{sec:motivation} and \ref{sec:statement_of_the_pesa} are adapted from excepts of \Sec 2 of \cite{Gryb:2021qix}.
    \item \Sec\ref{sec:explanatory_target} to \ref{sec:the_dilemma} and \ref{sub:conc_gen-part_impasse} are based on a manuscript in preparation with Simon Friederich.
    \item \chap\ref{ch:against_PH} is adapted from \cite{ph_takedown}.
\end{itemize}

\noindent This thesis is the result of my own work and includes nothing which is the outcome of work done in collaboration except as declared above and specified in the text.\medskip

\noindent It is not substantially the same as any work that has already been submitted before for any degree or other qualification except as declared above and specified in the text.\medskip

\begin{flushright}
{\makeatletter\itshape
    \@firstname\ \@lastname \\
    Groningen, 11 Sept 2025
\makeatother}
\end{flushright}


\mainmatter

\thumbtrue

\chapter{Introduction}
\label{ch:intro}

\ifchapcomp
    \tableofcontents
    \newpage
\else
    \cleardoublepage
\fi

\epigraph{
    To-morrow, and to-morrow, and to-morrow,\\
    Creeps in this petty pace from day to day,\\
    To the last syllable of recorded time;\\
    And all our yesterdays have lighted fools\\
    The way to dusty death. Out, out, brief candle!\\
    Life's but a walking shadow, a poor player\\
    That struts and frets his hour upon the stage\\
    And then is heard no more.
}{Macbeth (Act 5, Scene 5, lines 18–25)}

\section{Symmetry, possibility and Time's arrow}
\label{sec:general intro}

From the youngest child to the greatest poet, none can escape the advance of time. For some, this knowledge carries the burden of confronting the realities of age. For others, it highlights the wonderful fleetingness of the human experience. But for the physicist, time's directionality is a puzzle. While the direction of our experience in time is as undeniable as it is inseparable from our humanity, our best physical laws don't seem to explain it.

In this thesis, I will propose a new solution to this puzzle. I will develop a programme to account for the direction of processes in time despite the apparent time symmetries of our best fundamental theories. From now on, I will call this puzzle the problem of the \emph{Arrow of Time, (AoT)}. My approach will differ from standard attempts to explain the AoT in that it will be based on a symmetry argument. I will observe that, at the level of modern cosmology, the overall size of the universe --- but importantly \emph{not} its rate of change --- is devoid of empirical content. When the consequences of this simple observation are taken to their logical conclusion, a new understanding of our best fundamental physical theories will be revealed. In this new understanding, observers like us typically experience the pervasive and substantial time-asymmetries that comprise the AoT in our universe.

While this thesis is ultimately about explaining the AoT, it is also about understanding symmetry --- both what symmetry is and what its implications are for scientific theorising. To arrive at the proposed explanation of the AoT, I will first need to establish what it means for there to be more structure in a theory's models than in the phenomena it is trying to describe. In physics, this old problem involves understanding the nature of `gauge symmetry.' My first task will then be to give a clear definition of gauge symmetry and then use it to motivate rules for what to do when a gauge symmetry is present.

Applying those rules to the physical theories of cosmology will directly lead to my proposed explanation of the AoT. Consequently, the plausibility of my proposal hinges on the plausibility of my analysis of gauge symmetry. I will therefore dedicate significant effort to this analysis. There is, undeniably, an abundance of existing literature on the subject of gauge symmetry. One thing that will be different about my treatment is that it will be general enough to accommodate my case of interest, where the gauge symmetry in question has important formal differences from well-studied cases. In particular, the gauge symmetry relevant to cosmology, called \emph{dynamical similarity}, will be found to break the standard formal relations between configurations and their momenta: it treats scale as gauge but not scale velocity. It is this non-standard feature of dynamical similarity that will lead to a new understanding of the AoT. The philosophical foundations of my approach will, thus, require that my definition of gauge symmetry can handle all the well-studies cases of gauge symmetry and generalise naturally to dynamical similarity.

One important aspect of my proposal is that it requires a shift in convention. Some (or perhaps most) important problems associated with the AoT involve matters of degree because the puzzles at hand involve explaining the \emph{amount} of time-asymmetry in a system. For such questions, one needs to choose some method for counting possibilities, and such choices usually introduce a certain amount of convention. For almost two centuries, physicists have been in the rather fortunate position of having an obvious measure to choose from: the \emph{Liouville measure}, which is robustly time-independent and wildly successful as an explanatory tool in physical practice. Unfortunately, as we will show,\footnote{ See \Sec\ref{sub:measures_on_counting_solutions}. } the very nature of dynamical similarity makes it incompatible with the Liouville measure. This fact shatters long-held intuitions and opens a door to new ways of thinking about the counting of dynamical possibilities.

To be clear, I will not yet be able to settle on a single convention that can be used in all imaginable cases. But I will show that there exists a natural family of choices that can explain the AoT and share universal features rivalling the explanatory virtues of the Liouville measure. Moreover, this family includes the measures \emph{already} being used by modern cosmologists --- even though the correct usage of these measures is the subject of ongoing debate. My framework, then, severs as a way to give a clear conceptual understanding of how such measures should be used and interpreted.

My proposal for solving the AoT will be at the same time general and specific. Because it is leveraged on a symmetry argument, the proposal will have general features that reflect the structure of dynamical similarity. But it's one thing to provide a general argument for the existence of the AoT in general and another to predict the specific features of our world that comprise our experience of the AoT. To address the latter, I will study two specific models: one cosmological and one based on Newtonian gravity. Together, these models explain many of the properties of the AoT we experience in our universe.

This thesis can thus equally serve as a proposal for an explanation of the AoT, the first step in a more ambitious project to explain particular time-asymmetric features of the world, and a case study in understanding the implications of gauge symmetry. In the last regard, one may pose the question: what is it about gauge symmetries that makes them such powerful conceptual tools for theory construction? My tentative answer, which I will turn to at the end of my analysis, is that they provide strong constraints both on the space of states but also on the natural ways of counting dynamical possibilities over the space of states. Symmetries can thus radically reshape the inferential and explanatory structure of a theory by ensuring that the users of that theory count only those possibilities that really count in the world. When this is done for dynamical similarity in cosmology, the AoT is revealed.

\section{How to explain the Arrow of Time}
\label{sec:intro finding AoT}

\subsection{The problem of the Arrow of Time} 
\label{sec:intro AoT}

Let me now state more precisely the problem that I will aim to solve.\footnote{ My definition of the problem is based on the construction in \cite{price2002boltzmann} and is the subject of \chap\ref{ch:aot_prob}. } This problem is firstly based on the empirical observation that many physical processes in nature exhibit a significant and pervasive form of asymmetry \emph{in time}. This asymmetry is expressed as a numerically large temporal gradient in at least one quantity. Often, this quantity is some kind of entropy, but this is not essential --- a fact I will make use of later. Despite this large temporal gradient, our best fundamental physical theories are only able to explain a small amount of time asymmetry.\footnote{ The small amount of time asymmetry that can be explained is due to the lack of time-reversal invariance of the electroweak interaction of the Standard Model of particle physics. } The problem of the AoT is then to explain the large amount of time-asymmetry seen in physical processes despite the (near) time reversal symmetry of our best physical laws.

Several comments regarding this definition of the problem are now in order. First, I have not yet specified the particular processes for which I will seek an explanation. I will do so shortly below. Second, the formulation of the problem as posed above highlights the fact that I will be concerned primarily with finding explanations for particular empirical phenomena as they occur \emph{in} time. I will thus leave open metaphysical questions \emph{about} time itself, such as whether it has an intrinsic direction.

Third, while the numerical imbalance that needs explaining is often attributed to some type of entropy, restricting attention to particular notions of entropy is not essential except when one is specifically interested in the entropic arrows of particular thermodynamic systems. But defining entropy at the cosmological level is fraught with formal challenges resulting from the infinite dimensional nature of the state space.\footnote{ See, for example, the issues discussed in \cite{earman2006past}. } In my analysis, I will mainly be concerned with understanding the behaviour of physical quantities, such as the rate of expansion of the universe, that can be modelled by phase space functions. These can be given precise mathematical definitions even in the infinite dimensional case.

Finally, my phrasing of the problem suggests that the essential explanandum is the exceptional largeness of the numerical gradient. As such, I've identified the problem with a kind of counting problem: the universe is `far more' time-asymmetric than we could reasonably expect given our knowledge of our most fundamental laws. This clearly assumes, either explicitly or implicitly, a measure on the space of possible worlds. Insomuch as I will strive for mathematical rigour and epistemological clarity, I will try to avoid committing to any particular measure unless absolutely necessary. Thus, my general scenario for obtaining \emph{some} AoT does not involve the specification of a particular measure. It is only when I turn to specific questions, such as why some numerical imbalance is greater than a certain quantity, that the use of a measure is unavoidable. For the cases I'm interested in, it will suffice to give universality arguments that single out a \emph{class} of measures that can explain the broad quantitative features of the system.

With these comments in mind, let me now say more about the particular numerical gradients I will be interested in explaining. I will be concerned primarily with two problems. The first will be to explain the relative smoothness of the early state of the universe. This I will call the \emph{smoothness problem}. The second will be to explain the large and monotonic expansion rate of the universe.\footnote{ More specifically, the large monotonic \emph{decrease} of the Hubble parameter. } Because this expansion rate appears observationally as the red-shifting of waves as they propagate through space, I will call this the \emph{red-shift} problem.

I will give much more detail about and motivation for each of these problems later.\footnote{ See, for example, \Sec\ref{sec:explanatory_target}. } For now, let me note that while these problems are not meant to be exhaustive, they are considered by many to entail a significant portion of the commonly discussed puzzles associated with AoT. For example, an influential argument given in \cite[Ch 7]{Penrose:NewMin} states that our knowledge of the clumping properties of gravity is sufficient to explain the large reservoir of entropy produced by our Sun, and therefore the observed thermodynamic AoT seen on Earth, provided one can explain the early smoothness of the universe. This will form the basis for my motivations for considering the smoothness problem later. Additionally, \cite{Rovelli:2018vvy}, leveraging an argument made in \cite{wallace2010gravity}, argues for the importance of explaining the extreme expansion rates of the early universe to lock-in entropy in the form of the hydrogen atoms that fuel our Sun. This will later provide the basis for my motivations for considering the red-shift problem.

I will discuss the relative merits of the arguments above later. Note, however, that these empirical problems occur exclusively in the domain of classical physics. As a result, my explanation of these effects will be purely classical. I will thus leave open, for the moment, the question about the origin of time asymmetry in quantum mechanics. I will adopt the working hypothesis that the AoT can be explained without quantum theory --- notably, without solving its hard foundational problems.

Regardless of whether the smoothness and red-shift problems are the \emph{only} problems needing explanation, these problems certainly encompass a significant portion of the observed time asymmetries in the world. Finding explanations for these time asymmetries using a symmetry argument based on scale may even provide a new template for explaining time asymmetries more generally, including asymmetries relevant to the quantum formalism.

Support for this claim can be seen by noting that one of the primary virtues of my explanation for the AoT is that it is a very general explanatory scheme that can be applied to a wide class of physical situations. When applying my norms in systems with dynamical similarity to, for example, homogeneous and isotropic cosmologies, I will find a solution to the red-shift problem in those cosmologies (see \Sec\ref{sec:cosmological_models}). Similarly, when applying my norms to an $N$-body system of self-gravitating point particles, I will find a solution to the smoothness problem for those models (see \Sec\ref{sec:newtonian_gravitation_models}). This suggests that applying these norms to a more diverse set of models or to more realistic models of the universe will lead to solutions to more diverse aspects of the problem of the AoT.

\subsection{The Janus--Attractor scenario}
\label{sec:intro JA scenario}

Let me now give a sketch of the general scenario for obtaining an AoT in theories where dynamical similarity is a gauge symmetry.\footnote{ The full picture will be given in \Sec\ref{sub:the_janus_attractor_scenario}. } The idea will be to consider our two models of interest, namely the cosmological and $N$-body models, and first express them as dynamical systems on phase space. Then, we project the dynamical trajectories of these theories onto a smaller state space that is invariant under dynamical similarity.\footnote{ This projection can be either explicit, in which case the dynamics is projected onto a reduced system, or implicit, in which case the dynamics can be expressed on a fibre bundle with the fibres being the orbits of dynamical similarity. } My claim, which I will prove in detail in what follows, will be that the projected theories contains global structures, called \emph{attractors}, and local structures, that I will call \emph{Janus points}. It is these structures that define an AoT.

\begin{figure}[H]
    \centering
    \includegraphics[width=0.9\textwidth]{\pdots 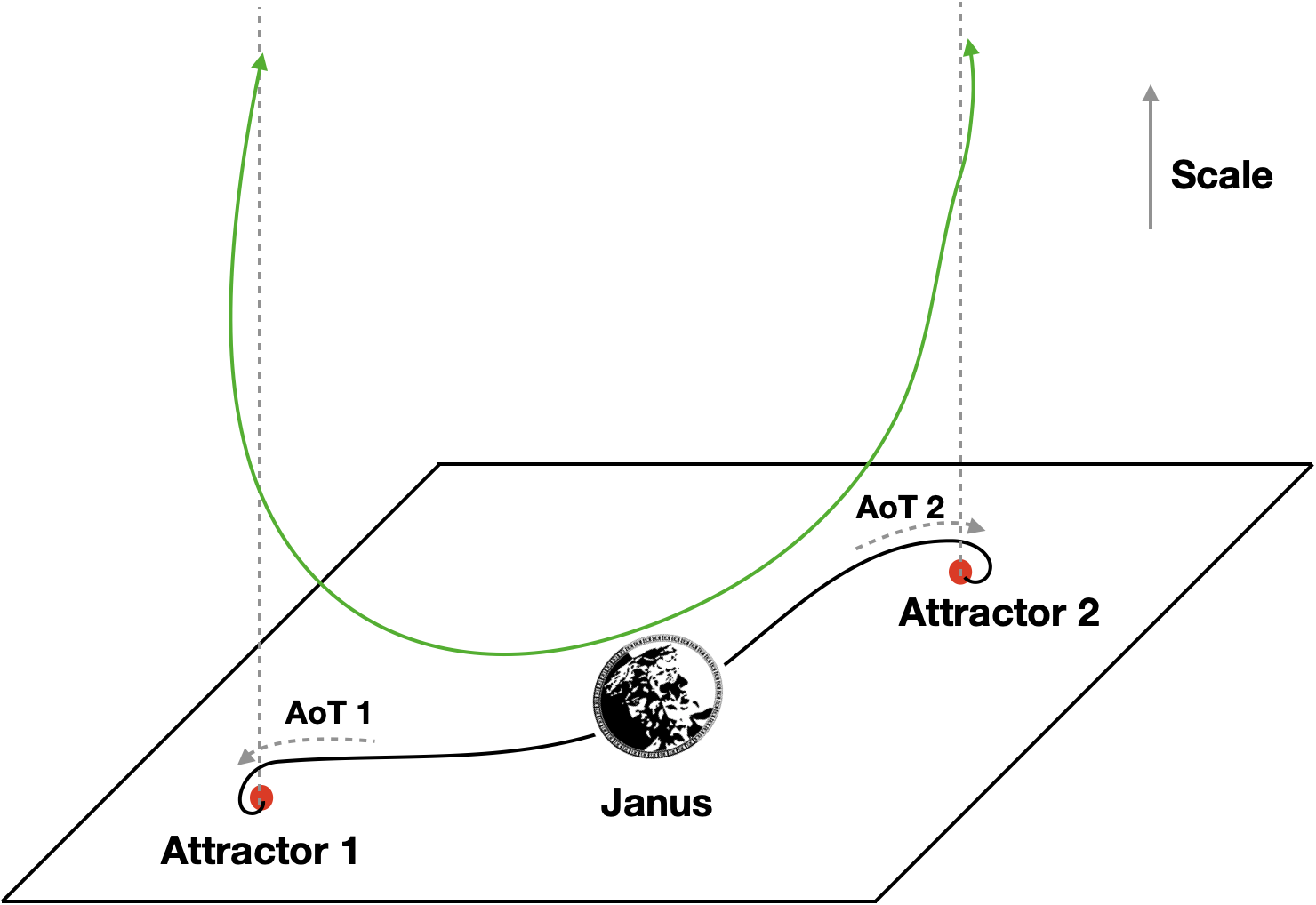}
    \caption{The Janus--Attractor scenario: an AoT pointing from a Janus point to an attractor is seen by an observer in a state close to that attractor.}
    \label{fig:intro JA scenario}
\end{figure}

More generally, I will argue that whenever attractors and Janus points occur along a dynamical trajectory of some dynamical system, there is an AoT seen by observers in states near a particular attractor that points from the Janus point to that attractor. The general situation is depicted in \fig\ref{fig:intro JA scenario}. Dynamical similarities are transformations that simultaneously rescale and reparametrize a dynamical trajectory.\footnote{ The word `similarity' refers to the change of scale in the geometric sense of a `similarity transformation' while the word `dynamical' refers to the change of time parameter. } Thus, the projection involves removing the directions on phase space that rescale unparameterized curves. Attractors in this scenario then arise when the dynamical evolution of the system before projection moves mostly along the directions of rescaling, as shown in \fig\ref{fig:intro JA scenario}. The existence of attractors therefore depends on the details of the dynamics as well as any dynamical conditions imposed on the solution space. For the models I will consider, the dynamical conditions required to guarantee the existence of attractors are well-motivated physically. For example, they require the cosmological constant to be positive and for the matter to not have negative energy. Once these conditions are satisfied, the existence of attractors follows from rigorous theorems.

Note that there can be many attractors in the state space --- even within a single dynamical trajectory. For that reason, the AoT in this scenario is not a global feature of a theory but something that is seen by particular observers in a specific kind of state --- namely, those that are \emph{close} to an attractor. Thus, different observers can see different, even incompatible, AoTs. For some observers, namely those \emph{close} to a Janus point, no AoT of this kind can even be assigned. A region around a Janus point then serves as a kind of transition region in the dynamics from which opposite AoTs emerge. The name \emph{Janus point}, introduced by Julian Barbour, is then aptly chosen since Janus is the two-faced Roman god of time, transitions and beginnings. Because this scenario for obtaining an AoT requires the existence of Janus points and attractors, I will call it a \emph{Janus--Attractor (JA) scenario}.

Let me describe in more detail the general features of the JA-scenario depicted above. I will assume, for simplicity, that the dynamical system in question is exactly time-reversal invariant, ignoring for now the small time-asymmetries that exist in realistic theories of the world because of the electroweak interaction. We will see that dynamical similarities preserve the time-orientation of a dynamical trajectory\footnote{ I.e., the time-reparametrisation resulting from the dynamical similarity is monotonic. } so that the invariant projection of the dynamics is time-reversal invariant. In this case, one might ask how it could be possible, even in principle, to have an AoT in such a theory?

The answer lies in the particular details of the construction. In \Sec\ref{sec:intro cosmo symmetry}, I will briefly motivate why one might expect attractors to arise when removing dynamical similarity. For now, let's discuss what they are in general. Attractors are defined to be stationary sets of the dynamics. That is, they are sets (not necessarily single points as they appear in \fig\ref{fig:intro JA scenario} due to the lack of dimensions on a page) reached from some dense set of trajectories in the asymptotic `forward' flow of the dynamics. However, if an attractor exists in the `forward' flow, another must also exist in the `backward' flow when the theory is time-reversal invariant. This is because any proof of the existence of an attractor in the forward direction can equally be applied to the same effect in the backward direction. So attractors can be consistent with time-reversal invariance provided such attractors bound trajectories in \emph{both} temporal directions. Note that nothing requires that these attractors be themselves symmetric to each other in any way.

This does not yet give rise to an AoT. The fact that it doesn't is a reflection of the underlying (near) time-reversal invariance of the dynamics and is the basis for the well-known \emph{temporal double-standard} objection to many proposed explanations of the AoT raised in \cite{price2004origins}. To get around this objection, I need to introduce an additional structure to break the time symmetry, at least locally, between the two attractors that bound a trajectory. This is achieved by a Janus point. In the theories I will consider, the choice of Janus point will involve some amount of convention. This convention essentially amounts to a choice of Riemannian metric on the state space. The Janus point can then be thought of as a kind of mid-point between the attractors.\footnote{ This is not a mid-point in the sense of equal distance, since such distances can be infinite, but in the sense that the trajectories are geodesics of the chosen metric at a Janus point. } Two AoTs will then ultimately point from the Janus point to each of the attractors.

To understand the temporal directions defined above as realistic AoTs, I can now provide a concrete notion of nearness to an attractor. Using the Riemannian metric defining the Janus points, a state is \emph{near} to an attractor when the distance defined by the metric goes to zero. An observer in a state near an attractor will then see a monotonic and numerically large gradient in the inverse of this distance, defining a generic AoT by our definition. Therefore, when there is a JA-scenario, we achieve our minimal goal of proving the existence of \emph{some} AoT.

If one wants to go beyond establishing the mere existence of an AoT and consider questions of degree, such as determining the \emph{amount} and \emph{nature} of the time-asymmetry seen by a particular observer, then one must choose a specific Riemannian metric, or family of metrics, with corresponding Janus points that can give precise answers to the questions of interest. Importantly, in order for that choice to be explanatory, it must be motivated by additional theoretical virtues that include more than just the ability to explain the AoT. I will discuss the virtues of my own choices in the concluding section of the thesis.

\subsection{Scale symmetry in cosmology}
\label{sec:intro cosmo symmetry}

\epigraph{Suppose that in one night all the dimensions of the universe became a thousand times larger. The world will remain \emph{similar} to itself, if we give the word \emph{similitude} the meaning it has in the third book of Euclid. Only, what was formerly a metre long will now measure a kilometre, and what was a millimetre long will become a metre. The bed in which I went to sleep and my body itself will have grown in the same proportion. When I awake in the morning what will be my feeling in face of such an astonishing transformation? Well, I shall not notice anything at all. The most exact measures will be incapable of revealing anything of this tremendous change, since the yard-measures I shall use will have varied in exactly the same proportions as the objects I shall attempt to measure.}{\emph{The relativity of space} from \emph{Science and method}\\Henri Poincar\'e, 1908 (translation \citeyear{poincare2003science}). }

When it comes to measurements in cosmology, it is commonly accepted that, while the global scale of the universe is not directly observable, \emph{changes} of scale are.\footnote{ The issues discussed in this section are covered in more detail in \Sec\ref{sec:cosmological_models}. }

Let us fist discuss the non-observability of global scale. This fact can be easily understood by imaging what would happen, as Poincar\'e does above, if one increased the size of everything in the universe a thousandfold. In that case, all relative measurements in that rescaled universe would give precisely the same value. Consequently, any law depending on the absolute scale would be underdetermined by all knowable facts about the world. It is thus noteworthy that our laws of cosmology do, in fact, not depend on the absolute size of the universe. As is well-known, the initial value of the scale factor, $a_0$, which sets the overall scale of the universe, can be rescaled without affecting the cosmological equations of motion. This symmetry is an example of the dynamical similarities discussed above. However, if one combines this fact with the time-reparametrisation invariance of general relativity, then one can see that the value of the scale factor is not empirically meaningful at \emph{any} time because one can always make a choice of clock where the dynamics of the scale factor is (nearly) arbitrary.\footnote{ The `nearly' here reflects the fact that the time evolution of the scale factor can be rescaled only up to an arbitrary \emph{but monotonic} function by redefining the lapse function. }

The interplay between dynamical similarity and time-reparametrisation invariance will be studied extensively in this thesis. It suggests that the universe's size at any given time is ultimately empirically irrelevant. In other words, spatial scale is a gauge degree of freedom. The connection between the freedom to choose a clock and the freedom to choose a spatial scale is easy enough to understand: a choice of clock sets a global scale at every instant for the velocities of the system. Fixing a scale then also requires fixing a convention for comparing configurations and their rates of change.

For consistency, different conventions for fixing the scale will similarly require different choices of internal clock. Of course, once a particular convention is made, a scale-free description of system can be given, in principle, using any scale-invariant choice of internal clock. But the explicit structure of the dynamical equations may depend on the choice of convention. This will be confirmed in the detailed investigations that will follow. What is important for now is to establish that global spatial scale is empirically irrelevant in cosmology even though there are choices in how to describe the system without scale.

On the other hand, \emph{changes} of scale, encoded in the Hubble parameter $H$, are not only measurable but essential to many important cosmological processes. These changes can be observed via the red shifting of waves, most notably radiation, propagating through space. One can easily measure the red-shift by comparing the wavelength of light coming from distant galaxies to light produced under similar conditions in the lab. It is this red-shift that causes cooling in the early universe and fixes, for example, the relative abundance of elements in our universe.

These considerations lead to an unusual situation: a particular variable, in this case the spatial volume of the universe, is determined to be a gauge variable while its corresponding momentum, in this case the Hubble parameter, is not. Consequently, it is possible to eliminate the spatial volume from the theory, keep the Hubble parameter, and retain empirical adequacy.

What is so unusual about this is that the laws of physics are generally second order differential equations in time.\footnote{ We will make the point in this paragraph much more concretely in \Sec\ref{sec:dynamical_similarity}.} If one performs the reduction described above, then one obtains equations that are first order in time for exactly one choice of variable: the Hubble parameter. Phase space is then no longer a viable state space for the theory because phase space assumes a pairing between configurations and momenta that is only guaranteed for second order systems. Instead, we must use the odd-dimensional analogue to phase space called \emph{contact space}. But while Liouville's theorem guarantees a universal time-independent volume-form on phase space, no such result holds on contact space. This dramatically changes the structure and interpretation of the dynamical equations.

The equations for dynamical flows on contact space look very similar to Hamilton's equations on phase space.\footnote{ See Equations~\ref{eq:Darboux contact}. } For the even-dimensional part of a contact space, they are virtually identical except for a term that looks like a friction term in Hamilton's second equation. The coefficient of this friction-like term depends on the variable along the odd dimension of the contact space. I will call this variable the \emph{drag} because of its formal resemblance to the drag coefficient of a theory with friction. We will see that the drag will also determine the time-dependence of any counting procedures (more specifically, measures) that one wants to define on state space. This is unsurprisingly related to its formal resemblance to the drag coefficient of a system with friction.

A cautionary note about the drag is that it will depend on the particular convention used to fix the global scale. In cosmology, the drag can be identified with the Hubble parameter using a natural representation of the dynamics. This choice is the one that arises by fixing the scale factor to be a constant. But it is also the choice that leads to a counting procedure widely used for counting solutions in cosmology.\footnote{ This counting procedure is given by the ``physical'' measured used by cosmological that is defined in \cite{Hawking:1987bi} and given in \eqn\ref{eq:GHS}. We will discuss the reasons why cosmologists regard it as physical in \Sec\ref{sub:dynamical_similarity_in_the_universe}. } Moreover, because the Hubble parameter appears, for example, as a friction term in the Klein--Gordon equation of a massive scalar field in a homogeneous cosmology, the Hubble parameter is perhaps naturally identified with a kind of drag variable.

I have thus sketched an argument, which I will make more carefully throughout the thesis, that cosmological systems can be treated as contact systems where the scale factor has been eliminated and the Hubble parameter behaves like a drag coefficient for the matter fields in the universe. This gives an alternative way to think about red-shift not as waves being dragged out because of the expansion of space but because of a universal friction-like force. In the new picture, the dynamics are characterised by dissipative-like phenomena. And when there is dissipation, there can be attractors. This is precisely how attractors will be seen to arise in our formalism. Moreover, the Janus points will be characterised by points of the dynamics where the dissipative terms are zero. The behaviour of the drag therefore determines whether a JA-scenario is possible. There will always be, for example, a JA-scenario if there is a way to represent the system where the drag is monotonic and passes through zero. This explains more generally why theories that exhibit dynamical similarity as a gauge symmetry can (but don't always) lead to a JA-scenario.

\section{How to define a gauge symmetry}
\label{sec:intro how to find a gauge symmetry}

\subsection{Symmetry and its problems}
\label{ssec:intro symmetry and its problems}

My solution to the problem of the AoT is motivated by a symmetry argument. In particular, it is based on the claim that dynamical similarity should be considered a gauge symmetry of cosmology. It is therefore essential to my argument to give a general definition of gauge symmetry and a set of rules for what do to when one is found. To do this, I will outline a general principle, called the \emph{Principle of Essential and Sufficient Autonomy (PESA)}, that will clearly identify the gauge structures of a theory and suggest several norms for how to deal with them. The statement and justification of the PESA will be the focus of Part~\ref{part:foundations} of the thesis.

The project of identifying the empirical core of a theory and its theoretical implications has independent philosophical interest outside considerations about the AoT. Debates about what constitutes a physically significant difference and how theories should handle such differences have been the subject of (at times intense) debate throughout the history of physics. In modern physics, the importance of gauge symmetries is exemplified by the central role played by the coordinate invariance of general relativity and the Yang--Mills symmetries of the Standard Model of particle physics despite lacking explicit methods of removing such symmetries.

Given that the concept of a gauge symmetry plays such an important role in physics, it is natural to ask what common principles define it? One philosophical challenge is to understand the extent to which the formal structures of a theory's models can inform epistemologically sound theoretical practice for identifying gauge structure. One may doubt whether a theory's formal structures should have \emph{any} relation to good epistemology but \cite{earman1989world} provides striking examples, stemming from the spacetime symmetries of Newtonian mechanics and general relativity, where considerations about spacetime symmetries \emph{do} inform good theoretical practice. \cite{belot:sym_and_equiv} asks whether this can be done in general and gives many physical examples where standard definitions of gauge symmetry are inadequate when applied to specific physical contexts. I will later formulate these worries in terms of a general challenge to find a good definition of gauge symmetry and will call it \emph{Belot's Problem}.

To illustrate some of the difficulties that Belot alludes to, consider general relativity. One of its fundamental principles is purported to be its invariance under arbitrary coordinate transformations. But early in the development of the theory it was pointed out by Kretschmann \citep{kretschmann1918physical} that any theory whatsoever could be written in generally covariant form, raising questions about the physical significance of this invariance.\footnote{ For a more modern discussion of this point, see \cite{norton2003general}. } Since that time, many notions have been introduced or (re-introduced) to try to identify what sort of invariance is truly meaningful and what is not. I will discuss many such notions as I develop my own proposal, and highlight the strengths and weakness of each as I go.\footnote{ See \Sec\ref{sec:narrow_proposals} for a list of different notions and a discussion of their strengths and weaknesses. }

For now, I will attempt to illustrate some of the most pointed aspects of the problem.\footnote{ One can find a version of the argument presented below in \cite{belot2018fifty}. The example discussed here illustrates the failure of what Belot calls \emph{Earman's Principle} discussed in \Sec\ref{sub:belot_s_problem}. } It is common practice in general relativity to regard general coordinate transformations as gauge transformations of the theory because they lead to underdetermination in the equations of motion. Consider, then, the Kerr solution used as a model of the exterior geometry of some astrophysical black hole. If general coordinate transformations are indeed gauge symmetries, then changing the location or speed of the black hole should lead to an undetectable change in the system. This is because translations and boosts can be implemented as coordinate transformations, which have been identified as gauge. But the location and speed of a black hole are certainly empirically relevant quantities. It would seem that there is a mismatch between the physical quantities modelled by the theory and the underdetermination in the equations of motion.

One approach to this problem is to distinguish between symmetries that act on isolated subsystems and symmetries that act on the system as a whole.\footnote{ Such an approach is advocated, for example, by \cite{greaves2014empirical}. See \Sec\ref{sec:other symmetry notions} for a more detailed list of references. } Perhaps our definition of a gauge symmetry should depend upon this distinction? Under such an approach, because a black hole is a subsystem of the universe, translating it rigidly relative to the rest of the universe could be empirically significant even if such transformations would be empirically \emph{in}significant had they acted on the whole system.

But this definition of a gauge symmetry depends on a lot of extra structure. One first needs to carefully define an isolated subsystem and then what it means to rigidly translate it.\footnote{ For an attempt to do so see \cite{gomes2021holism} and \cite{gomes2022gauge} for the more general construction. } Additionally, one must address the problem of having a definition of symmetry that applies differently to the `whole' system than to its parts given that there is no way to know empirically whether one has actually found the \emph{whole} system --- especially when that system is thought to be the universe.

To overcome these difficulties, I will require more specificity in the statement of a theory. In particular, I will insist that a theory state all the idealizations under which its models should be considered valid representations of the phenomena being studied. My definition of a gauge symmetry will then apply universally to any theory, but only when the conditions of the idealisations are satisfied. In this way, general relativity should be thought of as a different theory when using asymptotic boundary conditions to model an isolated subsystem, as is done when using the Kerr solution to model astrophysical black holes, than when considering a theory of cosmology. Clearly, these two applications of general relativity operate under different physical idealisations. It is then not inconsistent to have two different sets of gauge symmetries arising from these two different theoretical frameworks.

One advantage of this approach is that it is nicely compatible with notions of gauge symmetry that depend on the existence of isolated subsystems. In my case, the definition of an isolated subsystem and the notion of a rigid translation can be packaged into the idealisation conditions needed to apply the theory in the chosen context. This means that these definitions do not need to appear anywhere in the definition of a gauge symmetry, allowing my formulation to be simpler and more flexible. We will see explicit examples illustrating these advantages in \Sec\ref{sec:problems_solved} after developing my proposal.

A related advantage of this approach is that the theoretical structures required to distinguish different theoretical contexts (i.e., those that define the idealizations, approximations and other auxiliary hypotheses) can be conveniently contained in the model-description of the models used to define the theory. Then, it is possible to separate questions about what makes a good representation from questions about what makes a good dynamical definition of a gauge symmetry. 

Finally, let me note that I will be silent regarding the metaphysical status of gauge and other structures. My definition of gauge symmetry and the norms that it implies will depend on the empirical status of the representational structures of a theory's models. Further inferences about the ontological status of these structures will be left to future investigations. 

\subsection{The proposal for Gauge Symmetry}
\label{sub:intro PESA}

Let me now give a sketch of the PESA. At the core of my definition of gauge symmetry will be a particular kind of underdetermination of theory by phenomena. The motivation for my definition comes from a well-known proposal introduced in \cite{dirac2001lectures}. Dirac's proposal gives a specific set of criteria for identifying gauge symmetries in constrained Hamiltonian systems. While this proposal is not general enough for my purposes,\footnote{ Since I just argued above that I'll be interested in contact systems, which are not Hamiltonian. } its motivations are very general. My proposal generalises and refines Dirac's proposal by making use of its general motivations. The resulting formalism, articulated by the PESA in \chap\ref{ch:pesa}, can be roughly summarised by the following slogans:
\begin{enumerate}
    \item \textbf{Slogan-definition:} A gauge symmetry is present when the representations of a theory are underdetermined by the phenomena (in a way I will make more precise below).
    \item \textbf{Slogan-norm:} When this occurs, the underdetermination of the dynamical equations of the theory \emph{should} be arranged to exactly match the underdetermination of the representations by the phenomena.\footnote{ This norm will provide the basis for my generalisation of the well-known \emph{Gauge Principle}, which is discussed in \Sec\ref{ssub:the_gauge_principle}. }
\end{enumerate}
I will refer to these as `slogans' to highlight the fact that I have intentionally ignored many details for the sake of clarity --- details that I will make explicit in my more careful treatment later.\footnote{ For the full statement of the PESA, see \Sec\ref{sec:statement_of_the_pesa}. The slogan-definition and norm given here are based on the more complete definition of a gauge symmetry and first normative rule given in that section. } When taken together, these slogans can be seen to provide concrete dynamical and empirical criteria for identifying a gauge symmetry in a theory. 

Lets me now describe in a bit more detail the resulting picture. Consider that the models of a theory are written in terms of some set of representational structures $\mathcal A$ that can be used to describe some phenomena. Considerable effort will be put into stating precisely what it means for a model to faithfully represent a target system, and therefore what it means to have a basis for empirical fact. Once this is achieved, my claim will be that a gauge symmetry exists when there is some smaller set of observable structures $\mathcal O \subset \mathcal A$ that is both necessary and sufficient for describing the phenomena.\footnote{ In a way that will be made more precise throughout Part~\ref{part:foundations}.} That is, there is a gauge symmetry when there is no smaller subset of $\mathcal O$ that can be used to maintain the empirical adequacy of the theory.

The slogan-definition results from the fact that $\mathcal O$ is sufficient, which means that there are structures in $\mathcal A$ that are not necessary for describing the phenomena and, thus, cannot be determined from them. The gauge symmetries themselves are then defined to be all transformations on $\mathcal A$ that leave $\mathcal O$ invariant.

\begin{figure}[H]
    \centering
    \includegraphics[width=0.6\textwidth]{\pdots 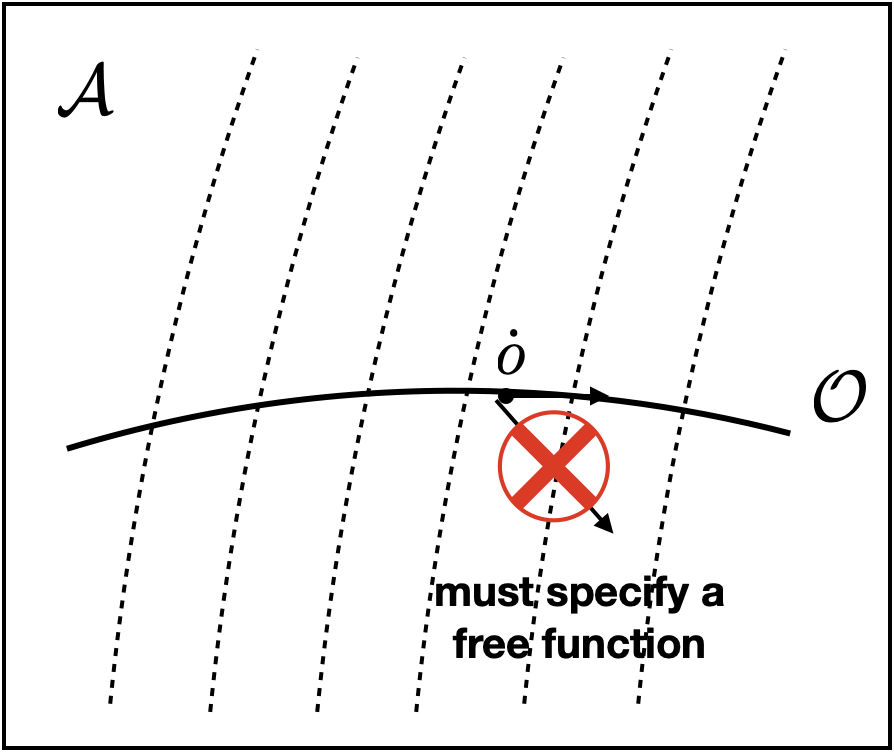}
    \caption{The general structure emerging from the PESA. Gauge symmetries are transformations on $\mathcal A$, the set of all the representational structures of a theory's models, that preserve $\mathcal O$, the set of all representational structures that are both necessary and sufficient to describe the phenomena. When a gauge symmetry exists, the dynamical equations are required to not be well-defined on $\mathcal A$ unless arbitrary values are assigned to gauge-dependent quantities. }
    \label{fig:intro PESA}
\end{figure}

My slogan-norm then requires that the theory has a formulation of its laws that is well-posed on $\mathcal O$ but \emph{not} on $\mathcal A$. I will motivate it below. Note that this criterion does \emph{not} restrict us to deterministic theories since the laws, while well-posed, may only determine states probabilistically. From a practical perspective, this means that the laws expressed on $\mathcal A$ simply can't be solved unless the user of the theory assigns arbitrary values to any quantities in $\mathcal A$ not determinable from $\mathcal O$. The overall picture is illustrated in \fig\ref{fig:intro PESA}.

The motivation for the slogan-norm is that it ensures is that a gauge symmetry cannot go undetected. In a formulation of the theory where the equations of motion on $\mathcal A$ are well-defined, one may not realise in using the theory that there is a variable whose value has no impact on empirical predictions. We will see that this can lead to reasoning errors that mislead intuitions, particularly when those intuitions involve typicality arguments. My norm can prevent such reasoning errors by identifying the structures of the theory that do not affect its empirical core.

Let me illustrate how to use my slogans with the simple example of a Newtonian universe. It is generally well-accepted that the position of absolute space in such a theory is not empirically accessible. Thus, in a simple $N$-body Newtonian system, the center-of-mass of the universe is underdetermined by the phenomena. The slogan-definition above then says that any transformation that changes the center-of-mass while leaving the center-of-mass coordinates invariant is a gauge transformation. 

At this point, one might worry that time-\emph{dependent} translations should qualify as gauge transformations under this definition. This appears to be problematic because Newton laws are not invariant under time-\emph{dependent} translations and predict complicated fictitious forces in non-inertial frames. This is where the slogan for my normative rule comes into play.

In this example, the norm requires that we \emph{reject} Newton's equations for this system because they are well-defined for the center-of-mass. Instead, my norm requires that we use an equivalent system of equations that is only well-defined for translation-invariant quantities. Fortunately, this is rather easy to do. All that is needed is to restrict Newton's equations to center-of-mass coordinates by imposing a constraint that enforces the vanishing of the linear momentum of the system. It is easy to see that this constrained system will be invariant under arbitrary time-dependent translations, and that the position of absolute space will be underdetermined.\footnote{ For a direct proof, see \Sec\ref{sub:bb_theory}. The constrained theory described here is equivalent to the Barbour--Bertotti theory defined in that section. } It is also easy to see that this new system is functionally equivalent to the previous one provided that the center-of-mass really is underdetermined by the phenomena.

Note, however, that we are now in an epistemically superior situation: we are secure in the knowledge that the centre-of-mass really is empirically irrelevant, as long as our theory continues to be empirically adequate, because the theory itself never makes use of the value of the center-of-mass to make any prediction whatsoever. On the other hand, if the new theory ever becomes inadequate for describing some phenomena of interest, we know that we can always reintroduce the center-of-mass into the dynamical system by returning to Newton's original theory.

\section{Roadmap}

The dissertation is divided into two parts. Part~\ref{part:foundations} motivates, formulates and justifies the PESA. This leads to my proposed definition of a gauge symmetry and the resulting consequences. Part~\ref{part:aot} makes use of this definition of a gauge symmetry to develop a proposal for explaining the AoT.

The first part of the thesis is divided into four chapters and the second into three. The first chapter of Part~\ref{part:foundations} (\chap\ref{ch:sym_probs}) establishes the main problem, which I call \emph{Belot's Problem}, that I aim to address with the PESA. Belot's Problem is motivated using several examples and then defined in \Sec\ref{sec:problems_with_symmetry}. I introduce the representational language used to articulate the PESA in \Sec\ref{sub:theory} and define several aspects of symmetry that will be useful in my analysis in \Sec\ref{sub:symmetry}. The chapter ends in \Sec\ref{sec:examples_of_symmetry} with a review of some standards notions of symmetry discussed in the literature and their relation to my analysis.

After defining the central problem in \chap\ref{ch:sym_probs} and developing the language for symmetry and representation that will be used throughout, \chap\ref{ch:en route gauge} describes several well-known attempts in the literature to define gauge symmetries. The emphasis is on describing the structural features of these proposals and the ways in which they fail to fully solve Belot's Problem. These failings are used to motivate my own solution. The chapter ends by introducing the notion of dynamical similarity, which is central to my analysis of the AoT, and explains how this symmetry differs from standard gauge symmetries (\Sec\ref{sec:a_new_kind_of_symmetry_dynamical_similarity}).

While \chap\ref{ch:en route gauge} introduces the conceptual foundation of different approaches to defining gauge symmetry, \chap\ref{ch:rep_sym} develops the mathematical machinery needed to model all the examples that will be relevant to the analysis later. Several important things are accomplished. First (although not in this particular order), I review standard approaches to gauge theory including Noether's theorems, Lagrangian constraints (\Sec\ref{sec:symmetries_of_the_variational_principle}), and the Dirac formalism (\Sec\ref{sec:the_dirac_algorithm}). Second, I develop a unifying mathematical formalism for treating general symmetries in physics (\Sec\ref{sec:the_initial_value_problem}) and show how this can be used to recover standard results. I illustrate the utility of this formalism by using it to shed light on a well-known problem, called the \emph{Frozen Formalism Problem}, affecting theories where the time parameter can be arbitrarily chosen (\Sec\ref{sub:reparametrisation_invariance}). Then, I apply the formalism to the case of dynamical similarity (\Sec\ref{sec:dynamical_similarity}). This important step will allow us to apply the norms resulting from the PESA to specific models of the universe to specifically address the red-shift and smoothness problems in the second part of the thesis. The mathematical formalism introduced in this chapter is thus essential to our analysis: it gives a general framework for treating symmetry that forms the basis of the PESA and provides a concrete way to implement the resulting norms.

In \chap\ref{ch:pesa}, the closing chapter of Part~\ref{part:foundations}, I state the PESA (\Sec\ref{sec:statement_of_the_pesa}) and then show how it can be used to solve the problems with symmetry introduced in \chap\ref{ch:sym_probs}. When treating the examples, I will base my analysis on the formalism developed in \chap\ref{ch:rep_sym}. This will allow me to compartmentalise the technical and conceptual analysis by giving the rigorous proofs that support the claims made in \chap\ref{ch:pesa} and then referencing them in \chap\ref{ch:rep_sym}.

Part~\ref{part:aot} of the thesis gives my proposed explanation of the AoT. The first chapter (\chap\ref{ch:aot_prob}), defines the problem. I begin by defining the smoothness and red-shift problems and show how the particular physical processes that characterise these problems are central to establishing the problem of the AoT (\Sec\ref{sec:explanatory_target}). I then review known approaches to explaining the AoT (\Sec\ref{sec:Price_taxonomy}) and illustrate their inadequacies (\Sec\ref{sec:the_dilemma}). The intention is to establish the need for a new approach based on fundamentally different assumptions about the character of the dynamical laws.

In \chap\ref{ch:against_PH} I take aim at the Past Hypothesis (PH), which is perhaps the leading approach to explaining the AoT. The purpose of this chapter is to review the mathematical and conceptual arguments typically employed for understanding the AoT in terms of a PH (\Sec\ref{sec:the_past_hypothesis}), describe and characterise the well-known criticism against it (\Sec\ref{sec:deconstructing_the_argument}), and add new criticism (\Sec\ref{sec:symmetries_and_measure_ambiguities}) making use of the PESA to argue for dynamical similarity as a gauge symmetry of cosmology (\Sec\ref{sub:dynamical_similarity_in_the_universe}). It is in this chapter that I will give my argument for why dynamical similarity should be considered a gauge symmetry. Upon accepting this, an advocate of the PH must choose between introducing a distinction without a difference by breaking a gauge symmetry or undermining the assumptions of the PH by accepting a time-dependent measure.

In \chap\ref{ch:new_aot}, the final chapter of Part~\ref{part:aot}, I introduce my new general scenario, the Janus--Attractor(JA) scenario, for explaining the AoT and use it to solve the smoothness and red-shift problems. The JA-scenario is defined in \Sec\ref{sub:the_janus_attractor_scenario}. The smoothness problem is addressed in \Sec\ref{sec:newtonian_gravitation_models}. Particular attention is paid to the properties of the chosen measure and the universality class that it is a part of. The red-shift problem is then addressed in \Sec\ref{sec:cosmological_models}. The universal nature of the assumptions required to solve this problem are highlighted.

I conclude the thesis in \chap\ref{ch:conclusions}. There, I assess the extent to which I have solved the problems I set out for myself (\Sec\ref{sec:conc problems_solved}), suggest further research (\Sec\ref{sec:prospectus}), and reflect on what my solution to the problem of the AoT could mean about the role of gauge symmetry in physical theory (\Sec\ref{sec:final_remarks}).

\cleardoublepage 
\part{Gauge Symmetry}
\label{part:foundations}

\chapter{Symmetry and its problems}
\label{ch:sym_probs}


\begin{abstract}
    In this chapter, I set the stage for my analysis of gauge symmetry. I develop a language for discussing symmetry in general and gauge symmetry in particular. I illustrate known problems for defining gauge symmetry using concrete examples and synthesize them into a problem I attribute to Belot. Then, I define the account of representation that I will later use to formulate my definition of gauge symmetry that solves Belot's Problem. Finally, I situate my analysis within the literature in terms of several standard gauge-symmetry concepts and review some well-known examples of gauge symmetries.
\end{abstract}

\ifchapcomp
    \tableofcontents
    \newpage
\else
    \cleardoublepage
\fi

\section{Introduction} 
\label{sec:introduction_sym_prob}

The purpose of Part~\ref{part:foundations} of the dissertation is to give a good definition of gauge symmetry that I can later use to motivate my explanation for the AoT. I will lay down the philosophical and physical foundations of my proposal in the first three chapters of Part~\ref{part:foundations}. In \chap\ref{ch:sym_probs}, I will set up the representational language that I will use throughout, define the main problem I want to solve (i.e., Belot's Problem), give my solution and illustrate how it applies to several examples.

One thing that distinguishes my definition of gauge symmetry from other proposals --- particularly those in the physics literature --- is that my definition will presuppose a detailed description of the modelling context when defining the theory's models. This means that my definition will require me to be explicit about my account of representation. I will do this in \Sec\ref{sub:theory} where I will adopt the account developed in \cite{frigg2020modelling}. Then, I will apply my construction in \Sec\ref{sec:problems_solved} and find that many puzzles usually associated with gauge symmetries are in fact common issues that arise when using models to represent target systems. Once these representational issues have been dealt with, the distinctive feature of gauge symmetries, namely that they generate underdetermination in dynamical evolution, becomes easier to identify. This leads directly to my proposal, which I will present in \chap\ref{ch:pesa}.

One important goal of \chap\ref{ch:sym_probs} will be to reconsider how a gauge symmetry is usually defined in the physics and philosophy literature. A well-known principle for doing so, dating back to Weyl, is the \emph{Gauge Principle}, which I will describe in detail in \Sec\ref{ssub:the_gauge_principle}.\footnote{See that section for a detailed list of references.} The Gauge Principle, as we will see, is a general prescription developed and widely used by physicists for defining a gauge symmetry. It will thus be important for my analysis to ask what motivates the Gauge Principle and what this principle achieves. The PESA and the resulting norms stated in \Sec\ref{sec:statement_of_the_pesa} will provide my answer to these questions.

My definition of a gauge symmetry will be very general and will allow me to motivate extensions of the Gauge Principle that apply to two important cases of symmetries: reparametrisation invariance and dynamical similarity. The first case will show how my proposal can be used to address an important conceptual problem, the so-called \emph{problem of time} in toy models of quantum gravity, that lies outside the proposal's original scope. The second case forms the basis of my proposed solution to the problem of the AoT. As a consequence, the strength of my solution hinges on the strength of the analysis of gauge symmetry provided by the PESA. Let me begin that analysis now.

\subsection{Theory and structure}
\label{sub:theory and structure}

In the first step of my analysis, I will clearly state the sense in which I will understand a `theory' and establish terminology to refer to a theory's structures. Throughout this thesis, I will be considering physics theories, where the dominant paradigm is to formulate theories in terms of mathematical models treated as representations of phenomena. My primary theory of interest will be general relativity, which is a field theory whose syntactic reconstruction is poorly understand and possibly ill-defined. For those reasons, I will be exclusively working within a semantic view of theories.\footnote{ The semantic view of theories was proposed in \cite{suppes:semantic_view} and advocated for in Chapter~3 of \cite{van1980scientific}. The version we will adopt here can be found in Chapter~9 of \cite{van1989laws}.} In brief, a theory will be seen as collection of models and rules for how to use those models to represent a target system in the world. In \Sec\ref{sub:theory}, I will be more precise about what I mean by a `model,' how a model can be used to represent a target, and what criteria need to be met for a theory to adequately describe phenomena. But for now, I will take for granted that this can be done and introduce some minimal terminology that will be helpful in defining a general notion of symmetry. 

One important role of a theory is to provide a set of laws that can reduce some large collection of possible models to a smaller subset of candidate models that might provide a faithful representation of the target system. I will call the large collection the space of \emph{Kinematically Possible Models (KPMs)} and the smaller space the space of \emph{Dynamically Possible Models (DPMs)}. The distinction between `kinematic' and `dynamic' models is traditionally used in physics to distinguish all possible worlds from those permitted by a theory's laws. The laws then define a projection from the space of KPMs to the space of DPMs.

To pinpoint how the laws accomplish this, let us first state, as is standard in physics, that we will be concerned with theories whose models are mathematical objects. We can then start our analysis with the notion of a mathematical structure, which I will broadly take to be a set of mathematical objects equipped with functions and relations. The models of a theory can then be understood to be composed of different mathematical structures playing different representational roles.

Adopting the terminology used in \cite[ch. 5]{gryb_thebault_book}, I will call the most basic structures of a theory the \emph{constitutive structures}. These structures are common to all KPMs and include things like the spacetime manifold, matter fields, and coupling constants of a theory. The laws are then implemented using a different set of structures that impose constraints on the constitutive structures. I will call these \emph{nomic structures}. Examples include differential equations to be satisfied or cost (e.g., Lagrangian) functions to be extremised by the DPMs. The laws then use the nomic structures to define a projection on the space of constitutive structures, which, in turn, induces a projection from the space of KPMs to the space of DPMs.

Before introducing the notion of symmetry, let me note that constitutive structures can be indexed by token in the sense introduced at the end of \Sec5.1 of \cite{gryb_thebault_book}. For example, a theory of particle mechanics could have two, three or $N$ particles representing different tokens of the same type of matter. Similarly, models of general relativity with different boundary conditions are different tokens of the same type of spacetime structure. I will generally consider models with different tokens in this sense to belong to different theories. Because my notion of gauge symmetry will require that the models of a theory include a specification of the idealisations and conditions under which the theory should apply, defining a theory in this narrow sense is reasonable. In this way, general relativity leads to different theories with possibly different gauge symmetries under different boundary conditions.


\section{Symmetry and its problems} 
\label{sec:problems_with_symmetry}

\subsection{Broad and narrow symmetries} 
\label{ssub:broad_and_narrow_symmetries}

With the terminology established in the previous section, I can now give the general definition of symmetry that I will work with throughout the thesis. Because constitutive structures are mathematical objects equipped with functions and relations and KPMs are collections of constitutive structures, the space of KPMs is equipped with automorphisms; i.e., transformations that map the space of KPMs to itself. Consequently, the space of DPMs is also equipped with automorphisms via the pullback of the projection from the space of KPMs to the space of DPMs. We now make the following definition:
\paragraph{Broad symmetry:}
\begin{quote}
    An automorphism on the space of Dynamically Possible Models of a theory.
\end{quote}
I call such symmetries \emph{broad} because they are the most general notion of symmetry used in the literature. This broad definition reflects a commonly held view that a symmetry is a map from one solution (i.e., DPM) to another. Note that sometimes symmetries are defined as transformations that preserve certain constitutive structures; e.g., spacetime structure; that are common to all DPMs. Such transformations, however, induce transformations on the space of DPMs via the appropriate pullback and, therefore, ultimately reduce to the definition above.

It is rather easy to see that the definition of symmetry given above is so broad that it is nearly devoid of physical content. Because the laws of a theory define the space of DPMs and, therefore, its automorphisms, the broad symmetries are little more than a particular way of stating the laws. After all, there is a broad symmetry that maps any DPM to any other. Thus, all DPMs are equivalent up to a broad symmetry, raising questions about what one can really learn from broad symmetries in general. For these reasons, the definition of a broad symmetry given above is referred to as a ``Fruitless Definition'' in \cite{belot:sym_and_equiv}.

To get a definition of symmetry that is more fruitful, one must give a more narrow definition. To do this, one can require that a symmetry additionally preserve certain specified nomic structures. I will call such proposals \emph{narrow symmetries} and consider several examples in \Sec\ref{sec:narrow_proposals}.\footnote{The broad-narrow distinction was introduced in \cite[\S 5.3]{gryb_thebault_book}, where it has been applied in a more general analysis of symmetry in terms of structure preserving transformations.} Narrow symmetries can therefore be used to characterise specific features of the laws and, thus, highlight certain aspects of the target system. In particular, narrow symmetries give information about the dynamically stable structures of a theory, like energy and momentum, and can have considerable conceptual and heuristic value. Note that, under this understanding, the broad symmetries are the broadest form of narrow symmetry in the sense that they preserve all the nomic structures of the theory.

\subsection{The Gauge Principle} 
\label{ssub:the_gauge_principle}

The narrow notion of symmetry defined in the previous section identifies dynamically stable structure using the criterion of invariance. This, however, is not quite how symmetry is commonly understood. Symmetry transformations are often thought to express identity relations, either approximate or exact, between models. In the case where the identity relations are considered exact, the symmetry is usually referred to as a \emph{gauge symmetry}. Two systems that are related by a gauge transformation are generally considered to be empirically indistinguishable.

Often the word `physically' is used instead of `empirically' when referring to indistinguishability in gauge theories. I will try to avoid metaphysical commitments and not try to make any inferences about what is physically real. Because my definition of gauge symmetry will require a specification of the empirical context, `empirical' indistinguishability is the more appropriate notion. Of course, two systems could be physically different and be empirically indistinguishable in a particular empirical context. I will take the view that, so long as the theory is empirically adequate in the given context, empirical indistinguishability is the notion of indistinguishability that is appropriate for making sound inferences.

The rough notion of gauge symmetry just sketched requires significant elaboration before it can be made philosophically precise --- not the least because of the well-known considerations involved in identifying the `empirical core' of a theory.\footnote{ See, for example, the distinction between theoretical \emph{terms} and \emph{entities} as made in \cite{Lewis:Ramsification} and the process of \emph{Ramsification} and the classic discussion in \cite{Hempel1958-HEMTTD}. } Before giving my own detailed proposal for how to understand a gauge symmetry, it is worthwhile to take note of existing efforts in the physics literature to define a gauge symmetry.

In physics, the common way to think about formulating a theory that contains a gauge symmetry is to start with a theory that doesn't have that gauge symmetry and implement a procedure that first locates and then functionally removes the parts of the theory's models that change under the gauge symmetry. Once the effects of the relevant features of the models have been eliminated, the theory is said to be \emph{gauged}. If the features are explicitly removed, the theory is said to be \emph{reduced} under the action of the gauge symmetry.\footnote{ Note that explicit reduction is often not possible.} Removal of the effects resulting from applying a gauge transformation is motivated by the belief that such effects are not empirically relevant to the system being modelled. A procedure that gauges a symmetry is an implementation of what is often called the \emph{Gauge Principle}, or sometimes the \emph{Gauge Argument}.

The first explicit implementation of the Gauge Principle was introduced in \cite{weyl1986elektron}.\footnote{ For an English translation see \cite[Ch. 5]{o1997dawning}. } Different implementations of the Gauge Principle typically follow the same basic procedure.\footnote{ The Gauge Principle is described in many standard physics textbooks such as \cite[Ch. 15]{Weinberg:1996kr} and \cite{ryder1996quantum} and in the philosophy literature: \cite{teller2000gauge,roberts2021gaugeargumentnoetherreason,earman2003tracking}. } First, one introduces auxiliary structures that transform in a well-known way under the symmetry group to be `gauged.' Then, one modifies the nomic structures of the theory so that the laws become, in a specified sense, independent of those auxiliary structures.

A simple example is an old procedure introduced by \cite{kretschmann1918physical} for making an arbitrary theory independent of the choice of coordinates that label the spacetime structures of a theory. A modern version of this procedure, called the \emph{St\"uckelberg mechanism} after the mechanism developed in \cite{stuckelberg1938interactionforces}, can be applied to the symmetries of particle physics.\footnote{ See, for example, \cite{K_rs_2005} for an illustration of how this is done. } In both procedures, one starts with some constitutive structures --- call them $K^i$ --- and then introduces so-called \emph{compensator} fields, $\phi_\alpha$, that `shift' the constitutive structures along an orbit defined by $K^i \to G({\phi_\alpha})_i^j K^j$. In this notation, $G({\phi_\alpha})^i_j$ is a representation of the symmetry group to be gauged and $\phi_\alpha$ is a parameter along the orbit.

The next step in the procedure is to impose corresponding conditions on the nomic structures of the theory. Usually, this involves putting constraints on the variation of an \emph{action}, which is a function on the space of KPMs. I will illustrate how such constraints can be imposed in \Sec\ref{sub:symmetries_of_the_variational_principle}. Their effect is typically to impose that the action be independent of $\phi_\alpha$. This results in the dynamics of the $\phi_\alpha$ being underdetermined by the equations of motion. This underdetermination is then taken to signal the presence of a gauge symmetry.

A slightly more elaborate version of the Gauge Principle involves introducing new constitutive structures that are only partially auxiliary. These are often referred to as \emph{gauge fields} and represented by a gauge connection $A^i$, which is required to take values in the Lie algebra of the symmetry group to be gauged. One can then replace all instances of partial derivatives, $\partial_\mu K^i$, in the nomic structures with the \emph{gauge-covariant} derivatives, $D_\mu K^i$ that depend on the gauge fields in such a way that the derivatives $D_\mu$ transform covariantly under the symmetry group. Similar the St\"uckelberg mechanism, this produces an action that is invariant under changes of the group parameters and a corresponding underdetermination of the group parameters by the equations of motion. This version of the Gauge Principle can be found in standard texts on gauge theory; e.g., \cite[Ch. 15]{Weinberg:1996kr}. An example of how this procedure works in a simple theory is given in \Sec\ref{sub:bb_theory}. One can also use the Gauge Principle in this form to construct general relativity and the gauge theories of the Standard Model of particle physics.

The procedures sketched above are easier to implement when the original action is already invariant under changes of the group parameters that are spacetime constants. For this reason, the Gauge Principle is often thought of as a way of making a global (i.e., constant over spacetime) symmetry local (i.e., arbitrarily dependent on the spacetime point). The naive implication of this is that local symmetries are gauge transformations, but global symmetries are not. In \Sec\ref{sec:narrow_proposals}, I will show why such an understanding of gauge symmetry is rather limited. Instead, I will propose that the gauge character of a symmetry is better associated with the underdetermination of the dynamical system introduced by the implementations of the Gauge Principle.

An important limitation of existing implementations of the Gauge Principle is that there are no known implementations that apply to dynamical similarity. I will use the norms of the PESA to motivate an implementation of the Gauge Principle for dynamical similarity in \Sec\ref{sub:gauge_principle_for_DS}. This will be central to my explanation of the AoT.

\subsection{Problems exemplified} 
\label{sub:examples1}

The difficulties alluded to above in our discussion of the Gauge Principle illustrate some challenges encountered in giving a universal definition of gauge symmetry. I will formulate this into a concrete problem in \Sec\ref{sub:belot_s_problem}. For now, I will start by illustrating the kinds of problems that arise when defining a gauge symmetry using simple examples. A more detailed analysis of these and other problems will be given in \Sec\ref{sec:narrow_proposals}.

\subsubsection{The Newtonian free particle} 
\label{ssub:the_newtonian_free_particle}

In Newtonian mechanics, a free particle is a model for a material body, idealised as a point particle with position $x^i(t)$ in a Cartesian space (with $i = 1,\hdots,3$) at time $t$, that is in motion in the presence of no external forces. The dynamics of the particle is given by Newton's law, which says that the particle has zero acceleration: $a^i(t) = \deby{^2 x^i}{t^2}=0$. This equation of motion has the general solution
\begin{equation}\label{eq:free particle}
    x^i(t) = x^i_0 + v^i_0 t\,,
\end{equation}
where $x^i_0$ is the initial position and $v^i_0$ is particle's initial (and constant) velocity. The curves \eqref{eq:free particle} then describe every the DPM of the theory.

It is easy to see that the infinitesimal Galilean transformations,
\begin{align}
    t &\to t + T \notag \\
    x^i &\to x^i + X^i + \epsilon\indices{^i_{jk}} \theta^j x^k\, \notag \\
    v^i &\to v^i + V^i\,, \label{eq:Gal trans}
\end{align}
send DPMs to DPMs, and are thus broad symmetries. In \eqref{eq:Gal trans}, $v^i = \deby{x^i}t$ is the particle's velocity, $(T, X^i, \theta^i, V^i)$ are group parameters, and $\epsilon$ is totally antisymmetric tensor with unit entries.

We immediately see the problem with such a broad definition of symmetry: the space of symmetries is larger than the space of DPMs (in the sense that it has a greater dimension) and is such that any DPM can be mapped to any other using a Galilean transformation.\footnote{ To see this, apply a translation to arbitrarily shift the initial position, $x^i_0 \to x^i_0 + X^i$, and a boost to shift the initial velocity \(v^i_0 \to v^i_0 + V^i\). } If one were to identify Galilean transformations as gauge symmetries, one would therefore trivialise the theory. This is a problem because there may be perfectly good empirically interesting systems that one might have reason to model as a particle moving with constant velocity. We will discuss one classic example, Galileo's ship, below. A more modern example would be a model of a free object travelling in a spaceship in an orbit around the Earth. It would be absurd to suggest that such models have no empirical content.

On the other hand, there are also good reasons to expect Galilean symmetries to be gauge symmetries of Newtonian mechanics in general. Newton himself was well-aware that Galilean transformations generate no detectable change within an isolated system. This was proved as early as Corollary V of the \emph{Principia}.\footnote{ For modern translations of the \emph{Principia}, see \cite{newton1999principia,Newton:principia1962}. See \cite{saunders2013rethinking} for an analysis of the status of Galilean transformations in the \emph{Principia}. } Relationalists might then argue, as is done in \cite{Barbour_Bertotti} and \cite{barbour:mach_before_mach}, that, since the Galilean transformations leave relational quantities invariant, the Galilean transformations of a system as a whole should be treated as gauge transformations. I will develop a formalism for doing this explicitly in \Sec\ref{sub:bb_theory}.

We now face a problem: it seems that it is impossible to devise a narrow proposal of symmetry that can make Galilean transformations gauge in a relational universe but not gauge for free objects in an orbiting spaceship.

\subsubsection{Galileo's ship} 
\label{ssub:galileo_ship}

In this section, I will introduce an example taken from Galileo's \emph{Dialogue concerning the two chief world systems} \citep[Second Day, pp 186-187]{galileo_dialogues} that will play an important role in justifying the PESA. This example will be studied in more detail in \Sec\ref{sub:solved_galileo_s_ship} after introducing the PESA.

In Galileo's example, a ship is moving smoothly and uniformly on water relative to a shore. Galileo imagines the ship to have a cabin with closed windows so that the motion of the ship relative to the shore is not visible or otherwise detectable within the cabin. He points out that all observable phenomena in the cabin unfold in the same way whether the ship moves relative to the shore or not.

The connection with the Galilean transformations introduced in the previous example is straightforward. Because the ship is said to be moving uniformly, we can model Galileo's ship using a free Newtonian particle with position $x(t)$ relative to the shore. Galilean invariance then states that the Galilean transformations acting on $x(t)$ relate empirically equivalent situations. Here, the ship can be thought of as a subsystem moving relative to its environment, which in this case is the shore. When the windows of the cabin are closed, there is no significant interaction between the ship's contents and its environment. It seems that the Galilean transformations should be thought of as gauge symmetries in this case.

On the other hand, if the windows of the cabin are opened, a passenger can detect relative motion between the ship and shore. Small interactions between the ship's contents and the shore have no noticeable effect on the motion of the ship but undermine the justification for treating the Galilean transformations as gauge symmetries.

The difference between the two situations is clear: the epistemic access of the ship's passengers has changed. Knowledge about the location of the shore affects whether motion with respect to it should be treated as gauge or not. Epistemic considerations seem to dictate whether the Galilean transformations should be treated as gauge or not.

But epistemic considerations alone, however, do not suffice to give an adequate understanding of gauge symmetry. Dynamical considerations are also essential. Consider the proposal made in \cite{Rovelli:2013fga}. There, it is argued that gauge symmetries are useful in physics because they entail specific dynamical criteria for coupling isolated subsystems to larger systems. In the example of Galileo's ship, the fact that Galilean invariance works as a gauge symmetry for the closed cabin restricts the kinds of possible interactions that could exist between the ship and the rest of the universe. For example, if you learn about the presence of the shore, you already know that the dynamics of the shore \emph{must} not be strongly coupled to the ship or its contents. This leads to specific constraints that must be satisfied by the relative motion of ship and shore. Such constraints can also be articulated for more sophisticated gauge symmetries (e.g., \cite{gomes2021holism}), illustrating the generality of this observation.

We see from this example that a good definition of a gauge symmetry must depend not only on the epistemic access of an observer to a particular system but also on the details of the dynamical interactions within that system. The challenge is then to articulate the precise connection between epistemic and dynamical criteria. This will be a central consideration when developing the PESA.


\subsubsection{The Kepler problem} 
\label{ssub:the_kepler_problem}

We now turn to the symmetries of the Kepler problem to illustrate some formal and conceptual problems that will be crucial to our analysis later. The particular symmetry we will consider in this section (Equation~\ref{eq:DS kepler}) has been discussed in \cite{belot:sym_and_equiv} and \cite{wallace2022isolated} as being particularly problematic for certain ways of defining a gauge symmetry. It results from the conserved magnitude of the so-called \emph{Runge--Lenz} vector.\footnote{ The explicit form of this vector is not necessary for this discussion. See Equation~$(5)$ of \cite{prince1981lie} for details. } I will show in \Sec\ref{sub:ds_in_Nbody} that this symmetry is actually a dynamical similarity and will indicate how to treat it using the PESA in \Sec\ref{sub:the_kepler_symmetries}.

The Kepler problem is a name for a model developed to study planetary motion in the solar system. A large central body (e.g., the sun) and a single satellite (e.g., a planet) are modelled as point particles with different masses. The large body is taken to be at the origin of a Cartesian plane. The motion of the satellite can then be represented in terms of a radial coordinate $r(t)$ and angular coordinate $\theta(t)$, which both evolve in time $t$. Newton's law of gravitation says that the acceleration of the satellite is proportional to the inverse of the second power of $r$. It is possible to show (see \cite{prince1981lie} or \Sec~3.2.2 of \cite{Gryb:2021qix}) that the transformation
\begin{align}\label{eq:DS kepler}
    t &\to a t & r &\to a^{2/3} r & \theta &\to \theta\,,
\end{align}
where $a$ is a dimensionless gauge parameter, takes DPMs of the Kepler problem to DPMs.

This symmetry is more difficult to visualise than the Galilean transformations because it involves a rescaling of both space and time by different powers of $a$. This means that one must rescale positions and velocities in different ways such that the angular momentum of the system is rescaled. In particular, these transformations have the strange property that the angular \emph{momentum}, $J = mr^2 \dot \theta$, transforms non-trivially (because of the non-trivial transformation of $t$ and $r$) even though the angular \emph{coordinate}, $\theta$, is invariant. We will see in \Sec\ref{sec:a_new_kind_of_symmetry_dynamical_similarity} that such transformations are not just conceptually odd but that they are mathematically different from most symmetries considered in the physics literature. As a consequence of this, these symmetries don't possess many of the standard properties of well-studied gauge symmetries. And because of the particular role played by time in these transformations, dynamical similarities will play a central role in our proposal for a solution to the problem of the AoT.

Two important questions now arise:
\begin{enumerate}
    \item Should symmetries of the form \eqref{eq:DS kepler} be identified as gauge symmetries?
    \item If so, what consequences would that identification have for the interpretation of the models of the Kepler problem?
\end{enumerate}

The considerations of our two previous examples would suggest that the answer to the first question will depend, in part, on epistemic considerations and that the answer to the second will have dynamical consequences. Indeed, standard use of the Kepler models would suggest that DPMs related by \eqref{eq:DS kepler} should be taken to represent different physical situations. In the solar system, the angular momentum of planets, which transforms under \eqref{eq:DS kepler}, is usually understood to be an observable quantity. In fact, Kepler's third law makes this explicit since the transformations \eqref{eq:DS kepler} are a subset of the transformations found in the third law.

Such a non-gauge interpretation of the transformations \eqref{eq:DS kepler} is natural for modelling the motion of planets in the solar system. The sun and single planet are treated as an idealised isolated subsystem of the solar system as a whole. The motion of the other planets, which is assumed to have no noticeable impact on the sun and planet, can provide rods and clocks that set an external scale for $r$ and $t$. Provided one has epistemic access to these rods and clocks, the transformations \eqref{eq:DS kepler} should therefore \emph{not} be treated as gauge.

On the other hand, in the absence of externally accessible rods and clocks, no external scale for $r$ and $t$ can exist. Under such circumstances, it would be justifiable to identify \eqref{eq:DS kepler} as a gauge transformation. In this case, one would have to contemplate what it would mean for the angular momentum to have no empirical consequences when the angular coordinate does. This involves providing an answer to the second question above. I will develop a prescription for doing this in general in \Sec\ref{sub:gauge_principle_for_DS}, where I will find that the resulting gauge theory has certain radically different features from the original Kepler theory. The key thing to note at this stage is that identifying a gauge symmetry in a theory can have non-trivial, and even unexpected, dynamical consequences for that theory.

The question that now arises is: even if we can find a definition of a gauge symmetry that works for standard symmetries like the Galileo transformations, will this solution also work for the less often considered symmetries of the Kepler problem and dynamical similarities in general?

\subsection{Belot's Problem} 
\label{sub:belot_s_problem}

In this section, I will identify the general difficulties encountered in the examples above and formulate a concrete problem that connects them all. I begin by motivating the discussion with a particular suggestion made by Earman. I will follow the discussion starting in Section~3.4 of \cite{earman1989world}.

First recall that the constitutive structures of a theory are typically composed of spacetime structures and matter fields (among other things such as coupling constants). Earman then begins by defining a \emph{spacetime symmetry} as a transformation that preserves the fixed spacetime structures of a theory. These fixed structures are structures that are the same for all KPMs. He then defines a \emph{dynamical symmetry} as a map between DPMs that is induced by some arbitrary shift of the spacetime structure where one drags along the matter fields of the theory. These transformations represent all the different ways one can rearrange the matter content of a theory by shifting it smoothly between spacetime points.\footnote{ Mathematically, these transformations are obtained by pulling back the matter fields by arbitrary diffeomorphisms of the spacetime. For scalar fields, this generates all possible scalar field configurations on spacetime --- but not for vector or tensor fields. } He then goes on to argue for the following two ``conditions of adequacy on theories of motion:''
\begin{itemize}
    \item [SP1] Any dynamical symmetry is a spacetime symmetry.
    \item [SP2] Any spacetime symmetry is a dynamical symmetry.
\end{itemize}

The justification for these conditions is the requirement that the theory have just enough, but no more, spacetime structure than is strictly necessary to represent the matter content of the theory. Throughout \cite{earman1989world}, it is argued that these ``conditions of adequacy'' exemplify good interpretive practice in physics. In particular, when these conditions are satisfied, Earman argues that one should identify the spacetime symmetries with gauge symmetries, although he does not use that terminology. \cite{belot:sym_and_equiv} notes that this is a normative principle where ``formal facts place interesting constraints on (good) interpretation.''[Original parentheses.] While many things have been called \emph{Earman's Principle}, we will follow Belot and associate this to Earman. Earman's Principle is then a proposal for identifying a certain class of gauge symmetries; i.e., those associated with spacetime symmetries; for the purpose of constraining good interpretive practice.

\cite{belot:sym_and_equiv} then poses the following question: is it possible to generalise Earman's Principle in a way that applies to any gauge symmetry? The answer given by \cite{belot:sym_and_equiv} is: almost certainly not. That is because the standard conditions used by physicists and mathematicians to define gauge theories lead to contradictory conclusions in different physical situations. We saw some examples of this already in the previous section. \cite{belot:sym_and_equiv} provides further examples. We will investigate some of these and others in \Sec\ref{sec:narrow_proposals}. To make matters worse, \cite{belot2018fifty} argues that even Earman's Principle fails for the asymptotic symmetries of general relativity. The example of the translated or boosted Kerr black hole given in the introduction (\Sec\ref{ssec:intro symmetry and its problems}) makes the point similarly.

For a good general definition of gauge theory, what we need instead is a replacement of Earman's Principle that does not place undue burden and either purely dynamic nor epistemic considerations. In other words, the challenge is to find a definition of gauge symmetry with minimal epistemic and dynamic ingredients that is broad enough to include all acceptable usages of gauge symmetries in physics but not too broad as to include cases that relate models that are obviously empirically inequivalent. I will refer to this challenge as \emph{Belot's Problem}:
\paragraph{Belot's Problem}
\begin{quote}
    To find formal conditions on the symmetries of a theory that are, under good interpretive practice, necessary and sufficient conditions for a gauge symmetry.
\end{quote}
Note that Belot himself does not give a formal statement of this problem --- even stating that: ``it isn't obvious how to give a precise and general formulation of the idea.'' Because of this, the bulk of my proposed solution in \chap\ref{ch:pesa} will involve defining the problem in a more precise and general way to give clear conditions for what is meant by `good interpretive practice.'

We will further investigate some difficulties that lead to Belot's Problem in \Sec\ref{sec:narrow_proposals}. Many of the examples discussed in that section are also treated in \cite{belot:sym_and_equiv}. Our discussion will clarify some points discussed in those examples and add new cases that are commonly treated in the physics literature on gauge theories. In particular, we will find that a proposal by Dirac, discussed in \Sec\ref{ssub:the_dirac_algorithm} and in more detail in \Sec\ref{sec:the_dirac_algorithm}, has sound representational motivations and gives a good solution to Belot's problem in many, but not all, cases. I will then discuss difficulties with each proposal and motivate our my proposed solution to Belot's Problem. What we will find is that there \emph{does} exist a definition of gauge symmetry that solves Belot's Problem in all the cases, and that this definition suggest good interpretive practices for gauge theories.

\section{Models and representation} 
\label{sub:theory}

\subsection{Account of representation}\label{sub:DEKI account}

In this thesis, I am working within a semantic view of theories and am, therefore, using models to represent phenomena. Models --- specifically DPMs --- have already played an essential role in my definition of broad symmetry and will continue to play an essential role in my analysis of gauge symmetry. My solution to Belot's Problem will involve giving specific prescriptive rules for good interpretive practice. This will require that I be explicit about my account of representation. More specifically, I will need to carefully explain how my account of representation encodes both the empirical context and the auxiliary assumptions that define a theory. This is because what counts as a gauge symmetry, in my view, can depend on these assumptions. Remarkably, although perhaps not surprisingly, we will see that some of the structures that play a central role in my analysis of symmetry also happen to be issues that are central to the literature on representation.

Instead of developing a theory of representation from scratch, I will borrow heavily from an existing account of representation developed in \cite{frigg2020modelling}. This account is called \emph{DEKI}, which is an acronym for the names of the account's defining features: Denotation, Exemplification, Keying-up, and Imputation. DEKI gives a general account of how a carrier object $X$ can represent a target system $T$. Let me give a brief description of this account and point out the structures that will be important for the considerations later.

In DEKI, the carrier object $X$ can be anything from a material to a fictional object. We will restrict attention to a particular kind of fictional object: a mathematical object, which requires somewhat special attention.\footnote{ The DEKI account is described for material objects in \chap 8 of \cite{frigg2020modelling} and for fictional and mathematical objects in \chap 9. } The goal of the account is to give a way to construct a model $M$ from $X$ that can give a representation of certain features $Q_a$ of a target system $T$ in terms of the features $X_\mu$ of $X$, where $\mu$ can range over a different (i.e., larger or smaller) set of integer values than $a$. The DEKI account requires a specification of the \emph{context} $C$ of the representation or, in this case, the scientific investigation and a \emph{key} $K$, which specifies, among other things, the auxiliary assumptions underlying the representation. These two structures; i.e., the context and the key; will have direct relevance to the original elements of my analysis of gauge symmetry.

The first step in the account is to provide an interpretation $I$ to $X$. Suppose that $X$ is a Newtonian spacetime equipped with some set of coordinates $y^i$. An interpretation, $I$, might specify that the coordinates $y^i$ be interpreted as the position of a particle in space. This introduces a structure, the structure $X$ is to be interpreted as, which we call $Z$. In our example, $Z$ is the set of particle positions. The interpretation $I$ then defines a set of functions $i^\mu: x^\mu \to z^\mu$ that map the values $x^\mu$ of some features $X^\mu$ of $X$ to the values $z^\mu$ of some corresponding features $Z^\mu$ of $Z$. In the Newtonian example, $I$ would define, among other things, a map from the values of the coordinates $y^i$ to the value of the particle positions $p^i$. We then say that $I$ defines a $Z$-representation of $X$.

In the next step, we define a model and use it to exemplify features of $Z$. In particular, a \emph{model} $M$ is defined as an ordered pair consisting of a carrier object equipped with an interpretation: $M = \mean{X, I}$. We then say that a model \emph{instantiates} a feature of $Z$ when $X$ instantiates the corresponding feature of $Z$ under the interpretation. Finally, $M$ is said to \emph{exemplify} a feature of $Z$ if it instantiates it and if the context $C$ selects this feature as relevant.\footnote{ In the full DEKI account, the relevant feature of $Z$ must also be epistemically accessible in $C$. However, this is only relevant when $X$ is a material carrier object. }

What's important for our purposes is that the context $C$ must be explicitly spelled out in order to specify which features of the bare mathematical object $X$ are relevant to the theory. Thus, what counts as a gauge symmetry will also depend on what features of $X$ are determined to be relevant to the context $C$. We already saw an example illustrating the importance of the context to the definition of gauge symmetry. In Galileo's ship, whether the ship's windows were opened or not determined whether the origin of the Cartesian coordinate system was relevant to the theory's models. In the DEKI account, we see that such considerations are frequently encountered in the general problem of representation and are not specific the considerations of gauge symmetry. In the more general case, context can be encoded in the notion of exemplification as defined above.

The last step of the DEKI account will also play an important role in my analysis of gauge symmetry. In this step, a \emph{key} $K$ is given and used to impute features of $Z$ to features $Q_a$ of a target $T$. To impute a feature to a target involves generating a hypothesis that the target has that feature. For a key $K$ to do this, however, it must give an explicit description of \emph{how} the particular features of $Z$ are to be mapped to the features $Q_a$ of the target. In general, this may involve an incredibly complicated procedure. For example, in particle physics experiments, a huge amount of experimental machinery, simulation and data processing are required to convert to the signals generated in the particle detectors to numbers that can be compared to the output of theoretical calculations. In general, as argued in \cite{doi:10.1093/bjps/axm013} and \cite{bogen1988saving}, this often involves a specification of complicated data models. The key must contain all of this.

Additionally, the key must also specify the approximations, idealisations and other auxiliary assumptions that must be satisfied for the features of $Z$ to correspond to the actual features of the target. This will often involve making judgements that make use of community norms for determining the validity of some idealisation or approximation, and may also rely on additional theoretical assumptions that need external validation. Putting all of this together, we see that the key $K$ is a very complicated structure indeed, and one that can easily be taken for granted. When we return to our problematic examples in \Sec\ref{sec:problems_solved}, we will see that unpacking the key is the main step in understanding appropriate role for gauge symmetry in the relevant system.

To understand why this will be the case, take, yet again, the example of Galileo's ship: the conditions under which the existence and dynamics of the shore can be ignored when modelling the motion of objects within the ship are given by the key. Thus, the key determines when it is acceptable to consider the ship to be an isolated subsystem with an approximate Galilean symmetry (or not). This will obviously affect whether a symmetry should be treated as gauge or not within a particular context. The extra representational structure of a key then gives us the resources to specify what a gauge symmetry should be without committing to any particular notion of subsystem, etc. This is because different keys correspond to different models which, according to my definition, lead to different theories. And one shouldn't expect different theories to have the same gauge symmetries.

In the final step of DEKI, one requires that the features imputed to a target also denote them. To achieve this, it is sufficient to describe each structure of the account --- specifically the carrier object $X$, the interpretation $I$ (and the corresponding structures of $Z$), the key $K$, as well as the target $T$ and its features $Q_a$ --- in some common language. In this case, the features of $X$ denote those of the target $T$ precisely when they denote them in the common language. In describing a mathematical object, the user of a scientific theory is engaging in a game of make-believe where the mathematical object itself is not a material object in the world but some fictional object in the scientist's mind.\footnote{ In DEKI, it is emphasised that the dynamical laws of theory should be further encoded into the model description in the form of \emph{generating principles}. This extra layer, however, is encoded for us in the distinction between a KPM and a DPM. The introduction of generating principles is then motivated by the fact that a KPM, on its own, is not a particularly useful model of anything. But since we have already made a distinction between KPMs and DPMs, we don't have a need for such generating principles.}

\subsection{Empirical adequacy}

In the previous section, I set up the representation theory that I will need later to give a precise definition of gauge symmetry. This discussion focused on what I will mean when I say that I have a model for some target phenomena. But to judge a scientific theory, we also need to know whether the theory is doing a good or bad job of representing the target system. The simplest criteria for doing so is Van~Fraassen's notion of \emph{empirical adequacy}.

The notion of empirical adequacy was first introduced in \cite{van1980scientific}. I will focus on the specific definition given on page 64. There, empirical adequacy is defined as a relationship between certain substructures of a theory's models and measurement reports. First, \emph{appearances} are defined as ``the structures which can be described by experimental and measurement reports.'' Then, the \emph{empirical substructures} of a theory are defined as ``candidates for the direct representation of observable phenomena.'' A theory is then said to be \emph{empirically adequate} ``if it has a model such that all appearance are isomorphic to the empirical substructures of that model.'' In other words, a theory contains representational structures that are hypothesised to describe observable phenomena, and the theory is empirically adequate when those structures are isomorphic to measurement reports.

It is clear, however, from this definition that Van~Fraassen's representation theory is different from the one we defined in the previous section. For instance, there is no mention of context or key in Van~Fraassen's definition. Instead, the focus is on the empirical substructures and isomorphisms between these and measurement reports.\footnote{ Van~Fraassen does present an updated picture in \cite{vanFraassen:Scientific_Representation} where the conditions of use of a model play a more central role in scientific theorising. While this is closer to our picture, we will stick to the DEKI account of representation in our analysis, which is uncommitted to any stance in the empiricism/realism debate. } While it is true that an interpretation $I$ and key $K$ will define an isomorphism between some features of a carrier object and certain independent features of a target system, the DEKI account requires more than a simple isomorphism.

Moreover, because Van~Fraassen introduced the notion of empirical adequacy as a tool for contrasting constructive empiricism with forms of scientific realism, empirical substructures play an important role in his definition. Here, I will not be engaging in a metaphysical debate about realism in science. The distinction between empirical and non-empirical substructures will then not play a central role in my analysis. Indeed, when using the word `observable' I will refer to the physics usage of the word,\footnote{ See, for example, the usage of the word `observable' in Chapter 1 of \cite{henneaux1992quantization}. } which I will make more precise in \Sec\ref{sec:statement_of_the_pesa}. Importantly, I will \emph{not} be referring to its usage in the philosophical literature when discussing the observable/unobservable distinction.

I will, however, be interested in defining a notion of empirical adequacy that fits with the account of representation I am using. Fortunately, this is relatively easy to do. Following \cite[p. 178]{frigg2020modelling}, I say that a model gives a \emph{faithful representation} of a target when the target does indeed have the features imputed by the representation. I will then define the following:
\paragraph{Empirical Adequacy}
\begin{quote}
    A theory is \emph{empirically adequate} when it has at least one model that gives a faithful representation of the intended target system in the DEKI sense.
\end{quote}
Note that the process of imputation is accomplished in DEKI by the key, which, as I have emphasised, can be a very complex structure that is often left implicit in theoretical practice. Thus, there is a lot of work to be done `under-the-hood' in giving a faithful representation of a system.

The definition above has the advantage that one only needs to specify how the models of a theory lead to representations in the DEKI sense in order determine whether a theory is empirically adequate. In particular, there is no need to establish an isomorphism between empirical substructures and measurement reports: the target system either has the features imputed by the key or not.

\section{Features of symmetry} 
\label{sub:symmetry}

\subsection{Symmetries over a history and at a time} 
\label{ssub:symmetries_over_a_history_and_at_a_time}

In this section, I will introduce several symmetry-related concepts that will be important for the analysis later. I will start by making a distinction that is not usually emphasised in the literature on symmetry but that I will find to be central to my insights about gauge symmetry.

The distinction I will make involves different ways of understanding and representing the action of symmetries in time. In this thesis, I will consider models that consist of matter distributions defined over spacetime. Let us call such a model a \emph{history}. We can then distinguish two notions of symmetry: those that act only at a particular time, which I will call \emph{at-a-time symmetries}, and those that act over an entire history, which I will call \emph{over-a-history symmetries}. On the one hand, over-a-history symmetries seem natural from a block-universe perspective, particularly when dealing with general relativity and its relativity of simultaneity. On the other hand, observers only ever have access to local information so that at-a-time notions of symmetry seem more natural from an epistemic perspective. The two notions, however, are not only conceptually distinct but, as we will see, can have different mathematical structure. I will argue that assuming, falsely, that these distinct notions are always interchangeable can lead to confusion. To avoid such confusion, I will take the view that the at-a-time notions are more fundamental because they alone are directly relevant for doing physics. My definition of gauge symmetry will then be based on an at-a-time notion from which the corresponding over-a-history notions can be recovered when appropriate.

To define these two notions more carefully, let us first define an \emph{instantaneous state}. Given the spacetime structures of a theory, an \emph{instant} is a collection of spatial structures all labelled by the same temporal structure. The archetypal example is a spatial manifold $\Sigma$ labelled by a single value of time $t$. An instantaneous state is then a particular function, or set of functions, of a KPM restricted to such an instant. An instantaneous state should contain enough information to specify the state at the next time using the dynamical laws.\footnote{ Note that nothing about this definition requires a KPM to consist of an ordered sequence of instants, that the state represent a unique state of affairs, or that the laws be deterministic. } The \emph{instantaneous state space} of a theory is then the space of all possible instantaneous states in the theory. A one-parameter curve on state space is then said to form a \emph{history}. More generally, I will take a history to be a collection of states that form a KPM. A theory therefore attempts to model a target system using a set of instantaneous states. While it is possible to define a state in more general terms without having to specify instants --- say by identifying states with entire histories --- I believe that it is more standard to speak of the state of a system at some instant of time. Unless otherwise specified, I will therefore drop the word `instantaneous' when talking about the `states' or `state space' of a theory.

Let us define an \emph{at-a-time symmetry} as a transformation between states. An \emph{over-a-history symmetry} is then a transformation that maps histories to histories.\footnote{ The \emph{over-a-history}/\emph{at-a-time}  distinction for symmetry was introduced in \cite{gryb_thebault_book} (\Sec 8.1 and \Sec 8.2 respectively). } This is a timeless notion of symmetry in the sense that it acts on all instants past, present and future. Note that, since an ordered sequence of states forms a history, at-a-time symmetries induce transformations over a history and therefore have corresponding over-a-history symmetries. Translating between the two notions, however, can sometimes be cumbersome so that an analysis in terms of one form of symmetry can be difficult to carry over to the other.

In contrast, a history may not correspond to an ordered sequence of states. In general relativity, instantaneous states depend on the simultaneity convention and are sometimes regarded as unnatural. Moreover, many models of general relativity of physical interest, such as the Kerr model of a spinning black hole, cannot be expressed as ordered sequences of instantaneous states. Over-a-history notions of symmetry are therefore essential for understanding all the models of general relativity. However, as we will argue for in \Sec\ref{sec:lagrangian_variational_principle}, at-a-time notions of symmetry are sufficient and, indeed, preferable for giving empirically relevant information. This is because all models of general relativity can be re-expressed as collections of models that \emph{are} ordered sequences of instants. Moreover, the question of underdetermination of the laws by the phenomena, which is central to our understanding a gauge symmetry, is best answered in an at-a-time setting.

Finally, quantum mechanics in its standard formulations takes observable operators to be defined at-an-instant. In fact, the difficulty of reconciling this fact with the relativity of simultaneity of general relativity is one of the great obstructions to formulating a conceptually and mathematically coherent theory of quantum gravity. This highlights the important physical role played by at-a-time notions.

\subsection{Symmetry, reduction and surplus structure} 
\label{ssub:reduction}

I have defined broad symmetries as automorphisms on the space of DPMs. As a result, all broad symmetries also form a group. Given this fact, one can define the \emph{orbit} of a symmetry transformation to be the set of all models related by that symmetry. The orbits of a symmetry can be used to construct a projection from the space of DPMs to a smaller space where all the models in a given orbit map to a single element. We can call this smaller space the \emph{reduced space} of the symmetry and the projected theory a \emph{reduction} of the theory by the symmetry. Note that reduction involves an elimination of the symmetry because the symmetry has no action on the reduced space.

The projection defined above gives the mathematical structure necessary to define a principal fibre bundle. In this case, the group orbits are the fibres and the reduced space is the base space of the fibre bundle. Reduction and the corresponding fibre-bundle structure take on a particularly striking role when the fibres are interpreted as relating empirically indiscernible states. In this case, the base space naively encodes all empirically relevant information while the fibres are interpreted as \emph{surplus structure} --- a notion introduced in \cite{redhead1975symmetry}.

For gauge symmetries, it might then seem appealing to perform a reduction and eliminate the surplus structure. Often, however, explicit reduction is a bad idea. This is sometimes because the reduced space cannot be explicitly constructed or cannot be expressed in a single coordinate chart; e.g., because of issues associated with the so-called \emph{Gribov Problem} \citep{gribov1978quantization}.\footnote{ This arises when the gauge orbits intersect a section of the fibre bundle more than once. Removing this ambiguity by selecting a particular region along individual fibres, or \emph{Gribov horizons}, can not be done globally in general. This probably typically affects gauge theories with compact non-abelian gauge groups. } In other cases, the group action of the symmetry might be broken by particular ways of representing the state. This occurs in general relativity for at-a-time representations of the spacetime diffeomorphisms, which have a group\emph{oid} rather than group structure. Finally, is it sometimes the case that the classical versions of the original and reduced theories lead to inequivalent quantizations. This is usually understood to be because the quantum state is sensitive to global properties of the state space. These global features can be part of the fibre-bundle structure but not the reduced space. This can occur in the Aharonov--Bohm effect \citep{ABeffect} or when there are anomalies.

A body of recent philosophical literature has emerged from such considerations and involves studying the role of surplus structure in gauge theories. See \cite{redhead2002interpretation}, \cite{healey2007gauging}, \cite{weatherall2016understanding}, \cite{nguyen2020surplus}, \cite{bradley2020representational}, and \cite{fletcher2020representational} for further explorations of the issues discussed above.

Even in cases where reduction is straightforward, there are often theoretical advantages to not performing the reduction. I will explore different reasons for doing so throughout the text. When explicit reduction is not advisable, it might be better to work with the original symmetric set of models. In that case, the full orbits of the symmetry are retained with an acknowledgement that all members of that orbit are symmetry-related. The consequences of acknowledging this will be explored extensively throughout this work, where I will establish strict norms for how to treat the members of an orbit when the symmetries are gauge.


\subsection{Degree-of-freedom counting} 
\label{ssub:degree_of_freedom_counting}

It is common in theoretical physics to talk of the `degrees of freedom' of a theory and different ways of counting such `degrees of freedom.' The concept of a `degree of freedom' is supposed to reflect some modal dimension of the world along which different possible states of affairs can be represented by different values of some variable. A \emph{degree-of-freedom count} is then some procedure for counting the different degrees of freedom of a theory. For example, a theory of a free Newtonian particle in three dimensions might be said to have three `degrees-of-freedom' that reflect the three different spatial directions in which the particle is allowed to move.

It's important to recognize, however, that degree-of-freedom counting is often done in different ways to reflect different modal aspects of a theory. In the free-particle example just given, one might choose to count the number of independent initial data required to fix a DPM of the theory instead of the number of spatial directions the particle is allowed to move in. These counts are not the same because both the position and the velocity of a free-particle (totalling 6 `degrees of freedom' and not 3) must be specified to fix a DPM using Newton's second law. One must therefore be careful to specify precisely what is being counted when determining the `degrees of freedom' of a theory.

Degree-of-freedom counts are often made in physics when considering gauge symmetries. Loosely speaking, the presence of a gauge symmetry should be expected to reduce the number of `empirically relevant' degrees of freedom of that system, where one must have some understanding of what `empirically relevant' means. In the case of a gauge symmetry, the counting is supposed to account for the number of structures that are not fixed by the laws alone. So if some theory is said to have $A$ representational structures and $B$ gauge symmetries, then it can represent at most $A-B$ features of the target.

Counting degrees of freedom in this way can be an extremely valuable conceptual tool for understanding the empirical content of a theory. That's why physicists put so much effort in trying to do it. But the ambiguities involved in counting `degrees-of-freedom' can lead to considerable confusion when the meaning of such counts are not spelled out clearly.

In this work, it will be important to understand the difference between the kinds of degree-of-freedom counts made using at-a-time and over-a-history notions of symmetry. At-a-time counts usually correspond to the number of variables whose values need to be specified at some time in order to fix a DPM by applying the dynamical laws. Over-a-history counts, in contrast, usually correspond to a count of the quantities that need to be specified in order to define a DPM as a whole. For example, if a symmetry transformation can be specified using a function $\epsilon(t)$ at \emph{all} values of $t$, then an over-a-history approach might count the values of $\epsilon$ at all values of $t$ as a single \emph{functional} degree of freedom. In contrast, an at-a-time approach might simply count the value of $\epsilon(t)$ and its derivatives at a \emph{particular} time $t$. This was the case in the free-particle example given above. Clearly, such counts will not always match. The over-a-history approach counts functions over a continuous domain while the at-a-time approach counts derivatives of that function at an instant. While both notions are undoubtedly important for general theoretical considerations, I will take the view that the at-a-time notion is more relevant for assessing what can be independently measured by physical observers.


\section{Examples of symmetry} 
\label{sec:examples_of_symmetry}

In this section, I will briefly introduce different kinds of symmetry studied in the physics and philosophy literature. The list is not exhaustive nor is it intended to be. The purpose of the list is to give a brief description of the different kinds symmetries that will be discussed later and to establish a consistent taxonomy to be used throughout the analysis.

\subsection{Discrete symmetries} 
\label{sub:discrete_symmetries}

Discrete symmetries are symmetries where the group parameters take discrete; i.e., countable; values. Simple examples are the parity (often called `$P$') and time reversal (often called `$T$') symmetries: $\vec x \to -\vec x$ and $t \to -t$, where the group parameter takes the binary values $\pm 1$. Discrete transformations in physics are usually global transformations that reshuffle entire kinematical structures in some countable, and often finite, way. These reshuffled structures don't have to be spatiotemporal like the $P$ and $T$ symmetries mentioned above. Instead, they can be permutations of particles or changes to the coupling constants of a theory. For example, the charge conjugation symmetry (often called $C$) reverses the sign of the charge of electromagnetic particles.

Discrete symmetries are hugely important in a variety of physics and engineering applications with philosophical and metaphysical implications. Parity symmetry, for example, was central to Kant's arguments in favour of the synthetic a priori and was mentioned among Leibniz' arguments against Newton's absolute space.\footnote{ See \cite[Ch. 11 and 8]{huggett1999space} for original texts and modern commentary on these points. } CPT symmetry; i.e., the symmetry transformations obtained by successively applying $C$, $P$, and $T$ operators; played an important role in the development of the standard model of particle physics. For the purposes of this thesis, time-reversal invariance, or $T$-symmetry, will obviously play a central role in our discussions about the AoT.

\subsection{Leibniz shifts} 
\label{sub:leibniz_shifts}

A \emph{Leibniz shift}, as introduced originally by Leibniz, is a constant translation, or `shift,' in time or space.\footnote{ See \cite[Ch. 8]{huggett1999space} for a nice introduction to the concepts of a Leibniz shift.} The apparent invariance of physical phenomena under such shifts, manifest in the translational invariance of Newton's laws, was used by Leibniz to argue against the use of Newton's absolute space and time as a way of determining true motion.

Modern debates about the nature of spacetime have evolved greatly since Leibniz' original conceptions. A more modern way of thinking about a Leibniz shift is as an active transformation of spacetime points by constant translations.\footnote{ See \Sec\ref{ssub:passive_and_active_symmetries} for a description of the difference between active and passive transformations. } Since these are a subset of the Galilean transformations, which are the full set of spacetime symmetries of Newtonian mechanics, it is sometimes helpful to generalise the notion of a Leibniz shift to the full group of Galilean transformations. This involves both the continuous symmetries of \eqref{eq:Gal trans} and the discrete $P$ and $T$ symmetries mentioned above. In this way, the Leibniz shifts can be taken as active versions of the global spacetime symmetries of Newtonian mechanics. The Galilean boosts; i.e., translations of the velocities; are time dependent translations of the spacetime points. This could be a reason to \emph{not} regard them as true Leibniz shifts. By default, we will not include boosts when referring to Leibniz shifts. For a more concrete mathematical definition, see \eqref{eq:NM glob sym} and the surrounding discussion in \Sec\ref{sub:leibniz_and_noether1}.


\subsection{Best-matching shifts} 
\label{sub:best_matching_shifts}

The Leibniz shifts can be generalised by allowing the group parameters of \eqref{eq:Gal trans} (excluding the boosts, which would be redundant) to be arbitrary functions of time $t$.\footnote{ For time translations, these should be required to preserve the temporal ordering. } We will call transformations of this kind \emph{best-matching shifts} for reasons we will state below. Here, the Galilean boosts can be seen as arising as a special case of time-dependent spatial translations.

These symmetries have played a prominent role in relational models of Newtonian mechanics such as those introduced in \cite{Barbour_Bertotti}. When thinking about a Newtonian theory as a model for the universe as a whole, relational considerations suggest imposing such symmetries as gauge symmetries of the theory. A procedure called \emph{best matching} was developed to modify the dynamics of a Newtonian system of particles in a way were the \emph{best-matching shifts} can be naturally interpreted as gauge symmetries.\footnote{ See \cite{barbour:mach_before_mach,Barbour:2010dp} and \cite[Ch. 2]{gryb2012shapedynamicsmachsprinciples} for introductions to best matching. } We will review the technical details of this procedure in \Sec\ref{sub:bb_theory}.

When restricted to translations in the context of a Newtonian universe, it was shown as early as Corollary VI of Newton's Principia \citep{newton1999principia,Newton:principia1962} that the best-matching shifts lead to no observable consequences for the system. This suggests treating such transformations as gauge symmetries of a Newtonian universe as argued in \cite{saunders2013rethinking} and \cite{knox2014newtonian}.

The motivation behind the best-matching procedure is similar to the one I will advocate in this thesis. In particular, the aim is to impose constraints on the dynamical equations that introduce an underdetermination that reflects the underdetermination of the models by the phenomena.\footnote{ We will discuss this point more carefully in \Sec\ref{sub:symmetries_of_the_variational_principle} and relate this to our proposal in \chap\ref{ch:pesa}.} While the final result is equivalent to applying standard versions of the Gauge Principle to the translation group of Newtonian mechanics, the intuition used to motivate this move is different from the justification given in standard physics texts (e.g., Chapter 15 of \cite{Weinberg:1996kr}). I will briefly explain how best-matching implements a version of the Gauge Principle in \Sec\ref{sub:bb_theory}.

\subsection{Maxwell and Yang--Mills gauge symmetries} 
\label{sub:maxwell_and_yang_mills_gauge_symmetries}

The gauge symmetries of Maxwell's theory of electromagnetism and those of the Yang--Mills fields of the Standard Model of particle physics are considered \emph{the} exemplars of gauge symmetry in the modern physics literature. In fact, Maxwell's theory is a special case of Yang--Mills theory but is worth consideration for its historical importance and relative mathematical and conceptual simplicity. Much physics literature implicitly refers to Yang--Mills theory (or generalisations thereof) when using the term `gauge theories.'

To understand these symmetries, it is easiest to first consider Maxwell's theory of the electromagnetic field. In this theory, redundancies in the representations of the electric and magnetic field, $E_i$ and $B_i$, are revealed by writing these fields in terms of the derivatives of the electromagnetic potential $A_\mu$ as
\begin{align}\label{eq:EB def}
    E_i &= \dot A_i - \partial_i A_0 & B_i &= \epsilon\du i {jk} \partial_j A_k\,.
\end{align}
The transformation
\begin{equation}\label{eq:max gauge}
    A_\mu \to A_\mu + \partial_\mu \phi
\end{equation}
thus leaves the electric and magnetic fields invariant under any choice of function $\phi(x,t)$. Since all electromagnetic phenomena can be expressed in terms of laws depending solely on $E_i$ and $B_i$ (see \Sec\ref{sub:electromagnetism} for an explicit illustration of this), the transformations \eqref{eq:max gauge} are almost always considered to be gauge symmetries of Maxwell's theory.\footnote{ In the presence of matter fields, the matter fields must transform appropriately under different choices of $\phi$. } These transformations are taken to be representational redundancies that can be eliminated by applying suitable conventions for fixing $\phi$. It is also possible to explicitly reduced Maxwell's theory by the action of \eqref{eq:max gauge} in terms of quantities that are invariant under these symmetries. But because such reduced representations are often not that easy to work with in practice, and retaining manifest gauge symmetry is usually preferable. Moreover, topological features of the gauge bundle (e.g., its cohomology) can describe non-local physical effects that are not present in the reduced theory. Upon quantization, such global features lead to observable effects such as the \emph{Aharonov--Bohm effect} \cite{ABeffect}. Thus, reduced and non-reduced theories can be empirically inequivalent.

The most elegant way to express the gauge symmetries of electromagnetism and to understand its generalisation to Yang--Mills theory is by using the language of differential forms. In this formulation, the electric and magnetic fields are written as components of a single $2$-form $F$. In this language, \eqref{eq:EB def} becomes
\begin{equation}\label{eq:F def}
    F = \de A\,,
\end{equation}
where $\de$ is the exterior derivative and $A$ is the electromagnetic potential  (a $1$-form). The exterior derivative is known to obey the identity $\de^2 = 0$. Thus, \eqref{eq:F def} is invariant under the transformation
\begin{equation}
    A \to A + \de \phi\,.
\end{equation}

In Yang--Mills theory, $F$ is generalised to take values in a Lie group $G$ usually taken to be semisimple. The \emph{gauge potential} $A$ is then taken to be valued in the Lie algebra $\mathfrak g$ of $G$ and
\begin{equation}
    F = \de A + [A,A]\,,
\end{equation}
where $[,]$ is the Lie bracket on $\mathfrak g$. For the group element $g \in G$ we then have that the gauge transformation
\begin{equation}\label{eq:YM gauge}
    A \to g^{-1} A g - g^{-1}\de g\,,
\end{equation}
sends $F \to g^{-1} F g$. Unlike in Maxwell's theory, $F$ itself is not invariant but covariant under the gauge transformations \eqref{eq:YM gauge}. The nomic structures of the theory are then required to define the DPMs independently of any value of $g$, and this puts some (mild) restrictions on the kinds of nomic structures allowed. Maxwell theory is recovered when $G$ is taken to be the one dimensional group of unitary transformations.

The symmetries \eqref{eq:YM gauge} then define orbits that can be interpreted as sets of empirically equivalent representations of the theory. A projection that identifies the elements of such orbits can be used to define a principal fibre bundle over spacetime. The gauge potential $A$ can then be interpreted as a connection on this principal bundle. This gives the paradigm case for understanding gauge symmetries in terms of fibre bundles. It is important to note that this notion of gauge orbit is formally different from the notion discussed in \Sec\ref{ssub:reduction}, which was in terms of the instantaneous states of a theory. The connection between the two notions is described in \cite{gomes2018unified,gomes2019unified}.

Unlike in Maxwell's theory, there is no known way to reduce Yang--Mills theories in general. Because of issues like the \emph{Gribov ambiguity} \citep{gribov1978quantization} discussed earlier, one can't construct a global coordinate chart on a regular section of the gauge bundle. Thus, for Yang--Mills theories, retaining a formulation in terms of redundant variables appears to be necessary.

The different proposals for narrow definitions of symmetry almost all have the example of Yang--Mills theories in mind. Many attempts to quantize gravity are based on reformulations of general relativity that bring it close to Yang--Mills form.\footnote{ See, for example, \Sec2 of \cite{ashtekar2004background} for a description of such formulations. } Much is known about the formal properties and quantization of Yang--Mills gauge theories. For this reason it serves as a useful guide to the characterization of gauge symmetries in general. However, Yang--Mills gauge symmetries preserve the form of Hamilton's equations, and are therefore importantly different from the dynamical similarity transformations that will be central to our analysis of the AoT.

\subsection{Coordinate and reparametrisation invariance} 
\label{sub:coordinate_invariance}

The laws of general relativity are famously invariant under arbitrary (smooth) invertible transformations of the spacetime coordinates $x^\mu$ of the form
\begin{equation}
    x^\mu \to f^\mu(x^\nu)\,,
\end{equation}
where the $f^\mu$ are smooth functions of the spacetime coordinates.\footnote{ The $f^\mu$ are required to obey $\det \lf( \partial_\mu f^\nu\rt) \neq 0 $. } Einstein initially saw the enforcement of this symmetry as a foundational principle for his theory of general relativity. As we have already noted, \cite{kretschmann1918physical} immediately argued that the mere presence of such a symmetry can have no empirical significance because it can be realised in any theory without changing that theory's empirical content.\footnote{ For a discussion of Kretschmann's argument, see \cite{norton2003general}.}

Since Kretschmann's observation, there has been much discussion about the empirical significance of coordinate invariance and the role it plays in general relativity. One more recent discussion involves a revival of the so-called \emph{hole argument} in general relativity.\footnote{For recent reviews of the hole argument see \cite{stachel2014hole,pooley2021:hole}.} I will not attempt a systematic study of the hole argument here. Instead, I will point out one important feature of these discussions that is relevant to my analysis.

Discussions about the hole argument centre around what ontological significance to give to spacetime points. A key observation is that the coordinates of spacetime points are underdetermined by the phenomena. While I am not interested in ontological questions here, I will be interested in questions about the direct empirical significance of spacetime coordinates. The fact that the spacetime coordinates are underdetermined by the dynamical equations in a way that mirrors the way that they are underdetermined by the phenomena implies that coordinate invariance is a gauge symmetry according to the slogans introduced in \Sec\ref{sub:intro PESA}. The fact that coordinate invariance is generally seen as a gauge symmetry of general relativity validates these slogans. A more detailed analysis of the hole argument using the PESA will be left to future investigations.

A noteworthy special case of coordinate invariance is when the $f$'s transform only the time coordinate
\begin{equation}
    t \to f(t)\,.
\end{equation}
Theories with this symmetry are called \emph{reparametrisation invariant}.\footnote{ Perhaps \emph{parametrisation invariance} would be a better term? But I will use the standard terminology. } In such cases, it is almost always assumed that $\dot f > 0$, so that the transformed temporal coordinates preserve the temporal order of instants within the theory (with the possible exception of discrete $T$-symmetry defined above, which reverses the order of \emph{all} temporal instants). Symmetries of this kind will be important to my analysis because they will provide an example, outside dynamical similarity and the examples discussed in \cite{belot:sym_and_equiv}, of how my proposal can clarify an important conceptual problem, the so-called \emph{frozen-formalism problem}, in physics. I will describe this problem in \Sec\ref{sub:reparametrisation_invariance} and my clarifications in \Sec\ref{sub:solved_reparametrisation_invariance}.


\section{Other notions of symmetry}
\label{sec:other symmetry notions}

I will end this chapter with a review of different notions of symmetry often encountered in the literature in relation to the topic of gauge symmetry. While these notions do not play a central role in my own analysis, it is nevertheless helpful to review these important concepts and indicate how they relate, or not, to my formalism.

\subsection{Global and local symmetries} 
\label{ssub:global_and_local_symmetries}

Because symmetries are transformations on the space of KPMs and KPMs are composed of spacetime structure, symmetries can depend on the spacetime structure. This dependence can be either global or local. A global dependence means that the symmetry transformation is constant along the structure in question whereas a local dependence means that the symmetry transformation is not constant. For example, a time-translation of the form $t \to t + a$, for some constant $a$, is global in time and space (it leaves spatial coordinates unchanged) whereas a time-reparametrisation of the form $t \to f(t)$, where $\dot f \neq $const, is local in time and global in space.

The global-local distinction has often been used as a tool for identifying gauge symmetries.\footnote{ See, for example, the motivations of the original paper, \cite{yang1954conservation}, introducing Yang--Mills theory and the general discussion of gauge symmetry in \cite{t1980gauge}. } Two influential theorems by Emmy Noether divide classes of symmetries along these lines. I will discuss these theorems in \Sec\ref{ssub:noether_symmetries} (see that section for references) and prove them in \Sec\ref{sub:noether_2_symmetries_and_noether_s_second_theorem} and \Sec\ref{sub:lagrangian_constraints_and_noether_s_first_theorem}. There, I will show that global symmetries are often not good candidates for gauge symmetries while local symmetries are. 

The global-local distinction is undoubtedly a useful tool for identifying gauge symmetries in certain systems. There is, however, no good a priori reason think that global symmetries should \emph{never} be gauge symmetries. We have already seen an example in \Sec\ref{ssub:the_newtonian_free_particle} where there could be a perfectly good reason to treat certain global symmetries as gauge symmetries. Other examples are amply discussed in the philosophical literature. See \cite{kosso2000empirical,brading2004gauge,wallace2022isolated,gomes2021holism} for a sample. So while the global-local distinction is important for classifying the different examples of symmetry that we will study, it is not a reliable criterion for identifying gauge symmetries in general.


\subsection{Passive and active symmetries} 
\label{ssub:passive_and_active_symmetries}

A common distinction used to identify gauge aspects of symmetry is the distinction between so-called \emph{passive} and \emph{active} symmetries. This terminology was introduced in \cite{bargmann1957relativity} in the context of Lorentz transformations in special relativity, but the general concept applies to arbitrary symmetry transformations.\footnote{ See \chap 3 of \cite{rosen2008symmetry} for a more recent explanation of the distinction for applications in physics. } In physics, the active/passive distinction is made possible by the concept of a reference structure, which is any structure that can be used to quantify some feature of material structure; i.e., a structure representing some material objects in the world. An \emph{active} transformation then acts on the material structure but leaves the reference structure invariant while a \emph{passive transformation} acts on the reference structure but leaves the material structure invariant.

A passive symmetry is interpreted as expressing a mere change of convention resulting from different possible choices of reference structure. A typical example would be the use of a different naming convention like calling the direction that the sun rises `west' instead of `east.' Contrastingly, an active symmetry is interpreted as a genuine shift in representational structure corresponding to some target system. Thus, a change in the rotation of the earth such that the sun travels from the American continents towards Asia while people get older is an active transformation. In this way, any invariance under an active symmetry is thought to signal the presence of genuine gauge symmetry.

While it is clear that the passive--active distinction does capture a difference between trivial (i.e., passive) and non-trivial (i.e., active) notions of symmetry, it does not provide a good definition of a gauge symmetry. In particular, it is not clear how to identify, in general, what counts as a `reference structure' and what counts as a `material structure.' Nothing about the distinction specifies how this can be done in general. Moreover, it is also not clear what the suitable notion of `invariance' should be, and whether any specific criteria can work for all situations. Finally, nothing about the notion of representational redundancy seems to \emph{require} the introduction of a reference structure. It thus seems unnecessary to introduce such extra structure when defining a gauge symmetry. I will thus try to avoid the use of such language and develop my proposal independently of this distinction.

\subsection{Subsystem symmetries} 
\label{ssub:subsystem_symmetries}

If a broad symmetry has a non-trivial action on some substructures of the KPMs but leaves the remaining structures invariant, we call that symmetry a \emph{subsystem symmetry}. The concept of a subsystem symmetry is important in physics because there are many applications where a particular transformation acts in a way that creates a contrast between two systems; e.g., a system and its environment. For example, the passive--active distinction made in the previous section can be recast in terms of subsystem symmetries with material structure playing the role of a subsystem and reference structure playing the role of the remaining part of the system. 

Subsystem symmetries have important implications when a particular subsystem is dynamically isolated. When this happens, the conditions for dynamical isolation imply that any symmetry of the larger system should imply the existence of a corresponding symmetry for the subsystem as illustrated in \cite{greaves2014empirical}. The new symmetry arises as a kind rigid transformation of the subsystem relative to the remaining system \citep{gomes2021holism}. Since dynamical isolation is never exact, such symmetries are typically used to model approximate symmetries; e.g., the approximate Galilean symmetries of the physics inside a smoothly running train.

Because approximately isolated subsystems are common in physics, subsystem symmetries have important physical relevance. For example, they seem to play an important role in explaining the ubiquity and methodological advantages of gauge theories in physics, as suggested in \cite{Rovelli:2013fga} and later in \cite{roberts2021gaugeargumentnoetherreason}. However, such discussions are tangential to the question of what defines a gauge symmetry in general. Moreover, just as with active symmetries, nothing about the concept of a gauge symmetry seems to \emph{require} the introduction of an isolated subsystem. I will therefore only require that my definition of gauge symmetry be consistent with the definition of gauge symmetry used to define a subsystem symmetry in the relevant literature. The examples discussed in \Sec\ref{sec:problems_solved} should be sufficient to show how this can be done in general.

\subsubsection{Subsystem recursivity} 
\label{ssub:wallace_s_proposal}

It is worth mentioning a property of a symmetry, called \emph{subsystem recursivity}, that was first defined in \cite{WALLACE2022239} and further illustrated in \cite{wallace2022isolated}. In those papers, subsystem recursivity is defined and shown to be a valuable methodological tool for constructing gauge theories.

A theory is said to be subsystem recursive if the approximate symmetries of isolated subsystems are also symmetries of the system as a whole. Galilean invariance in Newtonian mechanics is subsystem recursive because, as discussed in \Sec\ref{ssub:the_newtonian_free_particle}, Corollary V of Newton's Principia shows that any isolated subsystem will have approximate Galilean symmetries, and Galilean symmetries are also symmetries of a Newtonian universe.

The idea is that, in practice, there is no way to know whether a theory is really capturing all possible physical phenomena so that one can never be sure that one has actually identified the all the matter in the universe. As a result, it is advantageous, from a methodological perspective, to assume that one can indefinitely embed any particular system into a larger system without that larger system losing any of the symmetries of the original system.

While it is true that subsystem recursivity is a powerful methodological tool for judging theories with symmetry, it is not well justified as an epistemic or ontological principle for defining gauge symmetry. Poincar\'e invariance, for example, is not subsystem recursive in general relativity with closed spatial topology because isolated systems, such as black holes, have the approximate symmetries of asymptotically flat spacetimes, which include the Poincar\'e symmetries, but these are not symmetries of a spatial closed universe. But general relativity with closed spatial topology is observationally distinguishable from asymptotically flat general relativity \citep{ellis1986observational}, where Poincar\'e symmetry is subsystem recursive. Thus, it is possible and observationally meaningful for a theory to \emph{not} be subsystem recursive and to be empirically adequate.

I will, thus, not consider subsystem recursivity to be a necessary ingredient in defining a gauge symmetry. This observation will be relevant to my analysis because I will argue that dynamical similarity \emph{should not} be a gauge symmetry of the Kepler problem but \emph{should} be a gauge symmetry of cosmology.\footnote{ The potential failure of subsystem recursivity here, however, is not completely obvious since it is not the same dynamical similarity acting on both kinds of system. }
\chapter{En route to gauge symmetry}
\label{ch:en route gauge}

\begin{abstract}
    This chapter will serve two main purposes. Firstly, it will review a variety of different attempts to define gauge symmetries in the literature. These include attempts to define gauge symmetry using the spacetime dependence of the symmetry generators, the invariance properties of the variational principle used to define the laws, and the underdetermination of the equations of motion. Secondly, and more significantly for the purposes of this thesis, it will highlight various problems with each attempt. As the chapter progresses, I will identify elements that I find promising (or not) in each approach as I gradually motivate my own proposal based on underdetermination following the motivations of Dirac. Finally, I end the chapter by introducing the concept of dynamical similarity, highlighting the challenges and opportunities it poses as a new kind of gauge symmetry in physics.
\end{abstract}

\ifchapcomp
    \tableofcontents
    \newpage
\else
    \cleardoublepage
\fi

\section{Introduction}
\label{sec:en route intro}

While the previous chapter defined and motivated Belot's problem, this chapter and the next will motivate my proposed solution and illustrate its distinctive features. I will start by reviewing standard approaches to classifying symmetry in general and defining gauge symmetry in particular. Many of these approaches are discussed in \cite{belot:sym_and_equiv}. I will add several more, including a discussion of Dirac's definition of gauge symmetry. We will see that the motivations that he gives for his definition are similar to my own even if his proposal, which was only ever intended to give a sufficient condition for a symmetry being gauge, is more limited in scope. I will also describe Noether's famous theorems and contrast her methodology with Dirac's. Then, I will argue that Noether's methodology is more direct, and reflects more accurately the conditions under which physicists construct their theories in practice. I will then adopt a version of Noether's methodology when recovering Dirac's formalism in \chap\ref{ch:rep_sym}. 

The review in this chapter will serve two purposes. First, it will lay out a terminology and introduce a set of concepts that I will refer back to later when explaining the benefits of my proposal. Second, and more importantly, it will provide an opportunity to ask why one might expect any given definition of gauge symmetry to be fruitful or not. By identifying the reasons for the successes and failures of any given definition, we are given hints at what a better definition might look like. This analysis will favour definitions of gauge symmetry that involve a variational principle. In this case, the variation defining the dynamics can be engineered to match underdetermination of the equations of motion to underdetermination of the models by the phenomena. The ability to do this will be required by the normative rules that will result from my own proposal. Nevertheless, I will find that standard variational definitions are deficient in the sense that they cannot implement underdetermination with respect to dynamical similarity. To understand this last point, and to set up the formal manipulations of \chap\ref{ch:rep_sym}, I will end this chapter by giving a detailed description and definition of dynamical similarity.

Many of the explicit models, mathematical derivations, and important theorems referenced in this chapter well be detailed in \chap\ref{ch:rep_sym}. I link to relevant derivations when appropriate.

\section{Proposals for narrow symmetry} 
\label{sec:narrow_proposals}

In \Sec\ref{ssub:broad_and_narrow_symmetries}, I defined the notion of a \emph{broad} symmetry and found it to be too broad to be fruitful. In this section, I will describe more narrow definitions, in the sense defined in \Sec\ref{ssub:broad_and_narrow_symmetries}, and discuss their advantages and difficulties. By working through these examples, we will gain further insights into Belot's Problem and potential routes to solving it. I will refer to the different proposals for defining narrow symmetries as `narrow proposals' in what follows.\footnote{ Note that it's the symmetries that make up the proposal that are meant to be narrow not the proposals themselves. } 

Existing narrow proposals have a common aim of trying to identify formal features of symmetry that can plausibly be linked to the underdetermination of representations by phenomena. One approach is to base the definition on the functional dependence of the symmetry transformations on the spacetime structure. In particular, one can define narrow symmetries in terms of different locality conditions imposed on this dependence. I will say that narrow symmetries of this kind are based on \emph{locality-conditions} and will investigate them in \Sec\ref{ssub:symmetries_of_the_equations_of_motion}. I will find that these are not good proposals for gauge symmetry.

Another approach is to define narrow symmetries in terms of specific features of the nomic structure that are preserved under the symmetry transformations. This is most easily achieved when the laws are implemented by a variational principle. I will refer to these as \emph{symmetries of the variational principle} and investigate them in \Sec\ref{sub:symmetries_of_the_variational_principle}. I will show that such symmetries are better motivated than symmetries based on locality conditions because the underdetermination arising from a variational principle due to a symmetry can be matched to the underdetermination of representations by phenomena, as suggested by the first slogan of \Sec\ref{sub:intro PESA}.

The most promising existing approach, my my opinion, is a proposal made in \cite{dirac2001lectures}. This proposal does not focus directly on the properties of the variational principle itself but specifically on the underdetermination of the equations of motion. The drawback of Dirac's proposal, as we will see, is that its specific implementation is too narrow for my purposes. In particular, Dirac's proposal applies neither to the case of reparametrisation symmetry\footnote{ This runs counter to orthodoxy in the canonical quantum gravity community. We will defend this claim later. A more thorough defence is provided in \cite{gryb_thebault_book}. } nor to the all important case of dynamical similarity.

Let us start by briefly reviewing proposals for symmetry based on locality conditions so that we might highlight their basic aspects and limitations. Further details can be found in \cite{belot:sym_and_equiv}. A beautiful mathematical exposition of classical symmetries can be found in \cite{hydon2000symmetry}. More advanced discussions of symmetry can be found in \cite{olver2000applications}.

\subsection{Symmetries based on locality conditions} 
\label{ssub:symmetries_of_the_equations_of_motion}

\cite{belot:sym_and_equiv} reviews three different kinds of narrow proposals based on locality conditions that he calls: classical, generalised and non-local. These proposals are closely related. In fact, generalised and non-local symmetries are generalisations of classical symmetries.\footnote{ Other names for classical symmetries include \emph{point symmetries} or \emph{Lie (point) symmetries}. } I will review these notions here to provide context and vocabulary for our subsequent discussions.

\emph{Classical symmetries} are defined in terms of the properties of their infinitesimal generators, which will depend on the KPMs of the theory in question. Recall that the space of KPMs includes spacetime structure as well as matter and force fields. It is common to refer to the quantities representing the spacetime structure as \emph{independent variables} and to the matter and force fields as \emph{dependent variables}. But for our purposes, we need only note that a specification of a KPM is equivalent to a specification of the independent and dependent variables as well as the derivatives of the dependent variables (if necessary).

In \Sec\ref{ssub:reduction}, I noted that symmetries form groups. Continuous groups, or \emph{Lie groups}, have parameters that label their orbits. Using the group parameters, we can define the \emph{infinitesimal generator} of a group as the change along an orbit induced by a \emph{small} (i.e., infinitesimal) change in the group parameters about the group identity element.

One can learn a great deal about a group's structure by analysing the properties of its infinitesimal generators. These give all the local properties of the group. Many groups share the same local properties because they have the same infinitesimal generators even though they differ in their global properties. Moreover, discrete symmetries can't be made `small' in any sense and therefore have no infinitesimal generators. In many applications however, global properties are not relevant. All symmetries in this section are defined only in terms of the properties of their infinitesimal generators. These definitions thus ignore global properties and exclude discrete symmetries altogether.

We can use the definitions given above to define a \emph{classical symmetry} as a transformation on the space of KPMs that preserves the space of DPMs (i.e., is a broad symmetry defined at the kinematic level) and whose infinitesimal generator is independent of the derivatives of the dependent variables.

This extra locality condition on the form of the infinitesimal generator imposes a strong condition on the allowable transformations. For most theories, this reduces the symmetries to a tractable set of transformations that is often explicitly computable. For Newtonian systems with arbitrary potential energy (i.e., the potential energy has no additional symmetries), the classical symmetries are \emph{only} time translations (see \cite{hydon2000symmetry}). For general relativity in the absence of matter, the classical symmetries are arbitrary coordinate transformations and global rescalings (see \cite{torre1993gravitational,anderson1996classification}).

Classical symmetries are sometimes too narrow and sometimes too broad to reasonably be taken as gauge symmetries. Some examples of this are discussed in \cite{belot:sym_and_equiv}. But it is interesting to ask why one might expect them to be gauge symmetries in the first place. A simple answer might be that gauge symmetries should reflect the presence of \emph{representational baggage}; i.e., representational structure beyond what is strictly necessary to model the phenomena. If that is the case, then it is natural to expect that such excess structure should be kinematical: any dependence of the dynamics on this structure should imply some empirical consequence. One might try (misguidedly) to use this to motivate the converse implication that all representational structure that is independent of the dynamics should be considered excess structure. Then, maps between DPMs that depend only on the kinematical structures might be expected to relate empirically equivalent representations. Classical symmetries are transformations of this kind that depend, in the simplest possible way, on the kinematical state.

While such an argument may sound persuasive, it is not. There is no particular a priori reason that non-dynamical structure should be empirically irrelevant. Moreover, there is also no guarantee that classical symmetries will always be dynamically irrelevant. Indeed, the example discussed in \Sec\ref{ssub:the_newtonian_free_particle} of the Newtonian free particle with Galilean transformations as gauge and non-gauge symmetries illustrate both of these possibilities.

Generalisations of classical symmetries suffer from similar problems. A \emph{generalised symmetry} is a symmetry where the infinitesimal generator is allowed to depend on arbitrary derivatives of the dependent variables. Clearly, all classical symmetries are generalised symmetries, and there are interesting theories that have no generalised symmetries that are not classical (matter-free general relativity is an example). Nevertheless, there are important examples of non-classical generalised symmetries. We have seen one such example in \Sec\ref{eq:DS kepler} in terms of the dynamical similarity of the Kepler problem.

Finally, one can also allow symmetries where the generator is allowed to depend on non-local functions of kinematical structures. These are the \emph{non-local symmetries}. I will not discuss these in detail here but see \cite[\S4]{belot:sym_and_equiv} for references to examples. I will, however, make two short comments about such symmetries. First, one must be careful in allowing for a non-local dependence of the infinitesimal generator on the kinematical state to not broaden the definition of symmetry too much that one simply recovers the full set of broad symmetries. Second, matter-free general relativity has been shown to only have non-local symmetries aside from the pre-engineered symmetry under coordinate transformations and global re-scalings. Thus, such non-local symmetries may be indispensable in trying to understand a theory of quantum gravity.

The narrow notions of symmetry defined above are mathematically well-defined and can be used to identify several physically salient notions of gauge symmetry found in the physics literature. What I'd like to note here is that broadening or narrowing the notion of symmetry along these lines does not achieve much in capturing a formal criterion that can be used to identify gauge symmetries in general. The rough intuition that some non-dynamical principle could be used towards this end is, at best, questionable and runs into immediate problems when faced with explicit counter-examples.

One might hope that variational principles could provide a better-motivated approach to identifying gauge symmetries. I will now investigate the extent to which this might be achievable.


\subsection{Symmetries of the variational principle} 
\label{sub:symmetries_of_the_variational_principle}

The idea behind a variational principle is to define some mathematical quantity, usually called the \emph{action}, that depends functionally on the elements of a KPM and takes extreme values on the DPMs. The DPMs of the theory can then be found by computing the extreme values of the action. A difficulty arises when the extreme values are not unique; i.e, when there are many DPMs for which the action takes the same extreme value. The nomic structure defined by the action principle alone has no way to discern such DPMs. Formally, one often finds that the equations of motion generated by the action are ill-posed in the sense that they cannot be solved uniquely even with appropriate initial or boundary data. I will call variational principles of this kind \emph{irregular}, following standard nomenclature.

One might want to require that the mathematical underdetermination resulting from irregular variational principles could be arranged to reflect the physical underdetermination between representations and phenomena suggested by the presence of gauge symmetries. In fact, the way to resolve the underdetermination in the variational principle is to assign arbitrary conventions, called \emph{gauge fixings}, that assign a unique DPM to every extreme value of the action. We will see a very general way of doing this explicitly in \Sec\ref{sec:gauge_fixing_and_degree_of_freedom_counting}. But good interpretive practice should prevent arbitrary conventions from having observable implications. Thus, one might require that all gauge fixings be related by gauge symmetries.

One could then try to engineer the variational principle to be irregular in just the right way to reflect the features of the target system one is interested in modelling. In fact, we will see below that this is precisely how gauge symmetries are often understood in physics textbooks. Theorems and algorithms can be formulated for identifying the conditions under which a particular action will have particular irregularities, the gauge symmetries implied by these irregularities, and the set of gauge fixings one can use to remove the irregularities. The description of some of these tools can be found in standard textbooks on gauge theory such as \cite{sudarshan1974classical}, \cite{sundermeyer:1982}, or \cite{henneaux1992quantization}, and I will give a detailed description of many of them in \chap\ref{ch:rep_sym}.

Given these impressive tools, it is perhaps no wonder that the opening Chapter of Henneaux and Teitelboim's book on the quantization of gauge systems makes the claim that ``the action itself enables one to decide what are the observables'' \citep[\S1.5.2]{henneaux1992quantization}. Henneaux and Teitelboim take `observables' to be the structures of a theory's models that are invariant under all gauge symmetries.\footnote{Note that this definition is, importantly, \emph{not} equivalent to the sense used in the philosophical literature on the observable/unobservable distinction. We will give a precise definition of what we mean by an observable in \Sec\ref{sec:statement_of_the_pesa}. } As a result, if we take their statement at face-value, it would seem to close the discussion on how to identify the gauge symmetries of a theory. We will see, however, that the definitions and techniques used by this standard book, which reflect the techniques used in the broader physics community, have limited applicability. Moreover, it is often assumed by physicists that the theory in question is empirically adequate, and thus physics definitions of gauge symmetry appear to be silent on epistemic considerations. Finally, even within physics textbooks there are several methodologies used to `decide' on the observables of a theory, and these can be inequivalent when applied to different examples of symmetry. I will now briefly sketch some of these different approaches and highlight their differences.

\subsubsection{Variational and divergence symmetries} 
\label{ssub:variational_symmetries}

As described in the previous subsection, variational principles might plausibly be linked to gauge symmetries when they lead to DPMs where the action takes the same extreme value. Gauge transformations might then be defined by transformations that preserve both the extremality condition \emph{and} the value of the action. In general, actions are written in terms of integrals of functions of the matter and force fields over the spacetime structure. If we call the dependent variables $u^i$ and the independent variables $x$ then the action $S$ is an integral over the \emph{Lagrangian} function $\mathcal L(x, u^i)$ such that
\begin{equation}
    S[\gamma] = \int_M \mathcal L(x,u^i) \de x\,,
\end{equation}
where $\gamma$ is a history of the theory.\footnote{ We will define the action a bit more carefully in \Sec\ref{sec:lagrangian_variational_principle}. For a more rigorous definition of a variational symmetry, see \cite[definition 4.10 p. 253]{olver2000applications}. } The integration region is a manifold $M$ containing different values of the independent variables. We then define a \emph{variational symmetry} as a transformation $x \to \bar x$, $u^i \to \bar u^i$ and $M \to \bar M$ such that
\begin{equation}
    \int_M \mathcal L(x,u^i) \de x = \int_{\bar M} \mathcal L(\bar x,\bar u^i) \de \bar x\,.
\end{equation}
Clearly, transformations of this kind preserve the value of the action. Note that variational symmetries can also be classified into classical, generalised, and non-local by the dependencies of their infinitesimal generators.

While many variational symmetries do correspond to obvious gauge symmetries, many do not. One reason is that variational symmetries are, in a sense, too broad. Invariance of the action does not imply that it is irregular. It is possible to find variational symmetries that do not lead to any problems in solving the equations of motion. Time translations and global Euclidean symmetries; i.e., the Leibniz shifts of \Sec\ref{sub:leibniz_shifts}; are examples of this in Newtonian mechanics. More generally, any symmetry that is covered by Noether's first theorem, which we describe below, is a variational symmetry that doesn't obstruct the equations of motion. We will see the reason why in \Sec\ref{sub:global_symmetries}.

Another reason that variational symmetries should not be identified, in general, as gauge symmetry is that they are too narrow: invariance of the action is stronger than the preservation of the extremality condition. Galilean boosts, for example, are not variational symmetries in Newtonian mechanics.

A slightly broader category of symmetries is that of \emph{divergence symmetries}. These symmetries are transformations that preserve the Lagrangian up to a total derivative; i.e.,
\begin{equation}
    \mathcal L(x,u^i) = \mathcal L(\bar x, \bar u^i) + \text{div } \phi\,,
\end{equation}
where $\phi$ is some arbitrary function of $x$ and $\bar u^i$. By the Divergence Theorem, divergence symmetries can, at most, shift the action by a constant $S \to S + a$, where $a = \phi\big|_{\partial \bar M}$. Under a divergence symmetry, the extreme values of the action can be shifted by a constant so that DPMs remain DPMs. In this way, the variational principle, but not the action itself, has no way of discerning DPMs related by divergence symmetries. Famously, the Galilean boosts of Newtonian mechanics are divergence symmetries even though they are not variational symmetries.\footnote{ For a proof, see Example~4.35 (p. 297) of \cite{olver2000applications}.}

Divergence symmetries, however, still do not capture all the features one might want of a gauge symmetry. Since all variational symmetries are divergence symmetries, they suffer from the same too-broad criticism laid upon variational symmetries. But they are also too narrow as they don't include the dynamical similarities of the Kepler problem (under the right epistemic constraints) discussed in \Sec\ref{ssub:the_kepler_problem} or relatedly, the global scaling symmetries of matter-free general relativity (see \cite{anderson1996classification}). Still, it seems that the basic idea behind a variational and divergence symmetry is roughly on the right track because it highlights a basic source of underdetermination in the nomic principles used to define laws in physics. I will now explore several other related definitions of narrow symmetry that attempt to highlight this issue more explicitly.

\subsubsection{Noether symmetries} 
\label{ssub:noether_symmetries}

No discussion of symmetry would be complete without mention of Emmy Noether and her powerful approach to understanding symmetry. Noether proved two theorems that are of upmost importance to the understanding of the formal aspects of symmetry.\footnote{ Her theorems were published in a 1918 paper (see \cite{noether1983invariante} for a reprint and \cite{noether1971invariant} for an English translation). See \cite{kosmann2011noether} for an historical perspective, \cite{olver2000applications} for a formal treatment, and \cite{brading2003symmetries} for a philosophical analysis.} Her first theorem lays out certain conditions under which a particular variational principle will define DPMs that have quantities, called \emph{constants of motion}, that are constant along a history. I prove a modern formulation of this theorem together with a description of the necessary assumptions in \Sec\ref{sub:lagrangian_constraints_and_noether_s_first_theorem}. Noether's second theorem derives a set of identities satisfied by the theory when the action obeys certain conditions. A modern formulation of the second theorem with a proof is given in \Sec\ref{sub:noether_2_symmetries_and_noether_s_second_theorem}. For now, I would like to discuss the general attitude towards symmetry taken by Noether in deriving her theorems and how her second theorem leads naturally to a particular definition of gauge symmetry.

The basic idea behind Noether's approach is to first restrict to the set of symmetries that leave the action invariant. The invariance of the action is expressed as a mathematical definition from which various conclusions are then deduced. Normally, the definition of symmetry used is that of a generalised variational symmetry as given above --- but with more leniency regarding the action of the transformations on the boundary since these have no effect on the equations of motion. This is perhaps the most historically accurate definition. However, some modern texts, including \cite{henneaux1992quantization}, use a slightly different definition (see their \eqn 3.22) in terms of the second functional derivative of the action $S$.\footnote{We will give this expression in Equation~\ref{eq:N2}, which we explain in \Sec\ref{sub:noether_2_symmetries_and_noether_s_second_theorem}.} We find this definition enlightening because it can be more directly manipulated and because it has a nice geometric interpretation, which I will describe briefly below. For most applications, these definitions are equivalent.\footnote{ A full comparison of these definitions is beyond the scope of this text, but potential differences may involve the treatment of boundary conditions. } Regardless of one's preference, I will simply refer to the relevant symmetries as \emph{Noether-$2$ symmetries}, which one can take as a placeholder for either a generalised variational symmetry or the functional derivative definition. Noether-$2$ symmetries are then those symmetries that are relevant to Noether's second theorem.

The subset of Noether-$2$ symmetries that are constant functions of the independent variables are the symmetries relevant to Noether's first theorem. We will call those \emph{Noether-$1$ symmetries}. Let me now describe some consequences of these different symmetries and leave the mathematical details for the next chapter.

First, it is important to note is that Noether-2 symmetries are used to \emph{define} gauge symmetries in many physics texts including in Chapter~3 of \cite{henneaux1992quantization}. This definition is probably the most commonly accepted formal criterion for defining a gauge symmetry in the physics literature. Helpfully, the functional derivative definition of a Noether-$2$ symmetry can be used to derive a formal equation (see \eqref{eq:loc narrow sym def}) that can, in principle, be used to solve for these symmetries explicitly.

Second, the intuition behind identifying Noether-2 symmetries as gauge symmetries is the idea that an underdetermination in the variational principle should be linked to underdetermination of representations by phenomena. But the underdetermination implied by Noether-2 symmetries is often mistaken for a more pernicious form of underdetermination, emphasised by Dirac, where the variational principle is seen to produce equations of motion that are not well-posed. In many cases of interest, the two notions are equivalent --- but not always. For example, I will show at the end of \Sec\ref{sub:lagrangian_constraints_and_noether_s_first_theorem} that all the Noether-$1$ symmetries are Noether-$2$ symmetries that do not lead to any underdetermination in the equations of motion. Thus, Dirac's argument, which is often used to motivate Noether-$2$ symmetries as defining gauge symmetries, is actually inapplicable to many Noether-$2$ symmetries of physical interest. This is important because I will argue that underdetermination is the more reliable feature to associate with gauge symmetry.

Finally, it will be helpful to take note of Noether's methodology when deriving her theorems. First, she identifies some class of symmetries in terms of the formal properties of the action. Then she deduces the consequences of those symmetries. This fits well with how physicists build theories: they look for simple actions that they can engineer to have the gauge symmetries they think reflect the phenomena they are trying to study. In this way, one knows in advance at least some symmetries the action will have and can use Noether's theorems to deduce some consequences of those symmetries. This will be in stark contrast to the methodology developed by Dirac for identifying the gauge symmetries of a theory in which the symmetries are deduced by an algorithm designed to iteratively repair the equations of motion generated by the variational principle. I will describe Dirac's approach in \Sec\ref{ssub:the_dirac_algorithm} below.

\paragraph{More on Noether-2 symmetries} 
\label{ssub:noether_2_symmetries}


We can understand the formal features of the functional derivative definition of Noether-2 symmetries by considering in more detail the concept of a `variation' of a history. By `variation' I mean a small (i.e., infinitesimal) but arbitrary change of the independent variables for every value of the dependent variables. The variational principles used in physics consider arbitrary variations of this kind and then define the DPMs as those KPMs such that the resulting change in the action as a result of such variations is zero. Each small change in the independent variables defines a kind of derivative for every value of the dependent variables. The variational principle therefore requires that each of these derivatives is zero. Since, for most theories, the dependent variables can take a continuous range of values, there is an infinite number of such conditions. If everything works properly, these conditions will produce the equations of motion that define the theory. See \Sec\ref{sec:lagrangian_variational_principle} for a formal statement of the variation and the resulting equations of motion, \eqref{eq:EL eqs}.

A special case occurs when one considers particular variations \emph{of each of the derivatives} above and then requires those variations to be equal to zero. For all such variations, the action will retain its extreme value. Thought of as the infinitesimal generators of transformations on the space of KPMs, these variations match the intuitive notion outlined above for a proposal for the gauge symmetries of a theory because they preserve the extreme values of the action. These are the Noether-2 symmetries. Using this condition, it is a straightforward exercise using modern variational techniques to prove that Noether-2 symmetries imply certain identities on the DPMs of a theory. These are the identities of Noether's second theorem. An explicit mathematical derivation is given in \Sec\ref{sub:noether_2_symmetries_and_noether_s_second_theorem}.

\begin{figure}[tb]
    \centering
    \includegraphics[width=0.5\textwidth]{\pdots 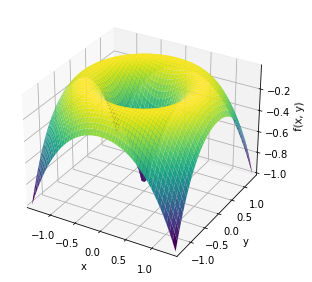}
    \caption{In a multidimensional space, a function can have a flat crest along a particular direction when both the first and the second derivatives are zero along that direction. In this example, the function has a maximum when $r = \sqrt{x^2 + y^2} = 1$ in both $r$ and $\theta$ directions, where $\theta = \tan \frac y x$. But it is also flat (i.e., the second derivative is vanishing) along the $\theta$ direction creating a flat crest. Directions such as that along $\theta$ point along the orbits of Noether-2 symmetries. }
    \label{fig:sym_min}
\end{figure}

The situation can be understood more intuitively by considering functions of only two (instead of infinite) variables. A maximum (or minimum) occurs when the derivatives of the function, in all directions, are equal to zero. But a special case occurs when the \emph{second} derivative in one particular direction is also equal to zero. Then the function has a flat crest along the direction in question as shown in Figure~\ref{fig:sym_min}. The Noether-2 symmetries, as expressed by Equations~\ref{eq:loc narrow sym def}, are then transformations along the flat directions in the infinite dimensional space of ways in which one can infinitesimally change a KPM.

\subsection{Symmetries in first order systems} 
\label{sub:symmetries_in_first_order_systems}

Most classical theories in physics are expressed in terms of second order partial differential equations, where, loosely speaking, accelerations are computed in terms of positions and velocities. The relevant kinematical structures are therefore the independent variables and at least two of their derivatives. At any given time, the independent variables and their derivatives have the ability to represent independent features of a target system. In analysing symmetry, a lot of leg work must go into keeping track of the fact that, while variables and their derivatives can be used to represent independent phenomena, they are nevertheless strongly intertwined by their mathematical definitions. Mathematical structures called jet bundles can be used to keep track of the constraints between these `independent' quantities that are not completely independent.\footnote{ For a simple illustration of how to do this, see the appendix of \cite{belot:sym_and_equiv}. For a comprehensive mathematical treatment, see \cite{saunders2008jet}. } The geometric picture, however, is sophisticated and explicit calculations can quickly become intractable.

An alternative is to reformulate the second order system as an equivalent first order system. We will be especially interested in doing this for the time dependence of the dependent variables. Spatial boundaries can complicate the issue but in general this can be done rather painlessly by following the general procedure outlined in \Sec\ref{sub:hamiltons_equations}. For most theories --- with the notable exception of general relativity --- splitting the temporal and spatial dependence allows for a relatively clean split of symmetry and dynamics. Thus, systems that are first order in time are convenient for analysing the symmetries of many theories.

First order systems come with extra constraints that must be satisfied in order for them to be equivalent to the corresponding second order system. At the end of the day, solving these constraints introduces all the extra complications of working with the second order system directly. Despite this, first order systems have remarkable geometric properties that can shed considerable light on the nature of symmetry. These properties are also important for constructing corresponding quantum theories. I will thus find it extremely valuable to consider first order systems here.

\subsubsection{Hamiltonian systems} 
\label{ssub:hamiltonian_systems}


A general class of first order systems are called \emph{Hamiltonian systems}.\footnote{ The canonical reference for the study of Hamiltonian systems in mechanics is \cite{arnol2013mathematical} or \cite{arnol2001dynamical} for a more advanced treatment. A more conceptual introduction is given in \cite[ch. VI]{lanczos2012variational}. For a review of the geometric methods and the link to symmetry, see Chapters 1-3 of \cite{woodhouse1997geometric}. } To describe such systems, it is customary to call the independent variables \emph{configurations} and their time derivatives \emph{velocities}. One then defines a map, called a \emph{Legendre transform}, between the space of configurations and velocities to a space of configurations and \emph{momenta}. That specifies the state space of the theory, which is called \emph{phase space}. To define the dynamics, one needs two additional structures: a function on phase space called the \emph{Hamiltonian} and a differential structure called a \emph{symplectic 2-form} that is antisymmetric in configurations and momenta. The Hamiltonian function and the symplectic 2-form can then be used to define a unique vector at all points in phase space. DPMs are then defined to be those curves on phase space that are tangent to these vectors.

The equations of motion of Hamiltonian systems are called \emph{Hamilton's equations}, which are a set of first order (in time) differential equations. These are split into two sets of equations. The first of these, called \emph{Hamilton's first equation}, is a constraint that reduces the first order system to an equivalent second order system. The second equation, called \emph{Hamilton's second equation}, contains the non-trivial part of the dynamics. As I just mentioned, Hamilton's equations assign a direction and magnitude (i.e., a vector) to every point in phase space. One then interprets the vector as giving the tangent to the DPM that passes through that particular phase space point. The DPMs can be defined by starting at a particular point in phase space and then \emph{flowing} to a neighbouring point by following the vector defined by the Hamilton function. Because this is similar to how fluid flows through a medium, Hamilton's equations are said to generate \emph{flows} on phase space. 

The utility of first order systems is that one can easily give mathematical criteria for when the equations of motion can be solved. The vectors produced by the flow equations define vector fields on phase space, and the existence, uniqueness and smoothness of these vector fields give precise constraints on the solvability of the system. Any obstructions to solvability can thus easily be identified and, in some cases, remedied. Moreover, it is usually also clear how many independently specifiable quantities are needed to define the vector fields in the first place. This is important for my definitions of gauge symmetry because I will focus on both the solvability of the equations of motion and on the number of independently specifiable representational structures. All of this is most cleanly achieved in a first order formalism.

First order systems are also ideal for studying symmetry transformations in general. Because the kinematical structure has been coded into the structure of curves on phase space, symmetry transformations with group structure can be generated by assigning vectors to every point on the curve defining a KPM. These vectors can be taken to represent the infinitesimal generators of an arbitrary symmetry transformation. In well-behaved Hamiltonian systems, the DPMs never intersect since they are defined as integrals curves of smooth vector fields. This means that all (smooth) maps between DPMs can be generated by vector fields and that one can represent an arbitrary (smooth) broad symmetry in terms of an arbitrary (smooth) vector field on phase space.\footnote{ This can be true even if that symmetry-generating vector field is the same vector field generating the dynamics. This is the case of reparametrisation symmetry. We will argue in \Sec\ref{sub:solved_reparametrisation_invariance} that such symmetries should \emph{never} be treated as gauge symmetries. } Thus, symmetries can be thought of as arbitrary (non-dynamical) flows on phase space. In this way, all the nice geometric conditions that can be used to understand the dynamical quantities of Hamiltonian systems can also be used to understand the symmetric quantities.

One important class of narrow symmetries are those symmetries of Hamiltonian systems that preserve the value of the Hamiltonian. These are called \emph{Hamiltonian symmetries}. This is a slightly smaller space than the broad transformation described above, which are only required to preserve the dynamical flow and not also the value of the Hamiltonian. One can narrow the definition of Hamiltonian symmetries further by requiring that they be generated by vector fields that satisfy locality conditions on phase space. Here, the locality conditions usually involve restrictions on the number of derivatives of the phase space quantities required to build the vector field in question the same way that classical, generalised and non-local symmetries were defined earlier. Locality conditions of this kind usually lead to a significant narrowing of the class of symmetries because a vector field needs to ``know'' something about the dynamical flow in order to preserve it. That usually requires more information than can be provided locally on phase space.

A narrowing of the Hamiltonian symmetries along these lines in order to identify gauge symmetries might seem plausible for the same reasons that motivated classical and/or generalised symmetries. Any Hamiltonian symmetry that depends locally on phase space points could be argued to be non-dynamical in the sense that they have no way of gaining information about the non-trivial dynamical implications of the Hamiltonian function. This could motivate them to be identified with gauge symmetries. However, such arguments are again vulnerable to the criticism levelled against classical and generalised symmetries above.

As a final observation, we note that while the physics and philosophy literature almost exclusively focuses on first order system that are Hamiltonian, these are not the most general first order systems because such systems are required to admit a symplectic $2$-form. More generally, one can define first order flow equations without introducing a symplectic $2$-form. An example of such a system is a \emph{contact system}. We will see that contact systems are the natural by-product of treating dynamical similarities as gauge symmetries in \Sec\ref{sec:dynamical_similarity}.

\subsubsection{Variational principles in first order systems} 
\label{ssub:variational_principles_in_first_order_systems}


Hamiltonian flows can be generated by variational principles. To achieve this, it is usually sufficient to slightly modify the action of the corresponding second order system along the lines laid out in \Sec\ref{sub:hamiltons_equations}. Just as in second order systems, the variational principle can lead to irregularities. And just as in second order systems,
one might try to match the irregularities of the variational principle with the underdetermination of representations by phenomena.

It is a significant advantage of first order systems that such an analysis can be done directly in terms of the geometric structures and vector fields on phase space. In \Sec\ref{sub:hamiltons_equations} I will derive a simple expression, Equation~\ref{eq:S var}, that will allow me to clearly identify all the notions of symmetry applicable to Hamiltonian systems discussed throughout this work and investigate their consequences. This expression provides a remarkable unification of most of the different aspects of symmetry discussed in this chapter.

The basic idea that I will follow is to use Noether's methodology and start by assuming that one knows that the action has symmetries with certain formal properties. The advantage of first order systems is that one can then easily match the presence of those symmetries with degeneracies in the geometric structures of phase space. Simply knowing that an action has symmetries of a particular form is sufficient to prove many immediate consequences --- including Noether's theorems. These consequences reproduce many results familiar to different approaches to gauge symmetry that have been developed in the literature. I will show this explicitly in \Sec\ref{sub:explicit_representation}. Before doing this, let me first review some of the central features of the most prominent of those approaches.

\section{The Dirac algorithm} 
\label{ssub:the_dirac_algorithm}

In this section, I will describe a definition of gauge symmetry introduced by Dirac. The motivations for this procedure, which I will describe in detail below, are also an important motivation for the definition I will give in \Sec\ref{sec:statement_of_the_pesa}.

Dirac's definition applies to Hamiltonian systems subject to constraints, and therefore further illustrates the utility of first-order approaches. Dirac's intuitions are often invoked in physics textbooks\footnote{E.g., Chapter~1 of \cite{henneaux1992quantization} or Chapter~2 of \cite{woodhouse1997geometric}} to motivate the interpretation of gauge symmetry for constrained Hamiltonian systems. Such systems form the basis for the BRST\footnote{ Named for Carlo Becchi, Alain Rouet, Raymond Stora and Igor Tyutin.} quantization methods used to quantize the Standard Model (e.g., Chapters~13 and 14 of \cite{henneaux1992quantization} or \chap 15 of \cite{Weinberg:1996kr}) and canonical attempts to quantize gravity (e.g., Chapters~4 and 5 of \cite{rovelli2004quantum} or Part~I of \cite{thiemann2008modern}). Among practising physicists working on gauge theories, Dirac's is, thus, the orthodox view of a gauge symmetry.\footnote{ A similar conclusion was reached in \cite{earman2003tracking}. } Let us describe that view now.

In a short series of lectures \citep{dirac2001lectures}, Dirac developed an algorithm that identifies the transformations of a theory that, in his words, ``lead to changes'' in the phase space quantities ``that do not affect the physical state.''\citep[p.20]{dirac2001lectures} Dirac seems to understand the `physical state' to be the state describing the physical state of affairs of the world. He then insists (p.20) that this state not depend on any freely specifiable parameters that may appear in the theory's models. It is clear that Dirac's notion of transformation-that-does-not-affect-the-physical-state has since been interpreted as a notion of gauge transformation, in our terminology, when a theory is empirically adequate.\footnote{ See, for example \S1.2.1 of \cite{henneaux1992quantization} that uses Dirac's terminology of ``transformations that do not change the physical state'' and explicitly identifies these with ``gauge transformations.'' } It is thus justified to treat Dirac's proposal as a proposal for defining gauge symmetry.

I believe that Dirac's proposal is generally well-motivated. When extended to include certain minimal epistemic considerations, his motivation reflects that of my own proposal. The applicability of Dirac's proposal, as we will see shortly, is unfortunately too narrow to serve all my purposes. To understand why, I will now briefly describe the key components of Dirac's algorithm. A more detailed technical account is given in \Sec\ref{sec:the_dirac_algorithm}. A modern account of the Dirac method is given in Chapters~1 to 5 of \cite{henneaux1992quantization} with a more complete description in \cite{sundermeyer:1982} and Chapter~2 of \cite{woodhouse1997geometric} for a more geometric approach.

As stated above, Dirac's proposal is formulated on phase space using the Hamiltonian formalism. Central to his reasoning is the claim that, in Hamiltonian systems, irregularities result in constraint equations on phase space. These constraint equations are necessary for the consistency of the dynamical equations and reflect the fact that the Legendre transform between configuration-velocity space, also called \emph{velocity phase space}, and phase space is many-to-one. Attempting to invert the Legendre transform then leads to underdetermination in the equations of motion on velocity phase space. This underdetermination is precisely the underdetermination that Dirac wants to associate with a gauge symmetry.

To understand the origin of the constraints, Dirac notices that the image of the Legendre transform is defined by a constraint equation on phase space. This is what Dirac called the surface of the \emph{primary constraints} of the theory. These primary constraints can be read-off directly from the action using the definition of the momenta. Consistency of the theory then requires that the dynamical flow preserve the primary constraint surface. If this is not automatically true, then new constraints, which Dirac calls \emph{secondary constraints}, must be imposed. Consistency then further requires that the dynamical flow also preserve the secondary constraint surface. This, in turn, can generate new constraints. The process of identifying new consistency conditions and their corresponding constraints continues until one finds that all consistency conditions are automatically satisfied. If this occurs, the Dirac algorithm is said to close.

The output of the Dirac algorithm is then a set of constraints (primary, secondary, etc) on phase space. If these constraints over-constrain the system then the original variational principle is said to be ill-defined. Otherwise, one can use the outcome of the algorithm to repair the irregularities of the original system. The first step of this repair process it to identify the constraints of the system that are responsible for the underdetermination in the equations of motion. This can be done using the form of the constraints and the geometric structures of phase space. The details are technical and are not important for understanding the repair procedure. I give a more detailed account in \Sec\ref{sec:the_dirac_algorithm}. Dirac calls the constraints responsible for the irregularities \emph{first class}, and the remaining constraints \emph{second class}.

The repair procedure involves introducing arbitrary functions, $v^\alpha$, as Lagrange multipliers for all first class constraints $\pi_\alpha$. These are then added to the original Hamiltonian, $H$, of the system to produce the \emph{extended Hamiltonian} of the theory $H_\text{ext} = H + v^\alpha \pi_\alpha$. Hamilton's equations for the extended Hamiltonian then lead to well-defined equations of motion. This resolves the underdetermination in the system at the cost of introducing the arbitrary functions $v^\alpha$.

Dirac then proceeds to argue for a conjecture that states that any change in the value of the arbitrary functions $v^\alpha$ should be thought of, in my language, as a gauge transformation. The motivation for his proposal is the idea that gauge transformations should reflect underdetermination in the equations of motion, and that this underdetermination is precisely reflective of the underdetermination between representations and phenomena (as described by the physical state). To see this, it is helpful to quote Dirac directly: \cite[p.20]{dirac2001lectures}
\begin{quote}
    $\hdots$ the $q$'s and $p$'s at later times are not uniquely determined by the initial state because we have the arbitrary functions $v$ coming in. That means that the state does not uniquely determine the set of $q$'s and $p$'s, even though a set of $q$'s and $p$'s uniquely determines the state. There must be several choices of $q$'s and $p$'s which correspond to the same state. So we have the problem of looking for all the set of $q$'s and $p$'s that correspond to one particular physical state.
\end{quote}
He then proceeds to argue that this set is exactly the orbit, in my language, of a gauge transformation.

Note that the same intuition persists in more modern treatments of gauge theory. From \cite[\S1.2.1,p.16]{henneaux1992quantization}:
\begin{quote}
    The presence of arbitrary functions $v^a$ in the total Hamiltonian tells us that not all the $q$'s and $p$'s are observable. In other words, although the physical state is uniquely defined once a set of $q$'s and $p$'s is given, the converse is not true --- i.e., there is more than one set of values of the canonical variables representing a given physical state. To see how this conclusion comes about, we notice that if we give an initial set of canonical variables at the time $t_1$ and thereby completely define the physical state at that time, we expect the equations of motion \emph{to fully determine the physical state at other times}. Thus, by definition, any ambiguity in the value of the canonical variables at $t_2 \neq t_1$ should be a physically irrelevant ambiguity.[Original emphasis.]
\end{quote}
Here, the word `observable' is to be understood in the sense defined at the end of \Sec\ref{sub:theory}. This notion of gauge symmetry clearly reflects Dirac's understanding in terms of underdetermination of the dynamics.\footnote{Interestingly, while they use this to motivate their definition of gauge symmetry, they later (Chapter 3, \S3.1) adopt a definition based on Noether-$2$ symmetries, which are not always associated with underdetermination because they include Noether-$1$ symmetries.}

Note that, in both of the quotes above, the emphasis is not just on the capacity of the representations to contain the phenomena but on their ability to \emph{uniquely} predict the phenomena based upon knowledge of the initial data. This is the general idea I wish to advocate in this thesis.

There are, however, several limitations to Dirac's analysis of symmetry. Dirac's starting point requires the theory be formulated as a Hamiltonian system. This requires that the theory be written as a system that is first order in time. Since most theories in physics are second order in space, this requires a split between space and time leading to a state space representation `at-an-instant.' Dirac-style definitions of symmetry involve transformations on states defined at an instant, and in this way are fundamentally at-an-instant notions of symmetry. This leads to two noteworthy complications.

First, at-an-instant symmetries require simultaneity conventions that are arbitrary in relativistic theories. Requiring that these conventions have no empirical consequences is often very difficult in the quantum formalism. In the standard model of particle physics, Dirac's proposal becomes intractable and is used only to motivate a more sophisticated procedure, the BRST procedure mentioned above, that does not require explicit simultaneity conventions. In canonical approaches to quantum gravity, which follow Dirac's proposal, implementing the arbitrariness of simultaneity conventions is probably the most important outstanding open problem. But even with a solution of this problem at hand, there are classical spacetimes that can't be represented as simple curves on phase space. Dirac's proposal seems ill-equipped to define gauge symmetries in such spacetimes. Note, however, that I will nevertheless defend the use of at-an-instant symmetries in this thesis using the arguments given in \Sec\ref{sub:pesa_applicability}.

A second complication arises when the Dirac-symmetry generators are also generators of evolution. This occurs in reparametrisation-invariant theories.\footnote{ In general relativity, the evolution generator is specified uniquely by the simultaneity convention. } When this happens, an argument first given in \cite{barbour2008constraints} and elaborated upon at the end of \Sec\ref{sec:the_dirac_algorithm} shows that Dirac's original demonstration is invalid. Conceptually, this can be understood in terms of a failure of the equivalence of at-a-time and over-a-history notions of symmetry, as I will now describe.

The evolution generator takes the instantaneous state of a system and produces the instantaneous state at a later time. In general, such states will \emph{not} be physically equivalent and should therefore \emph{not} be identified with at-a-time symmetries. However, if one applies such transformations along all points of a DPM taken as a curve in state space, then the resulting curve is invariant. Thus, transformations of this kind \emph{should} be regarded as gauge symmetries over-a-history. Whether one approves of Dirac's proposal in this context therefore depends on whether one takes the physically salient notion of symmetry to be at-a-time or over-a-history. Because the at-a-time notion is relevant to local observers, we will take this to be the salient notion in our definition of gauge symmetry. See \cite{gryb_thebault_book} for a more detailed defence of this position.

There is one additional formal limitation of Dirac's proposal that will be especially relevant to the considerations of my solution to the problem of the AoT. This is the fact that Dirac symmetries are associated with phase space constraints, which are defined in terms of phase space functions. To get the infinitesimal generators of these symmetries one needs to use the differential structure; i.e., the symplectic $2$-form; of phase space in addition to the constraints. But not all symmetries that preserve the dynamical flow can be produced in this way. One notable exception are the dynamical similarities that will be described in \Sec\ref{sec:a_new_kind_of_symmetry_dynamical_similarity}. Dirac's procedure is silent on the gauge status of such symmetries.\footnote{ Note that Dirac describes his criterion for a gauge symmetry as sufficient but not necessary. } The proposal we will introduce in \chap\ref{ch:pesa} will, thus, need to generalise Dirac's idea in a way that can handle this case.

I end this section with a comparison between the Noether and Dirac methodologies regarding symmetry. Recall that Noether starts with symmetries that obey certain formal criteria when applied to the action and then investigates the consequences. Dirac, starts with certain irregularities in the variational principle and then uses consistency conditions to identify and classify all the irregularities present in the variational principle. The basic idea, in contrast to Noether, is then to start with some action and try to figure out the ways in which it is problematic. Once this is done, Dirac develops a procedure for curing the irregularities and producing well-defined equations of motion. This procedure then motivates a proposal for defining the gauge symmetries of a theory because it clearly identifies the arbitrary functions required to evolve the system.

In \chap\ref{ch:rep_sym}, I will follow Noether's methodology but use Dirac's insight to identify a formal criterion for identifying the symmetries I want to investigate. In this way, I will assume that physicists can engineer their variational principles to possess certain gauge symmetries rather than guess blindly at an action and hope that this action can be remedied in such a way that it reflects the desired gauge symmetries.


\section{A different kind of symmetry: dynamical similarity} 
\label{sec:a_new_kind_of_symmetry_dynamical_similarity}

One of the main claims of this work will be that dynamical similarity is a gauge symmetry of modern cosmology. In this section, I will elaborate upon the general idea behind dynamical similarities and show that they are broad symmetries of most systems whose laws can be written using a variational principle based on an action. I will also give some an epistemological argument for treating dynamical similarities as gauge symmetries in certain contexts.

For the purposes of our general discussions about symmetry, the significance of dynamical similarity is that it is problematic for many standard definitions of gauge symmetry discussed in this Chapter. While I consider definitions based on underdetermination to be on the right track, standard definitions assume that the symmetries have formal properties that dynamical similarities do not.

In this section, I will describe the features of dynamical similarity that make it unique. I will then explain how my new principle, the PESA (developed in \chap\ref{ch:pesa}), can treat these unique features and solve many standard puzzles typically associated with gauge symmetry. To get started, let me give an epistemic argument that will help illustrate the features of dynamical similarity and the motivation for treating it as a cosmological gauge symmetry. For a general discussion of dynamical similarity in this context, see \cite{Sloan:2018lim} or further references in \Sec\ref{sub:generating_dynamical_similarity}.

\subsection{The dynamics of similarity} 
\label{sub:dyn of sim}

Consider the passage by Poincar\'e from \emph{Science and Method} \citep{poincare2003science} quoted in \Sec\ref{sec:intro cosmo symmetry} of the Introduction. There,  we're invited to image a universe like our own that is identical in every respect except that it is a thousand times larger. Such a universe, he argues, would be indiscernible from our own because any reference standard that could be used to measure the length of a body would have grown in exact proportion the length of that body, and therefore any measurement using this standard would be unaffected. He calls such a transformation a \emph{similarity} in reference to the same geometric symmetry discussed as early as Euclid.

Poincar\'e's argument for empirical indiscernibility under similarity can be applied more generally to measurements of temporal duration. If duration is measured using relative changes within a system --- say by recording the number of oscillations of the pendulum of a clock --- then, by Poincar\'e's argument, the transformed system should be indiscernible from the original, and therefore no change in duration can be measured.

But compared to length, representations of duration are made considerably more difficult by dynamical considerations. Useful standards of duration are defined by their ability to bring the dynamical laws into a particular form. For example, absolute time in Newtonian mechanics is defined in terms of inertial motion, which in turn is determined by the form of the dynamical laws. The precise way that a similarity transformation acts on the representations of a dynamical system is therefore affected by how the temporal standards depend on the laws. This requires a more comprehensive approach.

The most important consideration in defining such an approach is the way in which the form of the laws affects the transformation properties of the velocities. In theories based on variations of an action, length and time standards are related by the conventions used to define the unit of the action. Heuristically, this unit defines a standard of angular momentum that can be used to convert lengths to velocities using some convention for inertial mass. I will call similarity transformations that take into account the dynamical considerations involved in defining velocity \emph{dynamical similarities} because of their relation to the transformations \eqref{eq:DS kepler} of the same name studied in the context of the Kepler problem. Specifically, they will involve a rescaling of the unit of the action that can act differently on coordinates and their momenta because of how the temporal unit must be rescaled in order to correctly implement the conventions described above.

If Poincar\'e is correct that a rescaled world would be indiscernible from our own, then dynamical similarities, which transform spatial and temporal scales in a way that preserves the dynamical laws of a theory, should be treated as gauge symmetries. I will show that the PESA can be used to make this argument more precise.

\subsection{Dynamical similarity} 
\label{sub:dynamical_similarity}


In the previous section, I described how a similarity transformation intuitively acts on the representations of a dynamical system. I highlighted the subtle role of the dynamical laws in determining the transformation properties of the velocities using the example of the Kepler symmetries of \Sec\ref{ssub:the_kepler_problem}. I will now make these general comments more precise by giving a more formal definition of dynamical similarity. This definition will provide the starting point for the analysis of the remainder of this work.

The definition of dynamical similarity must involve a statement of the laws themselves because these define the convenient temporal standards for the theory. It will thus be helpful in my analysis to restrict to \emph{Lagrangian systems}, which are defined by an action principle --- although applications to Hamiltonian systems are straightforward.\footnote{ See \Sec\ref{sec:dynamical_similarity} for a general treatment in terms of Hamiltonian systems. } A Lagrangian system is a system whose DPMs, $\gamma_\text{DMP}$, are stationary points of the action, $S[\gamma]$, such that the variations $\delta S[\gamma]|_{\gamma_\text{DPM}}$ about $\gamma_\text{DPM}$ are zero. Given such a system, we define a \emph{dynamical similarity} as a transformation, $T: \gamma \to \gamma' = T(\gamma)$, that rescales the action by a constant:
\begin{equation}\label{eq:DS gen}
    S[\gamma] \to c S[\gamma']\,.
\end{equation}
Under this definition, if $\gamma$ is a DPM then $\gamma'$ is also a DPM because \eqref{eq:DS gen} preserves the extrema of $S[\gamma]$. Dynamical similarities, when they exist, are therefore broad symmetries of a theory.

To understand how such transformations match the intuitions of the previous section, consider that the units of $S$ are the same as the units of angular momentum, and therefore that the effect of a dynamical similarity is to actively rescale the global standard of angular momentum for the system. Different choices of $c$ therefore represent different ratios between the standards of length and the standards of velocity. Using these transformations, one can then determine how to rescale the standards of both length and time, as in \eqref{eq:DS kepler}, to produce a similarity transformation. Aside from the symmetries \eqref{eq:DS kepler} of the Kepler problem, another important example of a dynamical similarity is the global rescaling symmetry of general relativity mentioned in \Sec\ref{ssub:variational_symmetries}. The fact that dynamical similarities are symmetries of general relativity will be the basis for our cosmological arguments about the origin of the AoT in \Sec\ref{sub:dynamical_similarity_in_the_universe}.

We will embark on a more thorough investigation of dynamical similarities and their formal properties in \Sec\ref{sec:dynamical_similarity}. For the moment, however, I would like to describe in a bit more detail some features of dynamical similarity, alluded to the introduction (\Sec\ref{sec:intro cosmo symmetry}), that make dynamical similarities different from many other notions of symmetry. I pointed out above that because dynamical similarities rescale the action, they must also rescale the unit of angular momentum. But the unit of angular momentum fixes the differential structure of phase space. In particular, rescaling the unit of angular momentum causes a rescaling of the symplectic $2$-form, introduced in \Sec\ref{ssub:hamiltonian_systems}, that is used for finding a first order representation of the DPMs of a system using Hamilton's equations. Almost all gauge symmetries in physics are assumed to preserve this structure.

What's important for our purposes is that many basic theorems about the properties of gauge symmetries assume that the symplectic structure of phase space is preserved by that symmetry. Since dynamical similarities don't do this, they change many of the rules of the game.

For the purposes of Part~\ref{part:foundations} of the thesis, which is concerned with giving a general definition of gauge symmetry, dynamical similarities simply violate the formal conditions of Noether-$2$ and Dirac symmetries. Dynamical similarities then provide an important test of any good definition of gauge symmetry. I will test my definition of gauge symmetry on the dynamical similarities of the Kepler problem in \Sec\ref{sub:the_kepler_symmetries}.

For the purposes of Part~\ref{part:aot}, which is concerned with giving a new solution to the problem of the AoT, Liouville's theorem states that Hamilton's equations preserve the Liouville volume-form. Because the Liouville measure, which integrates over the Liouville form, is directly used to compute certain notions of entropy in classical mechanics,\footnote{ See \Sec\ref{sec:preliminaries} for more details and references. } Liouville's theorem is of central importance to attempts to understand popular explanations of the AoT.\footnote{ More specifically, those explanations that rely on a \emph{Past Hypothesis}, which will be explained in \Sec\ref{sec:Price_taxonomy}. } But, as we will see in \Sec\ref{sub:measures_on_counting_solutions}, dynamical similarities rescale the Liouville measure. For this reason, treating dynamical similarity as a gauge symmetry causes serious problems for these explanations of the AoT, where entropy considerations are central. This argument will be presented in detail in \chap\ref{ch:against_PH}. On the other hand, the breakdown of Liouville's theorem in the presence of dynamical similarity forms the basis of the new explanation of AoT developed in \Sec\ref{sub:the_janus_attractor_scenario}.
\chapter{Representing symmetry in dynamical systems}
\label{ch:rep_sym}

\begin{abstract}
    In this chapter, I derive many of the formal results used to justify claims made throughout the thesis. I develop a general formalism for understanding gauge symmetries in Lagrangians and Hamiltonian systems. For Lagrangian systems, this formalism follows standard treatments. I use it to derive many known results including Noether's theorems and other constraints that arise in the Lagrangian formalism. For Hamiltonian systems, I present a new view based on velocity, rather than conventional, phase space and show both how it can recover the standard Dirac analysis for constraint systems and shed light on the Frozen Formalism problem of canonical quantum gravity. I apply my construction to several examples that I refer to throughout the thesis. Finally, I develop a new set of tools for treating dynamical similarity. I first give a more complete definition of dynamical similarity and then show that quotienting by its action generically leads to flows on contact space. This result, in particular, forms the basis for my analysis of the Arrow of Time in Part~\ref{part:aot} of the thesis.
\end{abstract}

\ifchapcomp
    \tableofcontents
    \newpage
\else
    \cleardoublepage
\fi

\section{Introduction} 
\label{sec:intro_rep_sym}


In \chap\ref{ch:sym_probs}, I gave a general description of the concept of symmetry and its role in theory. I formulated Belot's Problem and illustrated it through several examples. In \chap\ref{ch:en route gauge}, I found that, while some definitions of gauge symmetry based on variational principles, particularly Dirac's proposal, had more compelling motivations, none provided a completely adequate solution to Belot's Problem. Dirac's proposal was seen to shed light on some puzzles regarding Noether-$1$ symmetries but was ill-equipped to handle reparametrisation symmetry and dynamical similarity. The discussion in the previous chapters was kept as non-mathematical as possible in order to focus on the logical structure of the arguments without getting lost in the technical details of particular proposals. But to give a good solution to Belot's Problem, it is necessary to understand the particular formal features of reparametrisation invariance and dynamical similarity that make them different from other gauge symmetries. These features will help explain how dynamical similarity can shed light on the problem of the AoT. Finally, it is also necessary to better understand the formal aspects of existing proposals that are responsible for their promising features in order to find an adequate solution to Belot's Problem.

In this chapter, I will investigate the detailed mathematical characteristics of different standard gauge symmetry proposals. One goal will be to develop a general unifying framework in which the main results of a variety of different approaches can be derived from a single equation (the resulting equation is Equation~\ref{eq:S var}). The formalism I will develop in this chapter is thus interesting in its own right --- independently of how it advances our understanding of the AoT. But the unifying nature of the formalism will also serve to give a deeper understanding of the role of gauge symmetry in theory.

The geometric properties of \eqn\ref{eq:S var} illuminate the formal mechanisms behind orthodox definitions of gauge symmetry. First-order differential equations will be seen to be general enough to model a vast array of systems using geometric flows on state space. Irregularities in the variational principle will be identified with the breakdown of these flows due to the degeneracy of the geometric structures on velocity phase space. This will be seen to lead to a specific form of underdetermination in the equations of motion associated with the need to introduce arbitrary functions in order to specify the models of the theory.

While this overall picture is not new, the specific implementation in terms of manipulations of a single equation on velocity phase space is simple and unifying. Moreover, the treatment of reparametrisation invariance gives an example, independent of dynamical similarity, of how my proposal can be used to clarify an important conceptual problem in quantum gravity: the so-called \emph{frozen formalism} aspect of the Problem of Time, which I will define at the technical level in \Sec\ref{sub:reparametrisation_invariance} and explain in more detail in \Sec\ref{sub:solved_reparametrisation_invariance}.\footnote{ While much of what is needed for this discussion was developed in \cite{gryb_thebault_book}, here I will study how it is possible to understand the frozen-formalism problem using the PESA. } In all of these examples, reliable degree-of-freedom counts can be given under an at-a-time conception of the state. These counts can be used to match the formal requirements of the theory with minimal epistemic expectations. This will pave the way for my proposed solution to Belot's Problem, which I will give in the next chapter.

As a final application of this formalism, I will develop a Gauge Principle for dynamical similarity. I will then illustrate how the unique formal features of dynamical similarity can generate time-dependent flows on state space. These ingredients will be essential for formulating my solution to the problem of the AoT. The application of this new Gauge Principle will produce results that are equivalent to the direct reductions performed in  \cite{Sloan:2018lim,bravetti2022scaling}. However, by implementing dynamical similarity as a gauge symmetry rather than eliminating it, it will be possible to consider arbitrary gauge-fixings, which, in some cases, can lead to conceptual and mathematical simplifications.

I will aim for the material of this chapter to be at the level of rigour of the theoretical-physics community. This is because my results have interest in-and-of-themselves to this community and because my proposed solutions to Belot's Problem and the problem of the AoT rely on mathematical results that have not been presented together elsewhere. I hope that my framework provides conceptual insight as well as mathematical clarity to the analysis of gauge symmetry.

\section{Lagrangian theories} 
\label{sec:lagrangian_variational_principle}

In this section, I will give a general framework for representing dynamical systems whose laws are specified using a \emph{Lagrangian} function.\footnote{ While I won't follow any particular treatment here, most of this can be found in standard texts on analytic mechanics. See \cite{lanczos2012variational} for a classical introduction, \cite{goldstein2000classical} for a standard treatment, and \cite{arnol2013mathematical} for an advanced text.} A general class of classical field theories over arbitrary spacetime manifolds can be represented in this way. The formalism is powerful enough to implement coordinate invariance in the independent variables of a theory and can be adapted to arbitrary numbers of derivatives of any kind of tensor or spinor field. In this way, one can use the framework to formulate classical models of general relativity and the standard model as well as a variety of field theories and particle models with applications to physics and engineering. In these theories, DPMs are selected on the basis that they extremise a particular action functional, $S$, also called \emph{Hamilton's principal function}. The output is a set of local equations of motion and boundary conditions that can be used to represent the nomic constraints of the theory.\footnote{ In quantum mechanical generalisations of these theories, the laws can be specified in terms of the action using a path integral rather than an extremization procedure. }

The variational principle is constructed by first defining a KPM as a history $\gamma: I \to \mathcal C$, which defines an embedding of a temporal manifold $I\subset \mathbbm R$ into the configuration space $\mathcal C$; i.e., a curve in $\mathcal C$. The temporal manifold consists of temporal points that label instants across space, and contains a natural ordering structure inherited from the real line. The configuration space $\mathcal C$ represents the matter and field content of the theory; i.e., the dependent variables; and is constructed from general field configurations at an instant. For notational convenience, we will write the history $\gamma$ in terms of the components of its image $q^i(t)\in \mathcal C \, \forall t\in I$. The action, which is a functional of the history $q^i(t)$, then takes the general form
\begin{equation}\label{eq:S}
    S[q^i(t)] = \int_{t_1}^{t_2} L\lf(q^i, \dot q^i, \hdots, \tfrac{\de^k }{\de t^k} q^i \rt) \, \de t\,,
\end{equation}
where $L$ is the Lagrangian function that depends on $k$ derivatives of $q^i$. I will allow for the possibility that $i$ range over a continuous spatial index $x$ of $n$-dimensions. In that case, the Lagrangian is itself a functional over instantaneous spatial configurations. I will show how this can be treated in full generality below.

For most applications, it is sufficient to consider theories whose spatiotemporal structure is that of a Lorentzian spacetime, $M$, of the form  $M = (\mathcal M, g)$, where $\mathcal M$ is a smooth manifold and $g$ is a spacetime metric; i.e., a bilinear map $g: T\mathcal M \times T\mathcal M\to \mathbbm R$. It is then most convenient to use the natural volume-form $\text{vol}_g = \sqrt{-|g|} \de^{n+1} x$ induced by $g$ as the integration measure for $S$, where $\{\de x^{\mu}\}$ is some basis on the exterior algebra of $\mathcal M$. In order to write our variations in a convenient form below, it will be helpful to perform integration by parts. This can be done most conveniently when the Lagrangian can be expressed in terms of the metric compatible covariant derivative $\nabla_g$.\footnote{This simplifying assumption can easily be avoided but won't be significant in our discussions below.} Using these ingredients, the action functional can be written as
\begin{equation}\label{eq:cov S}
    S[q^i(x)] = \int_{\Omega} \text{vol}_g\, \mathcal L(q^i, \nabla_\mu q^i, \hdots, \nabla^{(k)}_{\mu \cdots \nu} q^i) \,,
\end{equation}
where $\Omega$ is some connected spacetime region and $\mathcal L$ is the \emph{Lagrangian density}. Note that the $q^i$ stand for a general set of fields and, for instance, could also include the components of the spacetime metric $g$.

I will be interested in formulating the equations of motion generated by the variational principle in regions $\Omega \subseteq \mathcal M$ where $(\Omega,g)$ is globally hyperbolic so that I can understand the conditions under which they can be expressed in terms of a well-posed Cauchy problem. Note that this does not necessary require $M$ itself to be globally hyperbolic nor does it provide a significant restriction on $M$ since arbitrary $M$ can be obtained by suitably gluing together globally hyperbolic patches.\footnote{ This is always possible provided $M$ has an atlas. } If we consider a foliation of $\Omega$ into spacelike (Cauchy) hyper-surfaces $\Sigma_t$ of constant time $t$, then \eqref{eq:cov S} reduces to \eqref{eq:S} when
\begin{equation}
    L(t) = \int_{\Sigma_t} \text{vol}_{\bar g} N \mathcal L\,,
\end{equation}
where $\bar g$ is the restriction of $g$ onto $\Sigma_t$ and $N$ is the \emph{lapse} function such that $\sqrt{-|g|} = N \sqrt{|\bar g|}$ for $t$-coordinates adapted to $\Sigma_t$.\footnote{ One can also view the lapse function of the normal component of the deformation vector normal to $\Sigma_t$. } The discussion below therefore applies generally to any field theory with variational derivatives adapted accordingly and with boundary terms supplemented by the appropriate contribution from the time-like or null boundary of $(\Omega,g)$. A full treatment of these boundary contributions can be done straightforwardly but is beyond the scope of this work. I will, however, give a careful treatment of any contributions to the variation due to the space-like boundary of $(\Omega,g)$.

DPMs are specified by determining the extrema of $S$. This involves computing the variation of $S[q^i(t)]$ in response to arbitrary variations of $\delta q^i(t)$ and setting this variation to zero. Doing this, we find
\begin{align}\label{eq:EL eqs}
    \delta S[q^i(t);\delta q^i(t)]\Big\lvert_{q^i = q^i_\text{cl}}  =& \int_{t_1}^{t_2} \lf[ \diby L {q^i} \delta q^i +  \diby{L}{\dot q^i} \frac {\de }{\de t} (\delta q^i) + \hdots +  \diby{L}{\lf(\tfrac {\de^k}{\de t^k} q^i\rt)} \frac{\de^k \delta q^i}{\de t^k} \rt]\de t \nonumber\\
    =& \int_{t_1}^{t_2} \lf[ \diby L {q^i} - \frac{\de}{\de t} \lf( \diby L{\dot q^i} \rt)+ \hdots + (-1)^k \frac{\de^k}{\de t^k} \lf( \diby L{ \lf(\tfrac{\de^k}{\de t^k}q^i \rt)} \rt) \rt]_{q^i = q^i_\text{cl}} \delta q^i(t)\, \de t \nonumber \\
    & + \lf[  \diby L {\dot q^i} \delta q^i  + \diby{L}{\ddot q^i} \frac {\de }{\de t} (\delta q^i) - \frac{\de}{\de t} \lf( \diby{L}{\ddot q^i}  \rt) \delta q^i + \hdots \rt]_{t_1}^{t^2} = 0\,
\end{align}
after $k$ applications of integration by parts. This equation must be satisfied for \emph{all} smooth variations $\delta q(t) \in C^{k}[t_1,t_2]$ that are arbitrary except (possibly) on the boundary. The DPMs are the histories $q^i_\text{cl}$ satisfying this equation for a particular set of boundary conditions. I will adapt a notion where the first argument of the variation $\delta S$ indicates its functional dependence on histories and the second argument specifies the smearing functions used to perform the variation.\footnote{The index structure of the arguments is mostly for illustrative purposes.} The equations \eqref{eq:EL eqs} are called the \emph{Euler--Lagrange (EL)} equations. The set of three dots in the boundary term indicates an increasingly complicated set of boundary contributions whose explicit expression we will not use. If needed, it can be straightforwardly obtained by integrating the first line by parts.

The EL equations can be usefully split into a local piece, in the integrand, and a boundary piece. For the local piece, we define the EL `vectors'
\begin{equation}
    \alpha_i(t) \equiv \ddiby{S}{q^i(t)} = \diby L {q^i} - \frac{\de}{\de t} \lf( \diby L{\dot q^i} \rt)+ \hdots + (-1)^k \frac{\de^k}{\de t^k} \lf( \diby L{ \lf(\tfrac{\de^k}{\de t^k}q^i \rt)} \rt)\,,
\end{equation}
which can be obtained by smearing the variation $\delta S$ with a Dirac $\delta$-function. Strictly speaking, such a definition would only hold for $t$ inside the interval $(t_1,t_2)$ of variation because the $\delta$-function is not well-defined on the boundary. The local equations then reduce to
\begin{equation}\label{eq:loc EL}
    \alpha_i(t) = 0\,.
\end{equation}
These local equations must be supplemented by a boundary condition of the form
\begin{equation}\label{eq:bound EL}
    \lf[  \diby L {\dot q^i} \delta q^i  + \diby{L}{\ddot q^i} \frac {\de }{\de t} (\delta q^i) - \frac{\de}{\de t} \lf( \diby{L}{\ddot q^i}  \rt) \delta q^i + \hdots \rt]_{t_1}^{t^2} = 0\,,
\end{equation}
which can either be seen as a restriction on the boundary variation (Dirichlet boundary conditions) or on the quantities $\diby L{ \lf(\tfrac{\de^k}{\de t^k}q^i \rt)}$ (Neumann boundary conditions).

The boundary conditions above must be satisfied on the time-like, null or space-like surfaces that form the boundary of $\Omega$. As stated above, I will only explicitly treat the space-like terms.\footnote{ The time-like part of the boundary, if it exists, will imply additional conditions usually referred to as `boundary' conditions. Null boundaries can also be used to specify initial or boundary conditions. We leave these cases as an exercise to the reader. } These terms should be thought of as specifying the conditions that must be satisfied by the independent initial data that must be used for solving the $(k+1)^\text{th}$ order differential equations \eqref{eq:loc EL}. Only initial data satisfying the boundary equation \eqref{eq:bound EL} can be well-posed, although this condition is not sufficient for the existence or uniqueness of solutions.

On a temporal slice $\Sigma_t$, all functions and their derivatives should be considered to be \emph{independent} functions obeying certain dynamical consistency conditions. We see immediately that, in order to understand what the independently specifiable initial data of the theory is, it is helpful to reformulate this $k^\text{th}$-order system as an equivalent first-order system where functions and their derivatives are treated independently, and where the dynamical consistency conditions can be explicitly imposed. This will allow us to more easily asses the integrability of the EL equations because first order systems can be interpreted as geometric flows. We will do this explicitly in \Sec\eqref{sec:the_initial_value_problem}. There we will obtain an explicit procedure for identifying the independently specifiable data, and therefore obtain an at-a-time degree of freedom count using the straightforward integrability conditions of first-order systems. This count will form the basis of my proposal for identifying observables and gauge symmetries.

\section{Symmetries and Noether's approach} 
\label{sec:symmetries_of_the_variational_principle}

After outlining concrete representations for general Lagrangian systems, we are in a position to define and then assess different standard narrow proposals for gauge symmetry. I will now present a formalism that begins with the variations defined in the previous section and leads naturally to Noether's second theorem and the (functional derivative) definition of the Noether-$2$ symmetries introduced in \Sec\ref{ssub:noether_symmetries}. This formalism follows, with minor modifications, the standard treatment of symmetry presented in Chapter~3 of \cite{henneaux1992quantization}, and constitutes the paradigmatic definition of gauge symmetry in theoretical physics orthodoxy.\footnote{ See also (in order of mathematical sophistication) \cite{sudarshan1974classical}, \cite{sundermeyer:1982}, \cite{olver2000applications} for similar presentations. } This definition will also cover the Noether-$1$ symmetries, which feature in Noether's first theorem, as a special case that I will treat in \Sec\ref{sub:leibniz_and_noether1}. The differences between these symmetries and the systems to which they are applicable will be used to highlight the limitations of this approach for solving Belot's Problem and to motivate my proposed solution.

The attitude taken towards symmetry in these derivations follows Noether's methodology: one assumes that certain symmetries of the action are known --- either because they are known in advance or because they have been pre-engineered into the form of the action --- and the consequences of these symmetries are then investigated. Because the action is a functional of an entire history, the relevant notion of symmetry and corresponding degree-of-freedom counts will be over-histories.

\subsection{Noether-2 symmetries and Noether's second theorem} 
\label{sub:noether_2_symmetries_and_noether_s_second_theorem}


To begin, I identify some trivial transformations that do \emph{not} lead to useful notions of symmetry. For variations of the form
\begin{equation}
    \delta q^i = M^{ij} \alpha_j\,,
\end{equation}
where $M^{ij} = - M^{ij}$ are antisymmetric functions of the $q^i$, the local term of the EL equations is automatically satisfied:
\begin{equation}
    \delta S_\text{loc} = \int_{t_1}^{t_2}\de t\, \alpha_i M^{ij} \alpha_j = 0 \,.
\end{equation}
The vanishing of the boundary term requires $M_{ij}=0$ on the boundary for unrestricted Lagrangians and when $\alpha_i \neq 0$. Note that, while these transformations are formally symmetries of the action off-shell (i.e., for arbitrary KPMs), they are \emph{trivially} zero on-shell (i.e., when $\alpha_i = 0$). Such transformations are thus trivial broad symmetries that exactly vanish when the classical equations of motion are satisfied. I include them here for completeness, and because they are useful for considering symmetries in quantum mechanics (see \cite[\S 3.1.5]{henneaux1992quantization} for a more complete discussion of their characterization). But for the present purposes, I will be mainly interested in how symmetries act on DPMs so that such symmetries can be ignored.

Define an infinitesimal broad over-a-history symmetry as a transformation of the form
\begin{equation}\label{eq:broad trans}
    q^i(t) \to q^i(t) + \delta_{\epsilon} q^i(t);
\end{equation}
where $\epsilon$ stands for a freely specifiable ``gauge''\footnote{ Note that it is standard physics nomenclature to call the group parameter of a transformation that will ultimately be considered a gauge transformation as a `gauge' parameter. We will use this standard terminology while remaining open to the interpretation of the transformation in consideration. } parameter, $\epsilon_\alpha(t)$, defined over an entire history; such that if $q^i$ is a solution to the EL equation then so is $q^i+ \delta_{\epsilon} q^i$. For our analysis, we have assumed (see the form of \eqref{eq:EL eqs}) that the infinitesimal generator $\delta_{\epsilon} q^i(t)$ depends at most on $k$-derivatives of the independent variable $t$.

Noether-$2$ symmetries can be defined as the subset of the infinitesimal broad symmetries that exactly preserve the value of the action functional along \emph{any KPM} and that do not exactly vanish on-shell. Such symmetries leave the action invariant even \emph{off-shell} (modulo the trivial transformations discussed above). Formally, we can write this condition as
\begin{equation}\label{eq:N2}
    \delta S[q^i, \delta_\epsilon q^i] = 0\,,
\end{equation}
with $\delta_\epsilon q^i$ unconstrained on the boundary and $q^i$ not necessarily satisfying $\alpha_i = 0$. So far, this relation is simply a definition. Nothing guarantees that such symmetries exists or whether the action possessing such symmetries has any non-trivial solutions. However, since we are following Noether's methodology, we assume that one knows in advance at least some of the symmetries that obey this criterion.

When the Noether-$2$ symmetries are not known in advance, it is possible to formulate \eqref{eq:N2} in terms of a formal condition on $\delta_\epsilon q^i$. Taking the functional derivative of \eqref{eq:N2} and using the EL equations $\alpha_i = 0$ to eliminate the term depending on the functional derivative of $\delta_\epsilon q^i$, we find the on-shell relation
\begin{equation}
    \delta^2 S[q^i_\text{cl}; \delta q^i, \delta_\epsilon q^j] = 0\,.
\end{equation}
We have used a notation where the expression is to be evaluated at the value of the first argument; i.e., $F[q_0;\hdots] = F[q;\hdots]\big \rvert_{q = q_0}$ for some functional $F$ of $q$. The on-shell vanishing of the second variation above is equivalent to the local expressions:
\begin{align}\label{eq:loc narrow sym def}
    \int_{t_1}^{t^2}\de t'\, \frac{ \delta^2 S}{\delta q^i(t) \delta q^j(t')} \delta q^i(t) \delta_{\epsilon} q^j(t') &= 0  & \alpha_i &= 0\,,
\end{align}
where the variations $\delta q^i(t)$ are arbitrary except at the boundary, where they vanish. Given this arbitrariness, the Noether-$2$ symmetries are found to be the null eigenvectors --- in the functional sense above --- of the second functional derivative of $S$ when the EL equations are satisfied. This matches the conceptual picture used to describe Noether-$2$ symmetries in \Sec\ref{ssub:noether_symmetries}. Note that the fact that $\delta_\epsilon q^i(t)$ is defined over an entire history highlights the over-a-history nature of this notion of symmetry. Each symmetry of this kind thus reduces the solution space of $\delta S = 0$ by one \emph{functional} degree-of-freedom. In \Sec\ref{sec:gauge_fixing_and_degree_of_freedom_counting}, I will compare this functional degree of freedom with the independently specifiable initial data obtained from the at-a-time degree-of-freedom count.

A generating set for the Noether-$2$ symmetries can be systematically found by using a basis for the on-shell kernel of $\delta^2 S$. The definition \eqref{eq:loc narrow sym def}, though mathematically elegant, is usually not of much practical use because it is a functional eigenvalue equation subject to a constraint. Solving this constrained second-order functional differential equation is almost completely intractable in general unless particular solutions are known in advance. That is why knowledge of particular solutions, as assumed in Noether's methodology, is so powerful. Note that the difficulty of solving \eqref{eq:loc narrow sym def} means that finding \emph{all} the Noether-$2$ symmetries of a theory is a very difficult problem in general.\footnote{ Another possibility is to try to `discover' these symmetries by identifying the degeneracies in the resulting equations of motion. This is the idea behind the \emph{Dirac algorithm} discussed below. }

The Noether-$2$ symmetries, when they are known to exist, indicate a strong form of degeneracy in the nomic structure of the theory because there is a family of DPMs, spanned by the function $\epsilon_\alpha(t)$ at all values of $t$, that cannot be distinguished by the nomic structure. This indistinguishability is completely determined by the form of $S$ alone. As we will see shortly, in many, but importantly not all, cases this degeneracy will cause the variational principle to produce underdetermined equations of motion.

Before seeing how this arises, I will first demonstrate how the existence of a Noether-$2$ symmetry immediately implies certain identities satisfied by the KPMs of the theory in question. This will lead us to Noether's second theorem. In fact, using the functional notation of this section, expression \eqref{eq:N2} already contains all the information required for Noether's second theorem. It is, however, enlightening to separate \eqref{eq:N2} into its local pieces, which are relevant to the second theorem, and the global pieces, which are relevant to the at-a-time constraints that will shed light on the problem of underdetermination.

To do this, let us write the variations $\delta_\epsilon q^i$ in terms of a $k^\text{th}$ order differential operator $T^i_\alpha(\epsilon^\alpha)$ of the form
\begin{equation}\label{eq:T def}
    \delta_\epsilon q^i = T^i_\alpha(\epsilon^\alpha) \equiv \sum_{a=0}^{k} T^i_{(a)\alpha} \frac {\de^a}{\de t^a} \epsilon^\alpha\,.
\end{equation}
We then define the adjoint $\tilde T^i_\alpha(f_i)$ such that
\begin{equation}\label{eq:T ad def}
    \int_{t_1}^{t^2} \de t\, f_i T^i_\alpha (\epsilon^\alpha) = \int_{t_1}^{t^2} \de t\, \epsilon^\alpha \tilde T_\alpha^i (f_i) + (\text{boundary terms } 1) \,,
\end{equation}
for all continuously differentiable functions $f_i \in C^{k}(t_1, t_2)$ and $\epsilon^\alpha \in C^{n+k}(t_1, t_2)$. Normally when defining the adjoint, the vanishing of the boundary terms appearing in its definition is used to define its domain. But because we are using the adjoint to write Noether's second theorem, which is an identity we would like to hold off-shell, we have no particular reason to restrict the domain of $\tilde T^i_\alpha$ in any way. In this sense, we will use a non-standard definition of the adjoint and will need to explicitly compute the relevant boundary terms for any given application.

Using this definition of the adjoint, the expression \eqref{eq:N2} now reads
\begin{align}
    \delta S[q^i; \delta_\epsilon q^i] &= \int_{t_1}^{t^2} \de t\, \alpha_i T^i_\alpha (\epsilon^\alpha) + (\text{boundary terms } 2) \notag \\ &= \int_{t_1}^{t^2} \de t\, \epsilon^\alpha \tilde T_\alpha^i (\alpha_i) +  (\text{boundary terms } 1+2) = 0\,.\label{eq:N2 var}
\end{align}
Because the gauge parameters $\epsilon^\alpha$ can be chosen arbitrarily, this yields the local equation
\begin{equation}\label{eq:N2 loc}
    \tilde T^i_\alpha(\alpha_i) = 0\,,
\end{equation}
which is Noether's second theorem in its standard form.\footnote{See for example \cite[ch 5]{neuenschwander2017emmy} and \cite[\S 2.4]{logan1977invariant}. } Clearly, this is a necessary condition for the existence of a Noether-$2$ symmetry.

Note that Noether's second theorem is usually presented in a slightly broader context than what we have done here. In addition to transformations of the form \eqref{eq:broad trans}, one usually also considers transformations of the independent variables. However, for theories like the ones we are considering where the Lagrangian depends only on the independent variables through the implicit dependence of the dependent variables, these additional transformations are redundant. Any effect of transforming the independent variables can be achieved by pulling back the dependent variables by that transformation. In any case, it is a straightforward exercise to generalise our presentation when necessary.

\subsection{Lagrangian constraints and Noether's first theorem} 
\label{sub:lagrangian_constraints_and_noether_s_first_theorem}

The local equation \eqref{eq:N2 loc} is necessary but not sufficient for the vanishing of \eqref{eq:N2} and, therefore, the existence of a Noether-$2$ symmetry. To obtain a necessary and sufficient condition, which would give all the logical consequences of the existence of a Noether-$2$ symmetry, we must explicitly compute the boundary terms of \eqref{eq:N2 var} and require them to vanish. Because the conditions obtained in this way arise from the boundary terms of the variational principle, we can think of them as constraints on the class of histories for which the action has well-defined extrema. Restoring consistency of the variational principle will be the primary motivation behind the Dirac algorithm described below in \Sec\ref{sec:the_dirac_algorithm}. While space-like boundary terms are usually not the subject of discussions about Noether's theorem, I discuss them here because they provide an efficient tool for studying the underdetermination problem in the equations of motion and the connection with the Dirac formalism.

It is important to observe at this point that because the conditions I will derive here result from boundary terms, they must hold at-a-time on the space-like surfaces that bound the variation. On the other hand, because the choice of boundary surfaces is arbitrary, the conditions we will find must also hold at \emph{any} time. In this way, they will generate at-a-time constraints on the state space of the theory. A more direct, though equivalent, treatment will be given in \Sec\ref{sub:narrow_symmetries} using a first-order approach, which is adapted to an at-a-time analysis. It is nevertheless useful here to understand how the same constraints arise as boundary conditions in an over-a-history approach. The constraints I will derive here can be obtained through other standard means in texts such as \cite{sundermeyer:1982}. Many aspects of the derivation below can be found in Chapter~8 of \cite{gryb_thebault_book}.

General expressions for the boundary terms we will need can be given using the expansion of the operator $T^i_\alpha$ as defined by \eqref{eq:T def}. Explicit expressions are more illuminating if we stick to the lowest order non-trivial case of $k = 1$.\footnote{ I.e., when the action is first order in the time and, therefore, when the EL~equations are second order in time. } Then, integration by parts of the left-hand side of \eqref{eq:N2 var} gives
\begin{equation}\label{eq:T ad explicit}
    \tilde T^i_\alpha = \lf( T^i_{(0) \alpha} - \dot T^i_{(1) \alpha} \rt) - T^i_{(1) \alpha} \frac{\de}{\de t} 
\end{equation}
and the boundary term
\begin{equation}\label{eq:L bdy term}
    \lf[ \lf( T^i_{(0)\alpha} \diby L {\dot q^i} + T^i_{(1)\alpha} \alpha_i \rt) \epsilon^\alpha + T^i_{(1)\alpha} \diby L {\dot q^i} \dot \epsilon^\alpha  \rt]_{t_1}^{t_2} = 0\,.
\end{equation}
The adjoint defined in \eqref{eq:T ad explicit} leads to the explicit form of Noether's second theorem, $\tilde T^i_\alpha \alpha_i = 0$ when $k = 1$. Because the boundary terms must hold at-an-instant, the functions $\epsilon^\alpha$ and their time derivatives $\dot\epsilon^\alpha$ should be treated as arbitrary independent functions. We thus get two constraints. The first:
\begin{equation}\label{eq:Lagrangian c}
    T^i_{(1)\alpha} \diby L {\dot q^i} = 0\,,
\end{equation}
arises from the vanishing of the $\dot \epsilon^\alpha$ term and is called the \emph{primary constraint} of the theory. It holds off-shell. A second constraint, which holds only on-shell (i.e., when $\alpha_i = 0$), results from the $\epsilon^\alpha$ term and is given by
\begin{equation}\label{eq:further c}
    T^i_{(0)\alpha} \diby L {\dot q^i} \approx 0\,,
\end{equation}
where we have used the `$\approx$' sign to indicate an equation that is only required to hold on-shell. The condition above doesn't historically have a specific name since versions of it were originally derived using the Dirac methodology where it can play different roles. Note that, in deriving \eqref{eq:further c}, I have assumed that $\epsilon_\alpha$ has non-trivial time dependence --- an assumption I will revisit below when deriving Noether's first theorem.

Let us investigate the consequences of the degeneracies of the variational principle that result from Noether-$2$ symmetries for the solvability of the equations of motion. This will be a central part of our analysis of gauge symmetry later. To do this, note that $\deby{}{t} = \dot q^i \diby{}{q^i} + \ddot q^i \diby{}{\dot q^i}$ when $k = 1$. Thus,
\begin{equation}
    \alpha_i = \diby{L}{q^i} - \deby{}t \lf( \diby L {\dot q^i} \rt) = \diby{L}{q^i} - \dot q^j \diby{^2L}{q^j \partial \dot q^i} - \ddot q^j \diby{^2 L}{\dot q^i \partial \dot q^j}\,.
\end{equation}
The EL~equations $\alpha_i = 0$ can then be seen as differential equations relating the accelerations $\ddot q^i$ to the velocities $\dot q^i$ and configurations $q^i$ at any given time $t$. To solve them, the quantity
\begin{equation}\label{eq:hessian def}
    W_{ij} = \diby{^2 L}{\dot q^i \partial \dot q^j}
\end{equation}
must be invertible as a matrix.\footnote{ We will see later, when we derive the Legendre transform, that $W_{ij}$ is the Hessian of the Legendre transform. } The primary constraints \eqref{eq:Lagrangian c}, however, immediately imply that $W_{ij}$ is \emph{not} invertible because they imply that the $\T 1 i$ lie in the kernel of $W_{ij}$:
\begin{equation}\label{eq:null primary}
    \T 1 i W_{ij} = 0\,.
\end{equation}
This means that the primary constraints prevent the variational principle from generating well-posed equations of motion in the precise sense that the accelerations cannot be solved uniquely in terms of the configurations and their velocities at a given time. It is this form of underdetermination that, I will argue, should be part of a good definition of gauge symmetry --- although I will argue that this can arise in a more diverse set of circumstances than what we have seen here.

The condition \eqref{eq:null primary} can be used in combination with the identity $\T 1 i \alpha_i = 0$, which is trivially satisfied on-shell, to produce the on-shell constraint
\begin{equation}\label{eq:L constraints}
    \T 1 i K_i \approx 0\,,
\end{equation}
where we have defined
\begin{equation}
    K_i = \diby{L}{q^i} - \dot q^j \diby{^2L}{q^j \partial \dot q^i}\,.
\end{equation}
The constraints \eqref{eq:L constraints} are usually called the \emph{Lagrangian constraints} of the theory, which follow from the primary constraint and the equations of motion.

Normally the Lagrangian constraints are derived directly from the Lagrangian using Dirac's methodology as in \cite{sundermeyer:1982}. In this approach, one starts by reading off the primary constraints directly from the form of the Lagrangian and then derives additional constraints by consistency with the equations of motion. If one were to do that here, one would want to impose that the time derivative of \eqref{eq:L constraints} vanish when $\alpha_i =0$. Using the fact that the $\T 1 i$ are in the kernel of $W_{ij}$ off-shell, Noether's second theorem, which is also valid off-shell, then tells us that
\begin{equation}
    \deby{}{t}\lf( \T 1 i K_i \rt) = \deby{}{t}\lf( \T 1 i \alpha_i \rt) = \T 0 i \alpha_i\,.
\end{equation}
The condition $\T 0 i \alpha_i \approx 0$ then arises as a closure condition for the Dirac algorithm.

From the point of view taken here, the Noether methodology pre-assumes the existence of a Noether-$2$ symmetry. In this case, the on-shell condition \eqref{eq:further c} can be used, in combination with $\alpha_i = 0$ to derive the two on-shell conditions
\begin{align}\label{eq:extra L const}
    \T 0 i W_{ij} &\approx 0 & \T 0 i K_i &\approx 0\,,
\end{align}
in the same way that the analogous conditions were derived for $\T 1 i$.\footnote{ For simplicity, we have assumed here that $\diby{\T 0 i}{\dot q^i} = 0$ so that these expressions are valid only for classical symmetries that are not generalised symmetries. The form of the primary and Lagrangian constraints, however, does \emph{not} require such an assumption. } Together, these conditions imply $\T 0 i \alpha_i \approx 0$, which closes the Dirac algorithm. We thus see that the assumptions of the Noether methodology are mutually consistent: when a Noether-$2$ symmetry exists the Dirac algorithm is guaranteed to close. We also see that this closure condition can be expressed as resulting from an off-shell identity; namely Noether's second theorem. The various distinctions between `primary' and `Lagrangian' constraints arise simply from the formal distinctions between $\T 0 i$ and $\T 1 i$. In the Noether methodology, these arise as one tight, self-consistent package.

Let us note here that the amount of underdetermination in the equations of motion depends explicitly on the form of the infinitesimal generator of the Noether-$2$ symmetry in question. This can be read-off directly from the dimension of the kernel of $W_{ij}$. When $\T 1 i$ is non-zero there are off-shell null vectors of $W_{ij}$ for each value of $\alpha$.

The same is \emph{nearly} true when $\T 0 i$ is non-zero, although only on-shell. The arguments leading to \eqref{eq:further c} assume that the gauge parameter $\epsilon_\alpha$ is a non-trivial function of $t$. If this is \emph{not} the case (i.e., if the gauge parameters are constant functions of the independent variables), then $\T 1 i = 0$ (because $\dot \epsilon = 0$ definition) and \eqref{eq:L bdy term} has the solution
\begin{equation}\label{eq:N1}
    \deby{} t \lf( \T 0 i \diby{L}{\dot q^i} \rt) = 0\,.
\end{equation}
This implies, therefore, that $\T 0 i \diby{L}{\dot q^i}$ need only be a constant of motion and is not required to be equal to zero. This is Noether's first theorem restricted to the case where the action is not allowed to have explicit dependence on the independent variables. We see now why the presence of Noether-$1$ symmetries, whose infinitesimal generators are constant functions of the independent variables, leads to constants of motion. Importantly, the Noether-$1$ symmetries are still required to keep the action invariant but \emph{do not} lead to underdetermination in the equations of motion. They thus provide an important example of why invariance of the action is, in general, not a reliable way to implement a notion of gauge symmetry based on matching underdetermination in the equations of motion with underdetermination of representations by phenomena.

\subsection{Examples} 
\label{sub:examples2}

In this section, I will illustrate the general framework developed in the previous section using concrete examples. This serves at least two purposes. First, it illustrates how the general formalism develop above can be used in concrete terms. And second, it shows how the formal narrow symmetry proposals based on variational principles can be applied to the standard (non-discrete) symmetry notions introduced in \Sec\ref{sec:examples_of_symmetry}. Coordinate invariance can be treated using similar techniques but is beyond the scope of this work. The interested reader is encouraged to consult \cite[\S 14.3]{gryb_thebault_book} or \cite[Ch 1]{thiemann2008modern}.

\subsubsection{Leibniz shifts and Noether's first theorem} 
\label{sub:leibniz_and_noether1}

The Leibniz shifts, introduced in \Sec\ref{sub:leibniz_shifts}, correspond to the global (i.e., time independent) symmetries of Newtonian mechanics. These are the Euclidean symmetries $\text{ISO}(3) = \text{SO}(3) \ltimes \mathbbm R^3$ that have infinitesimal representations on $\mathbbm R^{3N}$ as
\begin{equation}\label{eq:NM glob sym}
    q^i_I \to q^i_I + \epsilon\ud i {jk} q^j_I \theta^{k} + a^i\,,
\end{equation}
in terms of the Euclidean coordinates $q^i_I \in \mathbbm R^{3N}$ and the time-independent parameters $a^i$ for translations and $\theta^i$ for rotations. The index $i$ ranges over the three spatial coordinates and $I$ ranges from $1$ to $N$, which is the total number of particles. 

We can match these generators to the general expansion \eqref{eq:T def} by splitting the $\alpha$ index of $T^i_\alpha$ into a translational piece, which we will call $T^i_{Ij}$, and a rotational piece, which we will call $R^{iI}_{j}$. Note that, because the gauge parameters are time independent, the generators only get non-zero contributions from the $0^\text{th}$ order terms labelled $(0)$. These generators can be read off from \eqref{eq:NM glob sym} and give
\begin{align}
    T^i_{(0)Ij} &= \delta^i_j &  R^{iI}_{(0)j}&= \epsilon\indices{^{i}_{kj}} q_I^k \,.
\end{align}
Noether's first theorem can then be read-off from \eqref{eq:N1} giving
\begin{align}
    \frac{\de}{\de t} \sum_I \diby L {\dot q^i_I} &= 0 & \frac{\de}{\de t} \sum_I \epsilon\ud j {ki} \diby L {\dot q^i_I} q^k_I = 0\,,
\end{align}
which, after identifying $\diby L {q^i_I}$ with the generalised momenta of the system, express the conservation of global linear and angular momentum respectively.

We see that, in Newtonian mechanics, the presence of Leibniz symmetries implies the existence of constants of motion and no underdetermination in the equations of motion. It is clear that, because such constants have the potential to represent real measurable quantities in the world, the Newtonian setting is not a natural one for treating Leibniz shifts as gauge symmetries. Depending on the kinds of target systems to be modelled, this could lead to the kind of mismatch between the formal definitions of gauge symmetry and the epistemic expectations indicative of Belot's Problem as illustrated in \Sec\ref{ssub:the_newtonian_free_particle}. We see here that not classifying Leibniz shifts as gauge symmetries in Newtonian mechanics is consistent with their presence not leading to underdetermination in the equations of motion. The next example will show how to modify Newtonian mechanics so that the Leibniz shifts may be matched to underdetermination in the equations of motion if the empirical context warrants it.

\subsubsection{BB theory} 
\label{sub:bb_theory}

We will call \emph{Barbour--Bertotti (BB)} theory the theory that modifies Newtonian $N$-particle mechanics in such a way that the best-matching shifts, defined in \Sec\ref{sub:best_matching_shifts}, are underdetermined by the equations of motion of the system. This provides a plausible way to treat Leibniz shifts, and indeed their generalisation as best-matching shifts, as gauge symmetries. For simplicity, we will restrict to the case of translations, which illustrate the general principles. The more general theory was introduced in \cite{Barbour_Bertotti}.\footnote{ See \Sec\ref{sub:best_matching_shifts} for a more detailed list of references. }

The underdetermination in BB theory is designed to match the epistemic considerations resulting from relational arguments of \Sec\ref{sub:best_matching_shifts}. This match is achieved by introducing into the usual Newtonian variational principle a \emph{translation shift} field, $w^i(t)$, whose role is to arbitrarily shift the Newtonian particle configurations $q^i_I(t)$ in a time-dependent way. To do this, define the shift-invariant time derivative
\begin{equation}
    D_t q^i_I \equiv \dot q^i_I - w^i
\end{equation}
that is invariant under the best-matching translational shifts
\begin{align}\label{eq:BB gauge}
    q^i_I &\to q^i_I + a(t) & w^i \to w^i + \dot a(t)\,.
\end{align}
We then substitute this derivative for the usual time derivative appearing in the Lagrangian of Newtonian mechanics to obtain the Barbour--Bertotti Lagrangian
\begin{equation}
    L_\text{BB} = \lf(\sum_I \frac {m_I} 2 \delta_{ij} D_t q^i_I \, D_t q^j_I\rt) - V(q^i_I)\,,
\end{equation}
where $m_I$ are the particle masses and $V(q^i_I)$ is a translation-invariant potential function. The Barbour--Bertotti action is then
\begin{equation}
    S_\text{BB}[\gamma] = \int_\gamma L_\text{BB} \de t\,,
\end{equation}
where $\gamma$ is a configuration space curve parametrized by $t$.

The BB-action is, by design, invariant under the time-depend shifts of \eqref{eq:BB gauge}, which are therefore variational symmetries of the theory. Because the gauge parameter $a(t)$ is allowed to be an arbitrary time-dependent function, the arguments leading to the constraints of \Sec\ref{sub:lagrangian_constraints_and_noether_s_first_theorem} apply. To write these explicitly we adapt our notation so that the $\alpha$ index of $T^i_\alpha$ splits into a piece acting on the variation of the $w^i$ variables, which we will call $U^i_j$, and one acting on the variation of the $q^i_I$ variables, which we will call $R^i_{Ij}$. Then, the non-zero components of the operators $U$ and $R$ can be read off from \eqref{eq:BB gauge} giving
\begin{align}\label{eq:BB Ts}
    U^i_{(1) j} &= \delta^i_j & R^i_{(0)Ij} &= \delta^i_j\,.
\end{align}
The adjoint operators $\tilde U^i_j$ and $\tilde R^i_{Ij}$ can then be computed from this using \eqref{eq:T ad explicit}. The result is
\begin{align}
    \tilde U^i_j &= -\delta^i_j \frac {\de} {\de t} & \tilde R^i_{Ij} &= \delta^i_j\,.
\end{align}

Calling $p^I_i = \diby{L_\text{BB}}{q^i_I}$, the $w^i$-components of $\alpha$ are
\begin{equation}
    \alpha^{(w)}_i = -\sum_I p^I_i\,
\end{equation}
while the $q^i_I$ components are
\begin{equation}
    \alpha^{(q)I}_i = - \diby{V}{q^i_I} - \dot p^I_i\,.
\end{equation}
The Noether identities \eqref{eq:N2 loc} are then
\begin{equation}
    -\sum_I \diby{V}{q^i_I} = 0\,,
\end{equation}
which simply express the fact that the potential $V$ has been required to be invariant under translations of $q^i_I$.

The primary constraint \eqref{eq:Lagrangian c} only receives a contribution from the $U$-component of $T$. This gives
\begin{equation}
    \diby L {\dot w^i} = 0\,,
\end{equation}
which says that the $w^i$ are Lagrange multipliers of the theory. The resulting condition on the $w$-components of the Hessian
\begin{equation}
    \diby L {\dot w^i} W^{(w)}_{ij}= 0
\end{equation}
implies that the accelerations $\ddot w^i$ are underdetermined by $w^i$ and $\dot w^i$ in the equations of motion. This was the desired effect because the shift fields $v^i$ were introduced as arbitrary functions into the variational principle. This confirms their role as Lagrange multipliers.

The Lagrangian constraints are given by computing the $w$-component of $K$:
\begin{equation}
     K^{(w)}_i = \sum_I \frac {m_I}2 \delta_{ij} D_t\, q^j_I\,,
\end{equation} 
which leads to the well-known \emph{Barbour--Bertotti} constraint
\begin{equation}\label{eq:BB const}
    U^j_{(1)i} K^{(w)}_j = \sum_I p^I_i \approx 0\,,
\end{equation}
where we have used the explicit expression for $p^I_i$ using the BB Lagrangian. This constraint expresses the on-shell vanishing of the total linear momentum of the system. A similar constraint would have been obtained for the angular momentum had we introduced a shift field for rotations instead. What we find is that the constants of motion, computed in the previous example for the Leibniz shifts, that result from Noether's first theorem are no longer freely specifiable: they are constrained to be zero. This means that the modifications of the variational principle introduced in BB-theory imply that the constants of motion are fixed by the dynamical principles of the theory and therefore cannot be used to model features of a target system where their measured value would need to be determined empirically. This matches the relational arguments that say that the total linear momentum of a system is not measurable because it depends on the linear velocity of the system, which is not observationally accessible through relational quantities.

These considerations are directly connected to further underdetermination in the equations of motion. Using \eqref{eq:BB Ts}, a short calculation shows that the on-shell condition \eqref{eq:further c} is equivalent to the Barbour--Bertotti constraint. The first condition of \eqref{eq:extra L const} then implies that the vector $R^i_{(0)Ij}$ is an additional null direction of the Hessian on-shell. This represents underdetermination in the part of $q^i_I$ that is transformed under \eqref{eq:BB gauge}; i.e., the location of the origin of the Cartesian coordinate system used to write the $q^i_I$. The Noether-$2$ symmetries of BB theory thus imply underdetermination in both this origin (on-shell) and the shift fields $w^i$. Those are precisely the variables needed to match underdetermination in the equations of motion with the underdetermination of representations by phenomena in a relational setting. We will return to this point in \Sec\ref{ssub:galilean_transformations_as_gauge_symmetries} after introducing the PESA.

The final on-shell condition, the second condition of \eqref{eq:extra L const}, is automatically satisfied because of the vanishing of the $q$-components of $K$ due to the translational invariance of the Lagrangian. Alternatively, this condition could be read as requiring translational invariance in the Lagrangian in order for \eqref{eq:BB gauge} to be a Noether-$2$ symmetry.

I end this section by noting that the procedure used in BB theory to promote the Leibniz-shift symmetries of Newtonian mechanics, which are of the Noether-$1$ type, to the best-matching-shift symmetries of BB theory, which generate underdetermination in the equations of motion, is an implementation of the Gauge Principle introduced in \Sec\ref{ssub:the_gauge_principle}.

\subsubsection{Electromagnetism} 
\label{sub:electromagnetism}

In this section, I treat Maxwell's theory of electromagnetism to illustrate how the various theorems and constraints I derived can be generalised to the field theory case. The generalisation to non-Abelian fields is straightforward so that this example also illustrates how the concepts presented in this section apply to well-studied notions of gauge symmetry in general.

Using the notation of \Sec\ref{sub:maxwell_and_yang_mills_gauge_symmetries} and the fact that the components of the electromagnetic field tenor can be written as $F_{\mu\nu} = \partial_\mu A_\nu - \partial_\nu A_\mu$, we note that the action for matter-free electromagnetism is the functional
\begin{equation}
    S[A^\mu] = \int_\Omega \de^4 x\, F^{\mu\nu} F_{\mu\nu}
\end{equation}
of the vector potential $A^\mu(x)$, which depends on the spacetime point $x$, with $\mu$ and $\nu$ spacetime indices. For simplicity, I have used a flat Minkowski metric in Minkowski coordinates so that the volume form is trivial and indices can be lowered and raised straightforwardly. The theory is manifestly invariant under the Noether-$2$ symmetries
\begin{equation}\label{eq:EM gauge}
    A^\mu \to A^\mu + \partial^\mu \epsilon = \begin{pmatrix} A^0  - \dot \epsilon \\ A^a + \partial^a \epsilon \end{pmatrix} \,,
\end{equation}
where $a$ is a spatial index.

Using the general notation introduced in \Sec\ref{sub:lagrangian_constraints_and_noether_s_first_theorem}, the $i$ ranges over all the fields and therefore becomes a spacetime index $\mu$ and a continuous spatial coordinate $x$. Moreover, there is only one gauge parameter indexed over space. We then have $T^i_\alpha \to T^\mu(x)$. The non-zero components of $T^i_\alpha$ can be read off from \eqref{eq:EM gauge}, and in this notation are
\begin{align}\label{eq:max Ts}
    T^0_{(1)} &= -1 & T^a_{(0)} = \partial^a\,.
\end{align}
The adjoint can then be calculated using \eqref{eq:T ad explicit} and takes the form
\begin{equation}
    \bar T^\mu = \begin{pmatrix}
                    \tfrac {\de}{\de t} \\ \partial^a
                 \end{pmatrix}\,.
\end{equation}
The EL vectors for electromagnetism are
\begin{equation}
    \alpha_\mu (x) = \partial^\nu F_{\mu \nu} = \begin{pmatrix}
        \partial^a E_a \\ - \dot E_a + \epsilon_{abc} \partial^b B^c
    \end{pmatrix}\,.
\end{equation}
We thus find that the Noether identities are
\begin{equation}
    \tilde T^\mu \alpha_\mu = \frac{\de}{\de t} \lf( \partial^a E_a \rt) - \partial^a \dot E + \epsilon_{abc} \partial^a \partial^b B^c = 0\,,
\end{equation}
which is trivially satisfied and equivalent to the covariant equation $\de F = 0$.

The non-zero expansion coefficients of $T^i_\alpha$ in \eqref{eq:max Ts} lead to primary constraints
\begin{equation}
    \T 1 \mu \diby{L}{ \dot A^\nu} = - \diby{L}{ \dot A^0} = 0\,,
\end{equation}
which say that $A^0$ is a Lagrange multiplier in the theory, and the Lagrangian constraint (after using the primary constraint)
\begin{equation}
    T^\mu_{(1)} K_\mu = \alpha_0 = \partial^a E_a \approx 0\,,
\end{equation}
which is the so-called \emph{Gauss constraint} of the matter-free theory. From \eqref{eq:further c}, we obtain the Gauss constraint more directly because
\begin{equation}
    \T 0 \mu \diby{L}{\dot A^\mu} = \partial^a E_a \approx 0\,.
\end{equation}
The second condition of \eqref{eq:extra L const}, $\T 0 \mu K_\mu = \partial^a K_a \approx 0$, then says, after a straightforward calculation, that the Gauss constraint is propagated in time $\partial^a \dot E^a = 0$, guaranteeing the closure of the Dirac algorithm.

Note the similarity between the overall structure of these constraints and those obtained for BB theory. This is perhaps unsurprising since Maxwell theory can be obtained by applying the Gauge Principle. BB theory thus serves as a useful toy model of standard gauge theories in general. The equations of this section, however, illustrate how our formalism can accommodate $T$ operators that are differential operators in space.


\section{The Hamiltonian formalism} 
\label{sec:the_initial_value_problem}


The notion of gauge symmetry developed in the first part of this chapter (above) is fundamentally an over-a-history notion because its definition relied on the invariance of a mathematical object, the action, that is a functional of an entire history. I already pointed out, however, a difficulty with this approach as a solution to Belot's Problem because it would include Noether-$1$ symmetries as gauge symmetries. I will also show, in \Sec\ref{sub:reparametrisation_invariance}, that such notions of symmetry have been interpreted as leading to the rather strange conclusion that time evolution is unphysical in theories that are independent of the choice of time parameter.

I will suggest that considerable insight can be gained on these and other issues by adopting an at-a-time notion of symmetry. With such a notion, I will give a definition of gauge symmetry that will be based on the number of freely specifiable initial data in the equations of motion of the theory. We have already seen that such a notion is adequate for pinpointing underdetermination in standard gauge theories and explains why Noether-$1$ symmetries should \emph{not} be considered gauge. Later, we will also see how such a notion can be used to clarify the case of time reparametrisation invariance.

I will now introduce a standard first-order formalism, the \emph{Hamiltonian formalism}, applicable to general systems that is well-adapted to an at-a-time analysis of symmetry. This is because a first-order formalism treats configurations and velocities, which are independent variables from an at-a-time perspective, on an equal footing. Moreover, dynamical laws and symmetries can both be represented as flows generated by vector fields on state space in a first order formalism. The integrability of these flows and questions about underdetermination can then easily be assessed by the smoothness of the relevant vector fields. This gives a clean and powerful way of counting at-at-time degrees of freedom and identifying invariant structures. What we will find is that all the interesting theorems and constraints derived in the previous sections can be extracted from the analysis of a single equation following Noether's methodology. Additionally, the conceptual framework suggested by the formalism provides natural ways of handling problematic cases. Thus, while an equivalent analysis can be obtained in a second order formalism, the first-order formalism provides a superior toolkit for understanding gauge symmetry.

The relationship between the following three different ways of representing a system are particularly important. The first is a description of the system in terms of fields over spacetime. The second is a description on the tangent bundle over configuration space, which we called \emph{velocity phase space} in \Sec\ref{sub:symmetries_in_first_order_systems}. The third is a description on the cotangent bundle over configurations space, which we called \emph{phase space}. While most discussions of gauge symmetry focus either on descriptions in spacetime or on phase space, I will suggest that the description on velocity phase space is especially helpful in characterising gauge symmetries.

\subsection{Hamilton's equations} 
\label{sub:hamiltons_equations}


Let us start this section by giving a general procedure for converting a second order dynamical system to a first order system. This procedure can straightforwardly be generalised to higher order systems. See, for example, \cite[\S~2.3, pp 24]{woodhouse1997geometric} for details.\footnote{ Our analysis of constraints on phase space is inspired by Chapter~2 of \cite{woodhouse1997geometric}. For more standard treatments of constrained Hamiltonian systems on phase space, see \cite{henneaux1992quantization} and \cite{sundermeyer:1982}. } Our analysis in terms of a first order system can therefore be applied to a general system of arbitrary order. As I've done throughout this chapter, I will restrict my attention to globally hyperbolic regions of spacetime. More complicated spacetimes can be obtained by suitably patching together such regions.

The main idea is to double the variables of the second-order system by introducing the velocity variables $v^i$ and then impose additional constraints on these variables so that the resulting first-order system is equivalent to the old one when $v^i = \dot q^i$. An elegant way of achieving this is to start with some second-order action with Lagrangian $L(q^i, \dot q^i)$ and modify it in the following way:
\begin{equation}
    S_1 = \int_{t_1}^{t_2} \de t\, \lf( \diby{L(q^i, v^i)}{v^i} (\dot q^i - v^i ) + L(q^i,v^i)  \rt)\,,
\end{equation}
where we replace $\dot q^i \to v^i$ in the Lagrangian $L$. When the quantities $\diby{L}{v^i}$ are linearly independent (i.e., they don't satisfy constraints), then the extra term will enforce $\dot q^i = v^i$ on-shell. As we have already seen in \Sec\ref{sub:lagrangian_constraints_and_noether_s_first_theorem}, this requirement on $\diby{L}{v^i}$ is precisely the requirement that the resulting EL equations of the corresponding Lagrangian system are well-posed. We will find a similar result below for Hamiltonian systems.

To unpack these claims more carefully we can take advantage of the first order form of the equations of motion of the system to express the formalism in terms of more geometric quantities. First, we define the velocity phase space $\Gamma = T\mathcal C$ as the tangent bundle over $\mathcal C$ equipped with local coordinates $(q^i, v^i)$. Then, we define the 1-form
\begin{equation}
    \theta_L = \diby{L}{v^i} \de q^i\,
\end{equation}
on $\Gamma$ and the Hamiltonian function 
\begin{equation}
    H = v^i \diby L {q^i} - L\,.
\end{equation}
Consider a curve $\gamma: I \to \Gamma$ on $\Gamma$, where $I \in \mathbbm R$ is the temporal manifold, and the tangent vector $X\in T\Gamma$ to $\gamma$, which in coordinates is
\begin{equation}\label{eq:X def}
    X = \dot q^i \diby{}{q^i} + \dot v^i \diby{}{v^i}\,.
\end{equation}
The derivatives above are with respect to the time parameter $t \in I$ in the domain of $\gamma$. Using these geometric quantities, the action $S_1[\gamma]$ then reads
\begin{equation}
    S_1[\gamma] = \int_{t_1}^{t_2} \de t \lf( \iota_X \theta_L - H \rt) = \int_\gamma \lf( \theta_L - H \de t \rt)\,,
\end{equation}
where $\iota$ denotes the interior product on the exterior algebra of $\Gamma$. In a slight abuse of notation, we define $\de t \equiv \gamma^* \de t$ as the differential obtained by the pullback of the differential $\de t$ by $\gamma$. We thus have that $\iota_X \de t = 1$ and $\iota_{Y_i} \de t = 0$ for all $Y_i$ such that $[Y_i,X] \in \text{Span}(Y_i)$. This can be taken as an alternative geometric definition of the vector $X = \frac{\de}{\de t}$.\footnote{ I.e., $X$ is the dual vector to $\de t$. }

As in our second-order treatment, we will restrict to the case where the action has no explicit dependence on the independent variables. Because of this, any variation due to a change in the independent variables can be expressed as a variation of the dependent variables by pulling these back by $\gamma$. Without loss of generality, we can thus express the variational derivative of the action $S[\gamma]$ in terms of infinitesimal variations generated by a vector field $u \in T\Gamma$. Using our previous notation, we can express such variations as $\delta q^i \to \mathfrak L_u q^i$. For our analysis later, it will be convenient to split these variations into those that lie tangent to $\gamma$, and therefore could correspond to variations of $t$ and are parallel to $X$, and those that are transverse to $\gamma$, and therefore satisfy $\Lie_u t = 0$ and are spanned by the $Y_i$ vectors above. But for a general $u$, the variation of $S_1$ can be written in terms of the Lie drag of $S$ by the vector field $u$:
\begin{align}
    \delta S_1[\gamma;u] \equiv \mathfrak L_u S_1[\gamma] &= \int_\gamma \Big( \iota_u \de \theta_L - (\mathfrak L_u H) \de t - H \mathfrak L_u \de t + \de \lf( \iota_u \theta_L \rt)  \Big)\nonumber \\
    &= -\int_{t_1}^{t_2} \Big( \iota_u \lf( \iota_X \omega_L + \de H  \rt) \de t + H \mathfrak L_u \de t \Big) + \iota_u \theta_L\Big\rvert_{t_1}^{t_2}\label{eq:S var}\,,
\end{align}
where we have defined the closed 2-form $\omega_L = \de \theta_L$ which is the \emph{symplectic 2-form} of the theory. Note that while $\omega_L$ is closed, we will see below that it will often, but not always, be degenerate when $S_1[\gamma]$ has Noether-$2$ symmetries. When $\omega_L$ is degenerate, it technically does not fit the mathematical requirements of a symplectic 2-form and is sometimes called a \emph{pre-symplectic} form instead.

Equation~\eqref{eq:S var} is the main unifying equation of this chapter. It expresses the variation of the action in terms of flows generated by the vector field $u$ on velocity phase space. By interpreting $u$ in different ways, one can therefore use \eqref{eq:S var} to extract all interesting information about the symmetries and dynamics of the theory using such flows.

Each term is significant. First let us consider \eqref{eq:S var} as an equation for fixing the classical solutions (i.e., the integral curves of $X$) for arbitrary $u$. For arbitrary variations $u$ transverse to $X$, $\mathfrak L_u \de t = 0$ and the integrand is only zero if the following equations
\begin{equation}\label{eq:Ham eqns}
    \iota_X \omega_L + \de H = 0\,
\end{equation}
are satisfied. These are \emph{Hamilton's equations}. When $\omega_L$ is non-degenerate, there is a unique vector field $X$ over $\Gamma$ that solves this equation. Using
\begin{equation}
    \omega_L = \de \theta_L = \de \lf( \diby{L}{v^i} \de q^i \rt) = \frac{\partial^2 L}{\partial q^i \partial v^j} \de q^i \wedge \de q^j + \frac{\partial^2 L}{\partial v^i \partial v^j} \de v^i \wedge \de q^j \label{eq:omegaL}
\end{equation}
and
\begin{equation}\label{eq:dH}
    \de H = \de \lf( v^i \diby{L}{v^i} - L \rt) = \lf( v^j \frac{\partial^2 L}{\partial v^j \partial q^i} - \diby{L}{q^i} \rt) \de q^i + \frac{\partial^2 L}{ \partial v^j \partial v^i} v^j \de v^i
\end{equation}
as well as \eqref{eq:X def}, Hamilton's equations \eqref{eq:Ham eqns} become
\begin{equation}
    \iota_X \omega_L + \de H = \lf[ \lf( v^j - \dot q^j \rt) \frac{\partial^2 L}{\partial v^i \partial q^j} + \frac{\de}{\de t} \lf( \diby L {v^i}\rt) - \diby L {q^i} \rt] \de q^i + \lf( v^j - \dot q^j \rt) \frac{\partial^2 L}{\partial v^i \partial v^j} \de v^i\,.
\end{equation}
The $\de v^i$ term enforces $v^i = \dot q^i$ (for non-degenerate $\frac{\partial^2 L}{\partial v^i \partial v^j}$) and the $\de q^i$ term enforces the EL equations, $\alpha_i = 0$, as expected. The variational principle \eqref{eq:S var} is therefore equivalent to Hamilton's principle -- at least when solutions exist and are well-posed.

Hamilton's equations \eqref{eq:Ham eqns} are an example of a general way of defining a vector field $v_f$ in terms of some function $f$ on $\Gamma$ and a symplectic $2$-form $\omega$. The unique solution to the equation $\iota_{v_f} \omega + \de f = 0$ then implicitly defines $v_f$ when $\omega$ is non-degenerate. The flow of the vector $v_f$ is then called the \emph{symplectic flow} (or sometimes the \emph{Hamiltonian flow}) of $f$ on the symplectic manifold defined as the duple $(\Gamma, \omega)$. Thus, Hamilton's equation define the DPMs of a theory as the integral curves of the symplectic flow of the Hamiltonian.\footnote{ Note that $v_f$ is importantly not the same as the dual vector $w_f$, which obeys $w_f(\de f) = 1$, that is defined using the duality between vector fields and different forms.}

Note that we have excluded variations tangent to $\gamma$. This means that, in deriving Hamilton's equations, we have --- at least for the moment --- only considered theories where the parametrisation of $t$ on $\gamma$ is fixed. The case where arbitrary parametrisations of $\gamma$ are allowed will be treated separately in Section~\ref{sub:reparametrisation_invariance}.

When $\omega_L$ is non-degenerate, $\theta_L$ will have no kernel so that the vanishing of the boundary terms requires $u(t_1) = u(t_2) = 0$. This means that the variation should be performed with fixed temporal boundary conditions. A DPM is then uniquely specified by fixing the initial and final values of the curve $\gamma$. We will explore the significance of these boundary conditions in the degenerate case below.

\subsection{Noether symmetries and constraints} 
\label{sub:narrow_symmetries}

In this section, I will follow Noether's methodology and explore the consequences of a theory having Noether-$2$ symmetries. To do this, I treat \eqref{eq:S var} as an equation for the variation of $S_1$ for \emph{particular} vector fields $u_i$ about \emph{arbitrary} trial curves generated by $X$ (i.e., not necessarily the solutions of \eqref{eq:Ham eqns}). Noether-$2$ symmetries are then generated by all $u_i$ satisfying $\Lie_{u_i} S_1[\gamma] = 0$ for all $\gamma$.

Let us now deduce some consequences of having such $u_i$. Again, we begin by considering the case where the $u_i$ are transverse to $X$. In this case, we again have $\mathfrak L_{u_i} \de t = 0$, and for arbitrary $X$ we must have separately that
\begin{align}\label{eq:loc u cond}
    \mathfrak L_{u_i} \theta_L &= 0 & \mathfrak L_{u_i} H &= 0\,.
\end{align}
The first of these equations can be written using Cartan's formula as
\begin{equation}\label{eq:ICVs can trans}
    \iota_{u_i} \omega_L + \de \lf( \iota_{u_i} \theta_L \rt) = 0\,.
\end{equation}
Using the definition of symplectic flow given above, this means that the $u_i$ are determined by the symplectic flow of the velocity-phase-space functions $\iota_{u_i} \theta_L$. Let us call these functions $h_i$. The general conditions for transverse Noether-$2$ symmetries is then that they be generated by the symplectic flow of the $h_i$ (although we will see the limitations of this direct reading below) and that they leave the Hamiltonian function invariant.

We can get a bit more insight about these conditions and, in particular, the significance of the functions $h_i$ by considering the implications for the DPMs of the theory. To do this, we use Hamilton's equations \eqref{eq:Ham eqns} from which we can immediately deduce that
\begin{equation}
    \iota_{u_i} \iota_X \omega_L \approx - \iota_{u_i} \de H = 0\,,
\end{equation}
because of the second independent equation of \eqref{eq:loc u cond}. On-shell, the vanishing of \eqref{eq:S var} therefore implies that we must have separately that
\begin{align}\label{eq:gen constraints}
    \iota_{u_i} \omega_L &\approx 0 & \mathfrak L_{u_i} H &= 0 
\end{align}
and
\begin{equation}\label{eq:IVCs}
    h_i = \iota_{u_i} \theta_L \approx 0\,.
\end{equation}
Thus, the $u_i$ generate transverse Noether-$2$ symmetries of $S$ on-shell when they are in the kernel of both $\omega_L$ and $\theta_L$ and are invariances of the Hamiltonian. These coordinate-free expressions tell us, in principle, all the consequences of transverse Noether-$2$ symmetries. It is useful, however, to investigate these conditions in more detail to see how they can recover the standard results of gauge theory.

To make contact with the standard conceptions, it is convenient to define a map called the \emph{Legendre transform} $\mathbbm L: T \mathcal C \to T^* \mathcal C$, which maps the velocity phase space $T\mathcal C$ (i.e., the tangent bundle over configuration space) to the phase space $T^* \mathcal C$ (i.e., the cotangent bundle over configuration space). In coordinates, this map takes the instantaneous configurations and velocities $(q^i, v^i)$ to the instantaneous configurations and \emph{momenta} $(q^i, p_i)$, where the momenta are defined as
\begin{equation}\label{eq:p def}
    p_i \equiv \diby{L}{v^i}\,.
\end{equation}
Using the new coordinates $p_i$, $\omega_L$ can be written in Darboux's compact form
\begin{equation}\label{eq:omegaL darboux}
    \omega_L = \de \theta_L = \de \lf( p_i \de q^i \rt) = \de p_i \wedge \de q^i\,.
\end{equation}
From the definition \eqref{eq:p def}, we see immediately that $W_{ij}$, defined in \eqref{eq:hessian def}, is the part of the Hessian of the Legendre transform associated with the transformation of $v_i$ to $p_i$. Because of its importance it is sometimes called \emph{the} Hessian of the Legendre transform, although the full Hessian would involve more components. If $W_{ij}$ has a non-trivial kernel then the definition \eqref{eq:p def} implies that $\mathbbm L$ will be non-invertible. We will now see how the non-invertibility of $\mathbbm L$ arises from the non-trivial kernels of $\theta_L$ and $\omega_L$.

Let us refer to $h_i = \iota_{u_i} \theta_L \approx 0$ as the \emph{initial value constraints (IVCs)} because their vanishing results in the vanishing of the boundary term of \eqref{eq:S var}, which, as we have shown, must hold at all times on-shell. The ICVs can thus be interpreted as at-a-time constraints on the instantaneous states --- including, of course, the initial data. Using the explicit definition of $\theta_L$, we find
\begin{equation}\label{eq:IVCs explicit}
    h_i = \iota_{u_i} \theta_L = p_j \iota_{u_i} \de q^j \approx 0\,.
\end{equation}
Let us assume that the theory has been set up in such a way that the $\de q^i$ form a linearly independent generating set for the exterior algebra on $\mathcal C$.\footnote{ This assumption can be lifted by introducing so-called \emph{second-class} constraints into the Lagrangian. This extra analysis, while valuable, requires considerable extra technical machinery and is tangent to the discussion of symmetry. We will thus not consider this generalization in this work. } The constraints \eqref{eq:IVCs explicit} thus must put constraints on the $p_i$ because the $\de q^i$ are linearly independent. These constraints define the image of $\mathbbm L$. All degeneracies of $\omega_L$ must therefore be due to the degeneracies in $\de p_i$ resulting from the constraints \eqref{eq:IVCs explicit}. This can also been seen directly from the form of \eqref{eq:omegaL darboux} and the completeness of the set of $\de q^i$. We conclude that the kernel of $\theta_L$ results in IVCs in the form of on-shell constraints on the momenta $p_i$ and the kernel of $\omega_L$ results from the non-invertibility of $\mathbbm L$ due to these constraints. This reproduces the orthodox view that phase space constraints lead to underdetermination in velocity phase space due to the non-invertibility of the Legendre transform. We note however, that this underdetermination is already manifest on velocity phase space because of the kernel of $\omega_L$ without ever having to define a Legendre transform in the first place. We will investigate this further below.

Because $h_i \approx 0$ it is not possible to treat \eqref{eq:ICVs can trans} as equations defining the $u_i$ in terms of the symplectic flow of $\omega_L$ on-shell because the $u_i$ must lie in the kernel of $\omega_L$ on-shell. It is, however, possible to define the $u_i$ in terms of a genuine symplectic flow using the \emph{canonical symplectic form}
\begin{equation}
    \omega = \de \tilde p_i \wedge \de q^i
\end{equation}
defined for \emph{unconstrained} $\tilde p_i$. Mathematically, $\omega$ is obtained by canonically extending $\omega_L$ off the image of $\mathbbm L$ over the \emph{extended} phase space where $q^i$ and $\tilde p_i$ are Darboux coordinates with the same dimensions. Note that the extended phase space inherits any non-trivial global structure from $\omega_L$. Confusingly,\footnote{Perhaps this is because physicists usually work in coordinates where the extended phase space and, in particular, its Poisson bracket, is trivial to write.} the extended phase space obtained in this way is usually referred to as \emph{the} phase space of the theory. Using $\omega$, we can then define $u_i$ implicitly through $\iota_{u_i} \omega + \de\lf( \iota_{u_i} \theta_L \rt) = 0$.

The most reliable way to define the $u_i$ vectors, however, is to stick closely to Noether's methodology and assume that they are known in advance. Then, one can simply use them to determine the null directions of $\omega_L$. Moreover, while phase space is important historically and is a useful computational tool (owing to the fact that $\omega_L$ takes a simple form in Darboux coordinates), the canonical constraints $h_i \approx 0$ can project out the action of some  $u_i$ as we will see in some examples below. This obscures, rather than clarifies, the role of the $u_i$ in regard to underdetermination of the equations of motion. For conceptual clarity over computational efficiency, I therefore recommend an analysis on velocity phase space where all $u_i$ have a non-trivial action and where all quantities are unconstrained. This will help us to more directly assess the amount of underdetermination in the equations of motion.

Towards this end, consider a particular solution $X_1$ of \eqref{eq:Ham eqns}. We can then define a new solution $X_2$ by arbitrarily following the flow of the $u_i$ according to
\begin{equation}
    X_2  = X_1 + \xi^i u_i\,,
\end{equation}
for \emph{arbitrary} functions $\xi^i$ of the independent variables and velocity phase space. This follows trivially from the fact that the $u_i$ are in the kernel of $\omega_L$. The $u_i$, however, are not just ordinary generators of broad symmetries. The complete arbitrariness of the $\xi^i$ suggests that the symplectic structure has \emph{no way} of distinguishing the solutions $X_1$ from any other $X_2$. This will become even more apparent below when we investigate more directly the invertibility of the equations of motion using more explicit expressions for the $u_i$. It is worth noting, however, that these conclusions were all derived under the assumption that the $u_i$ have a non-trivial $t$-dependence. If this is not the case; i.e., if the $u_i$ generate Noether-$1$ symmetries; then \eqref{eq:S var} simplifies considerably. In \Sec\ref{sub:global_symmetries}, I show that one no longer obtains degenerate directions in $\omega_L$ but instead recovers Noether's first theorem in accordance to what was found in the second order analysis.

Let us now investigate the final condition $\Lie_{u_i} H = 0$. Consider that \eqref{eq:ICVs can trans} implies $\iota_X \iota_{u_i} \omega_L = -\iota_X \de\lf( \iota_{u_i} \theta_L \rt) = - \frac{\de}{\de t} \lf( \iota_{u_i} \theta_L \rt)$. We thus have, using Hamilton's equations, that
\begin{equation}\label{eq:LC coord free}
    \mathfrak L_{u_i} H \approx - \frac{\de}{\de t} \lf( \iota_{u_i} \theta_L \rt) = - \dot h_i\,.
\end{equation}
The condition $\mathfrak L_{u_i} H = 0$ can then be seen as a condition for the propagation in time along DPMs of the IVCs. Clearly the IVCs, the degeneracies of $\omega_L$ and the invariance of the Hamiltonian are not all independent. The precise relationship between these conditions will depend on the precise form of the $u_i$. We can get a better understanding of these relationships by restricting to vectors $u_i$ that are induced by over-a-history transformations of the action. This will put constraints on the nature of the $u_i$ and will allow us to recover the entire formalism developed in the second order approach.

Before doing this, let us end this section by remarking on the geometric interpretation of the $u_i$ vectors. We have assumed throughout this section that such vectors are transverse to the vectors $X$, which generate the DPMs. By imposing the IVCs, we have assumed that the vectors $u_i$ generate no variation of the action even on the boundary. Because they are transverse, this means that their action is to re-shuffle the initial (or final) data at any instant of the theory. Recall that, in \Sec\ref{sub:hamiltons_equations}, when deriving Hamilton's equations, we initially required all boundary variations to be fixed. Satisfaction of the IVCs means that this requirement can be lifted along the directions generated by the $u_i$. This provides a clean split between the data that must be fixed in the initial value problem and the data that can be freely specified. The freely specified data are exactly the data that are underdetermined by the equations of motion due to the degeneracies of $\omega_L$ in the $u_i$-directions. Thus, the full set of IVCs gives us a precise means to determine the at-a-time degrees of freedom that are gauge and those that are not when using the definition of gauge symmetry that I will advocate in \chap\ref{ch:pesa}.

\subsection{Recovering the second order formalism} 
\label{sub:explicit_representation}


In the previous subsection, I derived the general coordinate-free conditions \eqref{eq:gen constraints} and \eqref{eq:IVCs} resulting from the presence of transverse Noether-$2$ symmetries that are not global (i.e., that \emph{do} depend on $t$). In this section, I will give more explicit restrictions on the $u_i$ that will allow us to make contact with all the expressions from the second order formalism. To do this, I will rewrite the $u_i$ vectors in terms of the expansion coefficients, defined in \eqref{eq:T def}, for a general over-a-history symmetry. Because such symmetries involve functional degrees of freedom, a single over-a-history symmetry can generate several independent $u_i$ vectors arising from the independent derivatives of the gauge parameter at a particular time.

To see how this arises, let us first note that for a general vector $u$ to generate a symmetry of a second order system for a gauge parameter $\epsilon$, we must have that
\begin{equation}\label{eq:u T}
    u = \delta_\epsilon q^i \diby{}{q^i} + \delta_\epsilon \dot q^i \diby{}{v^i}\,.
\end{equation}
As we have argued extensively, the terms of $u$ depending on different derivatives of $\epsilon$ should be considered independent vectors from an at-a-time perspective. This gives an expansion of the form
\begin{equation}
    u = \sum_{a=0}^{k+1}  u^a_\alpha \frac{\de^a}{\de t^a} \epsilon^\alpha\,.
\end{equation}
Let us then use \eqref{eq:T def} for $k = 1$ to write $\delta_\epsilon q^i$ and $\delta_\epsilon \dot q^i$ in terms of the expansion coefficients $\T 0 i$ and match these to the corresponding $u^a_\alpha$ in the expansion above. The result is:
\begin{align}
    u^0_\alpha &= \T 0 i \diby{}{q^i} + \dT 0 i \diby{}{v^i} \nonumber\\
    u^1_\alpha &= \T 1 i \diby{}{q^i} + \lf( \T 0 i + \dT 1 i \rt) \diby{}{v^i}\nonumber\\
    u^2_\alpha &= \T 1 i \diby{}{v^i} \label{eq:u expansion} \,.
\end{align}
In the rest of this section, I will insert the expansion above into the general expressions \eqref{eq:gen constraints} and \eqref{eq:IVCs} to obtain all the independent constraints that arise.

To begin, let us rewrite the ICVs $h_i \approx 0$. Inserting \eqref{eq:u expansion} into \eqref{eq:IVCs} we find two independent conditions
\begin{align}\label{eq:can constraints}
    \T 0 i p_i &\approx 0 &  \T 1 i p_i &\approx 0\,.
\end{align}
This immediately gives us the primary constraints\footnote{ In the first order formalism, weak equalities require Hamilton's first equation $\dot q^i = v^i$ to be satisfied. This makes the first order formalism equivalent to the second order formalism. Thus, weak equalities have a different meaning in the first order formalism. A careful examination of the primary constraint shows that it is a strong equality when Hamilton's first equation is satisfied. } \eqref{eq:Lagrangian c}, and the constraints \eqref{eq:further c} since $p_i = \diby L {q^i}$. The vector $u^3$ has no $\diby{}{q^i}$ component and therefore adds no independent constraint.

To obtain further conditions, let us rewrite \eqref{eq:gen constraints}. Since we are working with coordinate expressions, we must use \eqref{eq:omegaL} and \eqref{eq:dH} for the explicit form of $\omega_L$ and $\de H$. The simplest expressions to calculate are those resulting from $u^2$ because these only involve contractions with $\de v^i$. Both $\iota_{u^2} \omega_L \approx 0$ and $\mathfrak L_{u^2} H = 0$ lead to the same constraints (although the second equation is a strong equality)
\begin{equation}
    \T 1 i  \frac{\partial L}{\partial v^i \partial v^j} = \T 1 i  W_{ij} = 0\,,
\end{equation}
where we recall that $W_{ij}$ is the Hessian of the Legendre transform. This recovers our previous result \eqref{eq:null primary}, which states that the $\T 1 i$ are null directions of $W_{ij}$.

The next easiest terms to calculate are those due to $\iota_{u^0} \omega_L = 0$ and $\mathfrak L_{u^0} H = 0$. From $\iota_{u^0} \omega_L = 0$, one obtains a 1-form equation with independent $\de q^i$ and $\de v^i$ components. The $\de v^i$ component is easily seen to imply
\begin{equation}
    \T 0 i  W_{ij} \approx 0\,,
\end{equation}
which says that $\T 0 i$ are further independent null vectors of $W_{ij}$ on-shell. This reproduces the first term of \eqref{eq:extra L const}. Together, the two null vectors $u^0$ and $u^2$ of $\omega_L$ are therefore seen to be partially due to the two null directions of $W_{ij}$. We have already seen how the kernel of $W_{ij}$ contributes to the non-invertibility of the equations of motion produced by the variational principle. Thus, the degeneracies of $\omega_L$ are also directly connected to non-inversion of the equations of motion when the Noether-$2$ symmetries are transverse.\footnote{ We will see that this is not necessarily the case for the non-transverse symmetries of reparametrisation invariant theories. }

Because we are in a second order formalism, $\omega_L$ has more non-trivial components than $W_{ij}$. These can be combined with the symmetries of $H$ to derive the remaining identities from the second order formalism. The $\de q^i$ term of $\iota_{u^0} \omega_L \approx 0$ gives\footnote{ Note that the contraction of this equation with $\T 0 j$ is zero and thus not all components of $\dT 0 i$ are fixed by this equation. This is immaterial, however, for the argument that follows. }
\begin{equation}\label{eq:dT def}
    \dT 0 i W_{ij} \approx \T 0 i F_{[ij]}\,,
\end{equation}
where we have defined $F_{ij} = \diby {p_i}{q_j} = \diby{^2L}{v^i \partial q^j}$. When inserted into $\mathfrak L_{u^0} H = 0$, equation \eqref{eq:dT def} can be used to eliminate all terms depending on $\dT 0 i$. This resulting condition is
\begin{equation}
    \mathfrak L_{u^0} H \approx \T 0 i K_i \approx 0\,.
\end{equation}
This is the second condition of \eqref{eq:extra L const}.

The final condition is obtained by inserting $\iota_{u^1} \omega_L \approx 0$ into $\mathfrak L_{u^1} H = 0$ following the procedure used for $u^0$ above. Doing this and using $\T 0 i W_{ij} \approx \T 1 i W_{ij} = 0$ we find that
\begin{equation}\label{eq:Lag sec}
    \mathfrak L_{u^1} H \approx \T 1 i K_i \approx 0\,.
\end{equation}
This is the Lagrangian constraint \eqref{eq:L constraints}.

A dynamical relationship between the IVCs and the invariances of the Hamiltonian can be obtained by noting that, for $a = \{0,1\}$ the above considerations tell us, using $\iota_{u^a} \omega_L \approx 0$, that
\begin{align}
    \iota_{u^a}\theta_L =& \T a i p_i \equiv h^{(a)}_\alpha &  \Lie_{u^a} H &\approx - \T a i K_i\,.
\end{align}
We can then use \eqref{eq:LC coord free} to find
\begin{equation}\label{eq:prop constraint}
    \T a i K_i \approx \dot h^{(a)}_\alpha\,.
\end{equation}
There is no corresponding condition for $a = 2$ because the equations for $u^2$ are trivially satisfied. When $a = 1$, \eqref{eq:prop constraint} says that the Lagrangian constraints are the time derivative along a DPM of the primary constraints --- in line with standard results (see, \cite[p. 55]{sundermeyer:1982}). We get a similar condition for $a=0$. The role that the $a=0$ conditions plays in the formalism depends slightly on the theory in question. But, as we have seen, this condition usually express the closure of the Dirac algorithm when considered in conjunction with Noether's second theorem.

To complete our analysis, we note that the identities of Noether's second theorem can be derived directly from our variational principle \eqref{eq:S var} by using the expansion $u = \sum_a u^a_\alpha \frac{\de^a}{\de t^a}\epsilon^\alpha$ and performing integration by parts on the integrand. The resulting local identity is
\begin{equation}
    \sum_a (-1)^a \frac{\de^a}{\de t^a} \lf( \iota_{u^a_\alpha} \alpha \rt) = 0\,,
\end{equation}
where now $\alpha$ is the 1-form $\alpha = \iota_X \omega_L + \de H$. It is straightforward to show that this reproduces the explicit form $\tilde T^i_\alpha \alpha_i = 0$ of Noether's second theorem upon inserting the expansion \eqref{eq:u expansion} for $u$ and using \eqref{eq:T ad explicit}. We have now reproduced all the expressions obtained for non-global symmetries in the second order formalism.

\subsubsection{Global transverse symmetries} 
\label{sub:global_symmetries}

To complete the analysis, I will now consider global symmetries whose variations are transverse to the generator $X$ of the DPMs of the theory. These symmetries are relevant for Noether's first theorem. To get all the symmetries relevant to this theorem, we must also treat the case where the global variations are proportional to $X$. This will be done as a special case at the end of \Sec\ref{sub:reparametrisation_invariance}.

For global transverse variations, $u$ takes a particularly simple form. Again I will treat transverse and tangential variations separately. Because the variations are transverse, we have
\begin{equation}
    u = \T 0 i \epsilon^\alpha \diby{}{q^i}
\end{equation}
for some constants $\T 0 i$ and group parameters $\epsilon^\alpha$. The variation $u$ does not have a $\diby{}{v^i}$ component because the velocities $\dot q^i$ are invariant under global transformations. This dramatically simplifies the variation of $S$ because
\begin{equation}\label{eq:Global Lu theta}
    \mathfrak L_u \theta_L = \mathfrak L_u \lf( \diby{L}{v^i} \de q^i \rt) = \T 0 j \epsilon^\alpha \frac{ \partial^2 L }{\partial q^j \partial v^i} \de q^i + \diby{L}{v^i} \de \lf( \T 0 i \epsilon^\alpha \rt) = \T 0 j \epsilon^\alpha \frac{ \partial^2 L }{\partial q^j \partial v^i} \de q^i\,,
\end{equation}
where, in the second term, I have used the fact that $\T 0 i$ and $\epsilon^\alpha$ are constants. The vanishing of the second term implies that there is no longer any total divergence term in $\mathfrak L_u \theta_L$, so that usual boundary term is no longer present. This indicates that there is cancellation between the usual boundary term and the term proportional to $\iota_u \omega_L$. For this reason, transverse global symmetries neither imply IVCs nor degeneracies in $\omega_L$. Thus, \emph{they are not directly related to a reduction of at-a-time degrees of freedom and should not be identified with gauge symmetries according to the definition proposed in \Sec\ref{sec:statement_of_the_pesa}}. Global symmetries can be converted to gauge symmetries following, for example, the Gauge Principle described in \Sec\ref{ssub:the_gauge_principle} and illustrated in BB theory in \Sec\ref{sub:bb_theory}.

To derive Noether's first theorem for transverse $u$ we insert \eqref{eq:Global Lu theta} into the variation \eqref{eq:S var} to obtain
\begin{equation}
    \delta S[\gamma;u] = \int_{t_1}^{t_2}\lf[ \lf( \T 0 j \epsilon^\alpha \frac{ \partial^2 L }{\partial q^j \partial v^i} \dot q^i - \mathfrak L_u H \rt)\de t + \mathfrak L_u \de t\rt]\,.
\end{equation}
For transverse $u$, the last term is zero and, using \eqref{eq:dH}, the integrand becomes
\begin{equation}
    \epsilon^\alpha \T 0 j \lf( \frac{\partial^2 L}{\partial q^j \partial v^i} (\dot q^i - v^i) - \diby{L}{q^i} \rt) = 0\,.
\end{equation}
For arbitrary non-zero $\epsilon^\alpha$ and when Hamilton's first equation, $\dot q^i = v^i$, has been applied, we obtain to the identity
\begin{equation}
    \T 0 i \diby{L}{q^i} = 0\,.
\end{equation}
This condition requires that the Lagrangian be invariant under the global symmetries generated by $u$. But using Hamilton's second equation, which tells us that $\diby L {q^i} = \deby{}t \lf( \diby L {v^i} \rt)$, and the constancy of $\T 0 i$ we obtain Noether's first theorem \eqref{eq:N1} (with $\diby L {v^i} = p_i$) for transverse symmetries:
\begin{equation}
    \deby{}t \lf( \T 0 i p_i \rt) = 0\,.
\end{equation}

\subsection{Gauge-fixing and degree-of-freedom counting} 
\label{sec:gauge_fixing_and_degree_of_freedom_counting}

The analysis of the previous section allows us to make a clean at-a-time degree-of-freedom count on velocity phase space for Lagrangian systems with $k=1$ (i.e., first-order Lagrangians with second order equations of motion) in the presence of non-global, transverse Noether-$2$ symmetries. In this case, the vectors $\{u^a_\alpha\}_{\alpha = 0}^{\alpha = 2}$ give the independent null directions of $\omega_L$. Depending on the theory, not all values of $\alpha$ may lead to non-zero vectors (as can be seen, for instance, in the BB-theory in \Sec\ref{sub:barbour_berttoti_shifts} below). However, it is possible to read-off all non-trivial null vectors of $\omega_L$ directly from the form of the Noether-$2$ symmetry.

Each of the null vectors of $\omega_L$ is an obstruction to solving Hamilton's equation, which involve solving for $X$ in $\iota_X \omega_L + \de H = 0$. To find invertible equations, one can restrict to surfaces, called \emph{gauge-fixed surfaces}, on velocity phase space that are everywhere transverse to the non-zero components of $u^a_\alpha$. This means finding surfaces $\mathcal G_a^\alpha: \{ x | \Lie_{u^b_\beta} G_a^\alpha \not\approx 0\, \forall x \in \Gamma\}$ for some smooth functions $G_a^\alpha: \Gamma \to \mathbbm R $. One therefore needs to specify one independent function $G^a_\alpha$ for each non-zero null vector $u^a_\alpha$ of $\omega_L$. The restriction of $\omega_L$ to such a gauge-fixed surface is then non-degenerate. Gauge-fixed solutions can then be computed on this surface.

Following Dirac's notion of gauge-symmetry, we can then do an at-a-time degree-of-freedom count by taking the number of independent velocity phase space quantities and subtracting the number of independent gauge-fixing conditions required to solve the system. In this way, the gauge-fixed surface itself is a candidate representation of the physical state. However, a gauge-fixing procedure of this kind will lead to a parametrisation of this surface in terms of coordinates on the full velocity phase space. A parametrisation in terms of intrinsic coordinates on the gauge-fixed surface is equivalent to a gauge reduction of the theory and is, in many cases, intractable as discussed in \Sec\ref{ssub:reduction}.

Note that in theories where the non-primary constraints are non-zero, these constraints may represent non-trivial surfaces in velocity phase space. In this case, the consistency of the variational principle in the presence of a particular Noether-$2$ symmetry requires a restriction to such surfaces. This could entail relationships between the gauge-fixing functions $G_a^\alpha$ (see for example \Sec\ref{sub:barbour_berttoti_shifts}). This fact, however, does not change the at-a-time degree-of-freedom count, which simply counts the number of phase space functions that must be fixed by arbitrary functions regardless of how these functions might depend on each other.



\section{Reparametrisation invariance} 
\label{sub:reparametrisation_invariance}

In this section, I will investigate an important application of this formalism to the case of theories that are reparametrisation invariant, in the sense defined at the end of \Sec\ref{sub:coordinate_invariance} (which I will restate below). Reparametrisation invariance is important because of the special role played in this formalism by variations that are tangent to the DPMs $\gamma_\text{cl}$, which are intimately connected to reparametrisation symmetry, as we will see. In orthodox treatments of gauge symmetry, which usually start from Dirac's approach, the interpretation of the outputs of Dirac's algorithm is indifferent to whether the variations resulting in those outputs are transverse or tangential. Our geometric analysis on velocity phase space will suggest that this is a mistake. While the existence of transverse variations leads to underdetermination in the equation of motion, I will show that the existence of tangential variations \emph{defines} the equations of motion and leads to no underdetermination. This is a significant formal difference between these kinds of variations, and this formal difference has important interpretational consequences.

Two such consequences regard the classical at-a-time degree-of-freedom count and the quantization of reparametrisation invariant theories. In the orthodox view, time evolution is formally \emph{and} interpretationally indistinguishable from a gauge transformation. This ultimately leads to a timeless quantum formalism. Recovering a meaningful notion of time evolution in the quantum theory is often called the \emph{frozen-formalism problem}, which is an important aspect of the so-called \emph{Problem of Time} in canonical quantum gravity. I will describe this problem in more detail in \Sec\ref{sub:solved_reparametrisation_invariance}, where I will show how the PESA can address it. For now, I will focus on the mathematical details arising from treating reparametrisation invariance using the formalism developed above.

In the view I will present here, time evolution is neither formally nor intepretationally a gauge transformation. Tangential variations, unlike transverse variations, will not be seen to lead to IVCs. They will lead to degeneracies in $\omega_L$, but these degeneracies will \emph{define} the DPMs rather than provide obstructions to solving them. Reparametrisation invariance will be seen to constitute a tangential transformation, which I will identify as evolution, and a particular transverse transformation, which I will identify as a gauge symmetry. Velocity phase space will play a central role since this is the only space where both of these transformations have a non-trivial action.

A clear demarcation between the gauge and evolution aspects of reparametrisation invariance suggests possible solutions to the frozen-formalism problem. A particular proposal along these lines, called \emph{Relational quantization}, has been developed in \cite{gryb_thebault_book}. I will not describe this procedure here. Instead, I will use reparametrisation invariance as a non-standard example of a symmetry that illustrates how my notion of gauge symmetry can be used to clarify an important existing problem: the Problem of Time. This will be done in \Sec\ref{sub:solved_reparametrisation_invariance}.

\subsection{Reparametrisation symmetry} 
\label{sub:reparametrisation_symmetry}

Let us start the analysis by giving a more precise definition of a reparametrisation-invariant theory and then investigating some general consequences.\footnote{See \cite{sundermeyer:1982} or Chapter~4 of \cite{henneaux1992quantization} for standard treatments of reparametrisation-invariant theories.} A theory is said to be \emph{reparametrisation invariant} when its equations of motion are invariant under the symmetry
\begin{equation}\label{eq:RI def}
    t \to \bar t(t)\,,
\end{equation}
where $t$ is a time parameter on the domain, $I$, of $\gamma$ and $\bar t$ is some \emph{monotonic} smooth function $\bar t: I \to \mathbbm R$ on this domain such that
\begin{equation}
    f \equiv \deby{\bar t} t > 0\,.
\end{equation}

It will be important for our considerations below that this definition is given in terms of an invariance of the equations of motion rather than the properties of an action. That is because the properties of the equations of motion generated by a variational principle depend only on the \emph{local terms}, and not the boundary terms, of that variation principle. In other words, a theory will be reparametrisation invariant if the \emph{local form} of the action; i.e.,
\begin{equation}\label{eq:RI loc}
    \de t\, L\lf(q^i, \deby {q^i} t\rt) = \de \bar t\, L\lf(q^i, \deby {q^i} {\bar t}\rt)
\end{equation}
 is invariant under \eqref{eq:RI def}. Note that we have already restricted ourselves to second-order theories (i.e., theories with first-order Lagrangians) with no explicit $t$-dependence in $L$. Importantly, the local condition \eqref{eq:RI loc} is \emph{not} equivalent to invariance of the action. This is because, in general,
\begin{equation}
     \int_{t_1}^{t_2} \de t\, L\lf(q^i, \deby {q^i} t\rt) \neq \int_{\bar t_1}^{\bar t_2} \de \bar t\, L\lf(q^i, \deby {q^i} {\bar t}\rt)\,,
\end{equation}
even if \eqref{eq:RI loc} holds. This claim may seem strange at first, but its truth is obvious once one realises that the quantity $\de t\, L$ is not, in general, a constant along a DPM. While the functional form of the integrands of both sides of the expression above are identical, their regions of integration, $(t_1, t_2)$ versus $(\bar t_1, \bar t_2)$, are not identical unless $f(t_1) = f(t_2) = 0$. Evaluating these integrals over different endpoints will thus, in general, lead to a different result.\footnote{ A free particle is a notable exception. } Reparametrisations are, thus, not strict variational symmetries unless one imposes the condition $f(t_1) = f(t_2) = 0$.\footnote{ To see this another way, reparametrisation invariance would be an empty concept if the relevant notion of invariance was a passive relabelling of the domain of integration. Instead, one must actively shift this domain relative to the integrand to get a non-trivial notion of symmetry. If one does this, however, one must also actively shift the region of integration in a corresponding way. } This point will complement my analysis and interpretation of reparametrisation symmetry below.

Before giving my analysis using the first-order formalism, let us derive some immediate consequences of the invariance \eqref{eq:RI loc}. A sufficient condition for \eqref{eq:RI loc} is that the Lagrangian $L(q^i, \dot q^i)$ be homogeneous of degree one in the velocities $\dot q^i$. In that case, $\de t \to \frac {\de \bar t} f$ and $L(q^i, \dot q^i) \to L(q^i, f \deby{q^i}{\bar t}) = f L(q^i, \deby{q^i}{\bar t})$ so that $\de t\, L$ is invariant. A necessary and sufficient condition for $L$ to be homogeneous of degree one is that $\dot q^i \diby{L}{\dot q^i} = L$.\footnote{ This is an application of Euler's homogeneous function theorem. } This is equivalent to the vanishing of the Hamiltonian function in the second order formalism
\begin{equation}\label{eq:Weierstrass}
    H = \dot q^i \diby{L}{\dot q^i} - L = 0\,.
\end{equation}
This is called the \emph{Weierstrass condition} and is valid \emph{off-shell}. The vanishing of the classical Hamiltonian is often considered the origin of the frozen-formalism problem. In addition to the vanishing of $H$, if we differentiate \eqref{eq:Weierstrass} with respect to $\dot q^i$ we obtain
\begin{equation}\label{eq:RI null vec}
    \dot q^i W_{ij} = 0\,,
\end{equation}
which says that the velocities are in the kernel of $W_{ij}$. This suggests that reparametrisation invariance is associated with some underdetermination in the equations of motion. We will see in our first-order analysis that this underdetermination is associated with a particular null vector of $\omega_L$ and involves the freedom to arbitrarily choose the time parameter along $\gamma$. We will also see, however, that there is an additional null vector of $\omega_L$, due to the vanishing of $H$, which is \emph{not} associated with any underdetermination in the equations of motion but instead \emph{defines} the classical solutions. I will interpret the second null vector as the generator of evolution.

Before proceeding, let me note that homogeneity of degree one in velocities is not a necessary condition for reparametrisation invariance. In many theories, including the ADM form of general relativity, reparametrisation invariance occurs because of the transformation properties of a certain Lagrange multiplier function called the \emph{lapse}. Because it is a Lagrange multiplier, the lapse does not enter the theory as a velocity but must transform as such in a well-defined variational principle. The vanishing of the Hamiltonian thus occurs more generally when a theory is homogeneous of degree one in all variables that transform as velocities in the variational principle.

\subsection{Reparametrisations as gauge and evolution generators} 
\label{sub:tangential_variations}

An infinitesimal reparametrisation can be written in terms of the pullback of $q^i(t)$ by an infinitesimal transformation of the form \eqref{eq:RI def} using a Taylor expansion as
\begin{equation}
    \delta_f q^i(t) = f \dot q^i\,,
\end{equation}
where $f$ plays the role of a gauge parameter. Matching this to \eqref{eq:T def}, we find that the only non-zero component of $T$ in the first order formalism is $T_{(0)}^i = v^i$. Inserting this into \eqref{eq:u expansion} gives
\begin{align}
\bar u^0 &= v^i \diby {}{q^i} + \dot v^i \diby{}{v^i} & \bar u^1 &= v^i \diby{}{v^i}\,.
\end{align}
Thus, reparametrisations are generated at-a-time by the two independent vectors $u^0$ and $u^1$.

The $u^1$ vector can be treated as the generator of a standard transverse variation. Importantly, it leads to no ICV because it has no $\diby{}{q^i}$ component so its contraction with $\theta_L$ is automatically zero. The conditions $\Lie_{u^1} H = 0$ and $\iota_{u^1} \omega_L \approx 0$ both lead to
\begin{equation}\label{eq:RI null first order}
    v^i W_{ij} = 0\,,
\end{equation}
which is the first order form of \eqref{eq:RI null vec}. This can be understood by the fact that a redefinition of the time parameter corresponds to a rescaling of the velocities according to $ \dot q^i = f\deby {q^i}{\bar t}$. Thus, at any given time the overall size of the velocity vector $v^i$ should be underdetermined by the equations of motion. This is precisely what is implied by the condition \eqref{eq:RI null first order}. We therefore identify $u^1$ as the generator of an at-a-time reparametrisation reflecting the freedom to arbitrarily choose a time parameter in a reparametrisation invariant theory. Interestingly, the flow of $u^1$ is trivial on the image of the Legendre transform since it is precisely this null direction that makes the Legendre transform many-to-one. The flow of $u^1$ can therefore \emph{only} be understood on velocity phase space, where it is non-trivial.

To fully understand the representation of reparametrisation symmetry, we must also understand the consequences of the presence of the vector $u^0$. When Hamilton's first equation is satisfied, $u^0 = X$, which generates a tangential variation. Let us then study the general consequences of such variations.

We wish to evaluate \eqref{eq:S var} when the variations are of the form $u = f X$ for some positive function $f: T\mathcal C \to \mathbbm R^+$ on velocity phase space. We are looking for the off-shell consequences of tangential symmetries and, thus, we make no particular assumptions about $X$. Inserting $u = f X$ into \eqref{eq:S var} we find that the skew symmetry of $\omega_L$ (due to it being a 2-form) implies
\begin{equation}\label{eq:rep inv variation}
    \delta S_1[\gamma;f X] = -\int_\gamma \lf( \mathfrak L_X H f \de t + H \de f \rt) + f\iota_{X} \theta_L \Big\rvert_{t_1}^{t_2}\,.
\end{equation}
The vanishing of this variation requires the separate vanishing of the local and boundary terms since the restriction of $f$ to $\gamma$ is an arbitrary function of $t$ at all $t$. Because, in general, $f$ and $\de f$ are independent functions, both local terms must independently vanish.  Except in the case of global reparametrisations (i.e., time translations), which will be treated separately below, the vanishing of the $\de f$ term implies the Weierstrass condition:
\begin{equation}
    H = 0\,.
\end{equation}
This is an \emph{off-shell} relation.\footnote{ In the Dirac formalism, which will be described below, the vanishing of the Hamiltonian is usually considered to be valid only \emph{on-shell}. The difference between this and the \emph{off-shell} result on velocity phase space occurs after performing the Legendre transform and extending the phase space beyond the constraint surface. By doing the extension, the restriction to the constraint surface must be added as an extra on-shell condition. } The $f$ term implies $\de H = 0$ for arbitrary $X$, but this is automatically satisfied when the Hamiltonian itself vanishes off-shell. We thus recover the usual off-shell vanishing of the Hamiltonian in a reparametrisation invariant theory.

The vanishing of the Hamiltonian, however, does not imply that the DPMs are trivial or that there is no dynamical evolution. Hamilton's equations, which result from the vanishing of \eqref{eq:S var} under transverse variations, imply
\begin{equation}\label{eq:RI eoms}
    \iota_{X_\text{cl}} \omega_L = 0\,,
\end{equation}
where $X_\text{cl}$ is now taken to be the generator of a classical solution. Because of the explicit definition, \eqref{eq:omegaL}, of $\omega_L$ as a function of $L$ and its partial derivatives, the equations \eqref{eq:RI eoms} are non-trivial partial differential equations. These equations \emph{are} simply the equations of motion of the theory, and their solutions generate DPMs. Thus, unlike in the transverse case, the vanishing of the local terms of \eqref{eq:S var} implies that one of the null-directions of $\omega_L$ \emph{defines} the classical equations of motion. Finding solutions to the equations of motion is then tantamount to solving these eigenvalue equations. Crucially, in the tangential case, the existence of null-directions of $\omega_L$ has no bearing on the invertibility of the equations of motion. In fact, the equations of motion \emph{must} be invertible in order for \eqref{eq:RI eoms} to have solutions at all, which is an assumption in the Noether methodology. If solutions don't exist, it's because the action must have been poorly chosen for its saddle points and not its symmetries. We will see an explicit example of how the classical equations of motion are constructed from \eqref{eq:RI eoms} and how to resolve the underdetermination due to $u^1$ in \Sec\ref{sub:jacobi_theory}.

The vanishing of the local terms is sufficient for our definition of reparametrisation invariance, which is defined in terms of invariances of the local equations of motion. It is, nevertheless interesting to see why the boundary terms of \eqref{eq:rep inv variation} cannot be made to vanish without trivialising at-a-time notions of symmetry. Using the off-shell vanishing of $H$, we can rewrite the boundary term of \eqref{eq:rep inv variation} as
\begin{equation}\label{eq:ev gen}
    f \iota_X \theta_L \Big\rvert_{t_1}^{t_2} = f \iota_X L \de t\Big\rvert_{t_1}^{t_2} = f L \Big\rvert_{t_1}^{t_2}\,.
\end{equation}
But for non-trivial theories, $L$ can't vanish. This means that the only way for a tangential symmetry to be a strict variational symmetry is if
\begin{equation}
    f(t_1) = f(t_2) = 0\,,
\end{equation}
as we saw earlier from our more general considerations. But since the choice of boundary is arbitrary from an at-a-time perspective, $f$ must be zero at \emph{all} times, which would trivialise the symmetry. We thus see that for reparametrisations that act at-a-time, the boundary terms can never be made to vanish, in agreement with our previous considerations. Note that, from an over-a-history perspective, it is possible to have non-trivial reparametrisations of curves whose infinitesimal generators vanish on the boundary and are thus strict variational symmetries. But this fact depends on singling out some initial and final instant and has no bearing on the dynamical or epistemic considerations of local observers. It therefore should also not be relevant in the identification of gauge symmetries in general.

The fact that the boundary variations cannot be made to vanish for tangential symmetries implies that there are no IVCs associated to them. This is consistent with the fact, which I have shown above, that tangential variations imply no underdetermination in the equations of motion. While the transverse variations reshuffle the data on the boundary, tangential variations actually \emph{move} the boundary. In this way, they can be seen to generate evolution rather than underdetermination in the equations of motion.

In fact, the non-zero boundary terms due to tangential variations can be used to distinguish these variations in a theory that also has transverse symmetries. Consider a theory where the vectors $u^I_\alpha$ cause the local terms of \eqref{eq:S var} to vanish off-shell. If we define the change of basis
\begin{equation}
    \bar u^I_\alpha = \sum_{J,\beta} \lambda_{IJ}^{\alpha\beta} u^J_\beta 
\end{equation}
such that
\begin{equation}
    \iota_{\bar u^I_\alpha} \theta_L = \delta^I_0\delta^1_\alpha L\,,
\end{equation}
then we can simply define $\bar u^0_1$ as the tangential symmetry; i.e., the generator of evolution when the ICVs are satisfied.\footnote{ On the image of the Legendre transform, the evolution generator $\bar u^0_1$ has an interesting geometric interpretation using the formalism of \emph{contact geometry}, which will be the primary tool used in \Sec\ref{sub:generating_dynamical_similarity}. If we assume that all degeneracies due to the transverse components have been eliminated (and restricting to the image of the Legendre transform eliminates the degeneracy due to $\bar u^1_1$), then $\de \theta_L$ has only a one dimensional kernel spanned by $R = \frac 1 L \bar u^0_1$ with $\iota_{R} \theta_L = 1$. This means that $(\de \theta_L)^{\wedge (N-1)}$ is a contact form on this space. Moreover, the normalisation $\iota_{\bar u^0_1} \theta_L = 1$ implies that $R$ is simply the Reeb field on this contact space. The dynamical evolution can then be interpreted as the Reeb flow on the image of the Legendre transform. Finally, the condition $\iota_{R} \theta_L = 1$ can be used to identify a conservation law for the Reeb flow. In this case, this simply expresses the usual \emph{Hamiltonian constraint}. }

\subsection{Global tangential symmetries} 
\label{sub:global_tangential_symmetries}

The last case of variations to consider, which lead to the vanishing of the local terms of \eqref{eq:S var}, are the global tangential variations. These variations have the form $u = a_0 X$ for some constant $a_0$. Since $X = \deby{}t$, the variations $u$ of this kind generate global time translations of the form $t \to t + a_0$.

To study the consequences of these variations, it is best to use the variation of $S$ in the form of the second line of \eqref{eq:S var}, where the new term, $H \mathfrak L_u \de t = H \de a_0 = 0$ is now vanishing because $\de a_0 = 0$. We therefore \emph{do not} get a condition of the form $H = 0$ for global time translations as we did for local time reparametrisations. This also means that Hamiltonian's equations no longer imply that $u$ is a null direction of $\omega_L$ --- even on-shell. We find that global tangential $u$, like global transverse $u$, lead to no underdetermination in the equation of motion.

To recover Noether's first theorem for this case, note that the term $\iota_X \iota_u \omega_L = a_0 \iota_X \iota_X \omega_L = 0$ still vanishes automatically for all off-shell trajectories. The vanishing of the integrand of \eqref{eq:S var} then implies
\begin{equation}
    \mathfrak L_X H = 0\,.
\end{equation}
This says that the total energy $E$, which is the value of the function $H$ along a DPM, is constant along the DPM generated by $X$. This is the usual result of Noether's first theorem as applied to global time translations.

As in a general reparametrisation invariant theory, the boundary term
\begin{equation}
    a_0 \iota_{X} \theta_L\Big|_{t_1}^{t_2}
\end{equation}
is not zero. Again this expresses the fact that the local form of the action, and therefore the equations of motion, is invariant even though the value of the action itself can change because global time translation shift the region of integration by the constant $a_0$. It also means that global time-translations are not strict variational symmetries even thought this has no consequences for Noether's theorems.

\section{Examples} 
\label{sec:examples}

\subsection{Barbour--Bertotti shifts} 
\label{sub:barbour_berttoti_shifts}

In this section, I will redo the BB-theory example introduced in \Sec\ref{sub:bb_theory} using the first-order formalism. Recall that BB-theory was a theory of $N$ Newtonian particles in which time-dependent spatial translations have been made variational symmetries using a version of the Gauge Principle. Using the notation of that section, the instantaneous particle configurations are $\dot q^i_I$ and the shift-fields, which implement the translational gauge symmetry, are $w^i$. The variational symmetries \eqref{eq:BB gauge} tell us that the $T$ operator splits into a $q$-component, which we called $R^i_{Ij}$, and a $w$-component, which we called $U^i_j$. Inserting \eqref{eq:BB Ts} into the general expression for the $u$-vectors, \eqref{eq:u expansion}, gives four non-zero components
\begin{align}
    u^{0}_{(q)i} &= \sum_I \diby{}{q^i_I} & u^1_{(q)i} &= \sum_I \diby{}{v^i_I} \notag \\
    u^1_{(w)i} &= \diby{}{w^i} & u^2_{(w)i} &= \diby{}{v_{(w)}^i}\,, \label{eq:BB null directions}
\end{align}
where the $(q)$ and $(w)$ subscripts indicate the $q$ and $w$ components of $u$ respectively and $(v^i_I,v_{(w)}^i)$ are the velocities of $(q^i_i, w^i)$. This tells us that there are four independent sets of null directions of $\omega_L$.

We can now explicitly compute all the constraints arising from the first-order formalism. The primary constraints, $\T 1 i p_i$, in this notation (note the $i$ in the original notation ranges over particle indices as well as spatial indices for $q$ and $w$, and the $\alpha$ ranges over spatial $q$ and $w$ indices) only have the $w$-component:
\begin{equation}
    p_{(w)i} = 0\,,
\end{equation}
which says that the momenta $p_{(w)i}$ conjugate to $w^i$ vanish. This means that the $w$-component of the Hessian is zero so that its kernel is spanned by the vectors $\delta^i_{(j)}$ (one for each spatial dimension indexed by $j$) in line with expectations. In this notation, the Lagrangian constraints, $\T 1 i K_i$, become
\begin{equation}\label{eq:BB const 2}
    \sum_I p_i^I \approx 0\,,
\end{equation}
where $p_i^I$ are the momenta conjugate to $q^i_I$. This is the standard BB constraint that says that the total linear momentum of the system is zero. It is straightforward to see that the condition $\T 0 i p_i$ leads to the same BB constraint.

We can solve the BB constraint on velocity phase space using the definition
\begin{equation}
    p_i^I = \diby{L_\text{BB}}{v^i_I} = m_I \delta_{ij} \lf( v^i_I - w^i \rt)\,.
\end{equation}
Inserting this into \eqref{eq:BB const 2} gives
\begin{equation}\label{eq:vcm}
    w^i = v^i_\text{cm} \equiv \frac 1 M \sum_I m_I v^i_I\,,
\end{equation}
where we have defined the centre-of-mass velocity, $v^i_\text{cm}$, with $M = \sum_I m_I$. On the surface where this holds, a short calculation shows that the condition $\T 0 i K_i \approx 0$ reduces to the condition
\begin{equation}
    \sum_I \diby V {q^i_I} = 0\,,
\end{equation}
which is the requirement that the potential is translationally invariant as expected. One can also see that, on this surface, the vectors $\delta^i_{(j)}$ are in the kernel of the $q$-component of the Hessian. This explicitly confirms all the constraint equations we derived in \Sec\ref{sub:explicit_representation}.

We can go a bit further in our analysis by finding explicit gauge-fixing conditions for all the different independent null directions of $\omega_{L_\text{BB}}$ given in \eqref{eq:BB null directions}. The $w$-components of $u^1$ and $u^2$ suggest that $w^i$ and $v_{(w)}^i$ should be arbitrary functions. Indeed, the fact that $u^1_{(w)i}(w^j) = \delta^i_j$ and $u^2_{(w)i}(v^j_{(w)}) = \delta^i_j$ implies that setting $w^i$ and its velocity to arbitrary functions is a valid gauge-fixing procedure for these null directions.

For the $q$-components of $u^0$ and $u^1$ we can use \eqref{eq:vcm} to suggest a gauge-fixing procedure that involves setting the centre-of-mass position, $q^i_\text{cm}$, and velocity, $v^i_\text{cm}$, to arbitrary values. Indeed, it's straightforward to verify that $u^0_{(q)i}(q^j_\text{cm}) = \delta^i_j$ and $u^1_{(q)i}(v^j_\text{cm}) = \delta^i_j$. This tells us that we can think of the shift variables $w^i$ and their velocities as well as the centre-of-mass positions and their velocities as arbitrary functions at any given time in BB-theory. This satisfies Dirac's criterion, discussed in \Sec\ref{ssub:the_dirac_algorithm}, for treating the transformations \eqref{eq:BB gauge} as gauge transformations. Note that, because of \eqref{eq:vcm}, the free functions assigned to $w^i$ must match those assigned to $v^i_\text{cm}$ in order for the BB constraint to be satisfied. All that this means is that the freely specifiable functions cannot be chosen independently.


\subsection{Jacobi theory} 
\label{sub:jacobi_theory}

An important non-trivial example of a reparametrisation invariant theory is that of Jacobi's original variational principle for Newtonian mechanics.\footnote{ See, for example, \cite[\S V.7]{lanczos2012variational} for an introduction to Jacobi's principle. } From a modern perspective, Jacobi's principle can be understood as a geodesic principle on configuration space. This is the version of Jacobi theory that we will study here. As noted in \cite{Barbour_Bertotti}, Jacobi's theory not only covers virtually all theories of particle mechanics but is closely related to the Baierlein--Sharp--Wheeler (BSW) action for general relativity, introduced in \cite{baierlein1962three}.

Jacobi's theory, as a geodesic principle on a configuration space $\mathcal C$, can be implemented by a variational principle with an action of the form
\begin{equation}\label{eq: S jac}
     S_{\text{Jac}}[\gamma; q_1, q_2) = \int_{q_1}^{q_2} \de t \sqrt{ g_{ab} \dot q^a \dot q^b }\,,
\end{equation}
where $g_{ab}$ is a metric on $\mathcal C$. The DPMs of the theory are then the geodesic curves of extremal length $S$ computed with the metric $g_{ab}$ and with endpoints $(q_1, q_2)$. It is straightforward to see that $S_\text{Jac}$ is reparametrisation invariant owing to the fact that it is homogeneous of degree one in the velocities.

The EL equations resulting from this action lead to the geodesic equation:
\begin{equation}\label{eq:geo eqn}
     \ddot q^a + \Gamma\ud a {bc} \dot q^b \dot q^c = \kappa \dot q^a\,
\end{equation}
in terms of the metric compatible connection
\begin{equation}\label{eq:chritoffel}
     \Gamma\ud a {bc} = \frac 1 2 g^{ad} \lf( \partial_b g_{dc} + \partial_c g_{db} - \partial_d g_{bc} \rt)\end{equation}
and $\kappa = \frac \de {\de t} \lf( \log L \rt)$, where we recall that the Lagrangian is $L = \sqrt{ g_{ab} \dot q^a \dot q^b }$. For an affine parameter, $\kappa = 0$ and we obtain the more familiar form of the geodesic equation: $\dot q\ud a {;b} \dot q^b = 0$.\footnote{ In Jacobi's original formulation, $g_{ab} \dot q^a \dot q^b = 2 (E - V) T$, where $E$ is the total energy of the system, $V(q^a)$ is the potential energy function and $T(\dot q^a)$ is the kinetic energy. The equations of motion are then a reparametrisation invariant version of Newton's laws where the increment of Newtonian time $\de \tau$ is the choice of time parameter where $T + V = E$. In terms of an arbitrary time coordinate $t$, we have $\de \tau = \sqrt{ \frac {T(\dot q^a)} {E - V} }\,\de t $.  }

Given that Jacobi theory is reparametrisation invariant, the considerations of \Sec\ref{sub:tangential_variations} tell us that the vectors
\begin{align}
     u^0 &= \dot q^a \diby{}{q^a} + \dot v^a \diby{}{v^a} = X & u^1 &= v^a \diby{}{v^a}\,,
\end{align}
generate variations of $S_\text{Jac}$ that preserve the equations of motion. We can now explicitly illustrate the consequences of this in Jacobi's theory to confirm the general considerations of \Sec\ref{sub:tangential_variations}.

To do this, we compute the pre-symplectic form, $\omega_{L_J}$, for Jacobi theory using \eqref{eq:omegaL} and \eqref{eq: S jac}:
\begin{equation}
     \omega_{L_J} = \lf( \hat v^c \partial_a g_{bc} - \frac 1 2 \hat v_b \hat v^c \hat v^d \partial_a g_{cd} \rt) \de q^a \wedge \de q^b
               + \frac 1 {|v|} \lf( g_{ab} - \hat v_a \hat v_b \rt) \de v^a \wedge \de q^b\,,
\end{equation}
where we have defined the covectors $v_a = g_{ab} v^b$, the norm $|v| = \sqrt{ g_{ab} v^a v^b}$ induced by the metric $g$, and the unit vectors $\hat v^a = v^a / |v|$. The coefficient of the second term gives the explicit form of the Hessian,
\begin{align}
     W_{ab}^J &= \frac{\delta^2 L_J}{\delta v^a \delta v^b} \nonumber \\
          &= \frac 1 {|v|} \lf( g_{ab} - \hat v_a \hat v_b \rt)\,,
\end{align}
of the Legendre transform. The geometric formulation of Jacobi's principle allows us to explicitly see that $W_{ab}$ is proportional to the orthogonal decomposition of the metric $g$ along the directions $\hat v^a$. We can then explicitly verify the condition
\begin{equation}
     v^a W_{ab} = 0\,.
\end{equation}
In this picture, the above condition results from the fact that $u^1$ is a transverse symmetry so that
\begin{equation}
     \iota_{u^1} \omega_L = v^a\lf(g_{ab} - \hat v^a \hat v^b \rt) \de q^b = v^a W_{ab} \de q^b = 0\,,
\end{equation}
which must vanish for all $\de q^b$. The vector $u^1$ is therefore the generator of at-a-time reparametrisations, which rescale $|v|$. This is related to the freedom to define the evolution in an arbitrary time parameter as discussed in \Sec\ref{sub:tangential_variations}.

The vector $u^0$ is a tangential symmetry. We thus expect its presence to imply that the Hamiltonian function, $H = v^a \diby {L_J}{v^a} - L_J = 0$, vanishes identically. This can easily be verified by direct computation by noting that $L_J = |v|$. Hamilton's equations then tell us that the eigenvalue equation $\iota_{u^0} \omega_{L_J} = 0$ should determine the DPMs. This is a one-form equation with two terms that can be computed and set to zero separately. The first is the term proportional to $\de v^a$:
\begin{equation}
     \omega_{L_J}\lf(\tfrac{\partial}{\partial v^a}, u^1 \rt) = \frac{ \dot q^a }{|v|} \lf( g_{ab} - \hat v_a \hat v_b \rt) = \dot q^a W_{ab} = 0\,.
\end{equation}
Since the kernel of $W_{ab}$ is $v^a$, this equation tells us that $\dot q^a = v^a$, which is Hamilton's first equation for this system. The second term is proportional to $\de q^a$ and leads to the non-trivial condition
\begin{equation}
     \omega_{L_J}\lf(\tfrac{\partial}{\partial q^a}, u^0 \rt) = \frac 1 {|v|} \lf( \Gamma_{abc} v^b v^c - \frac 1 2 \lf( v^b \hat v^c \hat v^d \partial_b g_{cd} \rt) v_a + \dot v^b  \lf( g_{ab} - \hat v_a \hat v_b \rt)\rt) = 0\,\label{eq:jac_eom}
\end{equation}
after using $\dot q^a = v^a$ and applying several simplifications. Note that this is equivalent to the geodesic equation \eqref{eq:geo eqn}.\footnote{ Note that to show this one needs to expand the definition of $\kappa$ in \eqref{eq:geo eqn} and use $\dot q^a = v^a$ to prove that $\frac 1 2 \hat v^b \hat v^c \hat v^d \partial_b g_{cd} = |v| \kappa - \dot v^a \hat v_a$. } As we can see in this explicit example, the null direction $u^0$ of $\omega_L$ does \emph{not} contribute to any underdetermination in the equations of motion but, rather, \emph{defines} these equations of motion in the sense that the solutions of the geodesic equations define one of the null directions of $\omega_L$.

Importantly, the only degeneracy in this system is due to the presence of the transverse vector $u^1$. In order to solve for the acceleration $\dot v^a$ in terms of the velocities $v^a$ and configurations $q^a$ one would need to invert the matrix $W_{ab} = (g_{ab} - \hat v_a \hat v_b)/|v|$, which is degenerate. One can overcome this issue by imposing gauge-fixing conditions on velocity phase space. Such gauge fixings are given by restricting to surfaces transverse to the flow of $u^1$. This can be done, for example, by defining the surfaces on velocity phase space satisfying
\begin{equation}
    |v| - f(q,v,t) = 0\,
\end{equation}
for $f$ such that $v^a \diby f{v^a} \neq |v|$. A simple choice is $f = 1$. On this surface, the equation \eqref{eq:jac_eom} reduces to the invertible equation
\begin{equation}
     \dot v^a + \Gamma\ud a {bc} v^b v^c = 0\,,
\end{equation}
which is the (well-posed) geodesic equation in an affine parametrisation. This explicitly confirms both the statement that $u^1$ generates reparametrisations and that $u^1$ is the only source of underdetermination in the Jacobi system.


\section{The Dirac algorithm} 
\label{sec:the_dirac_algorithm}

I will now give a more detailed introduction to the Dirac algorithm that was first described in \Sec\ref{ssub:the_dirac_algorithm}. I will state his proposal for identifying gauge symmetries and relate the outputs of his algorithm to the formalism developed in the previous sections. This will explain how my proposal is related to Dirac's and what makes his too restrictive for my purposes. I will loosely follow the presentation in \cite{dirac2001lectures}. For a more modern treatment, see Chapter 1 of \cite{henneaux1992quantization} or consult the references of \Sec\ref{ssub:the_dirac_algorithm}.

Dirac starts by considering an arbitrary theory defined by an action of the general first order form \eqref{eq:S} (i.e., of a form that leads to second-order equations of motion), where $k= 1$. He then performs a Legendre transform $\mathbbm L: T\mathcal C \to T^*\mathcal C$ by defining momenta as $p_i = \diby L {\dot q^i}$. He is interested in a theory where the definition of the Legendre transform leads to an immediate set of $N$ independent identities, which Dirac calls \emph{primary constraints},\footnote{Dirac attributes this terminology as being due to Bergmann.} of the form
\begin{equation}
    \{ \varphi_\alpha = 0\}_{\alpha=1}^{\alpha = N}\,.
\end{equation}
As we have seen, identities of this kind define the image of $\mathbbm L$. Dirac suggests treating these as constraint equations on the unrestricted `phase space' of the theory whose canonical symplectic form $\omega$ takes the Darboux form
\begin{equation}
    \omega = \de p_i \wedge \de q^i\,.
\end{equation}
In this way, Dirac is working in coordinates on what I have called the `extended phase space' of the theory, which is constructed by canonically extending the image of $\mathbbm L$ using the pullback of $\omega_L$ by $\mathbbm L$. On Dirac's phase space, we can define a Poisson bracket $\pb f g = \iota_{\omega^{-1}} \de f \wedge \de g$ using the inverse, $\omega^{-1}$, of $\omega$. In Darboux coordinates, this is
\begin{equation}
    \pb f g = \diby{f}{q^i}\diby{g}{p_i} - \diby{f}{p_i}\diby{g}{q^i}\,.
\end{equation}

Using the Poisson bracket, the set $\{ \varphi_\alpha \}_{\alpha = 1}^{\alpha = N}$ forms a regular surface locally on Dirac's phase space when
\begin{equation}
    \pb {\varphi_\alpha}{\varphi_\beta} = c\indices{^\gamma_{\alpha \beta}} \varphi_\gamma + M_{\alpha \beta}\,,
\end{equation}
where $c\indices{^\gamma_{\alpha \beta}}$ and $M_{\alpha \beta}$ can be arbitrary phase space functions (subject to the appropriate anti-symmetry). The first step in Dirac's procedure is then to reduce this system to a smaller set of constraints, which he calls \emph{first-class constraints}, by restricting to a surface of what he calls \emph{second-class constraints}. On the second-class constraint surface, the remaining set of first-class constraints $\{\phi_\alpha\}_{\alpha = 1}^{\alpha = M}$, where $M < N$ (and $N-M$ is even), is required to obey
\begin{equation}\label{eq:first class algebra}
    \pb {\phi_\alpha}{\phi_\beta} = c\indices{^\gamma_{\alpha \beta}} \phi_\gamma\,,
\end{equation}
where $\pb{\cdot}{\cdot}$ is now the restriction of the previous Poisson bracket to the second-class constraint surface.

For the purposes of this analysis, I will assume that this first step has been carried out any time second-class constraints happen to appear, and that these second-class constraints can be explicitly solved for and eliminated. I will then work directly on the second-class constraint surface, which from now on I will simply call \emph{phase space}, and coordinatize this by $(q^i, p_i)$ in a slight abuse of notation. This assumption will simplify our considerations by eliminating aspects of the construction that are not related to gauge symmetry. Second-class systems can safely be ignored in our analysis because the fact that they can be eliminated using Dirac's procedure ensures that they are not associated with underdetermination in the equations of motion.\footnote{The only exception is if a second-class system was obtained by imposing a constraint as a gauge-fixing condition for a first-class constraint. But in this case, one need only consider the amount of underdetermination in the original first-class system.}

The starting point is then a theory, possibly reduced from a theory with second-class constraints, with a Lagrangian $L(q^i,\dot q^i)$ that generates the $M$ primary first-class constraints
\begin{equation}
    \{ \phi_\alpha = 0\}_{\alpha=1}^{\alpha = M}\,
\end{equation}
satisfying \eqref{eq:first class algebra}. Since Dirac is working on an unrestricted phase space, he must ensure that the dynamical flow he defines on this space remains tangent to the surface defined by the primary constraints. To do this, he starts by adding the primary constraints to the original Hamiltonian, $H = \dot q^i p_i - L$, of the theory using Lagrange multiplier functions $\lambda^\alpha$. This defines what he calls the \emph{total Hamiltonian} function
\begin{equation}
    H_\text{tot} \equiv H + \lambda^\alpha \phi_\alpha\,.
\end{equation}
The total Hamiltonian can be seen as a way of building conditions into the variational principle that lead to well-defined solutions when the original action generated primary constraints. One can then tentatively propose the equations of motion
\begin{align}
    \dot q^i &= \pb {q^i}{H_\text{tot}} & \dot p_i &= \pb {p_i}{H_\text{tot}}
\end{align}
and check whether these equations of motion preserve the primary constraint surface. To do this, one needs to compute the time evolution of $\phi_\alpha$ such that
\begin{equation}
    \dot \phi_\alpha = \pb{\phi_\alpha}{ H_\text{tot} } \equiv c_\alpha^\beta \phi_\beta + \chi_\alpha\,,
\end{equation}
for some independent phase space functions $c_\alpha^\beta$. If any of the functions $\chi_\alpha$ are non-zero, then Dirac proposes that these should be added as additional constraints to the theory. He calls these \emph{secondary constraints}. He then tentatively defines an \emph{extended Hamiltonian},
\begin{equation}
    H_\text{ext} \equiv H_\text{tot} + \Lambda^I \chi_I\,,
\end{equation}
where $I$ ranges over the number of independent secondary constraints. One must then check that the new set of equations of motion generated by $H_\text{ext}$ preserve the surface defined by both the primary and secondary constraints. If this is not the case, new \emph{tertiary constraints} are generated. The process continues in this way until no new constraints are generated. If this happens, Dirac's algorithm is said to close. One can then compile all primary, secondary, tertiary, etc constraints into a complete set of $R$ first-class constraints, $\{\pi_\alpha\}_{\alpha = 1}^{\alpha = R}$, satisfying
\begin{align}
    \pb {\pi_\alpha}{\pi_\beta} &= C\indices{^\gamma_{\alpha \beta}} \pi_\gamma & \pb{\pi_\alpha}{ H_\text{tot} } \equiv C_\alpha^\beta \pi_\beta\,,
\end{align}
for $\{\alpha,\beta\} = 1,\hdots, R$ and for the arbitrary phase space functions $C\indices{^\gamma_{\alpha \beta}}$ and $C_\alpha^\beta$. The full extended Hamiltonian then takes the form
\begin{equation}
    H_\text{ext} = H + v^\alpha \pi_\alpha\,,
\end{equation}
in terms of the $R$ arbitrary Lagrange multiplier functions $v^\alpha$.

Dirac then points out that the equations of motion of the system obtained in this way:
\begin{align}\label{eq:Dirac eoms}
    \dot q^i &= \pb {q^i}{H_\text{ext}} & \dot p_i &= \pb {p_i}{H_\text{ext}}\,,
\end{align}
are only integrable when the functions $v^\alpha$ are arbitrarily assigned specific values. Since these functions can take any value, Dirac argues that they should not describe anything physical in the theory and that
changing them should have no effect on the `physical state.'\footnote{ We say what Dirac means by the `physical state' in our language at the beginning of \Sec\ref{ssub:the_dirac_algorithm}. } If one changes the values of the Lagrange multipliers $v^\alpha$ by an infinitesimal amount $\delta v^\alpha$, then an arbitrary phase space function $f(q^i, p_i)$ will see its velocity shifted by
\begin{equation}\label{eq:Dirac gauge transf}
    \delta \dot f = \delta v^\alpha \pb{f}{\pi_\alpha}
\end{equation}
according to \eqref{eq:Dirac eoms}. This means that the vector $\pb{\cdot}{\pi_\alpha}$ on phase space generates a transformation that only involves an infinitesimal shift in the arbitrary functions $v_\alpha$, and therefore should induce no change in Dirac's physical state. Dirac proposes that such changes are gauge transformations. Importantly, Dirac's argument involves variations that act transversely to the dynamical trajectories. His argument is therefore inapplicable to tangential variations and is consistent with my analysis that tangential symmetries generate evolution and not gauge transformations. This observation was first stated in a slightly different form in \cite{barbour2008constraints}.

As a final step, Dirac proposed that the physical state be defined in terms of a complete set of functions $O^I$, now called \emph{Dirac observables}, such that
\begin{equation}
    \pb{O^I}{\pi_\alpha} = g_\alpha^\beta \pi_\beta \approx 0\,,
\end{equation}
for the arbitrary phase space functions $g_\alpha^\beta$. On the constraint surface, $\pi_\alpha \approx 0$ and \eqref{eq:Dirac gauge transf} tells us that
\begin{equation}
    \delta \dot O^i \approx 0\,.
\end{equation}
Thus, the evolution of the Dirac observables is invariant under Dirac's gauge transformations.

Note that the Dirac observables form an algebra on the constraint surface since, as can be seen from the definition above, linear combinations of observables and products of observables are also observables. Counting the dimension of a complete generating set for this algebra therefore gives a natural at-a-time degree of freedom count on phase space. This is often the degree-of-freedom count offered by physicists in standard gauge theory analysis. Crucially, Dirac's definition of gauge symmetries involve transformations that only change the values of the arbitrary functions $v^\alpha$ and his definition of observable is just any phase space function whose evolution is independent of the $v^\alpha$. This gives a more refined statement of the general observation made in \Sec\ref{ssub:the_dirac_algorithm}.

\subsection{Recovering Dirac's formalism} 
\label{sub:recovering_dirac_s_formalism}


To recover Dirac's formalism from the geometric first-order approach developed in \Sec\ref{sec:the_initial_value_problem}, it is helpful to take note of the differences between these approaches. First, our formalism was expressed on an unrestricted velocity phase space, whereas Dirac's approach is based on restricting the Hamiltonian flow to a constraint surface on phase space. Second, Dirac's algorithm makes no initial assumptions about the symmetry properties of the original action. Instead, the symmetries are derived by imposing consistency conditions that keep the dynamical flow on the image of Legendre transform. From the point of view of Noether's methodology, the number of iterations of the Dirac algorithm is fixed by the order of the Noether-$2$ symmetry. And the closure of the Dirac algorithm is guaranteed by the assumption that the action has a particular Noether-$2$ symmetry (and well-defined saddle points).

I will restrict attention to transverse variations because, as we saw in \Sec\ref{sub:tangential_variations}, tangential variations generate over-a-history reparametrisations that are best studied on velocity phase space. Because of the arguments made in that section I will consider the tangential case to be problematic for the Dirac picture.

With these considerations in mind, we can start to connect the two formalisms by identifying the primary constraints of our formalism with those that arise in the first step (modulo second-class constraints) of the Dirac algorithm. Since Dirac is using explicit coordinate expressions, we can compare his structures to ours using the explicit form of the $u$ vectors from \eqref{eq:u expansion} and pullback all expressions by $\mathbbm L$. For the primary constraints (i.e., the second term of \eqref{eq:can constraints}) this is straightforward. We thus identify
\begin{equation}\label{eq:primary dirac recov}
    \phi_\alpha = \T 1 i p_i\,.
\end{equation}
In Dirac's approach, the secondary constraints are obtained by requiring that the time evolution of the primary constraints is zero on the primary constraint surface. In our formalism the time evolution is given, on-shell, by the $a = 1$ component of \eqref{eq:prop constraint}. This tells us that Dirac's secondary constraints are just given by the pullback of the Lagrangian constraints by $\mathbbm L$, or
\begin{equation}
    \chi_\alpha = \T 1 i K_i\Big|_{p_i = \diby{L}{\dot q^i}}\,.
\end{equation}
This reproduces the well-known connection between Lagrangian constraints and secondary constraints \cite[p. 55]{sundermeyer:1982}.

In Dirac's approach, the number of secondary constraints is not assumed to match the number of primary constraints. Moreover, there may be more constraints generated by the propagation of the secondary constraints. However, in my approach, which uses Noether's methodology and assumes that all second-class constraints have been eliminated, there will be a match between the primary and secondary constraints unless the secondary constraints happen to be trivial. Furthermore, Dirac's algorithm is guaranteed to close because of Noether's second theorem, which says that $ \deby{}t\lf( \T 1  i K_i \rt) \approx \T 0 i \alpha_i $ on the primary constraint surface, and the remaining constraints, which imply $\T 0 i \alpha_i \approx 0$ following the arguments of \Sec\ref{sub:lagrangian_constraints_and_noether_s_first_theorem}. This recovers the output of the Dirac algorithm for the kinds of theories I've restricted to.


\section{Dynamical similarity} 
\label{sec:dynamical_similarity}

\subsection{Generating dynamical similarity} 
\label{sub:generating_dynamical_similarity}


In \Sec\ref{sub:dynamical_similarity}, I defined dynamical similarity as a transformation on the kinematical structures of a theory that rescales the action $S \to c S$ for some non-zero constant $c$. I will now give a more precise definition in terms of a transformation generated by a vector field on velocity phase space that is a broad symmetry of a large class of Lagrangian theories. The first attempt to write dynamical similarities as generated by a vector field was in \cite{Sloan:2018lim}. This definition was further refined in \cite{bravetti2022scaling}. My definition will be slightly more general than that given in \cite{bravetti2022scaling}\footnote{ It will work for arbitrary values of parameter $a$ defined below. } and will be more geometric than in \cite{Sloan:2018lim}. Specifically, my construction will aim to develop a generalisation of the Gauge Principle that will work for dynamical similarity in contrast to the reduced view taken in these earlier works.

As argued in \Sec\ref{sub:dynamical_similarity}, any transformation that rescales the action will preserve the stationarity condition of the action since $\delta S = 0 \to c \delta S = 0$. These transformations therefore map DPMs to DPMs and are broad symmetries. Note that such transformations are only required to rescale the action on-shell for them to qualify as broad symmetries. We will see that the on-shell requirement is a necessary for dynamical similarities to be generated by vector fields on velocity phase space.

I will now give the conditions for the existence of a family of vector fields, which I will call $D_\phi$, on velocity phase space that are parametrized by a single velocity phase space function $\phi$ and rescale the action when evaluated along a classical solution. I will also show more explicitly that the flow of this vector field preserves the classical solutions up to a global reparametrisation. I will restrict to regular Lagrangian theories that have a well-defined symplectic form. This in no way limits the procedure since all of my arguments can be applied to the gauge-fixed evolution of an irregular Lagrangian theory. What I will find is that, for many theories, the conditions for the existence of $D_\phi$ put constraints on $\phi$ which fix $D_\phi$ uniquely. Thus, this formalism can be seen as general way of inferring the unique dynamical similarity present in a large class of dynamical systems.

My starting point, once again, will be Equation~\eqref{eq:S var} for the variations of the first order action $S_1$ in the direction of a vector field $u$ on $\Gamma$. I will then define a dynamical similarity to be a transformation generated by the vector field $D$ such that
\begin{equation}\label{eq:DS vec def}
    \Lie_D S_1[\gamma] = S_1[\gamma]\,,
\end{equation}
where $\Lie_u S_1[\gamma] = 0$ for all $\gamma$ and transverse $u$ that vanish on the boundary. As we've proved, the last requirement states that $\gamma$ is tangent to the vector field $X$ obeying Hamilton's equation: $\iota_X \omega_L + \de H = 0$. If $f$ is a point on the image of $\gamma$, this means that
\begin{equation}\label{eq:int curves}
    \dot f = \Lie_X f\,.
\end{equation}
Absorbing the boundary terms into the first line of \eqref{eq:S var} tells us that \eqref{eq:DS vec def} reduces to
\begin{equation}\label{eq:DS vec def2}
    \int_\gamma \lf( \Lie_D \theta_L - \lf( \Lie_D H \rt) \de t - H \Lie_D \de t \rt) = \int_\gamma \theta_L - H \de t \,.
\end{equation}
When $\theta_L$, $H$ and $\de t$ are independent, we find that \eqref{eq:DS vec def2} has solutions when
\begin{align}
    \Lie_D \theta_L &= \theta_L \label{eq:theta DS} \\
    \Lie_D H &= a H \label{eq:H DS}\\
    \Lie_D \de t &= (1-a)\de t \label{eq:t DS}
\end{align}
for some arbitrary constant $a$. Recall that we defined $\de t$ in terms of trial curves with tangents $X$ such that $\iota_X \de t = 1$ and, when Hamilton's equations are satisfied, the vector $X$ depends on $H$. Using the fact that $\Lie_D \lf( \iota_X \de t \rt) = 0$, we can invert Hamilton's equations to prove that, when $\Lie_D \theta_L = \theta_L$ and $\Lie_D H = a H$, $\Lie_D \de t = (1-a)\de t$. This means that the latter condition is a consequence of the former when $X$ generates classical solutions. Only the two independent conditions \eqref{eq:theta DS} and \eqref{eq:H DS} are therefore required to define $D$. 

I will now show that a vector field $D$ satisfying \eqref{eq:theta DS} and \eqref{eq:H DS} will preserve the integral curves of $X$ up to a constant time parameter when $X$ satisfies Hamilton's equations. An immediate consequence of \eqref{eq:theta DS} is that
\begin{equation}\label{eq:omega DS}
    \Lie_D \omega_L = \omega_L\,,
\end{equation}
where we recall that $\omega_L = \de \theta_L$. Since we have assumed that our Lagrangian is regular (or can be made regular by gauge-fixing), then $\omega_L$ is non-degenerate, and we can write $X = -\omega^{-1}(\de H)$. A straightforward computation then shows that
\begin{equation}
    \Lie_D X = (1-a)X\,.
\end{equation}
The integral curves defined by \eqref{eq:int curves} are therefore invariant under the flow of $D$ if we redefine $t$ such that $ \de \bar t = (a - 1) \de t$. We thus see that the dynamical similarities generated by $D$ do indeed map DPMs to DPMs (up to a global time rescaling).

We now turn our attention to a procedure for finding $D$ as a solution to \eqref{eq:theta DS} and \eqref{eq:H DS}. We can guess at a solution by replacing \eqref{eq:theta DS} by the weaker condition \eqref{eq:omega DS}. Using the fact that $\omega_L$ is non-degenerate, we find that $D$ can be obtained from \eqref{eq:omega DS} by solving
\begin{equation}\label{eq:D def}
    \iota_D \omega = \theta_L + \de \phi\,
\end{equation}
for the arbitrary phase space function $\phi$. Contracting this with $D$ and inserting the result into the stronger condition \eqref{eq:theta DS} tells us that $\phi$ must satisfy
\begin{equation}\label{eq:phi cond}
    \Lie_D \phi = \phi\,.
\end{equation}
This gives us a means to solve explicitly for a family of vector fields $D$ parametrized by the function $\phi$.

The second condition \eqref{eq:H DS} on $D$ depends on the explicit choice of the Hamiltonian function $H$, and therefore does not have a general closed-form solution. However, we can usefully re-write this condition using the solution of $D$ from \eqref{eq:D def}. First we note that, for a general velocity phase space function $f$, \eqref{eq:D def} implies
\begin{equation}
    \Lie_D f = \omega^{-1}(\theta_L, \de f) + \pb \phi f\,,
\end{equation}
where $\pb \cdot \cdot$ is Poisson bracket constructed from the inverse of $\omega_L$. The condition \eqref{eq:H DS} then reduces to
\begin{equation}\label{eq:H condition}
    \omega^{-1}(\theta_L, \de H) + \pb \phi H = a H\,.
\end{equation}
Using Hamilton's equations we can write $\omega^{-1}(\theta_L, \de H) = \iota_X \theta_L$, which gives the first term in a form that is independent of $H$. For most theories, the condition \eqref{eq:H condition} can be seen as a way of fixing the function $\phi$. In fact, we will see that, for the examples of theories that will be important for the analysis in this work, \eqref{eq:H condition} will fix $\phi$. Equation~\ref{eq:H condition} is therefore a strong constraint on $D$. In fact, there is no guarantee in general that \eqref{eq:H condition} has solutions at all. We will see, however, that solutions can be found for a large class of physically interesting theories.

Before doing this, it is helpful to write our coordinate-free expressions in a convenient set of coordinates. This will allow us to connect to previous results and will provide a useful computational tool for solving specific examples. Because we have assumed a regular Lagrangian, the Legendre transform is invertible. There is thus a one-to-one correspondence between velocity phase space and phase space. We therefore find it convenient to write our expressions using Darboux coordinates $(q^i, p_i)$ such that
\begin{align}\label{eq:omega Darboux DS}
    \theta_L &= p_i \de q^i & \omega_L = \de p_i \wedge \de q^i\,.
\end{align}
In these coordinates, \eqref{eq:D def} has the solution
\begin{equation}\label{eq:D def2}
    D = p_i \diby {}{p_i} + \pb{\phi}{\cdot}\,.
\end{equation}
The Lie drag of $D$ on a phase space function $f$ then takes the form
\begin{equation}
    \Lie_D f = p_i \diby{f}{p_i} + \pb \phi f
\end{equation}
and the condition \eqref{eq:H condition} becomes
\begin{equation}\label{eq:H cond2}
    p_i \diby{H}{p_i} + \pb{\phi}{H} = p_i v^i + \pb{\phi}{H} = a H\,.
\end{equation}

\subsubsection{Example: homogeneous potentials} 
\label{subs:example_homogeneous_potentials}

We can find explicit solutions for $\phi$, and therefore $D$, if we restrict to Hamiltonians of a particular form. The simplest case, which will also prove to be general enough for our considerations in this work, is when the Hamiltonian has the form
\begin{equation}
    H = \tfrac 1 2 g^{ab}(q) p_a p_b + V(q)\,,
\end{equation}
where the potential function, $V(q)$, is homogeneous of degree $n$ in $q^a$ and the inverse of the kinetic metric, $g^{ab}(q)$, is homogeneous of degree $m$ in $q^a$. Euler's homogeneous function theorem says that, under these conditions,
\begin{align}\label{eq:homo conds}
    q^a \diby{V}{q^a} &= n V & q^a \diby{g^{bc}}{q^a} &= m g^{bc}\,.
\end{align}
The ansatz
\begin{equation}
    \phi = k q^a p_a\,,
\end{equation}
where $k$ is a constant, solves the condition \eqref{eq:phi cond} for all $k$. The homogeneity conditions \eqref{eq:homo conds} further tell us that \eqref{eq:H cond2} has solutions for this ansatz when
\begin{align}
    a &= \frac {2n}{n-m+2} & k &= \frac {2}{m - n - 2}\,.
\end{align}
This fixes both the form of $\phi$ and the transformation properties of $H$ and $\de t$ under $D$. Inserting our solution into \eqref{eq:D def2} gives
\begin{equation}
    D = \lf(\frac{ 2 }{ n - m + 2}\rt) q^a \diby{}{q^a} + \lf(\frac{ n - m }{ n - m + 2 }\rt) p_a \diby{}{p_a}\,,
\end{equation}
which, as we can easily verify, has the required properties.


\subsection{A Gauge Principle for dynamical similarity} 
\label{sub:gauge_principle_for_DS}

In the previous section, we gave conditions that can be used to define a vector field (or family of vector fields) on phase space for generating dynamical similarities when they exist. The transformations generated by $D$ where seen to preserve DPMs up to a constant rescaling of the time parameter. Because we restricted to regular Lagrangian theories, the equations of motion for the system where assumed to be invertible for all DPMs --- including those related by a dynamical similarity. This means that, in such theories, there is no underdetermination caused by the dynamical similarities. I have so far hinted that my proposal will be to treat gauge symmetries at the formal level as transformations associated with underdetermination. It follows, therefore, that my proposal will regard the dynamical similarities in theories of this kind as non-gauge symmetries.

I will, however, argue later that there are good epistemic reasons to treat dynamical similarities as gauge symmetries in cosmology. For models of this kind, my proposal requires a procedure for converting a theory that treats dynamical similarity as a non-gauge symmetry into one that treats dynamical similarity as a gauge symmetry. I would thus like to apply something like the Gauge Principle to dynamical similarity.

Unfortunately, existing implementations of the Gauge Principle work only for symmetries that preserve the symplectic structure of the corresponding Hamiltonian system. This is because those implementations start from modifications of the Lagrangian, which is used to generate equations of motion using Hamilton's principle. But dynamical systems of this kind are always non-symplectic --- as we've seen from our general derivations in \Sec\ref{sub:hamiltons_equations}. The difficulty with dynamical similarity is the fact that such transformations rescale the symplectic potential and $2$-form according to \eqref{eq:theta DS} and \eqref{eq:omega DS}. These are immediate consequences of the definition of a dynamical similarity, as I've shown. I thus need to develop a version of the Gauge Principle that works for symmetries that do not respect the symplectic structure of the corresponding Hamiltonian system.

To do this, I will \emph{not} use modifications of the Lagrangian to generate new equations of motion via Hamilton's Principle. Instead, I will directly modify the corresponding Hamiltonian system of the original theory in a way that identifies all elements in the gauge orbits generated by $D$.\footnote{ An alternative approach is to construct a modified variational principle different from Hamiltonian's principle that implements dynamical similarities explicitly. This can be done using \emph{Herglotz's Principle} as in \cite{sloan2023herglotz,sloan2021new,sloan2021scale}. Showing the equivalence between the two approaches is interesting future work. } This will effectively treat phase space as a fibre bundle with one dimensional fibres that are the orbits of dynamical similarity. Because we are quotienting phase space by a one-dimensional group, the base space on which our invariant dynamics is defined cannot be a symplectic manifold. We will see that the base space is, in fact, a contact manifold with a natural contact form.\footnote{ For a general introduction to mechanical systems in contact spaces, see \cite{bravetti2017contact}. } The projection of the original symplectic flows onto this space will be seen to be contact flows with a contact Hamiltonian related to the Hamiltonian of the original symplectic system. Crucially, neither the value of the contact Hamiltonian nor the natural volume form on the contact space are guaranteed to be preserved by the contact flows. This observation will be the central component of my proposed solution to the problem of the AoT.

To implement dynamical similarity as a gauge symmetry, I will project the flow of $X$ onto surfaces that are everywhere transverse to the flow of $D$ in a time parametrisation that is invariant under the action of $D$. Restricting the flow in this way will pick a single element from each of the orbits of $D$ along the dynamical flow of $X$. This will ensure that all quantities that are invariant under $D$ have identical dynamics, up to a time parametrisation, and that the quantity that changes only under $D$ is arbitrary. This arbitrariness arises from the choice of transverse surface, which can be arbitrarily chosen anywhere along the integral curves of $D$.

The procedure above can be implemented for any choice of transverse surface $G_w = \{x\in \Gamma| w(x) = w_0  \}$ for some smooth phase space function $w: \Gamma \to \mathbbm R$ and constant $w_0 \in \mathbbm R$ such that
\begin{equation}\label{eq:w def}
    \Lie_D w = 1\,.
\end{equation}
We can define the projected flow, $X_\parallel$, of $X$ onto $G_w$ by removing the transverse component according to:
\begin{equation}
    X_\parallel = X - \lf(\Lie_X w \rt) D\,.
\end{equation}
We can use the function $w$ and the scaling properties of $\de t$ from \eqref{eq:t DS} to define a $D$-invariant time parameter $\tau$ using
\begin{equation}
    \de \tau = e^{(a-1)w} \de t\,,
\end{equation}
where we recall that $a$ is the scaling of $H$ under a dynamical similarity. The flow equation \eqref{eq:int curves} then projects to the invariant flow equation
\begin{equation}\label{eq:gf flow}
    \frac {\de f}{\de \tau} = \Lie_{\Xinv} f \equiv f'\,,
\end{equation}
where
\begin{equation}\label{eq:gf X}
    \Xinv = e^{(1-a)w} X_\parallel =  e^{(1-a)w} \lf( X - \lf(\Lie_X w \rt) D \rt)\,
\end{equation}
and where we have used primes to indicate derivatives with respect to $\tau$. It is straightforward to check that $[\Xinv, D] = 0$, where $[\cdot, \cdot]$ is the Lie bracket on $\Gamma$.

Equations~\ref{eq:gf flow} and \ref{eq:gf X} define an invariant flow on the gauge-fixed surface $G_w$. In principle, this is all that one needs in order to produce $D$-invariant equations of motion on phase space. Note that these equations of motion depend on the arbitrary values of $w$ that, by virtue of the transversality condition \eqref{eq:w def}, transform non-trivially under dynamical similarity. The theory defined by this projected flow therefore meets Dirac's criterion for calling dynamical similarity a gauge symmetry. This establishes my main goals for this section.

It is extremely valuable, however, to characterise the invariant flows we've obtained more geometrically. We can do this by using the gauge-fixing conditions, flow equations, and geometric structures on phase space to define intrinsic geometric structures on the gauge-fixed surfaces $G_w$. This will allow us to establish the geometric properties of the flows on the base space obtained by quotienting out by the action of the dynamical similarities.

Towards this end, note that we can define invariant structures in terms of $\omega_L$, $\theta_L$ and $H$ by using their transformation properties under $D$ and the transversality of $w$. This can be achieved by noting that $w$ is translated under a dynamical similarity so that one can construct invariant quantities, $\bar Q$, by forming products with exponentials that involve the appropriate scaling of $Q$ under $D$; i.e., $\bar Q = e^{-s w} Q$, where $\Lie_D Q = s Q$. Using this procedure, we can define the invariant $1$-form
\begin{equation}
    \eta_\Gamma = -e^{-w} \iota_D \omega_L = e^{-w} \omega_L(\cdot, D)\,\label{eq:eta def}
\end{equation}
on all of $\Gamma$ and its restriction, $\eta$, to $G_w$:
\begin{equation}
    \eta = \eta_\Gamma\big|_{G_w}\,.
\end{equation}
Because $\eta_\Gamma$ is $D$-invariant and $G_w$ is transverse to $D$, $\eta$ is independent of the choice of $w$. On phase space, the function $\de \eta_\Gamma$ has a two-dimensional kernel spanned by $D$ and the vector
\begin{equation}
    R_\Gamma = e^{w}  \omega_L^{-1}(\de w) = e^{w} \pb{w}{\cdot}\,.
\end{equation}
But, when restricted to the gauge-fixed surface, the kernel of $\de\eta$ is just the vector
\begin{equation}
    R = R_\Gamma\big|_{G_w}\,.
\end{equation}

To see that $R_\Gamma$ is indeed in the kernel of $\de \eta_\Gamma$, we can use the scaling of $\omega_L$ under a dynamical similarity to prove the useful relation
\begin{equation}\label{eq:d eta}
    \de \eta_\Gamma = - e^{-w}\omega_L - \de w \wedge \eta_\Gamma \,.
\end{equation}
Using the fact that $\Lie_{R_\Gamma} w = 0$, which follows from the anti-symmetry of $\omega_L$, and
\begin{equation}\label{eq:R norm}
    \iota_{R_\Gamma} \eta_\Gamma = 1\,,
\end{equation}
which follows from the definitions of $R_\Gamma$ and $\eta_\Gamma$, we get
\begin{equation}\label{eq:R kernel}
    \iota_{R_\Gamma} \de \eta_\Gamma = 0\,.
\end{equation}
Note that $\Lie_R w =0$ can also be used to show that $[R,D] =0$ using $\Lie_D w = 1$. This implies that the restriction $R$ to $G_w$ is $D$-invariant. The two conditions \eqref{eq:R norm} and \eqref{eq:R kernel} then tell us that $\eta$ is contact $1$-form with Reeb vector field $R$. (See, \cite{bravetti2017contact}, for definitions of the Reeb field and other contact structures.)

We can now use the contact structures; i.e., $\eta$ and $R$; defined by $D$ and $\omega_L$ to show that $\Xinv$ is a contact flow on the surface $G_w$. Since we've already shown that $\Xinv$, $\eta_\Gamma$, and $R_\Gamma$ are $D$-invariant, it is easier to do all calculations on $\Gamma$ and then restrict to $G_w$ when required. For this reason, we will work on $\Gamma$ below and suppress $\Gamma$ subscripts for convenience.

Following Definition~11 of \cite{bravetti2022scaling} and proofs given therein, we consider a contact manifold with contact $1$-form $\eta$ and Reeb vector field $R$. We then define a \emph{$\Lambda$-contact vector field} $Y$ generated by the function $y$ as the \emph{unique} vector field satisfying the two conditions
\begin{align}\label{eq:CVF def}
    \iota_Y \de \eta &= \de y - R(y)\eta & \iota_Y \eta &= -\Lambda y\,.
\end{align}
This generalises the notion of Hamilton vector field on a symplectic manifold. It also generalises the notion of a contact vector field for an arbitrary non-zero constant $\Lambda$. When $\Lambda = 1$, we recover a standard contact flow. We will see below that, for $\Lambda$-contact vector fields as for standard contact vector fields, the relation between $Y$ and $y$ is bijective (at least locally on the contact space) so that a $\Lambda$-contact vector field $Y$ also defines a unique function $y$ through \eqref{eq:CVF def}.

We will now show that $\Xinv$ satisfies \eqref{eq:CVF def} for the $\Lambda$-contact Hamiltonian
\begin{equation}\label{eq:contact H}
    H_c \equiv e^{-a w} H\,,
\end{equation}
for $\Lambda = a$. The second relation,
\begin{equation}\label{eq:Hc def}
    \iota_{\Xinv} \eta = - a e^{-a w} H = - a H_c\,,
\end{equation}
follows trivially from the definitions of $\eta$ and $\Xinv$ as well as Hamilton's equations. To show that the first relation holds, we can make use of the helpful result that
\begin{equation}\label{eq:friction}
    R(H_c) = e^{(1-a)w} \pb{w}{H} = e^{(1-a)w} \Lie_X w\,,
\end{equation}
which tells us that
\begin{equation}
    \Xinv = e^{(1-a)w} X - R(H_c) D\,.
\end{equation}
The quantity $R(H_c)$ is the Reeb flow of the contact Hamiltonian. It will play a central role in our discussions because, as can be seen from the form of $\Xinv$ above, it determines the amount of deviation in the projected system from a conservative symplectic system. Indeed, because $D$ generates a local rescaling of the angular momentum scale, $R(H_c)$ will play the role formally similar to a drag coefficient in a mechanical system. Following this analogy, I will call $R(H_c)$ the \emph{drag} consistent with our earlier terminology.

The formulas \eqref{eq:d eta} and \eqref{eq:friction} can then be used in combination with the definition of the contact Hamiltonian \eqref{eq:contact H} and Hamilton's equations to derive
\begin{equation}\label{eq:flow def}
    \iota_{\Xinv} \de \eta = \de H_c - R(H_c) \eta\,.
\end{equation}
Thus, when restricted to $G_w$, $\Xinv$ is indeed the $a$-contact flow generated by $H_c$.

It is useful to write the $D$-invariant flow equations \eqref{eq:gf flow} for the contact Hamiltonian, the contact form and the natural measure on $G_w$. Using the above definitions and, in particular, the relation \eqref{eq:friction}, we find
\begin{equation}\label{eq:Hc_eom}
    H_c' = \Lie_{\Xinv} H_c = - a R(H_c) H_c\,.
\end{equation}
Thus, the contact Hamiltonian is only preserved if $w$ was a constant of motion in the original symplectic system (i.e., when $\Lie_X w = 0 \Rightarrow R(H_c) = 0$). We see that the quantity, $R(H_c)$, which I have called the drag, is proportional to the decay (or growth depending on its sign) coefficient of the total energy in analogy to the drag coefficient of a mechanical theory. Note also that we get a slight modification to the usual evolution equation for the contact Hamiltonian in a standard contact system where $a = 1$. This difference with a standard contact system is more significant when we look at the $D$-invariant flow of $\eta$. In this case, we can use the defining relations, \eqref{eq:Hc def} and \eqref{eq:flow def}, of an $a$-contact flow to show that
\begin{equation}
    \eta' = \Lie_{\Xinv} \eta = (1-a) \de H_c - R(H_c) \eta\,.
\end{equation}
The extra term proportional to $\de H_c$ is zero for a standard contact system where $a = 1$.

A natural volume form can be constructed from the contact form $\eta$. It is straightforward to show that the kernel of $\eta$ is such that the top-form
\begin{equation}
    \rho \equiv \eta \wedge \de \eta^{n-1}\,,
\end{equation}
where $n$ is the dimension of the configuration space of the original symplectic system, is non-degenerate. The formula \eqref{eq:flow def} can then be used to show that
\begin{equation}\label{eq:mu evo gen}
    \rho' = \Lie_{\Xinv} \rho = (1-a) \de H_c \wedge \de \eta^{n-1} - n R(H_c) \rho\,.
\end{equation}
This reproduces the standard contact evolution when $a = 1$. Note that, in general, this measure will evolve along the dynamical flow. In particular, the quantity $R(H_c)$ that I have called the drag acts as a decay coefficient for the measure $\mu$. We see that the drag quantifies the amount of non-conservative behaviour in the system in the form of loss (or gain) of energy and focusing (or de-focusing) of solutions. The drag will thus be the central quantity I will consider in later discussions about the arrow of time.

As a final step, we can write our flow equations using special coordinates on $G_w$. Since the gauge-fixed surface is a contact space, it is possible (see, for example, \cite{bravetti2017contact}) to write the contact form locally in the Darboux coordinates $(S, Q^a, P_a)$ as
\begin{equation}
    \eta = \de S - P_a \de Q^a\,,
\end{equation}
where the index $a$ runs over one less value than the index $i$ used for the Darboux coordinates, \eqref{eq:omega Darboux DS}, on $\Gamma$. We note that the equations \eqref{eq:R norm} and \eqref{eq:R kernel}, which define the Reeb vector $R$, imply that
\begin{equation}
    R = \diby{}{S}\,.
\end{equation}
Inserting these explicit coordinate expressions into \eqref{eq:Hc def} and \eqref{eq:flow def} tells us that
\begin{equation}
    \Xinv = \lf( -a H_c + \diby{H_c}{P_a} \rt) \diby{}{S} + \diby{H_c}{P_a} \diby{}{Q^a} - \lf( \diby{H_c}{P_a} + \diby{H_c}{S} P_a \rt) \diby{}{P_a} \,.
\end{equation}
The contact flow equations, \eqref{eq:gf flow}, then become
\begin{align}
    Q^{a\prime} &= \Xinv(Q^a) = \diby{H_c}{P_a} \notag \\
    P^{\prime}_a &= \Xinv(P_a) = - \lf( \diby{H_c}{Q^a} + \diby{H_c}{S}P_a \rt) \notag \\
    S' &= \Xinv(S) = - a H_c + \diby{H_c}{P_a} P_a\,.\label{eq:Darboux contact}
\end{align}
Note the factor of $a$ in the equation of motion for $S$ that is equivalent to the standard contact case when $a = 1$. The system above differs from a symplectic system when the drag, $R(H_c) = \diby{H_c}{S}$, is non-zero and because of the extra coordinate, $S$, along the Reeb direction. We can see from the form above that the drag is indeed analogous to the drag coefficient of a mechanical system described by the symplectic sub-system $(Q^{a}, P_a)$. This is consistent with our general findings that the drag is proportional the decay coefficients for the contact Hamiltonian, $H_c$, and the density $\rho$.


\subsection{Measures and counting solutions} 
\label{sub:measures_on_counting_solutions}

When it comes to counting solutions in a symplectic or contact theory, a useful tool is the natural volume-form defined by the differential structures of these manifolds. This volume-form can be used to define a \emph{measure} $\mu$ on the relevant state space. A measure $\mu$ is a non-negative function $\mu:\Sigma \to \mathbbm R_{\geq 0}$ on a Borel $\sigma$-algebra, $\Sigma$, over a topological space, $X$, that is countably additive and satisfies $\mu(\emptyset) = 0$.\footnote{ More concretely, a Borel $\sigma$-algebra on a topological space $X$ is the collection of all open sets of $X$ that are closed under countable union, countable intersection and relative complement. Countable additivity is defined as $\mu\lf(\bigsqcup\limits_i R_i\rt) = \sum\limits_i \mu(R_i)$ for a \emph{countable} collection $\{R_i\}_{i=1}^\infty$ of disjoint sets. The set $\emptyset$ is the empty set. } In our case, $X$ is a symplectic or contact manifold so that $\mu(R)$ is a function of some open set $R\in \Sigma$ in $X$. A straightforward way to obtain a measure in this case is to take integrals of the natural density defined by the symplectic or contact structures over the regions $R$.

For a symplectic theory, this natural density is the Liouville form
\begin{equation}\label{eq:Liouville form}
    \rho_L = \omega^{n/2}\,,
\end{equation}
where $\omega$ is the symplectic $2$-form of the symplectic geometry in question and $n$ is its dimension (which must be even). The natural measure is then the Liouville measure
\begin{equation}
    \mu_L(R) = \int_R \rho_L\,.
\end{equation}
For a contact theory, the natural density is
\begin{equation}\label{eq:contact vol}
    \rho_c = \eta\wedge \de \eta^{(n-1)/2}\,.
\end{equation}
This can be used to construct a natural contact measure
\begin{equation}
    \mu_c (R) = \int_R \rho_c\,,
\end{equation}
which is, perhaps confusingly, often \emph{also} called the Liouville measure --- but of a contact space. I will try to reverse the term \emph{Liouville measure} for the natural measure of a symplectic theory.

Because of the defining features of a measure given above (particularly its non-negativity but also its additivity), a measure can be understood as attributing a size, usually called a weight, to a region $R$. This weight is often loosely thought of as a count of the number of elements in $R$. In this sense, the measures $\mu_L$ and $\mu_c$ give a counting of the states in a region $R$ of state space. This is not the same thing as a counting of solutions. But, as we will see, it can be used to construct one. Still, it is important in the discussion below to distinguish the properties of the natural measure on state space, which count states, with the measures used to count solutions constructed from them. Some of these properties are inherited but not all --- particularly in the case we'll be interested in where the measure is not translation-invariant.

The Liouville measure of a symplectic system has several important properties that make it a useful measure on phase space. In Darboux coordinates $(q^i,p_i)$, we have that $\omega= \de p_i \wedge \de q^i$ --- a fact that can be used to see that the Liouville measure is invariant under translations of the Darboux coordinates. In this sense, $\mu_L$ is a Lebesgue on $\Gamma$ in these variables. Sometimes, this fact has been used to motivate the Liouville measure using a principle of indifference. But perhaps a more mathematically and philosophically motivated justification relies on the invariance of the measure under \emph{any} symplectic flow. Because a symplectic flow generated by a vector field $Y$ satisfies $\iota_Y \omega = \de y$ for non-degenerate $\omega$ and some phase space function $y$, the invertibility of $\omega$ leads directly to
\begin{equation}
    \Lie_Y \omega = 0\,
\end{equation}
for \emph{any} function $y$. This means that the Liouville form of a symplectic theory is preserved under the dynamical flow of \emph{any} choice of Hamiltonian function:
\begin{equation}\label{eq:Liouville t-constant}
    \dot \rho_L(R) = \Lie_Y \rho_L(R) = 0,\, \forall y \,.
\end{equation}
This result is usually called \emph{Liouville's theorem} after a similar result first derived by Liouville \citep{liouville1838note}. It is an incredibly powerful result precisely because it implies an invariance property of a \emph{density} (not just a measure) on phase space under a very large class of transformations. One can think of Liouville's theorem as a universality argument for the dynamical invariance of the natural volume form under any choice of symplectic evolution.

\subsubsection{Counting solutions in a symplectic theory} 
\label{ssub:counting_solutions_in_a_symplectic_theory}


The Liouville measure can be used to count solutions in a symplectic theory. To do this, a surface transverse to the dynamical flow is needed. Such a surface can be obtained by finding a function $\tau$ such that
\begin{equation}\label{eq:t cond}
    \Lie_X \tau \neq 0\,.
\end{equation}
Such a surface will intersect the integral curves of $X$ exactly once. Its volume under some measure can then be used to count solutions. I will ignore global issues to keep the discussion as simple as possible. But note that it is, in general, difficult to explicitly find surfaces of this kind that will intersect \emph{all} the different branches of the solutions of a theory (e.g., solutions with different energies) and that stay well-defined throughout the entire evolution. Given this caveat, note that a simple application of Darboux's theorem allows us to show that there exists, at least locally on phase space, a class of functions satisfying
\begin{equation}
    \Lie_X \tau_e = 1\,.
\end{equation}
I will call clocks of this kind \emph{ephemeris clocks}. Given this relation, we can write $X = \deby{}{\tau_e}$ so that $\Lie_X H = 0$ tells us that the Hamiltonian is $\tau_e$-independent. Such clocks are only defined up to canonical transformations on the level surfaces of $\tau_e$ since the vector fields generating such transformations will be transverse to $X$ by definition.

A very useful and common way to count solutions uses the Liouville form and an ephemeris clock, restricts $\rho_L$ to a level surface of $\tau_e$, and then integrates:
\begin{equation}
    \mu_{\tau_e}(R) = \int_R \rho_L \big|_{\tau_e = \text{const}}\,,
\end{equation}
for some region $R$ on the level surfaces of $\tau_e$. In this case, the time invariance of $\rho_L$ and the normality of $\tau_e$ implies $\dot \mu_{\tau_e}(R) = 0$. A counting of solutions, defined in this way, is a simple choice in the sense that the counting is the same for all values of $\tau_e$.

Note that the time independence of measures on the solution space is, in general, \emph{not} true for an arbitrary clock choice $\tau$. While the Liouville measure itself is time-dependent, its restriction, $\rho_L \big|_{\tau = \text{const}}$, onto an arbitrary co-dimension one submanifold \emph{can} depend on time. This is because the flow of $X$ won't be symplectically normal to the level surfaces of $\tau$ if $\Lie_X \tau \neq 1$ so that the embedding of these surfaces into $\Gamma$ can pick up a time dependence. Thus, non-ephemeris clocks will induce measures on the space of solutions that are not dynamically preserved despite Liouville's theorem. 

We can think of the flow of $X$ projected onto the surfaces of constant $\tau$ as a new flow on these surfaces. Let us call the Hamiltonian generating this flow $H_\tau$.\footnote{ In general, one can solve for $H_\tau$ by treating the conservation equation for the Hamiltonian as an equation for the momentum, $p_\tau$, conjugate to $\tau$. } A similar argument to the one made above for the measure can be used to show that, in general, only ephemeris clocks will have an evolution that preserves $H_\tau$. A description of a symplectic system in terms of a non-ephemeris $\tau$-clock is therefore non-conservative and will, in general, contain dissipative or anti-dissipative behaviour depending on the value of $\Lie_X \tau$.

The time invariance of the Liouville measure can then be understood as a way of privileging a certain class of clocks --- namely the ephemeris clocks --- for counting solutions because they lead to conservative, time-independent descriptions in terms of internal clocks. Such systems are motivated, among other things, by the simplicity of a time-independent Hamiltonian and the universal time invariance of the Liouville form.

One can gain further insight into the preferred role of time-independent descriptions by inquiring into the physical nature of ephemeris clocks. The condition $\Lie_X \tau_e = 1$ can be seen as a partial differential equation on phase space for $\tau_e$. Such an equation is equivalent to finding a variable, $\tau_e$, that is conjugate to an integral of motion. These variables are clocks that are dynamically isolated from the system because they behave exactly as isolated free particles. But for non-trivial theories, such clocks can be extremely complicated functions of phase space. In most cases, one can only find explicit integrals of motion when there are known symmetries in the system. Fortunately, for most everyday situations, it is relatively easy to find (or build) clocks that are sufficiently isolated from their environment --- either because of contingencies or the presence of approximate symmetries. Thus, the relative abundance of such clocks in everyday situations allows for approximately conservative, time-independent descriptions of these systems in terms of internal clocks and justifies the use of the Liouville measure for counting solutions (and not what it is usually used for, which is counting states).

The assumption of the existence of readily available, dynamically isolated clocks is manifest in theories with a fixed time parameter. This time parameter is assumed to be external to the system and have no dynamical effect on it. In fact, the time parameter is often modelled mathematically as a parameter on an extended phase space that is defined to by dynamically uncoupled from the original system. There are, however, situations in which no such convenient clocks exist. For such situations, one requires a theory that is reparametrisation invariant. This is especially true in general relativity where, as we will discuss more extensively in \Sec\ref{sub:time_laws_and_conventions}, integrals of motion are always non-local functions on phase space (unless the space-time has asymptotic global symmetries).

In reparametrisation invariant theories, the non-conservative nature of most internal clock parametrisations is apparent. As we proved in \Sec\ref{sub:reparametrisation_symmetry}, the image of the Legendre transform is restricted to a \emph{Hamiltonian constraint} surface. This means that the Hamiltonian takes the standard form $H = N\Ham = 0$, where $N$ is an arbitrary positive definite Lagrange multiplier. Theories of this kind are invariant under changes of the normalization of $\Ham$ of the form $\Ham \to f \Ham$, where $f>0$. This can be seen more explicitly by noting that, because $\de H = 0$, Hamilton's equations simply state that $X$ is in the kernel of $\omega$. On phase space, this is usually computed by solving $\iota_X \omega = \de \Ham$ and then restricting to the surface $\Ham = 0$. But this procedure is clearly invariant under the transformation
\begin{align}
    X &\to f X & \Ham \to f \Ham\,
\end{align}
when $\Ham = 0$. This reflects the fact that the Lagrange multiplier $N$ is arbitrary and, in particular, says that a rescaling of the Hamiltonian constraint leads to a rescaling of the time parameter along the integral curves of $X$.

An immediate consequence of the ability to rescale $\Ham$ is that the normality condition $\Lie_X \tau_e = 1 \to \Lie_X \tau_e = f^{-1}$ that we previously used to fix ephemeris clocks, is not invariant under reparametrisation. Fortunately, because we have a fixed $2$-form in a symplectic theory, we can nevertheless define an ephemeris clock by requiring that $\dot \mu_{\tau_e} =  \Lie_X \mu_{\tau_e} =0$. Because $\mu_{\tau_e}$ is only defined in terms of the restriction of $\omega$ to the level surfaces of $\tau_e$, it is invariant under reparametrisation so that the condition $\Lie_X \mu_{\tau_e} = 0 \to f \Lie_X \mu_{\tau_e} = 0$ is invariant. Thus, the symplectic $2$-form of a theory selects a preferred clock choice $\tau_e$ where the laws take a time-independent form.

Nevertheless, gauge-invariance under reparametrisation suggests that all time-independent descriptions are physically equivalent to the descriptions in terms of non-ephemeris clocks, which will be time dependent in general. Moreover, the time-dependent descriptions are highly fine-tuned to the choice of $X$, and are only computationally tractable when dynamically isolated degrees of freedom can be found. For generic systems, this is not possible. Thus, the simplicity argument for time-independent descriptions fails, and the only reasonable justification of such descriptions comes only from the universality properties of the Liouville measure. This significantly weakens the argument for privileging time-independent descriptions. Next, we will see that, when quotienting by dynamical similarity, the universality argument based the properties of the Liouville measure evaporates. Thus, the only good reason to favour time-independent descriptions is that they are convenient when reliable clocks are available. And even this argument evaporates in cosmology.

\subsubsection{Counting solutions in a contact theory} 
\label{ssub:counting_solutions_in_a_contact_theory}

Just as the Liouville measure, $\mu_L$, for a symplectic system can be used to count solutions in a symplectic theory, the Liouville measure, $\mu_c$, of a contact theory can be used to count solutions in a contact theory. As in the symplectic case, one can look for level surfaces of some function $\tau$ on the contact space that obey $\Lie_X \tau \neq 0$ and intersect solutions exactly once. In this case, however, even if one can find a function $\tau_e$ that obeys $\Lie_ X {\tau_e} = 1 $, the time evolution of the induced measure $\mu_{\tau} = \mu_c\big|_{\tau = \text{const}}$ will not be time independent. This is because the time evolution of $\eta$ in a contact system is, in general, not zero because of the terms on the RHS of \eqref{eq:mu evo gen}.

One could reverse engineer a clock that would induce a time-independent measure by requiring $\Lie_X \mu_{\tau_e} = 0$ as was done in the symplectic theory. For such choices of clock, the time dependence of the embedding of the level surfaces of $\tau_e$ would be required to exactly cancel the time dependence of $\mu_c$ due to \eqref{eq:mu evo gen}. This choice would, however, be extremely fine-tune to the choice of contact Hamiltonian causing both the simplicity and universality arguments present in the symplectic theory to fail. We thus conclude that time-independent measures are no longer favourable in contact systems. This can explain why contact systems have traditionally been used to model dissipative systems, whose descriptions are generally time dependent. 

The situation becomes even worse for time reparametrisation invariant theories. Consider the case we are interested in where the contact theory is obtained by quotienting a symplectic theory by dynamical similarity. In this case, let us recall that a gauge choice that fixes the level surfaces of some function $w$ obeying $\Lie_D w = 1$ leads to a contact form and Hamiltonian such that (see equations \eqref{eq:eta def}, \eqref{eq:Hc def} and \eqref{eq:friction})
\begin{align}
    \eta &= - e^{-w} \iota_D \omega & H_c &= e^{-aw} H & R(H_c) &= e^{(1-a)w}\Lie_X w\,,
\end{align}
for some flow $X$ in the symplectic theory. Consider now what happens when we change the gauge-fixing. To be concrete, consider the most general smooth transformation that preserves the condition on $w$:
\begin{equation}
    w \to w - \log f\,,
\end{equation}
where $\Lie_D f = 0$ and $f > 0$. Under such a transformation, the contact form, Hamilton, and measure focusing factor transform as
\begin{align}\label{eq:contact rep_sym}
    \eta &\to f \eta & H_c &\to f^a H_c & R(H_c) &\to f(a-1) \lf( R(H_c) - \Lie_{\Xinv} \log f \rt)\,.
\end{align}
Because we are quotienting by dynamical similarity, we must require that our theory be reparametrisation invariant so that the contact Hamiltonian $H_c$ is constrained to be zero. Using this constraint, the $a$-contact equations, (see equations~\eqref{eq:CVF def})
\begin{align}
    \iota_{\Xinv} \eta &= -a H_c & \iota_{\Xinv} \de \eta &= \de H_c - R(H_c) \eta\,,
\end{align}
which define the invariant flow $\Xinv$, are invariant provided $\Xinv$ transforms as
\begin{equation}
    \Xinv \to f^{(a-1)} \Xinv\,.
\end{equation}

The rescaling of $\Xinv$ by a positive function on the contact space means that different gauge choices for $w$ lead to reparametrisations of the integral curves of $\Xinv$ and highlights the importance of having reparametrisation invariance as a gauge symmetry when quotienting by dynamical similarity. This is rather remarkable because both the contact form and the Hamiltonian are re-scaled under this transformation. For the contact Hamiltonian, this is not a surprise because it corresponds to a different choice of normalization of the Hamiltonian constraint. But for the contact form, this means that the measure density $\rho_c$ can be re-scaled up to a free positive function on the contact space --- saturating all freedom in the measure. Since these contact forms all lead to equivalent solutions, reparametrisation invariance implies that there is no geometrically preferred measure in the resulting theory.

We can work out some of the consequences of the previous argument in a bit more detail by noting that the inhomogeneous transformation properties of $R(H_c)$ under changes of $w$ imply that the time evolution of the measure density transforms as
\begin{equation}\label{eq:measure ambiguity}
    \Lie_{\Xinv} \rho_c = \rho_c' = - n R(H_c) \rho_c \quad \to \quad \rho_c' = - n \lf( R(H_c) - \Lie_{\Xinv} \log f \rt) \rho_c\,
\end{equation}
when $H_c$ is constrained to be zero. The new term involving $\Lie_{\Xinv} \log f$ is only restricted by $f > 0$, and can be used to produce just about any possible amount of measure focusing or dissipation. In particular, if $f$ is chosen so that $w' = w - \log f$ is an integral of motion in the symplectic theory; i.e., if $\Lie_X w' = 0$; then $\rho_c' = 0$ and the measure is conservative. We see that, as in the symplectic theory, integrals of motion can be used to construct time-independent descriptions of the system on the contact space. Again, such descriptions are highly fine-tuned to the choice of dynamics and are difficult to identify without the presence of simple, dynamically isolated degrees of freedom. But unlike the symplectic case, there is no universality argument to privilege the Liouville measure of a contact theory. We conclude that the choice of measure and, in particular, the amount of time-dependence in the dynamics is entirely conventional in the contact theory.

I will return to this general problem in \Sec\ref{sub:time_laws_and_conventions} where I will discuss the implications of this conventionality for my proposed solution to the problem of the AoT. For now, note that the arguments in favour of time-independent descriptions, which were already under threat in the symplectic theory because of reparametrisation invariance, completely evaporate in the contact theory when obvious integrals of motion are unavailable. And as I will explain in \Sec\ref{sub:time_laws_and_conventions}, general relativistic solutions generically lack simple integrals of motion --- particularly in cosmological applications.
\chapter{The principle of essential and sufficient autonomy}
\label{ch:pesa}

\begin{abstract}
    In this chapter, I give my definition of gauge symmetry, which I propose as a solution to Belot's Problem. I motivate this definition by formulating a principle, called the Principle of Essential and Sufficient Autonomy (PESA). Using the examples introduced in \Sec\ref{sub:examples1}, I illustrate how my definition of gauge symmetry can be used to address the difficulties encountered in those examples. My proposal generalises the physical insights of Dirac and adds the representational insights of the DEKI account of reparametrisation presented in \Sec\ref{sub:DEKI account}. I end the chapter by applying my proposal to two novel cases: the Kepler problem, which illustrates a simple application to dynamical similarity, and the Frozen Formalism problem in reparametrisation invariant systems.
\end{abstract}

\ifchapcomp
    \tableofcontents
    \newpage
\else
    \cleardoublepage
\fi

\section{Motivation} 
\label{sec:motivation}

In this chapter, I will give my proposed solution to Belot's Problem and apply this solution to several notable examples, including those used in \Sec\ref{sub:examples1} to motivate Belot's Problem. I will also apply my proposal to globally reparametrisation-invariant systems and show how it solves the so-called \emph{frozen-formalism problem} in those theories. This illustrates how the proposal can be extended beyond the context of the standard examples considered in the philosophy literature. An early version of this proposal was developed in \Sec 2 of \cite{Gryb:2021qix}. Here, I will considerably develop upon this early proposal and refine it using the representation theory outlined in \Sec\ref{sub:DEKI account}.

Before stating the PESA, it is useful to take stock of what has been developed so far and give context to some of the results of Chapters~\ref{ch:en route gauge} and \ref{ch:rep_sym}. In those chapters, I began by considering two different ways to characterise symmetry. The first was in terms of locality conditions on the symmetry generators and the second was in terms transformation properties of the action. In \Sec\ref{ssub:symmetries_of_the_equations_of_motion}, I gave reasons to flatly reject definitions of gauge symmetry based on locality conditions because these definitions provide no concrete way of matching epistemic expectations to the nomic structures of a theory. In \Sec\ref{sub:symmetries_of_the_variational_principle}, symmetries based on properties of the variational principle were found to be more promising because the variational principle itself could be engineered, using some version of the Gauge Principle, to achieve the matching required. It wasn't clear, however, why certain properties of an action \emph{should} have anything to do with epistemic constraints. Indeed, I gave examples, such as the Noether-1 symmetries, that fit the conditions for being variational symmetries but should not be understood as gauge symmetries.

This led to a proposal by Dirac. In this proposal, outlined in \Sec\ref{sec:the_dirac_algorithm}, two states are related by a gauge transformation if the time evolution of quantities that change under the transformation involves a choice of arbitrary function.\footnote{ Note that, in Dirac' proposal, gauge transformations act on instantaneous states and thus correspond to at-a-time symmetries unlike the over-a-history symmetries considered by Noether and treated in many physics textbooks. } In terms of the equations of motion, this means that solutions are underdetermined by the freely specifiable initial data. Dirac has shown that, in Hamiltonian theories, this underdetermination can be linked to certain kinds of constraints --- first-class constraints --- and has given a general procedure for defining gauge transformations using these constraints.

In \chap\ref{ch:rep_sym}, I showed in detail how the constraints relevant to Dirac's notion of gauge symmetry and the identities of Noether's theorems can arise from the properties of the variational principle. I showed how the underdetermination of the equations of motion resulting from Dirac's constraints could be matched to well-known examples of gauge symmetries. This suggested that Dirac's analysis underpins all successful attempts to use variational symmetries to define gauge symmetries. Finally, I gave examples of how to implement different versions of the Gauge Principle to engineer actions that would possess specific Dirac-style gauge symmetries.

This led to a promising strategy for solving Belot's Problem in many contexts. There are, however, important drawbacks to Dirac's proposal. Firstly, Dirac's proposal is rather unsophisticated regarding the relationship between theory and phenomena.\footnote{ This is not to say that Dirac himself was unsophisticated about the relationship between theory and phenomena but rather that much was left unsaid in his writing in this regard. } Throughout his analysis, Dirac does not clarify how representations are to be related to physical phenomena and, in particular, does not define what he means by the `physical state' of a system. Secondly --- and perhaps more restrictive for our purposes --- is the fact that his proposal applies only to Hamiltonian theories.

Regarding the second drawback: while it is true that Dirac's proposal is tied to Hamiltonian systems, his motivation is much more general. Underdetermination of a system of equations is a formal property of general dynamical systems, and can therefore be used to pinpoint gauge symmetries more broadly. In this chapter, I will take this fact as the basis of a general definition of gauge symmetry.

But as I have pointed out repeatedly, no purely formal definition of gauge symmetry will be able to cover every possible way that a particular mathematical model can be used to represent features of the world. To address the first drawback, I will thus need to combine formal dynamical criteria with a normative principle that takes into account the relationship between the theory in question and the particular phenomena being studied.

I will take guidance for the construction of such a principle from a proposal due to Caulton. In the proposal laid out in \cite{CAULTON2015153}, he begins by defining what he calls an \emph{analytic symmetry}. While he is using a slightly different formalism than mine for talking about the relationship between theory, representation, and phenomena, it is possible to loosely translate his concepts into the language I am using.\footnote{ See \S2.3 of \cite{CAULTON2015153} for the definition is his language. }

Recall from \Sec\ref{sub:DEKI account} that, for a given mathematical carrier object $X$, the interpretation $I$ and key $K$ used by a theory to define its representations will impute the interpreted features of $X$ to features, $Q_a$, of the target. In this language, an analytic symmetry in the sense of Caulton translates most closely to a transformation on the space of \emph{values}, $x_\mu$, of $X$-features that preserves all the \emph{values}, $q_a$, of the features imputed to the target.\footnote{Caulton leaves open whether these transformations should act at a time or over a history.} Caulton calls any function of $x_\mu$ that is invariant under all analytic symmetries a \emph{physical quantity} because these are the quantities that the theory hypothesises to correspond to physical features of the world. One of the main differences between the language used by Caulton and the DEKI account used here is in the role played by the context and key. This extra input, which is required in our model description, will play a central role in separating puzzles related to building good models in physics and puzzles related to defining gauge symmetry.

Dirac's proposal for gauge symmetry can be used to illustrate how Caulton's language can be used. In Dirac's proposal, Caulton's physical quantities are the Dirac observables and the analytic symmetries are the symmetries generated by first-class constraints. One can think of analytic symmetries then as corresponding to the transformations of a theory's models that are hypothesised to preserve all the physically salient features of the world that those structures are intended to represent. It is only by checking the empirical adequacy of that theory that one can judge the validity of such hypotheses. 

Empirical adequacy alone, however, only ensures that Caulton's physical quantities have the capacity to represent the phenomena, and does not eliminate the possibility that the physical quantities overdetermine the phenomena. Using the notion of analytic symmetry, Caulton then proposes a methodological principle for aligning theoretical descriptions with physical reality. This principle involves maximising a theory's analytic symmetries while preserving its empirical adequacy. The proposal is described in two phases:
\begin{quote}
    During the first phase we set up representational links between the theory and the observable portion of the physical world, under the assumption that the theory is empirically adequate (or similar). In the second phase, we maximise the theory's analytic symmetries, taking advantage of the representational links forged in the first phase so as not to compromise empirical adequacy. [\S 4]
\end{quote}
This two-phase procedure is designed to ensure that a theory has enough physical quantities to represent phenomena --- but no more than strictly necessary.

I will adopt a similar attitude in my own definition of gauge symmetry. In particular, I will insist on identifying as many analytic symmetries as possible while preserving empirical adequacy. In contrast to myself, however, Caulton is hesitant to commit to any particular formal condition for identifying analytic symmetries. This may be partly due to the fact that Caulton defines `gauge symmetries' as a ``cluster of closely related formal notions'' which are ``often defined purely formally, as a space-, time- or spacetime-dependent transformations in some internal space.'' He then goes on to say:
\begin{quote}
    It may indeed be claimed that all gauge symmetries are analytic and all analytic symmetries are gauge. But establishing that claim would be a significant philosophical achievement: one that could be clear only in a language that refers to them with different names. Besides, it seems to me that there are both analytic symmetries that are not gauge and gauge symmetries that are not analytic.
\end{quote}
The last comment seems to reflect the difficulty of solving Belot's Problem without a consistent or precise definition of gauge symmetry.

In my proposal, I aim to bring clarity by proposing a concrete definition of gauge symmetry. I will use Dirac's motivation as a guide for identifying a general formal condition for an at-a-time notion of gauge symmetry. This condition will involve identifying a set of structures that can be autonomously evolved uniquely using the equations of motion. As we have seen in \chap\ref{ch:rep_sym}, for the gauge symmetries of well-known symplectic systems, this involves identifying the rank of the dynamical equations. I outlined a general procedure for doing that. (Specifically, one counts the number of independent null vectors of the symplectic $2$-form that are not tangential to the dynamical flow.)


For more general systems, it is not possible to be as specific in one's definition. Instead, I will focus on the general character of the equations of motion. Instead of adopting Caulton's language of physical quantities and analytic symmetries, I will use the terms \emph{(candidate) observable quantities}, since I am concerned with modelling empirical features, and \emph{(candidate) gauge symmetries}, to match my earlier terminology, but will sometimes drop the word \emph{candidate} as this can be understood from context.

Using this terminology, I will require that the observable quantities form an algebra. Physicists often refer to this as the \emph{observable algebra} of a system, matching the terminology I have introduced. In this language, the elements of the observable algebra define the hypothesised set of features of a theory's representations that are essential for imputing all the features of the target system that are relevant to the modelling context. The symmetries that preserve the observable algebra are then the candidate gauge symmetries of the theory.

The choice of an algebra as the basic mathematical component of the representational structures is ultimately a pragmatic one. It provides a minimal set of operations (both addition and multiplication) that allow for the group structure I've assumed and the ability to express non-trivial laws (e.g., in terms of differential equations). The closure property of an algebra is particularly important for my purposes since I want observables to be defined in terms of a closed set of structures. Algebras also possess important substructures, \emph{generating sets}, that are minimal sets whose elements can be combined using the basic operations of the algebra to construct \emph{any} element of the algebra. Such sets will be useful for counting the basic representational degrees of freedom of the system. But perhaps the most important reason for choosing an algebra is that many, if not all, observable quantities used in scientific theories can be represented by quantities that \emph{do} form an algebra.\footnote{ In quantum mechanics, for example, these could be the elements of a C$^*$-algebra. } Nevertheless, it would not require much effort to extend the formalism to some other (closed) mathematical structure.

After specifying the observable algebra of a theory, I will require that the equations of motion be \emph{well-posed} and \emph{autonomous} in this algebra. The requirement that the equations be well-posed is motivated by Dirac's definition and requires that the solutions be determined uniquely in terms of an element of a generating set of the observable algebra.\footnote{ I will discuss the applicability of this formalism to stochastic and quantum systems in \Sec\ref{sub:pesa_applicability}. } The requirement that the equations be autonomous requires that no additional mathematical quantities need to be specified in order to solve the dynamical system. Not only this, but specifying the values of any \emph{other} variables in the theory should have no effect on the evolution of any element of the observable algebra. This is precisely how Dirac understood and defined his observables: for Dirac, the evolution of non-observable variables requires the specification of arbitrary functions. Thus, any non-trivial transformation of the instantaneous state of the theory that leaves the observable algebra invariant is then a candidate gauge symmetry according to my definition.

I will then proceed along the lines suggested by Caulton: increase the number of candidate gauge symmetries until empirical adequacy can no longer be maintained. This aspect of my proposal is also similar to the process of \emph{Ramsification}, advocated in \cite{Lewis:Ramsification}, for identifying the empirical core of a theory. Note that empirical adequacy, in my language, means that the target system does indeed possess the features imputed by the theory's representations. In general, this procedure will involve implementing some version of the Gauge Principle (discussed in detail in \Sec\ref{ssub:the_gauge_principle}) because the Gauge Principle can be used to add specific amounts of underdetermination to a system of equations. I have given examples for how to do this for various kinds of theories including the Kretschmann and St\"uckelberg procedures described in \Sec\ref{ssub:the_gauge_principle} and the best-matching procedure performed in \Sec\ref{sub:bb_theory}. I have even provided a general scheme for doing this for non-symplectic symmetries like dynamical similarity in \Sec\ref{sub:gauge_principle_for_DS}. While the details of such a procedure are not important for the general argument, the net result of applying such a procedure is now clear: one should obtain a new dynamical system where elements of the desired observable algebra are uniquely determined by the dynamical equations and the remaining representational variables can be assigned arbitrary values. Once we have obtained a formulation of the theory that maximises the candidate gauge symmetries in this way, and if the theory is empirically adequate, then we can identify the resulting \emph{candidate} gauge symmetries as \emph{actual} gauge symmetries.

By requiring empirical adequacy, I ask that the theory in question have sufficient mathematical structure to represent the relevant phenomena. By maximally increasing the candidate gauge symmetries while maintaining empirical adequacy, I require that the theory have no more than the essential (or necessary) amount of structure for representing the phenomena. Autonomy says that the remaining representational structures determining the instantaneous state can be specified arbitrarily without affecting the dynamics of the essential and sufficient structure. Together, these requirements form what I will call the \emph{Principle of Essential and Sufficient Autonomy (PESA)},\footnote{ Pronounced `PEA-zah'. } which I will state more concretely below. This forms the basis of my proposal for solving Belot's Problem.

In the remaining sections of this chapter, I will first give a more concrete statement of the PESA. Then, I will briefly compare and contrast this proposal with a similar proposal made by Barbour. Finally, I will give examples that will illustrate both how the PESA should be used in general and how it provides a solution to Belot's Problem for these situations.


\section{Statement of the PESA} 
\label{sec:statement_of_the_pesa}

Let me now state the PESA. Consider a general theory whose representations include an instantaneous state $\psi(t,b^I(t))$ at time $t$, labelling Cauchy surfaces, that depends on the time-dependent elements of a set $B = \{ b^I(t) \}$ that is a generating set of some algebra $\mathcal B$. Since the algebra $\mathcal B$ can be extremely general, this puts virtually no restriction on the state other than it depend only on the current time $t$. Since $t$ itself can be considered an element of $\mathcal B$, there is no loss of generality in writing $ \psi(t,b^I(t)) = \psi(b^I(t))$. The time evolution of $\psi$ can then be determined by pulling back the time dependence of the $b^I(t)$. Let me assume that this can be expressed in first-order form\footnote{I will justify this assumption in \Sec\ref{sub:pesa_applicability}.}
\begin{equation}\label{eq:red ev}
    \dot b^I(t) = f^I(b^I(t))\,,
\end{equation}
where dots are $t$-derivatives and the $f^I$ are automorphisms of $\mathcal B$ for which I won't specify any significant restrictions (in particular, they could depend on $\psi$ itself). In general, equations \eqref{eq:red ev} are not required to be well-posed in terms of an element of the generating set $B$. This means that a specification of initial data, $\{b^I(t_0)\}$ at some time $t_0$, does not necessarily uniquely determine the solution $\{b^i(t)\}$ at some later time $t$ for all $t>t_0$. I do, however, require that \emph{at least one} solution does exist for some choices of initial data in a non-trivial domain. This requirement excludes theories with inconsistent or otherwise poorly defined equations of motion.

Following the definitions of \Sec\ref{sub:DEKI account}, I identify the algebra $\mathcal B$ as the algebra of values, $b^I$, of the features of the mathematical carrier object $X$ used to define the theory's models. The theory must then assign an interpretation $I: \mathcal B \to \mathcal B$ that is a bijection on $\mathcal B$ that maps values of the $X$-features to values of \emph{interpreted} $Z$-features. The values of these interpreted features are then fed into a key that associates them with features of the target as candidates for imputation. If we denote the values of the target features as $q^a$ and the algebra they belong to as $\mathcal P$, then the key (among other things) defines a map $K: \mathcal B \to \mathcal P$ that is usually non-bijective. 

I now call $\mathcal A \subseteq \mathcal B$ the \emph{observable algebra} of a theory if it is the smallest subset of $\mathcal B$ such that the composition of $K$ with $I$ obeys
\begin{equation}
    K\circ I(\mathcal A) = K\circ I(\mathcal B)\,.
\end{equation}
That is, the algebra is observable when its elements span the minimal set necessary to impute all features of the target system deemed relevant by the interpretation and key. Note that the observable algebra, as noted above, is defined a priori in terms of the mathematical structures (specifically the maps $I$ and $K$) that are necessary to define a theory's models. The specification of an observable algebra therefore amounts to a hypothesis about what features of a target system can be imputed by a theory.

We can now understand the PESA to be a normative principle that gives the conditions for having a \emph{good} interpretation and key such that the dynamical laws, defined by \eqn\ref{eq:red ev}, and the observable algebra, defined by the maps $I$ and $K$, are aligned with the actual properties of the target system. To accomplish this, consider a generating set $A = \{a^i\} \subseteq B$ of $\mathcal A$ and the restriction of \eqref{eq:red ev} to $A$, i.e.,
        \begin{equation}\label{eq:gen ev}
            \dot a^i(t) = \tilde f^i(a^i(t))\,.
    \end{equation}
We then have:
\paragraph{The Principle of Essential and Sufficient Autonomy (PESA)}
\begin{quote}
    If the equations~\eqref{eq:gen ev} are
    \begin{enumerate}[(i)]
        \item well posed, and \label{crit:well-posed}
        \item autonomous in $A$,\label{crit:autonomy}
    \end{enumerate}
    and if
    \begin{enumerate}[resume*]
        \item The observable algebra $\mathcal A$ is \emph{empirically sufficient}; i.e., $\mathcal A$ has sufficient structure to faithfully represent the target system,\label{crit:sufficiency}
        \item The observable algebra $\mathcal A$ is \emph{empirically necessary}; i.e., $\mathcal A$ has the necessary amount structure to faithfully represent the target system,\label{crit:necessity}
    \end{enumerate}
    then $I$ and $K$ are \emph{good}.
\end{quote}
I will define well-posedness, autonomy, empirical sufficiency and empirical necessity more carefully bellow. I can now use the PESA to define a gauge symmetry as follows:
\paragraph{Gauge symmetry}
\begin{quote}
    A \emph{gauge symmetry} is a non-trivial automorphism of $\mathcal B$ that preserves $\mathcal A$ when $I$ and $K$ are good.
\end{quote}
Note that while gauge symmetries, when they exist, will be automorphisms by this definition, there is really nothing that guarantees, in general, that they be smooth or otherwise nice in any particular way. The fact that they \emph{do} often happen to be simple and well-behaved transformations in our best theories of physics is a compelling fact worthy of further investigation.

Let me now describe in more detail the formal features of the PESA to connect these with the motivations of \Sec\ref{sec:motivation}. As a definition of gauge symmetry, the PESA has two important components. The first is the dynamical requirements on the representational structures. These are expressed by the conditions~\ref{crit:well-posed} and \ref{crit:autonomy} on the equations~\eqref{eq:gen ev}. The well-posed condition requires that the dynamical equations be invertible in terms of any $A$. Condition~\ref{crit:autonomy} enforces the autonomy of the evolution equations. This ensures that changes to any element of $\mathcal B$ that is not in $\mathcal A$ will not affect the dynamical evolution of $\mathcal A$. In other words, autonomy requires that the empirical core of a theory is closed. Together, these dynamical conditions are in line with Dirac's conditions for observables.

The second important component of the PESA involves the empirical adequacy of the theory and is encoded in the sufficiency condition~\ref{crit:sufficiency} and the necessity condition~\ref{crit:necessity}. Empirical sufficiency requires that the observable quantities specified by the theory are sufficient for representing the relevant phenomena. This avoids that there are features of the target system that the theory should describe according to its key but cannot. In \Sec\ref{ssub:extending_galileo_s_ship} we will see an example of how this condition can be violated. Empirical necessity requires that the observable quantities of a theory only have the minimal representational structure necessary to represent the phenomena. This avoids situations where a theory purports to describe more features than what are actually relevant to the modelling context. Violations of this condition are particularly difficult to spot, and will be the subject of many of our considerations in Part~\ref{part:aot}.

Empirical adequacy is often implicitly assumed in many accounts of gauge symmetry. However, empirical considerations are essential for understanding the concept of gauge symmetry because gauge symmetries are usually understood to be statements about the relationship between theory and phenomena. Explicitly requiring that symmetries only be considered gauge when they satisfy certain formal requirements \emph{and} when the theory can adequately describe the phenomena in question implies that both dynamical and empirical considerations are relevant.

The PESA can then be used to decide what interpretive practice should be used to align these dynamical and empirical considerations. It is then clear how my definition of gauge symmetry corresponds to a proposed solution to Belot's Problem. The PESA gives conditions on good interpretive practice, through the specification of a good interpretation and key, that align dynamical and empirical considerations. When these are satisfied, gauge symmetries are then given an exact definition in terms of the formal properties of a theory's models.

Let me say a bit more about the requirement that $I$ and $K$ be `good'. Goodness, in my sense, requires that the theory's models provide a faithful representation of the target. This, in turn, requires that the context $C$ of the representation selects the appropriate features of the mathematical carrier object $X$ to be imputed to the target. This illustrates the dependence of my definition of gauge symmetry on the experimental context. Moreover, the key must specify the right idealisations, approximations, assumptions, judgements, etc that will lead to a faithful representation. This extends and clarifies what Dirac may have meant by gauge symmetries preserving the ``physical state'' of a system. By separating the representational aspects of the proposal (i.e., the specification of a modelling context as well as the necessary idealisations, approximations and other auxiliary assumptions) from the dynamical ones (i.e., the conditions on the amount of underdetermination), a clear definition of gauge symmetry is possible.\footnote{ Focus on the context and key also brings to light the role of modelling assumptions in shaping what should rightly be considered the empirical core of a theory.  }

Finally, the PESA leads to two clear prescriptive methodological rules:
\begin{enumerate}
    \item If a theory violates condition~\ref{crit:necessity}, then apply some version of the Gauge Principle to eliminate the non-essential elements of $\mathcal A$.\label{rule:gauge}
    \item If a theory violates condition~\ref{crit:sufficiency}, then extend the theory to accommodate new phenomena.\label{rule:extend}
\end{enumerate}
In \Sec\ref{sec:problems_solved}, we will see examples of how to implement these prescriptions. These examples will illustrate how Belot's Problem can be handled using the PESA and will relate these discussions to known ways of treating gauge symmetries.

\subsection{On the applicability of the PESA} 
\label{sub:pesa_applicability}

As presented above, the PESA assumes a theory that can be written as a dynamical system of differential equations that are first order in time. In \chap\ref{ch:rep_sym}, I went to great lengths to show how all of our most fundamental theories of physics, and indeed many non-fundamental and non-physics theories, can be cast into this form. One notable difficulty is general relativity --- particularly its non-globally hyperbolic DPMs. I will now defend my approach in light of this difficulty. My arguments will assume some technical expertise in general relativity.

The attitude I will take here will be to think of theories as tools that scientists can use to explain observable phenomena. In this regard, observers can only be assumed to have access to information on their past light-cone. Using such information, these observers can reconstruct a Cauchy problem for temporal evolution in their causal diamond.\footnote{ This can be done, for example, using a double-null decomposition of the spacetime in the causal diamond, as introduced in \cite{Sachs:double_null}. } This sort of usage of general relativity is perfectly compatible with the first-order formalism I am assuming because Cauchy problems can be expressed in terms of the kind of first-order differential equations at the basis of my proposal. Moreover, causal diamonds can be patched together with other causal diamonds to obtain any possible DPM of general relativity.\footnote{ This is because solutions in general relativity are Lorentzian space-times, which can be equipped with an atlas. By taking infinitesimally small patches of causal diamonds, which are locally Minkowski, one can always construct such an atlas. } Thus, my picture is not incompatible with a theoretician's world-view involving groups of different observers embedded in an arbitrarily complicated Lorentzian space-time --- even if this would involve imposing complicated non-local constraints between the different observable patches.

The attitude expressed above also justifies why I believe that an at-a-time notion of symmetry is the physically salient one at the cosmological level. Using data on their past horizons, observers can reconstruct the at-a-time state of a theory and check whether this state has specific properties in relation to some laws. But such an observer cannot `observe' an entire history --- particularly the non-globally hyperbolic ones of general relativity --- and check whether that history has the desired properties. I therefore conclude that, under my understanding of how a theory should be used, at-a-time notions of symmetry are more physically salient. The instantaneous states of first-order systems are ideally suited to studying these notions of symmetry.

One last comment about the applicability of the PESA is that, by only assuming an algebraic structure for observables that evolve according to differential equations, I can accommodate quantum theories (including quantum field theories and classical stochastic theories) whose observable algebras can be written in terms of operator algebras acting on a Hilbert space or, more generally, $C^\star$-algebras. Indeed, my attempt to represent the observable quantities of a theory in term an observable algebra was inspired by canonical attempts to quantize gravity where the goal is to do this explicitly.\footnote{ See, for example, \cite{rovelli2004quantum} or \cite{thiemann2008modern}. } Note that requiring that the evolution equations be well-posed in terms of an observable algebra doesn't require that those equations be deterministic in the instantaneous state. This extends the domain of applicability of my proposal to statistical and quantum mechanics.

\section{Barbour and ``Poincar\'e's principle''} 
\label{sec:barbour_and_poincare}

I have already commented on the motivations for the PESA stemming from Dirac's definition of gauge symmetry. Similar ideas have been expressed at least as early as Poincar\'e, and have been re-emphasised more recently by Barbour (in, for example, \cite[\S 5]{Barbour:2010dp}). Indeed, the motivation for the formal conditions of the PESA are similar to what Barbour has called \emph{Poincar\'e's Principle}. According to \cite{Barbour:2010dp}, the attribution of this principle to Poincar\'e results from an analysis in Chapter 7 of \emph{Science and Hypothesis} \citep{poincare_foundation} where Barbour claims that Poincar\'e was attempting ``to define relativity in terms of \emph{the amount of information} needed to be specified in coordinate-independent (gauge-invariant) form if the evolution is to be predicted uniquely.'' [p.1271, original emphasis.] 

In the relevant passage, Poincar\'e discusses the empirical consequences of the angular momentum of the spinning Earth. He notes that what is relevant for describing the data is the amount of independent data required to solve the evolution of the system: [p.114]
\begin{quote}
    ``Provided that the future indications of our instruments can only depend on the indications which they have given us, or that they might have formerly given us, such is all we want, and with these conditions we may rest satisfied.''
\end{quote}
In this case, he points out that if ``thick clouds hide the stars from men who cannot observe them, and even are ignorant of their existence''[p.109] then only the value of the angular momentum and not the absolute orientation of the Earth is required to describe the phenomena. In the language of the PESA, the prescriptive rule \eqref{rule:gauge} would suggest applying a version of the Gauge Principle to Newtonian mechanics where the angular momentum, but not the orientation, would be part of the observable algebra.\footnote{ See \cite{GOMES2021138} for an implementation of the Gauge Principle that achieves this in general, and that also has other appealing explanatory features. }

One can plausibly use the quote above to motivate the conditions \ref{crit:well-posed} - \ref{crit:necessity} used to define the observable algebra of the PESA. In any case, these conditions are closely aligned to how Poincar\'e's comments have been interpreted by Barbour. This match was used to motivate the naming of the elements of the observable algebra $\mathcal A$ the \emph{Poincar\'e observables} in \cite{Gryb:2021qix}. This naming is a nod to the \emph{Dirac observables}, which are generalized by the Poincar\'e observables. Note, however, that the account of representation I am using (i.e., the DEKI account) was designed to be as metaphysically neutral as possible. In this way, the PESA is not committed to Barbour's relationalism.

\section{Problems solved} 
\label{sec:problems_solved}

In this section, I will reconsider the examples discussed in \Sec\ref{sub:examples1} to illustrate Belot's Problem and show how they can be dealt with using the PESA. This will serve both to show how common illustrations of Belot's Problem can be dealt with using the PESA and as a template for how to apply the PESA more generally. Towards this end, I will also use the PESA to study the case of reparametrisation invariance and the resulting frozen-formalism problem described in \Sec\ref{sub:reparametrisation_invariance}. The conceptual problems related to reparametrisation invariance are not as well-known in the philosophical literature as, say, those of Galileo's ship. This example will then serve as a way to show that the PESA has general applicability outside the original scope of problems considered in much of the philosophical literature on symmetry. Finally, the example related to the symmetries of the Kepler problem will relate to our broader discussions of dynamical similarity and the applications of the PESA to cosmology.

An important aspect of the applications of the PESA we will consider is the way in which the methodological prescriptions \eqref{rule:gauge} and \eqref{rule:extend} appear in different contexts. The rule \eqref{rule:gauge} appears as a way of testing whether a particular theory has \emph{too many} observables while rule \eqref{rule:extend} occurs when the observables are \emph{too few}. Interestingly, we will see that \emph{both} cases can be associated with significant theoretical advances. This highlights the importance of both the sufficiency and necessity conditions (i.e., conditions~\ref{crit:sufficiency} and \ref{crit:necessity}) of the PESA.

\subsection{Galileo's ship} 
\label{sub:solved_galileo_s_ship}

\subsubsection{The empirical consequences of Galilean transformations} 
\label{ssub:the_empirical_consequences_of_galilean_transformations}


In \Sec\ref{ssub:the_newtonian_free_particle}, I described the Newtonian free-particle and noted its invariance under the Galilean transformations \eqref{eq:Gal trans}. I commented on the fact that treating the Galilean transformations as a gauge symmetry would trivialise the theory since all solutions, \eqref{eq:free particle}, of the free particle can be related by a Galilean transformation. This was seen as a generally unfavourable outcome because a free particle can be used as a model for many physical systems in the world that are not devoid of physical content.

One example of this kind is Galileo's ship, which was discussed in \Sec\ref{ssub:galileo_ship}. In this case, the position of the ship relative to a fixed point on the shore could be modelled as a free particle because, to a good degree of approximation, there are effectively no forces impeding the motion of the ship along the shore.

The PESA requires that the observable algebra of this system be generated by the minimal set of instantaneous representational structures required to model the system in terms of first order differential equations. The Hamiltonian for the Newtonian free particle is
\begin{equation}
    H_\text{free} = \frac{\vec p^2}{2m}\,,
\end{equation}
where $\vec p$ is the particle's momentum in the $\vec x$-direction and $m$ is its mass. Here the vector components range over the two-dimensional plane spanned by the surface of the water.\footnote{ We're ignoring the constraint that the boat is not allowed to move onto the shore, which would only complicate the model unnecessarily. } Hamilton's equations tell us
\begin{align}
    \dot {\vec x} &= \frac{ \vec p }m & \dot{\vec p} &= 0\,.
\end{align}
These equations are integrable for all values of $(\vec x(t), \vec p(t)) \in \mathbbm R^4$ and autonomous in these variables. 

We know from experience that this theory is empirically adequate as it correctly describes the instantaneous position and velocity of the ship relative to the shore given the approximations we are making. Moreover, removing any of the position or momentum variables would ruin the empirical adequacy of the theory because we would be constraining the ship's motion in an unphysical way. Thus, the PESA tells us that we have a good interpretation of the system with an observable algebra generated by the set $\{\vec x, \vec p\}$.

We are then justified in using this observable algebra for defining the gauge symmetries of the theory. Because all the elements of $\{\vec x, \vec p\}$ transform under Galilean transformations, the Galilean transformations are \emph{not} gauge symmetries of the theory according to my definition. This is precisely in line with the expectation that Galilean symmetries should have empirical consequences when applied to Galileo's ship.

This shows how the PESA deals with an earlier puzzle, noted in \Sec\ref{ssub:the_newtonian_free_particle}, regarding the gauge-status of the Leibniz shifts, which are the Euclidean subgroup of the Galilean transformations. As I showed in \Sec\ref{sub:leibniz_and_noether1}, the Leibniz shifts are Noether-1 symmetries of Newtonian mechanics in general. They are therefore variational symmetries of the free-particle theory. However, because such symmetries are not associated with underdetermination in the equations of motion and because the model we have given is empirically adequate, there is no need to treat these as gauge symmetries in this context according to the PESA.

\subsubsection{Galilean transformations as gauge symmetries} 
\label{ssub:galilean_transformations_as_gauge_symmetries}

In contrast to the previous section, one might be interested in analysing the role of Galilean transformations in the same system but in a slightly different context, and one that was discussed by Galileo in his original thought experiment. In his \emph{Dialogue Concerning the Two Chief World Systems}, Galileo considers how everything moving inside the contents of a large ship's cabin will move the same way no matter what the motion of the ship, provided that motion is uniform. It is illuminating (and amusing) to quote the wonderful passage directly: \citep[pp. 186-187 (Second Day)]{galileo_dialogues}
\begin{quote}
    Shut yourself up with some friend in the main cabin below decks on some large ship, and have with you there some flies, butterflies, and other small flying animals. Have a large bowl of water with some fish in it; hang up a bottle that empties drop by drop into a wide vessel beneath it. With the ship standing still, observe carefully how the little animals fly with equal speed to all sides of the cabin. The fish swim indifferently in all directions; the drops fall into the vessel beneath; and, in throwing something to your friend, you need to throw it no more strongly in one direction than another, the distances being equal; jumping with your feet together, you pass equal spaces in every direction. When you have observed all these things carefully (though there is no doubt that when the ship is standing still everything must happen in this way), have the ship proceed with any speed you like, so long as the motion is uniform and not fluctuating this way and that. You will discover not the least change in all the effects named, nor could you tell from any of them whether the ship was moving or standing still.
\end{quote}
In this passage, Galileo is clearly describing a form a gauge symmetry: the physical processes in the ship's cabin proceed in the same way regardless of the velocity of the ship relative to some reference point on the shore. And as long the cabin is closed off from the rest of the world, there is no empirical way to detect the ship's motion because the location of a reference point on the shore in empirically inaccessible.

Let us use the PESA to analyse the symmetries of this system. First, we will try to model the contents of the cabin as rigid objects with their locations approximated by their centre of mass. This can be done using a system of $N$ Newtonian point particles with positions $\vec x_I$ and velocities $\dot {\vec x}_I$ interacting with some potential, $V(\vec x_I)$, that we won't need to specify explicitly. In this model, one assigns coordinates to these $N$ particles by imagining that they are placed in some absolute Cartesian coordinate system. For observers in the cabin, this can be done by choosing an arbitrary reference point inside the cabin --- say the centre-of-mass of all the contents. Let us call this theory \emph{Naive Newtonian Mechanics (NNM)} because it makes the (naive) assumption that the Cartesian coordinates and their velocities are observable quantities. I will now give a more detailed model of the system and show that the PESA tells us that NNM does not provide a good description of the system. I will then use the prescriptive rules of the PESA to generate a new theory that does provide a good description.

Newton's laws can be expressed in first-order form using the well-known Hamiltonian
\begin{equation}
    H_\text{NNM} = \sum_I \frac{\vec p_I^2}{2m} + V(\vec x_I)\,,
\end{equation}
which leads to the Hamiltonian equations of motion
\begin{align}\label{eq:eoms NNM}
    \dot{\vec x}_I &= \frac{\vec p_I} m & \dot{\vec p}_I &= -\vec{\nabla} V\,. 
\end{align}
For suitably well-behaved potentials, which we will assume in this model, these equations are well-posed and autonomous in the set $A_\text{NNM} = \{\vec x_I, \vec p_I\}$. Using the obvious interpretation and key of NNM from our description above, the algebra $\mathcal A_\text{NNM}$ generated by $A_\text{NNM}$ is observable because NNM assumes that all Cartesian coordinates and their velocities can be measured. The algebra $\mathcal A_\text{NNM}$ thus meets all the dynamical criteria of the PESA. That should be unsurprising since, as a mathematical system, this model is nothing but a direct generalization of the free-particle model of the previous section.

Now though, the target system of this representation is different (it doesn't include any reference frame outside the cabin), and so we must test whether the empirical conditions of the PESA continue to be satisfied. I will assume that the Newtonian time can be reliably measured for this system using some suitably isolated external clock. Under this assumption, experience tells us that the sufficiency condition~\ref{crit:sufficiency} is satisfied because, by choosing a reference frame inside the cabin, one should be able to correctly predict the behaviour of this Newtonian system using Newton's laws in the form \eqref{eq:eoms NNM}. However, Galileo's passage above suggests that the necessity condition~\ref{crit:necessity} is no longer satisfied. This is because one could choose any reference point moving with uniform velocity relative to the cabin and predict all the same phenomena inside the ship's cabin --- provided one converts all coordinates to centre-of-mass coordinates inside the cabin. This suggests that there is more representational structure in $\mathcal A_\text{NNM}$ than is strictly necessary to represent the known features of the target. The PESA then tells us that NNM is not a good theory for describing this system because its interpretation and key are not good.

In this way, the PESA reproduces standard relationalist critiques of Newtonian mechanics. But the prescriptive rule~\ref{rule:gauge} allows us to go further. This prescription tells us that we should use some form of the Gauge Principle to introduce the right amount of underdetermination into our equations of motion to match the underdetermination of NNM by the phenomena. I have already given, in \Sec\ref{sub:bb_theory}, a procedure for doing this called \emph{best-matching}. I called the theory obtained from this procedure \emph{Barbour--Bertotti (BB)} theory and worked out the first order analysis and degree-of-freedom counting in \Sec\ref{sub:barbour_berttoti_shifts}. For the system considered here, the underdetermination is in the position of the origin of the Cartesian system used to write the coordinates of the $N$ particles. We can therefore exactly use BB-theory, which best-matches spatial translations.

At the interpretational level, Galileo's considerations suggest to us that the observable quantities of the new theory should be invariant under translations. This defines a particular interpretation and key of the theory where the observable algebra, $\mathcal A_\text{BB}$, of BB-theory is translation invariant. Indeed, our analysis of the equations of motion of the BB-theory first in \Sec\ref{sub:bb_theory} and later, in first-order form, in \Sec\ref{sub:barbour_berttoti_shifts} confirm this to be the case.

What we found in \Sec\ref{sub:barbour_berttoti_shifts} (see that section for details) was that there were exactly four directions (see Equations~\eqref{eq:BB null directions}) on velocity phase space for each translation (i.e., each spatial direction) in which the sympletic structure of the theory was degenerate. This means that there are four representational quantities per translation that cause the equations of motion to not be well-posed. Two of these correspond to the arbitrariness of the gauge field and its velocity, which we introduced as part of the Gauge Principle itself. It is then expected that the gauge theory be invariant under arbitrary changes of these variables.

More significantly, we found that the remaining two obstructing directions were associated with the centre-of-mass position and velocity. These latter directions generate arbitrary time dependent translations, which preserve the centre-of-mass coordinates. Thus, while the first-order equations are not well posed on the velocity phase space of Cartesian coordinates and their velocities, they \emph{are} well-posed and autonomous in terms of the centre-of-mass coordinates and their velocities. This is exactly a generating set for the observable algebra, $\mathcal A_\text{BB}$, identified by the interpretation and key of the BB-theory. Thus, the dynamical conditions of the PESA are satisfied.

Experience tells us that the sufficiency condition~\ref{crit:sufficiency} is also satisfied because the evolution of the centre-of-mass variables can be used to predict all the relevant empirical phenomena simply by choosing the right reference point within the cabin. But now, the necessity condition~\ref{crit:necessity} is also satisfied because the observable algebra $\mathcal A_\text{BB}$ is a maximal translation-invariant set for this system.\footnote{ To see this, note that the centre-of-mass coordinates are the result of a formal quotient of the Cartesian coordinates by the translation group. } The PESA then tells us that the interpretation and key used in BB-theory is appropriate for this system.

As a last step, we can identify gauge symmetries for the BB-theory as applied to this system. Because the Galilean transformations leave the observable algebra $\mathcal A_\text{BB}$ invariant, it follows from my definition that they are gauge symmetries. This exactly matches Galileo's intuition above, and shows that the PESA can provide a good tool for determining when and how to match the gauge symmetries of a theory with the physical features of the target system in question.

Note that the prescriptive rules of the PESA don't specify what version of the Gauge Principle to use. Rather, the PESA puts reasonable constraints on the output of the gauging procedure. Instead of using best-matching, one could have chosen to modify the geometry of a neo-Newtonian space-time. However, the advantage of the canonical analysis performed in \Sec\ref{sub:barbour_berttoti_shifts} is that it gives us a direct tool for assessing the amount of at-a-time underdetermination in the equations of motion.

\subsubsection{Extending Galileo's ship} 
\label{ssub:extending_galileo_s_ship}

There is one final context in which the PESA can illuminate our understanding of the symmetries of Galileo's ship. Consider what happens if the cabin of Galileo's ship has windows that can be opened to reveal a view of the shore. In this case, there is a small but negligible interaction between the shore and the contents of the cabin that gives observers in the cabin empirical means of detecting motion relative to the shore.

In this case, the BB-theory of the previous section will fail the sufficiency condition~\ref{crit:sufficiency} of the PESA so that it is no longer an appropriate theory for describing the system. This is because there is a new  phenomenon --- the motion of the ship relative to the shore --- that the BB-theory cannot represent. In this case, the prescription~\ref{rule:extend} tells us that we should extend the system by enlarging the observable algebra to accommodate the new features.

When the insufficient theory is a gauge theory, there is an obvious way of doing this extension: simply undo the Gauge Principle. It is easy enough to see that such an extension will work in this case. Undoing the Gauge Principle of BB-theory simply reproduces the NNM-theory of the previous section. But now the position and velocity of the centre-of-mass can be taken to be measured with respect to some fixed point on the shore as in \Sec\ref{ssub:the_empirical_consequences_of_galilean_transformations}. With this interpretation and key, the observable algebra $\mathcal A_\text{NNM}$ satisfies all the conditions of the PESA, and NNM is therefore an appropriate theory for describing this new system. Importantly, rigid Galilean transformations of the contents of the ship's cabin will change the position and velocity of the centre-of-mass of the cabin, which is now part of the observable algebra. These Galilean transformations should then no longer be considered gauge transformation according to the PESA in accordance with our expectations.

Of course there is no guarantee that this sort of extension procedure will work for every system. However, it is noted in \cite{Rovelli:2013fga} that this particular extension procedure\footnote{ The example that he uses involves two rocket ships travelling through outer space, but this is mathematically equivalent to the system of (navel) ship and shore that we've considered here. } is an example of a general principle that can be used in many contexts to take advantage of a gauge symmetry in order to describe the ways in which an isolated system can be coupled to an external system. What I'd like to note here is that the different physical situations of Galileo's ship help to serve as a template for how the PESA can be used in general to correctly identify the appropriate observable quantities and gauge symmetries of a particular theory and its target phenomena.

\subsection{The Kepler symmetries} 
\label{sub:the_kepler_symmetries}

In \Sec\ref{ssub:the_kepler_problem}, we studied the symmetries of the Kepler problem. We found that the DPMs of a Keplerian model were related by the dynamical similarity \eqref{eq:DS kepler},\footnote{ To see that these are dynamical similarities see, for example, Equations~\ref{eq:ds Nby explicit}. } which is a broad symmetry of that system. In \Sec\ref{sec:a_new_kind_of_symmetry_dynamical_similarity}, I studied the formal properties of dynamical similarity in general and gave an implementation of the Gauge Principle applicable to that case. The Kepler problem therefore provides us with an opportunity to test the applicability of the PESA to non-symplectic symmetries and gives us a simple way to test the PESA outside the domain of applicability of Dirac's proposal.

It is relatively straightforward to apply the PESA to the Kepler problem and, in particular, the symmetries \eqref{eq:DS kepler}. Let us consider the standard context of the Kepler problem in which the aim is to model the motion of planets in our solar system. In this context, Newtonian mechanics stipulates that we can measure the absolute values of $t$, $r$ and $\theta$ using external rods and clocks available to earth-based observers. The observable algebra $\mathcal A_\text{Kepler}$ is then generated by a set that includes the coordinates $(r,\theta)$ and their velocities: $A_\text{Kepler} = \{ r, \dot r, \theta, \dot \theta \}$. Because the equations of motion of the Kepler problem are just Newton's laws, we know that they are well-posed and autonomous in $A_\text{Kepler}$. The dynamical conditions of the PESA are therefore met.

The sufficiency condition~\ref{crit:sufficiency} of the PESA is also met because experience tells us that we can use Kepler's laws to model the motion of the planets in our solar system --- at least up to the approximations specified by whatever key Kepler might have used. Finally, the necessity condition~\ref{crit:necessity} of the PESA is also met. This is because removing extra structure from the Kepler model will prevent it from being empirically adequate. What one would like to know regarding the necessity of $\mathcal A_\text{Kepler}$ is whether re-scaling the time $t$ and spatial size $r$ according to \eqref{eq:DS kepler} leads to an empirically equivalent system. If this were the case, then one could reduce the size of the observable algebra without affecting empirical adequacy.

We know, however, from the standard use of Kepler's laws that this is not the case. There are external structures; e.g., other planets, the fixed stars, the spinning Earth; that provide reliable external clocks and rods that can be used as references scales for the system. The measurement procedures for using these clocks and rods are described in the key of the standard Kepler theory even if they aren't always explicitly stated. In light of these empirical and interpretive considerations, the PESA tells us that the standard Kepler theory is a good theory for describing the motion of planets in our solar system.

As a final consideration, the dynamical similarities \eqref{eq:DS kepler} act non-trivially on $\mathcal A_\text{Kepler}$, and are therefore \emph{not} gauge symmetries of the Kepler theory according to the PESA's definition. This is precisely what we would expect from the remarkable empirical success of Kepler's laws and Newton's theory of planetary motion. We should however note that, in the absence of external clocks and rods, the Kepler theory could fail the necessity condition~\ref{crit:necessity}. Then we would be forced to use the version of the Gauge Principle developed in \Sec\ref{sub:gauge_principle_for_DS} for dynamical similarity. A procedure for how to do this is given in \Sec 3.2.2 of \cite{Gryb:2021qix} In \Sec\ref{sub:dynamical_similarity_in_the_universe}, we will see that precisely this sort of condition will fail in modern theories of cosmology. This will be central to our analysis of the problem of the AoT. It will also be an important and novel application of the PESA.


\subsection{Reparametrisation invariance} 
\label{sub:solved_reparametrisation_invariance}

For the final section of this first part of the dissertation, I will apply the PESA to a conceptual problem related to symmetry that is not often considered in the philosophical literature on symmetry. This example will also serve to illustrate that the dynamical conditions of the PESA are non-trivial and have important consequences for identifying the gauge symmetries of a theory.

The problem I will consider is the \emph{frozen formalism} problem of classical reparametrisation invariant theories. The quantum analogue of this problem is the subject of much debate in the theoretical physics literature on canonical quantum gravity.\footnote{ For an introduction to the frozen formalism problem, see \Sec 3 of \cite{Anderson:2012vk}. } But the classical problem captures all the main conceptual difficulties. Here, I will consider the global problem, where the time parameter is constant across spatial slices.

The frozen formalism problem appears by combining Dirac's proposal for gauge symmetries with the fact, proved in \Sec\ref{sub:reparametrisation_symmetry}, that reparametrisation invariant theories have vanishing Hamiltonians. (Recall that this is the Weierstrass condition of Equation~\ref{eq:Weierstrass}.) In Dirac's proposal, the gauge symmetries of a theory are generated by the first class constraints of the Hamiltonian formalism (see \Sec\ref{sec:the_dirac_algorithm}). But because the Hamiltonian is generally represented on phase space by a first class constraint, the evolution generator and the gauge generator are taken to be the same thing. This results in the oft repeated slogan appearing explicitly in \cite[p. 103]{henneaux1992quantization}: ``motion is just the unfolding of a gauge transformation.'' The problem here is that, if this slogan is to be taken literally then past, present and future are all physically indistinguishable. This has led \cite{rovelli1991time} to question whether time exists at all in quantum gravity and \cite{barbour2001end} to proclaim ``The End of Time.'' The challenge is then seen to be to recover our perceived notion of time from a timeless formalism. We will now see that the PESA puts into question whether the formalism is really timeless in the first place.

One might suspect that the empirical conditions of the PESA would be enough to conclude that, given any reasonable key, states at different times should be considered empirically inequivalent in any theory. This is partly because the PESA regards at-a-time notions of symmetry to be more fundamental than over-a-history notions, where all instants differ only by a partial order on the real line. But given the extensive literature on trying to extract an emergent notion of time from a timeless formalism, it is clear that metaphysical preferences are playing a role in deciding what reasonable expectations there should be on a theory's key regarding temporal structure.

For this reason, it is noteworthy that even before these metaphysically fraught considerations, there are good reasons to believe that the \emph{dynamical} conditions of the PESA suggest that time evolution should not be identified with a gauge symmetry. This is because Dirac's proposal, which is purely dynamical, is based on assumptions that are violated in reparametrisation invariant theories. The reason why is technical and is explained at the end of \Sec\ref{sec:the_dirac_algorithm}, and is inspired by an argument made in \cite{barbour2008constraints}. What we will see now is that an analysis on \emph{velocity phase space}, rather than Dirac's extended phase space, can bring technical clarity to the discussion.

I have already treated the case of reparametrisation invariance in detail in \Sec\ref{sub:reparametrisation_invariance}, and there is no need to repeat the technicalities here. What we found there is that, on velocity phase space, the vector field generating an at-a-time reparametrisation was emphatically \emph{not} the generator of the dynamical evolution. And while the generator of at-a-time reparametrisations was responsible for underdetermination in the equations of motion, the evolution generator was not. Instead, the evolution generator \emph{defined} the classical solutions rather than providing any obstruction to solving them. In particular, when the time variable labelling instantaneous states along a DPM is included in the observable algebra of the system, the evolution equations \emph{are} well-posed and autonomous. Thus, including this time variable in the observable algebra satisfies both dynamical conditions of the PESA. This directly contradicts the expectations one gets from Dirac's proposal since the first-class Hamiltonian constraint changes the value of such time parameters. This leads to no inconsistency, however, due to the fact that the conditions of Dirac's theorem, on which his more general conjecture is based, are violated for reparametrisation invariant theories.

Finally, the extended observable algebra that includes the value of a time parameter (but not its velocity) labelling instantaneous states is also the algebra we concluded above should result from any reasonable key (modulo more exotic temporal metaphysics). Thus, a theory that uses this extended algebra would also satisfy the PESA's sufficiency and necessity conditions~\ref{crit:sufficiency} and \ref{crit:necessity}. We thus conclude that this extended algebra is appropriate for describing the system according the PESA, and that time evolution transformations are \emph{not} gauge symmetries. The main insight of this analysis is the observation that while the evolution generator is, in fact, a null direction of the symplectic $2$-form, the existence of this null direct does not make the equations of motion ill-posed but, rather, defines them. This solves the frozen formalism problem of the classical (global) theory. 

\cleardoublepage 
\part{The arrow of time}\label{part:aot}

\chapter{The problem of the arrow of time}
\label{ch:aot_prob}

\begin{abstract}
    In this chapter, I introduce and motivate the main problem that I will be concerned with in Part~\ref{part:aot} of the thesis: the problem of the Arrow of Time. I begin by defining the general problem and add two important sub-problems: the smoothness problem, which involves explaining extreme smoothness of the early universe, and the red-shift problem, which involves explaining the rapid cooling in the early universe. After motiving these problems, I discuss two standard approaches to these phenomena. The first involves explaining the Arrow of Time using a Past Hypothesis and the other using time-asymmetric laws. I identify several difficulties with these approaches and conclude that a new kind of solution is welcomed.
\end{abstract}

\ifchapcomp
    \tableofcontents
    \newpage
\else
    \cleardoublepage
\fi

\section{Introduction}\label{sec:aot_prob intro}

I will now shift attention to the problem of the Arrow of Time (AoT). In the introduction (\Sec\ref{sec:intro AoT}), I referred to this as the problem of finding an explanation for the large amount of time-asymmetry seen in physical processes given the (near) time-reversal symmetry of our best physical laws. In this chapter, I will try to justify the claim that there is considerable time-asymmetry in the universe and pinpoint precisely where that asymmetry lies. I will then review some standard ways of explaining the AoT and argue that no single explanation is completely adequate. This will motivate the need for a new way of thinking about the AoT, which I will provide in the last chapter (\chap\ref{ch:new_aot}).

The problem of the AoT has its origins in a debate about the recovery of the laws of thermodynamics, which describe many time-asymmetric processes, from the laws of statistical mechanics, which are fundamentally time-reversal invariant.\footnote{ See \cite{brush1976kind} for historical context, \cite{frigg2011guide} for a modern introduction to the central issues and \cite{sklar1993physics} for a philosophical analysis of the problem of recovering time asymmetry in thermodynamics from statistical mechanics. } A central figure in this discussion is Ludwig Boltzmann who championed an influential view for explaining the time-asymmetric behaviour of thermodynamic systems from statistical mechanics.\footnote{ This view was espoused, for example, in \cite{boltzmann1895certain} or \cite{boltzmann2012suicide}. } I will summarise this view using modern concepts in \Sec\ref{sec:preliminaries}. For now, let me note that, while Boltzmannian explanations of time-asymmetry were originally designed to address the specific time-asymmetries observed in thermodynamic systems, they have become a template for explaining time-asymmetries more generally.\footnote{ For some notable examples where Boltzmannian reasoning is used to explain general time-asymmetries in the universe, see \cite{price2002boltzmann}, \cite{lebowitz1993boltzmann}, \cite{goldstein2001boltzmann}, \cite{goldstein2004boltzmann}, and \cite{albert2009time}. } 

Renewed interest in the problem of the AoT is illustrated by an influential argument made by Roger Penrose in Chapter~7 of \cite{Penrose:NewMin} in which Boltzmannian explanations were argued to provide an account for a multitude, if not all, of the different time-asymmetric processes observed in the Universe.\footnote{ For different approaches with a similar goal, see \cite{Reichenbach1956-REITDO-2} and \cite{horwich1987asymmetries}. For a review of the relative merits and draws backs of these different approaches, see \cite{sep-time-thermo}. } This view has been notably advocated in \cite{albert2009time} and further developed in works such as \cite{price2004origins}, in a way I will elaborate upon below, resulting in an entire literature of explanations of the AoT in terms of the so-called \emph{Past Hypothesis (PH)}.\footnote{ For an introduction to the PH with a detailed list of references, see \chap\ref{ch:against_PH}. } Explaining the AoT using a PH is undoubtedly the dominant approach currently taken in the literature on the AoT. Acceptance of such explanations, however, is not universal. We will study the different reasons for that in \chap\ref{ch:against_PH}.

In this second part of the thesis, the goal will be to develop a new kind of approach for explaining time-asymmetry using a symmetry argument rather than a Past Hypothesis. This new approach is inspired by a related approach developed in \cite{Barbour:2014bga} and \cite{barbour2020janus}. In these approaches, explanations for the AoT are based on the same hypothesis about dynamical similarity;\footnote{ See \Sec\ref{sec:intro cosmo symmetry} and \Sec\ref{sec:a_new_kind_of_symmetry_dynamical_similarity} for an introduction to dynamical similarity and \Sec\ref{sec:dynamical_similarity} for a detailed treatment of the symmetry. } namely that it is a gauge symmetry of modern cosmology; and share common explanatory mechanisms such as Janus points. However, there are many details that differ between the two approaches, particularly regarding the definition and implications of the resulting AoT. In addition, the motivations presented here are based on the empirical principles of the PESA and not on the relational ontology of \cite{Barbour:2014bga} and \cite{barbour2020janus}.\footnote{ For a more complete discussion of the differences between the two approaches, see \Sec\ref{Nbody_motvation}. }

Before turning to my proposed explanation of the AoT, I will give more detail about the specific problem I will aim to solve. This will involve giving a progressively more detailed descriptions of the phenomena believed to define the AoT and, thus, what I think a good solution to the problem of the AoT should be able to explain. We will see that the AoT is, in fact, a multifaceted problem resulting from the interplay between specific physical processes during the evolution of the Universe. In \Sec\ref{sec:explanatory_target}, I will lump the resulting collection of problems into the broad headings of the \emph{smoothness} and \emph{red-shift} problems that were briefly introduced in \Sec\ref{sec:intro AoT} of the introduction. A good solution to the problem of the AoT must at least be able to solve both of these problems.

This leads to two different kinds of explanatory projects: a modest project, which involves giving a \emph{general} procedure for extracting \emph{some} AoT from nearly time-symmetric laws, and an ambitious project, which involves using that general procedure to explain the many \emph{specific} empirical features of our world that lead to the observed AoT.\footnote{ The modest/ambitious terminology has been adapted from a private exchange concerning \cite{ryder2022directed}. } Our goal in this second part of the dissertation can then be stated as giving a proposal that will complete the modest project and make significant headway towards the ambitious one.

My proposed solution to the modest project will involve the realisation of the Janus--Attractor scenario that was sketched in \Sec\ref{sec:intro JA scenario} of the introduction. I will give a more detailed account, including all the necessary definitions, of that scenario in \Sec\ref{sub:the_janus_attractor_scenario}. There, I will show how the existence of Janus points and attractors can formally define a general AoT for observers approaching an attractor. This will address the modest project.

I will then turn attention to the ambitious project. One way that a Janus--Attractor scenario can be seen to come about in a time-reversal invariant theory is by applying the Gauge Principle developed in \Sec\ref{sub:gauge_principle_for_DS} to dynamical similarity. By looking at two separate classes of models, the $N$-body Newtonian particle models of \Sec\ref{sec:newtonian_gravitation_models} and the cosmological models of \Sec\ref{sec:cosmological_models}, I will show that the gauge theory resulting from applying my implementation of the Gauge Principle will lead to a Janus--Attractor scenario and, therefore, an AoT under certain specified physical assumptions. I will then argue that the resulting AoT can separately solve the smoothness and red-shift problems in the corresponding model. This suggests a new and promising path towards solving the ambitious project by combining those models into a more realistic and comprehensive model of the Universe.

But before I can claim to have made progress in solving the ambitious project, I must first give a convincing reason to justify treating dynamical similarity as a gauge symmetry of the Universe. This is where my definition of gauge symmetry, carefully crafted in Part~\ref{part:foundations} of the thesis, will be necessary. The PESA will be used first in \Sec\ref{sub:dynamical_similarity_in_the_universe} and then in Sections~\ref{sec:newtonian_gravitation_models} and \ref{sec:cosmological_models} to argue that dynamical similarity should be treated as a gauge symmetry when using those models to represent phenomena in the Universe. For the purposes of the second part of the thesis, the first part can be seen as a way of developing the conceptual and mathematical tools necessary to make progress on the ambitious project.

Before describing my proposed solution, however, I will first motivate the need for a new programme by reviewing the known problems in the existing approaches. I will embark on this task in \Sec\ref{sec:the_dilemma} of this chapter. \chap\ref{ch:against_PH} will be dedicated to describing and critiquing explanations of the AoT in terms of a PH. What we will find is that no existing approaches to explaining the AoT are completely adequate. I will then contrast this with my own proposal in the concluding chapter (\chap\ref{ch:conclusions}).

\section{Identifying the explanatory target} 
\label{sec:explanatory_target}

In this section, I will try to identify the different physical processes that physicists have attributed to the AoT. This will give a more precise idea of what the AoT is and what is involved in the ambitious project of explaining the AoT. To do this, we must engage with various details of particle physics, cosmology, astrophysics, thermodynamics and even speculations about quantum gravity. One of the purposes of this analysis is to pinpoint the basic phenomena that need explaining. This is an important task because, as we will see, authors disagree about what those basic phenomena should be. What we will find is that this disagreement exists because there are at least \emph{two} distinct classes of phenomena that are responsible for many of the relevant aspects of the AoT.\footnote{ For a more comprehensive account of the different puzzling aspects of the AoT that may not be covered by these two problems, see \cite{sep-time-thermo}. }

Before describing these phenomena, let me say a few brief words about entropy. Entropy was first introduced as an extensive property of a thermodynamic system. Its tendency to increase in time, usually attributed to the second law of thermodynamics, is often identified with the \emph{thermodynamic AoT}, which, as I have said, was central to early debates about the problem of the AoT. The second law, however, has been argued to \emph{not} be sufficient for proving entropy \emph{increase} (rather than decrease), and requires at least one further assumption: the tendency of thermodynamic systems to evolve towards equilibrium. This is the so-called \emph{zeroth law of thermodynamics} advocated for in \cite{uffink2001bluff} and \cite{brown2001origins}.

As I have described, many attempts at explaining the AoT take statistical-mechanical accounts of the thermodynamic AoT as the basis for more general attempts at explaining time asymmetry. In such accounts, the statistical mechanical notion of entropy normally considered is that of the \emph{Boltzmann entropy}, which is a classical notion of entropy that I will define more carefully in \Sec\ref{sec:preliminaries} (see Equation~\ref{eq:entropy def}). The Boltzmann entropy can be generalised in various ways to be compatible with open classical systems using the \emph{Gibbs entropy} and for quantum systems using the \emph{von Neumann entropy}. For many considerations about the AoT, however, quantum effects can be ignored and the systems in question can be argued to be well-approximated by closed systems. The Boltzmann entropy is thus often tacitly assumed (or explicitly mentioned) in many discussions about the AoT. One advantage of the proposal given in \chap\ref{ch:new_aot} is that I won't need to define it directly in terms of entropy, avoiding having to make such distinctions. 

Let me now describe two distinct kinds of phenomena which together encompass key aspects of the AoT. The first regards the local behaviour of matter degrees-of-freedom in the observable universe --- particularly its homogeneity and isotropy. I will refer to the problem of explaining this behaviour as the \emph{smoothness problem} and use the word `smoothness' as a stand-in for approximate homogeneity and isotropy. This is the problem emphasised by \cite{Penrose:1979WCH}, \cite{price2002boltzmann}, \cite{albert2009time}, and others. As we'll see below, much of the discussion in the literature focuses on explaining the apparent low (Boltzmann) entropy of matter in the early Universe. Price is so convinced by Penrose's arguments that he states in \cite{price2004origins} that the extremely smooth distribution of matter in the early Universe ``is the \emph{only} anomaly necessary to account for the vast range of low entropy systems we find in the universe.'' [Original emphasis.]

In contrast, the second mechanism regards the global properties of the observable universe --- particularly the red-shift phenomenon normally associated with the rate of expansion of the Universe. I will call this the \emph{red-shift problem}. This problem is emphasised by \cite{Rovelli:2018vvy} and \cite{wallace2011logic}. Rovelli considers the behaviour of an important cosmological variable, called the \emph{scale factor}, that roughly describes the size of the Universe at any given time (see \Sec\ref{sec:cosmological_models} for a more precise definition). He then takes this to be the primary explanatory target: ``it is the scale factor \emph{and only the scale factor} that was [...] the ultimate source of low entropy.'' \citep{Rovelli:2018vvy} [Emphasis added.] We will see in \Sec\ref{sub:dynamical_similarity_in_the_universe} that the PESA will suggest that the scale factor should \emph{not} be considered part of the observable algebra of cosmology. Nevertheless, its rate of change is a measure of the amount of red-shifting at any given time in the observable universe. This information is contained in the so-called \emph{Hubble parameter} $H$ used by cosmologists. The PESA will say that the Hubble parameter \emph{should} be part of the observable algebra. Fortunately, all the significant points made by Rovelli actually carry through if one considers the Hubble parameter as the source of low entropy rather than the scale factor itself. We will study the reasons for this in \Sec\ref{sub:the_red_shift_problem}.

Given these two point of view, it is useful to define:
\begin{enumerate}[(1)]
    \item \textit{The smoothness problem:} Why was the Universe so remarkably smooth early in its history?\label{prob:smoothness}
    \item \textit{The red-shift problem:} Why was the Hubble parameter, which measures the rate of red-shifting in the Universe, so large and monotonic in the past?\footnote{ Note that the Hubble parameter is the \emph{relative} rate of expansion and is monotonically \emph{decreasing}, not increasing, towards what we normally call the `future' (and which I will later identify as the direction of the approaching de~Sitter attractor). } \label{prob:red_shift}
\end{enumerate}
I will argue that, to complete the ambitious project, one must at least solve both the red-shift and smoothness problems. Thus, Price and Rovelli are ultimately correct in identifying necessary conditions for explaining the AoT but wrong to think that those conditions are also sufficient.

Let me make a couple of comments before giving a more detailed description of the smoothness and red-shift problems. First, as I have already emphasised, there are many kinds of time-asymmetric phenomena (sometimes referred to as different arrows of time) that can be associated with the problem of the AoT. \cite{Penrose:1979WCH} highlights no less than seven. For the most part, reasonable arguments can be given to either show that these different time-asymmetric phenomena have a common cause or that some subset of them is irrelevant.\footnote{ See \cite{Penrose:1979WCH,price2004origins,Rovelli:2018vvy} and \cite[Ch 7]{Penrose:NewMin} for arguments of this kind.} But many open problems remain --- many of which won't be directly addressed in this work. For example, even the thermodynamic AoT itself is no longer a clear explanatory target if one follows the criticisms that I will describe more fully in \chap\ref{ch:against_PH}. If the entropy of the universe cannot be well-defined, which may be the case given the arguments of \chap\ref{ch:against_PH}, then it is no longer obvious that what actually needs to be explained is the very low entropy of the very early universe.

Second, in the discussions below, I will often refer to \emph{equilibrium} and \emph{non-equilibrium} states of different variables. In the study of thermodynamic systems, the notion of equilibrium can be made precise and represents a steady state of affairs in terms of a particular macroscopic description of the system. For my purposes, I will simply regard equilibrium states as states that do not noticeably vary in the natural time scales relevant to the process in question. This notion can apply to basically any variable in the theory. It will also be helpful to introduce the notion of a \emph{local}, which is sometimes also called \emph{metastable}, equilibrium. Local equilibrium states are states that reach an approximate equilibrium over some physically relevant timescale before exiting that state at some later time. Such local equilibria are common to thermodynamic systems and, as we will see below, can be used to describe many phases of the Universe's evolution.

\subsection{The smoothness problem} 
\label{sub:smoothness_problem}

Let me begin by defining the \emph{smoothness problem} and explaining its relevance to the AoT. The call to explain why the early universe was extremely smooth --- i.e. why the distribution the energy-momentum and geometric degrees of freedom was approximately homogeneous and isotropic --- is familiar from the Past Hypothesis literature. \cite{price2004origins} argues:
\begin{quote}
The crucial thing is that matter in the universe is distributed extremely smoothly, about one hundred thousand years after the Big Bang. 

... In effect, the smooth distribution of matter in the early universe provides a vast reservoir of low entropy, on which everything else depends. The most important mechanism is the formation of stars and galaxies. Smoothness is necessary for galaxy and star formation, and most irreversible phenomena with which we are familiar owe their existence to the sun.
\end{quote}
The motivation here is that smoothness leads to a very low-entropy state when the system is self-gravitating. Standard arguments from the statistical mechanics\footnote{ See \cite{PADMANABHAN:1990book} for detailed derivations of gravity's unusual thermodynamic properties and how the statistical mechanics of the gravitational $N$-body problem can be applied to galaxies. } of the $N$-body system suggest that, because the force of gravity is attractive and leads to the clumping of point masses, smooth states that are not clumpy are highly entropically suppressed.\footnote{ More specifically, clumpy states occupy large phase space volumes because the $-1/r$ gravitational potential is peaked on them. } The observation that the contents of the Universe were very smooth at a particular point in its evolution then implies that the Universe was in a low-entropy state at that time.

In reference to the arguments given above, \cite{price2004origins} goes as far as declaring that the ``discovery about the cosmological origins of low entropy is the most important achievement of late twentieth century physics.'' As I have already stated, according to him, this observation is the \emph{only} thing that needs explaining in order to account for the thermodynamic AoT because it entails ``most irreversible phenomena with which we are familiar.''

Let us evaluate this claim.

To begin with, it is true that, in order to even apply thermodynamic notions to the universe, we must already take advantage of the observational fact that it is, to a good approximation, relatively smooth  on large scales --- at least during the epochs to which we have direct observational access. To understand why, consider a patch of space containing smooth matter whose boundary evolves as if it were made of free dust particles. Such a patch of space acts as a kind of box through which there is effectively no heat flow because the homogeneity of the matter distribution implies that any heat flowing out of the box should be balanced by an equal amount flowing in. This allows us to define the thermodynamic properties of the matter in that box provided a variety of other physical conditions on the states of the matter itself are met.\footnote{ For example, the timescales of changes in the temperature and other thermodynamic quantities must be large compared to the mixing times of the matter. } If we take the box to be the spatial boundary of the visible universe; i.e., the finite part of the whole `Universe' that is causally accessible to us (which we will henceforth call the `universe' with no capitalization); then we have what we need to talk about the temperature and entropy of matter (or radiation or geometry). Because of this, smoothness is a necessary, but not sufficient, condition for being able to define the entropy of the universe in the first place.

We can now evaluate the evidence for the universe being `smooth' in its `early' state. The epoch of $10^5$ years after the Big Bang, mentioned by Price, is relatively late in the evolution of the early universe. This is roughly the time of \emph{recombination} in which the universe cooled to the point where the first electrically neutral hydrogen atoms could form, releasing the first visible light in the universe called the \emph{Cosmic Microwave Background (CMB)} radiation.\footnote{ More precisely, recombination occurred $3.78\times 10^5$ years after the Big Bang. It's not completely clear whether Price meant this period or some slightly earlier epoch. } The fact that any cooling occurred at all is an essential assumption of Price's claim, and one that I will criticise shortly in \Sec\ref{sub:the_red_shift_problem}. But putting this assumption aside for now, there is good evidence that the universe was relatively smooth at the time of recombination. This evidence comes from direct observations in radio-astronomy of the CMB, indicating temperature variations of $1$ part in $10^5$ over background levels.\footnote{ The details about this and other cosmology evidence discussed in this section can be found in any good cosmology text book such as \cite{dodelson2020modern}, \cite{mukhanov2005physical}, or \cite{weinberg2008cosmology}. My own treatment roughly follows \cite{baumann_2022}. See Chapter~7 of \cite{baumann_2022} for an introduction to the CMB observations.}

This confirms that the universe was smooth up to variations of $10^{-5}$ at the epoch of $3.78 \times 10^5$ years. But is this state `smooth' and `early' enough? While we don't have \emph{direct} observations of the large-scale structure of the universe before recombination, we do have reliable \emph{indirect} evidence that the universe was even smoother at earlier times. Some of this evidence comes from the inflationary paradigm of modern cosmology. It's important to note, however, that our knowledge of the evolution of the universe gets more speculative as one approaches the Big Bang. Fortunately, just how early the universe was smooth and the extent of this smoothness is not particularly relevant to the overall structure of the argument. At a minimum, one needs to explain the extremely smooth state of the CMB, and there is no good reason to suspect that the universe was less smooth until \emph{very} early in its evolution.\footnote{ A reasonable estimate for what `very early' might be is the epoch when inflation is thought to have ended (if it occurred at all). This is because inflation was introduced as a mechanism for smoothing out the contents of the universe. }

According to well-known arguments, clumping due to the self-gravitation of the over-dense regions of the CMB dramatically amplified the inhomogeneities that were small at recombination. This clumping led to the formation of the dense gas clouds that produced the first stars. These processes have been modelled by state-of-the-art numerical simulations, such as the Millennium Simulation \citep{springel2005simulations}, and the resulting matter distributions match the observed large scale structure of the universe. The early stars created in those gas clouds undoubtedly produced material that collapsed under gravity to produce later generations of stars including our Sun --- although the exact details of this process are still not completely understood. The Sun, in turn, is a vast reservoir of entropy that we understand to ultimately be the source of nearly all the time-asymmetric processes on Earth. As we will see in the next section, however, this vast reservoir of entropy is \emph{not} completely, or even mostly, due to the gravitational collapse of the CMB inhomogeneities.

The chain of arguments just described leading from the clumping of homogeneities in the CMB to the emergence of a local entropic AoT here on Earth are generally well-understood and uncontroversial. Further details are provided in Chapter 7 of \cite{Penrose:NewMin}. I conclude from this that there are strong arguments to suggest that the initial smoothness of the CMB guarantees that the conditions favourable to star formation happened at roughly the same time and in roughly the same way across the entire visible universe. As I will argue below, this combined with a solution to the red-shift problem ensures both that an observer near a particular star will see a strong entropic AoT \emph{and} that the direction of this AoT will be consistent, to a very high degree, across the visible universe. Solving the \emph{smoothness problem} is thus \emph{necessary} for explaining the AoT. But because of the caveats mentioned above, explaining smoothness is not sufficient. I will turn to these now.

\subsection{The red-shift problem} 
\label{sub:the_red_shift_problem}

\subsubsection{Red-shift and the cosmological arrow}

Most of the heat, and therefore the entropy, produced by stars is produced in nuclear fusion reactions. A simple calculation (see, for example, \cite{wallace2010gravity}) shows that the entropy generated by such nuclear reactions dwarfs the entropy generated by the gravitational collapse. The primary role of gravitational collapse is thus not to explain entropy gradients on the Earth but to trigger, as a kind of catalyst, the start of nuclear reactions in stars, which are the real driver of entropy increase. Thus, the time-asymmetric processes caused by the Sun and other stars can only be properly understood by giving an explanation for the existence of the entropy reservoir stored in the Sun's nuclear fuel.

The origin of this entropy reservoir is tightly connected to the large \emph{rate of expansion} in the very early universe, as I will show below. Firstly, when we talk of the universe as ``expanding'' we mean, from an observational perspective, that the spatial distance between bound systems free of external forces is increasing relative to the characteristic size of those bound systems. The fact that expansion in this sense is a relative notion will be central to the symmetry argument I will present in \Sec\ref{sub:dynamical_similarity_in_the_universe}. On cosmological scales, expansion is manifest in the form of light waves having increasingly larger wavelength the further they get from their source. This is the \emph{red-shifting} phenomena in terms of which the red-shift problem will be understood. The relative rate of change of the red-shift is Hubble parameter, $H$, which is a central quantity in observational cosmology.\footnote{ The Hubble parameter is usually defined in terms of the scale factor, $a$, as $H=\frac {\dot a}a$, which is related to the red-shift, $z$, by $z(t) = 1 - a(t)/a(0)$, where $t=0$ is some arbitrarily chosen reference time. We will see that the definition in terms of the scale factor is actually unnecessary and, even, misleading. } 

Central to my argument later is the fact that it is the relative rate of expansion, $H$, and not the absolute size of the universe itself that is responsible for creating the entropy reservoir stored in the Sun and other stars. To understand why, let us first recognise that the temperature of the early universe scales like $\sqrt H$ according to standard arguments.\footnote{ See, for example, Equation~3.55 of \cite{baumann_2022} and the surrounding derivation. } As a result, decreasing values of the Hubble parameter corresponds to decreasing temperature.\footnote{ Recall that the Hubble parameter is the \emph{relative} rate of the expansion, which is decreasing even as the red-shift itself is increasing. } Let us now see how decreasing temperatures lead to the entropy reservoir stored in stars like the Sun. Two timescales are relevant to the discussion: the average time between particle interactions $t_c$ and the characteristic timescale of expansion $t_H \propto 1/H$. When $t_c \ll t_H$, the rate of interactions is large compared with the rate of expansion, and the particles in the universe reach a local thermal equilibrium. But $t_c$ and $t_H$ usually scale differently with temperature so that $t_c$ grows relative to $t_H$ over time as the temperature decreases. When $t_c \gtrsim t_H$, universal expansion is too rapid for significant particle interactions to occur and the interactions effectively stop. Whatever relative abundances of different particle species exist at that time then get ``frozen-out.'' Because different particle species interact at different rates, this ``freeze-out'' time is different for different interaction types.

Many freeze-out events are believed to have occurred over the history of the universe. These freeze-out moments define the boundaries of different cosmological epochs, whose relevance to the local AoT on Earth we will discuss below in \Sec\ref{ss2:global to local arrow}. For a given freeze-out event, the relative abundances of particle species in the universe can be computed using different cosmological models, and the abundances predicted by these models can be compared to their observed values. The match between these predictions and observations is one of the great successes of modern cosmology, confirming our understanding of the physics of these freeze-out moments. It should be stressed, however, that the earlier the freeze-out moments, the more speculative our understanding of them becomes.

From the perspective of the cosmological AoT, what is mysterious about these freeze-out moments is the fact that they occurred at all. The freeze-out process is time-asymmetric, and thus can't be explained with time-symmetric laws alone. Importantly, if the universe had started out in a global equilibrium state, then the second law of thermodynamics suggests that it should stay that way on the timescales relevant to cosmology. No freeze-out moments could ever occur because the state would effectively be permanently frozen. The question then is: what could be responsible for the lack of global equilibrium we observe?

Before the freeze-out times of most cosmological epochs, the particles in the universe are not in global equilibrium. They are, however, in local thermodynamic equilibrium. When freeze-out occurs, the system is driven out of that local equilibrium and eventually settles to a new one on timescales determined by the $t_c$ relevant to the dominant interaction of that epoch. Crucially, this drive away from equilibrium is caused by a single phenomenon: the decreasing relative rate of expansion of the universe. This is only possible if the Hubble parameter was itself wildly out of equilibrium in the early universe and at each subsequent freeze-out time. Thus, the time-asymmetry created by the sequence of freeze-out moments that occurred in the universe can be directly linked to the monotonically decreasing Hubble parameter, which indicates a significant departure from equilibrium. I will then understand a good explanatory account of the extremely large initial Hubble parameter \emph{and} its monotonic decrease as a solution to the \emph{redshift problem}.

\subsubsection{From a cosmological to a local arrow}\label{ss2:global to local arrow}

Let me now give a chain of arguments explaining how the freeze-out moments in cosmology lead to the thermodynamic AoT observed on Earth. During the period starting at roughly $3$ minutes after the Big Bang, the light elements started to form in the epoch called \emph{Big Bang Nuclearsynthesis (BBN)}. During this epoch, cooling made proton-neutron pairs (deuterium nuclei) stable, and this in turn enabled a wave of early nuclear fusion reactions that used any available neutrons, binding most of them, together with protons, into the more stable helium nuclei. Given the stability of helium, if the universe had stopped expanding at this point nuclearsynthesis would have slowly continued and eventually --- although on a very large timescale --- helium (and other heavier elements) would have been produced until no hydrogen would have remained. This would have effectively burnt up all stellar fuel and prevented any stars from forming. Instead, the relative expansion rate continued to decrease and drastically reduced the rates of nuclear reactions leaving enough stellar fuel for our sun to form.

The connection between the red-shift and the origins of the large entropy reservoirs produced at the end of BBN has been emphasised by \cite{wallace2010gravity}. It is not, however, the whole story. In practice, the bottleneck that effectively stopped the production of further light elements during BBN was the availability of neutrons. Creating deuterium directly from protons, as is done in the Sun, is exceptionally slow at the temperatures and timescales of BBN. Thus, helium and other light elements needed to be formed from the neutrons produced in the earlier epochs of the universe. Before BBN, neutrons and protons were in local thermodynamic equilibrium and roughly equal in number through the process of $\beta$ and inverse $\beta$-decay:
\begin{align}\label{eq:beta decay}
    n + \nu_e &\leftrightarrow p^+ + e^-\\
    n + e^+ &\leftrightarrow p^+ + \bar \nu_e\,.
\end{align}
However, because of the slightly heavier mass of the neutron, protons gradually became more dominant once the universe cooled below the mass difference of the neutron and proton. Eventually, the reactions mediating these exchanges turned off and the neutron-proton ratios became ``frozen-in.'' Neutron decay further limited its numbers. Once the conditions for deuterium production became favourable, the fraction of neutrons was only about $1/8$ so that no more than $1$ helium atom could be formed for every $4$ atoms of hydrogen. This excess of hydrogen is the origin of the main source of stellar fuel in the universe.

While the details above are interesting and necessary for evaluating different cosmological models, what is important for this discussion is the striking role played by the various different freeze-out moments in the epochs discussed above. From the time the process of \eqref{eq:beta decay} turned off until the end of BBN, the universe passed between several phases with local equilibria dominated by different processes. At each stage these local equilibria are disrupted by the relentless decrease of the Hubble parameter. The entropy that got trapped at the end of these processes in the form of excess hydrogen is a direct result of the universe being forced out of its trajectory towards a global entropy maximum into metastable states in the new local equilibrium. These are all well-understood processes in modern cosmology even though the special role of the Hubble parameter in driving them is not always emphasised. Understanding the behaviour of the Hubble parameter in the universe is thus essential to understanding the origins of the AoT.

\cite{Rovelli:2018vvy} highlights the importance of the relatively long period of local equilibrium that the universe enjoyed while the processes in \eqref{eq:beta decay} were dominant, and he is clearly right about its relevance to the origins of the local thermodynamic AoT observed here on Earth. However, it is clear that even this period of relatively long-lived equilibrium was preceded by other epochs where different equilibria were present. For example, since free neutrons are metastable through $\beta$-decay, their abundance in the early universe represents a new kind of entropy reservoir similar the entropy reservoir of hydrogen atoms. Why was the early universe put in a low-entropy state with any neutrons to begin with? The answer can presumably be found by looking at the structure of the freeze-out moments in the even earlier universe; i.e., those that triggered by the beginning of the so-called \emph{hadron epoch} between around $10^{-5}$ s and $1$s after the Big Bang.

Before this was the \emph{quark epoch} where the universe was in a state of quark-gluon plasma, in which no hadrons, such as protons and neutrons, had yet formed as bound states. Again, the decreasing expansion rate of the universe caused cooling which allowed such bound states to form. The disruptive push of the decreasing Hubble parameter drove the universe into a new local equilibrium storing entropy in the form of new metastable bound states.

From our knowledge of particle physics experiments, we can predict several earlier freeze-out moments including the period of \emph{electroweak symmetry breaking}, where the electromagnetic force and the weak nuclear force are not yet distinct and the Higgs mechanism has not yet taken place to give rise to the masses of elementary particles. Earlier still we encounter less understood and somewhat controversial physics like the very rapid period of early expansion of inflation and the \emph{reheating} period that follows it.

Eventually, we reach the scale where quantum gravity effects are believed to be important. According to untested but widely accepted theories, the highest entropy state during this epoch is a black hole state. In Chapter 7 of \cite{Penrose:NewMin}, it is estimated that such a state would be $10^{10^{120}}$ times more entropically favoured than that of the current observed universe. While certain aspects of this estimate could easily be disputed, the shear size of this number indicates the scale of the problem. 

Interestingly, \cite{Penrose:1979WCH} uses an early version of this estimate to motivate the so-called \emph{Weyl curvature hypothesis}, which is a constraint on the smoothness of the geometry of the universe at the Big Bang.\footnote{ Specifically, the conjecture holds that the Weyl curvature of the spacetime metric is zero at the Big Bang. } The Weyl Curvature Hypothesis, however, has the potential to address the smoothness problem, but it has no way of addressing the red-shift problem because the Weyl curvature does not constrain the behaviour of $H$.\footnote{ To understand why, note that the Weyl curvature only depends on the conformal degrees of freedom of the metric while the scale factor caries units and is therefore non-conformal. }

The explanatory chain leading to various entropy reservoirs in the universe can, thus, be pushed back as far as our understanding of physics allows. At each link in the chain, the explanatory target is clear: the decreasing relative rate of expansion of the universe. The entropy reservoirs produced by the decreasing Hubble parameter create the conditions necessary for nuclear synthesis to begin in stars once enough gravitational collapse has occurred. The local AoT observed on Earth and other star systems then results from the combination of the gravitational clumping of early inhomogeneities and the presence of nuclear potential energy trapped in hydrogen. I conclude that any reasonable explanation of the AoT should thus solve \emph{both} the smoothness and red-shift problems.

\section{Price's taxonomy}\label{sec:Price_taxonomy}

In the previous section, I described several important phenomena that comprise the AoT. I outlined a series of processes that appear to have led from a smooth state far from equilibrium in the very early universe to a significant, spatially uniform, monotonic entropy gradient across our present observable universe. The problem of the AoT is then to provide an explanation for this universal observed gradient in terms of laws that have no apparent time asymmetry. The AoT is puzzling to the degree that it is puzzling to have a universal physical feature that is not in any way suggested by the known universal laws of nature. I will now discuss different approaches to finding a solution to this problem.

\cite{price2002boltzmann} provides a helpful taxonomy for distinguishing different existing approaches to explaining the AoT. While Price was mainly motivated by the smoothness problem, his classification applies to general explanations of the AoT, and is therefore applicable to the red-shift problem as well. The distinction he draws in between two different approaches he calls ``causal-generalism'' and ``acausal-particularism.'' Causal-generalism is characterised as follows:
\begin{quote}
    On one side are what I shall call \textit{Causal-General} theories. These approaches take the explanandum to be, at least in part, a time-asymmetric \textit{generalisation}---the general fact that entropy never decreases, or some such. Broadly speaking---perhaps taking some liberties with the terms \textit{causal} and \textit{dynamical}---they seek a causal explanation of this general fact in dynamical terms. Approaches I take to fall under this heading include `interventionism' and certain appeals to asymmetric initial microscopic independence conditions, as well as to suggestions grounded on law-like asymmetries in the dynamical laws themselves. What unifies these diverse approaches, in my view, is their sense of the nature of the project. All of them seek a \textit{causal-explanatory} account of a \textit{time-asymmetric generalisation} about the physical world as we find it.  (p.\ 90)
\end{quote}
According to the opposing, acausal-particularist views,
\begin{quote}
     All the time-asymmetry of observed thermodynamic phenomena resides in an existential or particular fact---roughly, the fact that physical processes in the known universe are constrained by a low entropy `boundary condition' in one temporal direction. Against the background of a \textit{time-symmetric} understanding of the normal behaviour of matter, this particular fact alone is sufficient to account for the observed asymmetry in thermodynamic phenomena. The task of explaining the observed asymmetry is thus the task of explaining a particular violation of contrast class (b)---a particular huge entropy gradient, in a world in which (roughly) none are to be expected. (p.\ 92)
\end{quote}
For the purposes of this thesis, I will sidestep \emph{causal} questions about the AoT and will refrain from using the terminology of cause and effect. Instead, I focus on whether any given approach hypothesizes time-reversal invariant \textit{laws} or not.

A complication arises from the fact that the laws of nature encoded in our best current theories are not completely time-reversal, or $T$, invariant but only ``nearly'' $T$-invariant. In particular, the Standard Model of elementary particle physics is notably \emph{not} $T$-invariant. For example, the time-reversed dynamics of a ``left-handed'' electron are those of a ``right-handed'' anti-electron (positron). There is some disagreement among philosophers regarding the appropriate metaphysical implications of this lack of $T$-invariance. Some; e.g., \cite[Ch 7]{earman1989world}; argue that this supports substantivalism about temporal orientation while others; e.g., \cite{pooley2003handedness,price1997book}; disagree.\footnote{For a discussion of how such $T$-violations can be understood and a summary of some of the issues involved, see \cite{roberts2022reversing} --- particularly Chapter 7.}

But when it comes to the quantitative empirical question of whether the \emph{amount} of time asymmetry resulting from $T$-violations in the Standard Model is sufficient for providing an adequate explanation of the cosmological AoT, the overwhelming consensus is an emphatic `no.' In particular, $T$-violations in the Standard Model are neither able to explain the shear amount of smoothness seen at recombination nor the dramatic red-shift in the early universe. Thus, while $T$-violation in the Standard Model does constitute evidence for time asymmetry in the known fundamental laws, this time asymmetry is not able to provide an adequate explanation for the AoT.

Given the numerical insignificance of the $T$-symmetry violations in the Standard Model, I will regard our fundamental laws as $T$-invariant \emph{For All Practical Purposes (FAPP)}. With this terminology in place, I propose the following taxonomy of ``generalist'' and ``particularist'' accounts of the AoT:
\begin{itemize}
    \item[]{\bfseries Generalism:} accounts of the AoT according to which the true laws of nature are time-asymmetric and make the thermodynamic AoT an expected feature.
    \item[]{\bfseries Particularism:} accounts of the AoT according to which the laws are FAPP time-symmetric while there is some particular, contingent, fact that makes the AoT expected.
\end{itemize}

Examples of attempts to provide generalist accounts of the AoT involve using spontaneous collapse models of quantum mechanics to introduce fundamental and significant time asymmetry into the laws. A well-known example, advocated in \cite{albert2009time}, makes use of the spontaneous collapse theory introduced by \cite{ghirardi1986unified}.

Alternatively, particularist theories make use of some version of the PH. The most common way to implement a PH is in terms of a low-entropy initial state. However, low-entropy is not the only way to frame a PH. The \emph{Weyl curvature hypothesis}, discussed near the end of \Sec\ref{ss2:global to local arrow}, is a PH that imposes a constraint  on the spacetime geometry of the initial state, and such a constraint has no obvious (or proven) connection to any notion of entropy. Additionally, while smooth states in self-gravitating systems are certainly low-entropy, it's not clear whether the same can be said about states with a rapidly decreasing Hubble parameter. Thus, there is no good reason to assume that low-entropy is the right kind of condition on a past state for solving the red-shift problem. Instead, what makes a PH is that the past state falls into a class of states (eg, smooth, highly red-shifted or low in Weyl curvature) that are deemed to be atypical according to some reasonable measure on the state space of the theory. We will see the reason for this in a moment.

Price argues in favour of acausal-particularism or, in our terms, particularism. The basic idea behind this type of approach is based on an old argument by Boltzmann (see, for example, \cite{boltzmann1895certain}) regarding the thermodynamic AoT in free gases.\footnote{ I will describe this in more detail in \Sec\ref{sec:the_past_hypothesis}.} This style of argument is then claimed to underpin a general explanation of how an atypical past state can explain the AoT. In \cite{price2002boltzmann}, the approach is described as follows:
\begin{quote}
    [T]he basic character of Boltzmann's statistical approach is well known. Consider a system not currently in equilibrium, such as a vial of pressurised gas within a larger evacuated container. We want to know why the gas expands into the larger container when the vial is opened. We consider what possible future `histories' for the system are compatible with the initial set-up. The key to the statistical approach is the idea that, under a plausible way of counting possibilities, almost all the available microstates compatible with the given initial macrostate give rise to future trajectories in which the gas expands. It is possible---both physically possible, given the laws of mechanics, and epistemically possible, given what we know---that the actual microstate is one of the rare `abnormal' states such that the gas stays confined to the pressurised vial. But in view of the vast numerical imbalance between abnormal and normal states, the behaviour we actually observe is `typical', and therefore calls for no further explanation. There is no need for an asymmetric causal constraint to `force' the gas to leave the bottle---this is simply what we should expect it to do anyway, if our expectations about its initial state are guided by Boltzmann's probabilities.(p.\ 92)
\end{quote}
The assumption of the initial non-equilibrium state of the system, in this case the ``vial of pressurised gas within a larger evacuated container,'' is the key ingredient of the proposed explanation of the observed behaviour of the gas. Because the Boltzmann entropy is the log of the phase space volume of a macrostate, low-entropy states are atypical under the natural measure on phase space. If Boltzmannian entropy is low for some time $t_0$ (i.e., the state at $t_0$ is atypical), then entropy increase is expected for times $t>t_0$ (i.e., the states after $t>t_0$ are expected to become more typical). Thus, the riddle of the thermodynamic AoT is removed.  The only explanatory task left, the idea goes, is to account for why the state at the early time $t_0$ is so atypical.

The framing of this argument in terms of the typicality of states is central to its generality: states of low-entropy, vanishing Weyl curvature and rapidly decreasing Hubble parameter are all understood to be atypical --- at least according to standard arguments. In this way, the particularist strategy claims to be able to provide an explanation for the AoT. I will revisit whether early smoothness and rapidly decreasing Hubble parameter are atypical states in light of the symmetry arguments presented in \Sec\ref{sub:dynamical_similarity_in_the_universe}. But for now, let us accept the general premiss and assess the relative merits of particularism and generalism.

Price defends particularism and opposes generalism. His main reason for doing so is that, as he sees it, generalists must also explain, in addition to the AoT, the highly unusual early state of the universe. As a result, the overall generalist explanation of the AoT is less economical:
\begin{quote}
     It is surprising that such a stark contrast --- the invocation of one temporal asymmetry in the latter [Acausal-Particular] approach as against two in the former [Causal-General] --- seems to have received little explicit attention in the literature. The contrast suggests that \emph{prima facie}, at least, the Acausal-Particular approach has considerable theoretical advantage. To the extent that asymmetry is a theoretical `cost', the Causal-General approach is a great deal less economical than its rival. (p.\ 99)
\end{quote}

Price here notes that the generalist is as burdened to explain the atypicality of the past state as the particularist.\footnote{ While Price's original argument focused on the smoothness problem, it could easily be generalised to include the red-shift problem as well. } The bulk of Price's paper is then devoted to an exorcism of the idea that a causal mechanism or ``engine'' must be specified in order to explain \textit{why} entropy rises even after the assumption has been granted that it starts out low. Once the idea that such a mechanism is needed has been debunked, the motivation for generalism is undermined, as he sees it, and it is clear that particularism is the only way forward.

I agree with Price that, inasmuch as generalists face the same challenge as particularists to account for why the very early universe is in a highly atypical state, generalism provides no added benefits. But I disagree that the path to victory for particularism is as clearly laid out as he takes it to be. In the next section, I will describe an impasse according to which both particularist and generalist accounts of the AoT appear to have highly unattractive features, seemingly leaving no good option available.

\section{An impasse}
\label{sec:the_dilemma}

In this section, I will argue that both generalist and particularist approaches to explaining the AoT face serious objections. All the lines of thought underlying these objections are familiar from the philosophy of physics literature, though only some have been harnessed explicitly as objections against particular views of the AoT. I will outline the objections in what follows, first for generalism, then for particularism.

\subsection{Objections to generalism} 
\label{sub:objections_to_generalism}

\begin{enumerate}[G.I]
    \item {\bfseries Objection from redundancy} This objection is just Price's primary criticism of generalism, which I have described in \Sec\ref{sec:Price_taxonomy}. It says that any generalist approach must posit special initial conditions \textit{in addition to} time-asymmetric dynamics. This makes the move of postulating time-asymmetric dynamics redundant and undermines the rationale for generalism.
    \item {\bfseries Objection from lack of independent motivation:} The fundamental known laws of physics, as currently encoded in the Standard Model of elementary particle physics and general relativity, are time-symmetric in the FAPP sense defined above. These laws are extremely successful in describing the physics and microphysics underlying all known processes in nature, including those which instantiate the thermodynamic AoT on a macroscopic level. The hints we have about where new physical laws are needed -- for instance the inability of the Standard Model of elementary particle physics to account for dark matter -- do not seem to be related to the time symmetry of the Standard Model and general relativity. There is thus no independent empirical motivation, beyond potentially the AoT itself, for believing that time-asymmetric laws would be correct instead. It is also unclear how any account that incorporates generalism could preserve the facts that are so well accounted for by time-symmetric laws while simultaneously accounting for the thermodynamic AoT in terms of its significant time-asymmetric aspects.
    \item {\bfseries Objection from historical progress} From a historical perspective, comparing the evolution of our understanding of the spatial dimensions with the time dimension, it would be surprising if the laws of nature turned out to be time-asymmetric after all. ``Naive'' physics, based on our everyday experience, suggests that there is a principled difference not only between opposite time-directions ``past'' and ``future'', but also between opposite spatial directions ``up'' and ``down.'' Progress in understanding the world around us has made it clear that the laws of nature make no principled distinction between two opposite space direction when it was recognised, already in antiquity, that the Earth is spherical. The distinction between ``up'' and ``down'' has since been understood in terms of a particular fact: the Earth, as the body that exerts the strongest gravitating pull on us, defines what counts locally as ``down.''

    Since Newtonian times, our best laws of physics --- from Newton's theory of gravitation to the Standard Model of elementary particle physics --- have treated the time-coordinate analogously to the space-coordinates in this respect. They do not treat the two time-directions qualitatively differently. It is very natural to believe that this feature is not accidental and that fundamental laws to be discovered in the future will treat the two time directions analogously ---  at least on the ``global'' level for which this is true for the spatial directions. The CPT-theorem of quantum field theory even mixes time-reversal with reversal of spatial coordinates and thus provides further support that this analogy is not accidental. It can of course not be ruled out that future physics might treat the past and future directions radically differently. But this would seem to be an odd turn in the historical development of our understanding of time and its role in the laws of nature --- not something that we would otherwise expect. 
\end{enumerate}

These objections to generalism have had a stinging impact on the philosophical community. While there are still attempts to conceive of explanations of the AoT by appealing to time-asymmetric laws, many modern approaches embrace particularism instead. Bus as we will see next however, particularism faces serious problems as well --- problems that have been glossed over by Price in his endorsement of it.

\subsection{Objections to particularism} 
\label{sub:objections_to_particularism}

Here I note three important categories of objections against particularism. I will describe these here only very briefly to balance the discussion. Later, in \Sec\ref{sec:deconstructing_the_argument}, I will expand on each of these points in greater detail. The categories of objections I will consider are:
\begin{enumerate}[P.I]
    \item {\bfseries Objections from mathematical and conceptual ambiguity:} These objections arise from the fact that the particular state required by particularist explanations is difficult to characterise mathematically and motivate physically. The PH, requires that the early state be atypical. This requires a reasonable typicality measure for ``counting'' the states of a theory. At early enough times in the universe's history, general relativistic degrees of freedom must also be taken into account. Unfortunately, recent research has brought to light several daunting difficulties involved in doing so.

    As discussed at length in \cite{earman2006past}, \cite{Wald:2012zf} and \cite{Curiel:2015oea}; even in homogeneous cosmology, which is a dramatically simplified version of general relativity, there are serious ambiguities in defining a measure on the space of possible models of the theory. In \cite{Ashtekar:2011rm}, it is noted that such ambiguities lead to estimates of the probability of inflation that differ by up to $85$ orders of magnitude (e.g., \cite{Kofman:2002cj} compared to \cite{Turok:2006pa}). Using more realistic models only adds to the problem. As is show in \cite{Wald:2012zf}, the infinite dimensional nature of the state spaces of perturbed cosmological models leads to considerable mathematical and interpretational problems.

    In addition to these arguments, the arguments I will present in \Sec\ref{sec:symmetries_and_measure_ambiguities} show that for a PH to have explanatory force, it must make a distinction without a difference. This distinction arises by over-counting empirically equivalent states that are simply global rescalings of each other.\label{obj:ambiguity}

    \item {\bfseries Objections from the breakdown of thermodynamic assumptions:} These arise from a lack of precision often involved in defining \emph{the} entropy of the universe or in connecting the typicality of the universe's states to the notions of entropy relevant to the local thermodynamic arrow. Since the universe is a complicated many-body self-gravitating systems, it is not clear whether there is a good notion of entropy that can be applied consistently to the universe as a whole. Moreover, even if such a notion were available, the particularities of the dynamics of these systems raise questions about whether thermodynamic concepts are even appropriate to use at various levels.\label{obj:thermo_breakdown}
    \item {\bfseries Objections from lack of explanatory force:} These arise from the absence of any good candidate explanations for the particular fact used to explain the AoT itself. Without such an explanation, it is not clear whether particularist accounts can have any real explanatory force. Even \cite{price2002boltzmann} admits that a ``solution to this new puzzle [of explaining why the PH holds] is not yet in hand. Indeed, it is not yet clear what a solution would look like.'' (p.118) As pointed out by \cite{callender2004measures}, it is difficult to see what \textit{kind} of explanation there could even be for such a particular fact and whether demanding such an explanation is even warranted at all.\label{obj:explanation}
\end{enumerate}

I conclude from this catalogue of objections against generalism and particularism that attempts to account for the AoT seem to be at an impasse: none of the available approaches seems promising, let alone clearly laid out.

\chapter{Against the Past Hypothesis}
\label{ch:against_PH}

\begin{abstract}
    In this chapter, I argue that explanations for time-asymmetry in terms of a `Past Hypothesis' face serious difficulties. After reviewing the basic assumptions of the Past Hypothesis, I strengthen grounds for existing objections by outlining three categories of objections that put into question essential requirements of the proposal. Then, I provide a new argument showing that dynamical similarity should be treated as a gauge symmetry of cosmology. Finally, I use this result to show that an advocate of the Past Hypothesis faces a dilemma: introduce a distinction without difference or lose explanatory force by introducing a time-independent measure. This further motivates the need for the new solution I will present in \chap\ref{ch:new_aot}.
\end{abstract}

\ifchapcomp
    \tableofcontents
    \newpage
\else
    \cleardoublepage
\fi

\section{Introduction} 
\label{sec:intro}

In the previous chapter, I introduced the problem of the AoT and presented two different approaches --- particularism and generalism --- to solving it. I noted that generalism faces severe challenges, and that particularist approaches have been favoured in the recent literature. But particularism faces its own set of difficulties. I outlined different categories of objections to particularism in \Sec\ref{sub:objections_to_particularism}. In this chapter, I will investigate these objections more carefully. Then, I will use the PESA to raise a new one. My conclusion will be that particularist accounts of the AoT face serious problems. This adds to the impasse introduced in \Sec\ref{sec:the_dilemma} and paves the way for a new solution, which I will provide in \chap\ref{ch:new_aot}.

My critique of particularism will be based on the observation that one of the key assumptions on which it is based --- particularly the assumption of a time-independent measure --- is incompatible with a straightforward application of the PESA to cosmology. Thus, the particularist must either reject the PH as a viable explanation of the AoT or reject the PESA. In this thesis, I will do the former and embrace the norms of the PESA. This will lead to my new solution, which I will present in \chap\ref{ch:new_aot}.

It will be essential for the motivations of my proposal to understand why dynamical similarity is a gauge symmetry of cosmology according to the PESA. The argument for this will be given in \Sec\ref{sub:dynamical_similarity_in_the_universe}. There, I will show that the PESA is in perfect alignment with standard practice in modern cosmology. In particular, the measure used by cosmologist (Equation~\ref{eq:prob inflation}), which they regard as physically acceptable, is time-\emph{dependent} and invariant under dynamical similarity. This is in contrast to another measure introduced by cosmologists, the Gibbons--Hawking--Stewart measure (Equation~\ref{eq:GHS}). This measure is modelled off the Liouville measure and, therefore, is time-\emph{independent} and \emph{not} invariant under dynamical similarity. Importantly, however, this measure is \emph{not} regarded as physically plausible by cosmologists for reasons we will discuss in \Sec\ref{sub:dynamical_similarity_in_the_universe}. Consequently, proponents of the PH, who make use of the time-independence of the Liouville measure, must not only reject the PESA but also standard practice in cosmology.

In order to better understand why this is the case, it is useful to recall from \Sec\ref{sec:aot_prob intro} that particularist explanations are modelled off Boltzmann's proposal for deriving time-asymmetric thermodynamic behaviour at the macroscopic level from time-reversal invariant statistical mechanical laws at the microscopic level. In this approach, the time-symmetry of the laws is broken by asymmetrically restricting one's theory to DPMs that have highly atypical initial (but not final) states. In this way, Boltzmann attempted to explain why one might readily expect a cup of coffee to fall and shatter onto the ground but would not expect a mess of coffee and shards of cup to reassemble themselves. Because the cup of coffee is a highly unusual state in the space of possible ways that the constituents of the cup and coffee could be arranged, it is more typical to see the pieces scatter haphazardly than to see them reassemble as a cup of coffee.

Particularist approaches to the AoT result from trying to extend Boltzmann's reasoning to the Universe as a whole. But while this kind of explanation works reasonably well for simple thermodynamic systems, complications arise when attempting to apply this strategy to our actual Universe. It is not clear whether Boltzmann's reasoning can be used to iteratively provide an explanation for why nested subsystems of the Universe --- such as a coffee cup in a room in a city on a planet etc --- should individually be expected to start off in atypical states. For the rest of this chapter, I will be considering the viability of particularist approaches and, specifically, the use of a PH.

Early versions of the PH date back to Boltzmann himself \citeyearpar{boltzmann2012suicide} and comprehensive improvements making use of modern lessons from cosmology have been advanced mostly notably by Roger Penrose \citeyearpar{Penrose:1979WCH,penrose1994second}, Joel Lebowitz \citeyearpar{lebowitz1993boltzmann}, Shelly Goldstein \citeyearpar{goldstein2001boltzmann,goldstein2004boltzmann} and Huw Price \citeyearpar{price1997book,price2002boltzmann,price2004origins}. A well-known formulation has been advocated in \citet{albert2009time} where the phrase `Past Hypothesis' was coined after an initial proposal by Richard Feynman \citeyearpar[p.116]{feynman2017character}.

Despite the dominance of particularist proposals in the literature, the status of the PH remains controversial: it is not difficult to find both glowing appraisals and scathing criticism. Barry Loewer rates the problem of the AoT as ``among the most important questions in the metaphysics of science'' \citep{loewer2012two} and the PH as ``the most promising approach to reductive accounts of time's arrows''. Huw Price rates the discovery of the low entropy past as ``one of the most important [achievements] in the entire history of physics''\citeyearpar{price2004origins}.

Despite these grand claims, criticism abounds. John Earman \citeyearpar{earman2006past} puts it bluntly:
\begin{quote}
    This dogma [the Past Hypothesis], I contend, is ill-motivated and ill-defined, and its implementation consists mainly in furious hand waving and wishful thinking. In short, it is (to borrow a phrase from Pauli) not even false.
\end{quote}
\cite{Wald:2012zf} deliver a scathing critique of the basic technical premises of the idea identifying ``a number of serious difficulties in'' attempting to formulate concrete implementations of the proposal.

In this chapter, I will first provide a comprehensive analysis of existing objections to the PH for the purpose of assessing its status. The three broad categories of objections introduced in \Sec\ref{sub:objections_to_particularism} will be expanded upon at the beginning of \Sec\ref{sec:deconstructing_the_argument}. These categories provide a formal scheme for describing and evaluating different objections to the PH that have been advanced in the literature. To add precision to this process, I will start in \Sec\ref{sec:the_past_hypothesis} by giving a modern presentation of the arguments motivating the PH and identify a list of important conditions (in \Sec\ref{sub:key_assumptions_of_the_past_hypothesis}) that underlie these arguments. I will then analyse several particular objections, taken as exemplars, in each category by identifying the specific conditions that each objection puts into question. One important conclusion I will draw is that the time independence of the counting procedure used by advocates of the PH is essential for justifying its use. This observation will play an important role in the considerations of \chap\ref{ch:new_aot}.

While the list of objections I will consider is not meant to be exhaustive and no single objection may be seen as providing grounds to reject the entire proposal, when taken together these objections are sufficient to raise serious concerns regarding the PH. The resulting analysis already paints a rather grim picture for the prospects of formulating a PH in an unambiguous way using sound mathematical and physical principles.

One common response to such objections is that they amount merely to an unreasonable insistence on technical rigour given the immense mathematical difficulties associated with defining measures in general relativity. In response to such objections, I will show in \Sec\ref{sec:symmetries_and_measure_ambiguities} that the PH encounters a troubling dilemma that persists even if all such technical concerns are removed. This dilemma is an uncomfortable choice between a loss of explanatory power --- the \emph{first horn} (see \Sec\ref{sub:the_origin_of_measure_ambiguities_in_cosmology}) --- and violation of the PESA and standard practice in cosmology --- the \emph{second horn} (see \Sec\ref{sub:symmetry_and_ambiguity}).

To establish this dilemma, I will begin by using the analysis of \Sec\ref{sec:the_past_hypothesis} and \ref{sec:deconstructing_the_argument} to describe the first horn. In \Sec\ref{sec:the_past_hypothesis} I will show that it is essential to the arguments of the PH to provide a justification for the measure used in the required typicality argument. Then in \Sec\ref{sec:deconstructing_the_argument} and \Sec\ref{sub:the_origin_of_measure_ambiguities_in_cosmology} I will argue that the existence of a unique time-independent measure on the cosmological state space is essential to the explanatory claims of the PH. In \Sec\ref{sub:dynamical_similarity_in_the_universe}, I will show that this unique measure is not gauge-invariant. Using this, I will establish the second horn of the dilemma, in \Sec\ref{sub:symmetry_and_ambiguity}, by arguing that a failure of the measure to be gauge-invariant according to PESA introduces a distinction without difference by over-counting empirically indistinguishable states. This leads to the following dilemma: either reject a time-independent measure and undermine the explanatory basis for the PH (horn 1) or introduce a distinction without difference by breaking a gauge symmetry (horn 2). Since violation of the second horn is a rejection of the norms of the PESA and standard practice in cosmology, the only reasonable option is to reject the first horn and abandon the PH. This opens the door to a new approach, which I will present in \chap\ref{ch:new_aot}.

\section{The Past Hypothesis} 
\label{sec:the_past_hypothesis}


In this section, I will first provide a modern outline of Boltzmann-style explanations of time-asymmetry (\Sec\ref{sec:preliminaries}) and then use this framework to illustrate the basic logic of the PH (\Sec\ref{sub:the_past_hypothesis}). I will compile a list (\Sec\ref{sub:key_assumptions_of_the_past_hypothesis}) of conditions necessary for the arguments of the PH collected from \Sec\ref{sub:the_past_hypothesis}. In \Sec\ref{sec:deconstructing_the_argument}, we will see that are good reasons to question the validity of many of the conditions identified below when applying this style of explanation in general relativity.

\subsection{Boltzmannian explanations of time-asymmetry}
\label{sec:preliminaries}

In the Boltzmannian reasoning, the ultimate goal is to explain within a given system the time-asymmetry of some macroscopic processes from the fundamentally time-symmetric microscopic processes that underlie it. The main formal ingredients of this procedure therefore involve a specification of the macro- and microstates of the system, a particular reductive map between them, and a way to describe their behaviour.

This is usually achieved using the Hamiltonian formalism introduced in \Sec\ref{sec:the_initial_value_problem}. In this formalism, the microstates of the systems in question are represented by points in phase space $\Gamma$. The Liouville measure, $\mu_L$, defined in \Sec\ref{sub:measures_on_counting_solutions}, then gives a privileged translation-invariant measure on $\Gamma$ when the phase space is finite dimensional. Moreover, as illustrated by Liouville' theorem in Equation~\eqref{eq:Liouville t-constant}, the flow of the density corresponding to this measure (i.e., the Liouville form) is also time-independent for \emph{any} choice of Hamiltonian. An immediate caveat of this result is that, up to a constant, the Liouville form is the unique (smooth) measure preserved by any choice of Hamiltonian.\footnote{Proof: Formally Liouville's theorem implies $\mathcal L_{\chi_H} \rho = 0\,,\forall H: \Gamma \mapsto \mathbbm R$ where $\rho = \omega^n$ and $\omega$ is the symplectic 2-form on $\Gamma$ and the vector field $\chi_H$ is determined via $\de H = \iota_{\chi_H} \omega$. Writing an arbitrary smooth volume-form as $v = f \rho$, where $f$ is some arbitrary smooth positive function $f:\Gamma \mapsto \mathbbm R^+$, then Liouville's theorem and the condition $\mathcal L_{\chi_H} v = 0$ immediately lead to $f =$ constant.} This fact is doubly useful for Boltzmann's reasoning: it provides, at the same time, a potential justification, via uniqueness, for the choice of measure $\mu$ and a consistency argument, via the invariance property under evolution, for being able to use the same measure at different times.

The considerations above lead to a host of difficulties in finding meaningful measures when trying to follow Boltzmann's programme in realistic theories of the Universe. In particular, the programme assumes that there exists a mathematically precise and unambiguous measure $\mu$ (not necessarily the Liouville measure) on $\Gamma$. I will identify both of these as conditions; Conditions-\ref{assumption:rigor} and -\ref{assumption:uniqueness} respectively; that are necessary to the PH.

With an appropriate measure one can assign weights to arbitrary regions of phase space. These weights can be used to define a notion of \emph{typicality} for these regions. For example, one can say that a particular region $A$ is \emph{typical} on phase space if its weight as determined by $\mu$ is sufficiently large with respect to the weight of phase space itself:
\begin{equation}
    \frac{\mu(\Gamma) - \mu(A)}{\mu(\Gamma)} \ll 1\,.
\end{equation}
In general, a set $S$ is said to be typical with respect to some property $P$ and measure $\mu$ if its weight according to $\mu$ is large as compared with all other sets that possess the property $P$ \citep{frigg2009typicality}. Clearly, any notion of typicality requires some interpretation for the weights provided by $\mu$ in order to have any meaning. For the purposes of Boltzmann's argument, we will see below that it will be necessary to interpret the weight $\mu(\Sigma)$ as the relative likelihood of finding the system in a particular region $\Sigma$ (as opposed to somewhere else in $\Gamma$) at any given time.\footnote{ This is sometimes called the \emph{statistical postulate}. } I identify this as an additional requirement (Condition-\ref{assumption:Boltzmann 2}) of the formalism. Note also that in order for the notion of typicality to be non-trivial, $\Gamma$ must have finite Liouville volume if $\mu(A)$ is finite. Such considerations are also relevant to Conditions \ref{assumption:rigor} and \ref{assumption:uniqueness}.

The next formal step is to define the macro-states of a system. Physically these correspond to macroscopic states of the system such as temperature, volume, pressure, etc. Formally they are represented by some macro-state space $M$ which must have a (much) smaller dimension than $\Gamma$. Because Boltzmann was usually considering closed systems where the total energy $E$ is preserved, it is customary to consider states restricted to constant energy surfaces $\Gamma_E = \Gamma|_{E = \text{constant}}$ (i.e., the micro-canonical ensemble).

In general, many microscopic states will be indistinguishable from each other at the macroscopic level. This indistinguishability is modelled as a projection from $\Gamma_E$ to $M$. The microstates identified under this projection define a partitioning of $\Gamma_E$ into the partitions $\Gamma_m$, where $m \in M$ ranges over all macro-states in $M$. These partitions represent equivalence classes of macroscopically indistinguishable microstates. In order for these to be meaningful physically, there must exist some epistemologically motivated coarse-graining procedure that realizes this projection. For example, if the macroscopic variable in question is the temperature, then the temperature must be a well-defined quantity. I identify this requirement with a further condition (Condition-\ref{assumption:epistomology}). With these ingredients in hand it is now possible to define the \emph{Boltzmann entropy} (from now on called the `entropy' unless otherwise stated) of a particular macro-state $m$ as the logarithm of the Liouville weight of the partition $\Gamma_m$:\footnote{$k_B$ fixes the units of $S_\text{B}$.}
\begin{equation}\label{eq:entropy def}
    S_\text{B} = k_\text{B} \log[ \mu_L(\Gamma_{m}) ]\,.
\end{equation}

\begin{figure}
    \begin{center}
        \includegraphics[width=0.618\textwidth]{\pdots 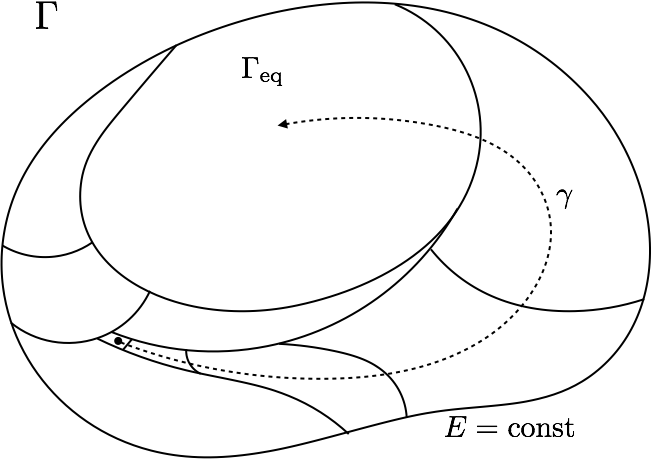}
        \caption{\label{fig:phase space} A small, atypical initial state will typically spend most of its future in a large equilibrium state $\Gamma_\text{eq}$.}
    \end{center}
\end{figure}


We are now equipped to give a modern synthesis of Boltzmann's reasoning. First one must show that for the system in question there exists an exceptionally large macro-state $\Gamma_\text{eq}$ that takes up most of the phase space volume of the system. I take this to represent a further requirement that $\Gamma_\text{eq}$ be a typical state in $\Gamma_E$ (Condition-\ref{assumption:Boltzmann 1}). The relevance of Condition-\ref{assumption:Boltzmann 1} can be seen by the interpretation given to the weights of $\mu$ given Condition-\ref{assumption:Boltzmann 2}. If $\mu(\Gamma_\text{eq})$ gives the relative likelihood of finding the system in $\mu(\Gamma_\text{eq})$ then for all practical purposes $\Gamma_\text{eq}$ is a steady or \emph{equilibrium} state of the system because the system will almost always be found there. More significantly, if an equilibrium state exists, then a system that starts in a small macro-state will typically spend most of its future time in $\Gamma_\text{eq}$. The basic picture is depicted in Fig~\ref{fig:phase space}. This picture is plausible because the counting suggested by the required interpretation of $\mu$ immediately suggests that a system starting outside $\Gamma_\text{eq}$ has little option but to quickly wander into $\Gamma_\text{eq}$, where it will remain for a very long time. But now there is a puzzle. Applying the same reasoning backwards in time suggests that a state finding itself in a small macro-state will also typically spend of all its \emph{past} in equilibrium. Because this apparently violates our knowledge that the past entropy of the universe was low, we are faced with the so-called \emph{second problem} of Boltzmann (see \cite{brown2001origins}). To solve this problem, one can posit an extremely \emph{atypical} condition on the earliest relevant state of the system. Under this condition, the system will typically find that it will approach the equilibrium state in the future. Note the temporal significance of the measure (Condition-\ref{assumption:Boltzmann 2}) and its central role in grounding the explanation of time asymmetry.

Before ending this section, I will mention one further requirement on the measure that is motivated by the representational considerations of Part~\ref{part:foundations}. Because a \gauge symmetry relates states that describe empirically equivalent situations according to my definition, a measure that is not gauge-invariant will count empirically equivalent states as distinct, introducing a distinction without a difference. 

To see this more explicitly, consider a region $R$ that lives in the domain $\mathcal D(\mu)$ of some general measure $\mu$ and a transformation $T: \mathcal D(\mu) \to \mathcal D(\mu)$ that maps this domain onto itself. Assume that $T$ map states of a system to empirically indistinguishable states. The set of states in the region $R$ is therefore empirically indistinguishable from the set of states in the transformed region $R' = T(R)$. In general, the non-invariance of $\mu$ under $T$ implies that the weight of the transformed region is not necessarily equal to the weight of the original: $\mu(R) \neq \mu(R')$. But if this is true then the weights $\mu(R)$ and $\mu(R')$ provide a distinction at the representational level between the regions $R$ and $R'$. Given our assumptions, this distinction cannot represent any empirical difference. In this sense, the measure $\mu$ therefore introduces a representational distinction that can't be captured by the empirical properties of the world.

This general argument is reinforced by standard practices in particle and statistical physics whereby the physical measure is required to be invariant under all gauge symmetries. In the standard model of particle physics, the gauge invariance of the path-integral measure is a central foundational principle of the theory. More generally, the Faddeev--Popov determinant, which enforces the gauge invariance of the path-integral measure, is considered a necessary ingredient in gauge theory (see \cite[Chap 15]{Weinberg:1996kr} for an overview and defence of this standard practice), in statistical physics, \cite{jaynes1973well} has argued influentially that measures should be invariant under transformations that relate indistinguishable states of a system. I therefore conclude that there are strong reasons for requiring that the measure be invariant under all \gauge symmetries, in agreement with the norms suggested by the PESA. I will refer to this as Condition-\ref{assumption:invariance}.

\subsection{The Past Hypothesis} 
\label{sub:the_past_hypothesis}

The main idea behind the PH is to invoke the Boltzmann-style reasoning of the previous section to explain time asymmetry in the actual Universe. The system in question is then taken to be the entire Universe and the PH itself translates into a special condition on the earliest relevant state of the Universe. All the mathematical quantities discussed above --- phase spaces, measures, macro-states, etc --- are then taken to represent aspects of the Universe as a whole. The proposed explanation is given in terms of a typicality argument: Universes that obey the appropriate PH, it is claimed, will typically evolve towards an equilibrium state in the future. Time-asymmetry arises by asymmetrically applying the special condition to past, rather than future, states. That the Boltzmann reasoning, whose empirical success is traditionally realised in closed sub-systems of the Universe, can provide explanatory leverage when applied to the Universe \emph{as a whole} is then taken as a further condition (Condition-\ref{assumption:typicality}) for the PH. Empirical support for the extreme atypicality of the initial state of our Universe is taken to be implied by abundant cosmological evidence for a low-entropy (e.g., near-thermal CMB power spectrum) and large red-shift early Universe. I take the existence of this empirical evidence to be a final condition (Condition-\ref{assumption:observations}) for the viability of the PH.

\subsection{Requirements of the Past Hypothesis} 
\label{sub:key_assumptions_of_the_past_hypothesis}

I will now state all conditions identified in \Sec\ref{sec:preliminaries} (this list of conditions is \emph{not} intended to be exhaustive).
\begin{enumerate}[A]
    \item There exists a measure, $\mu_\text{universe}$, on the phase space of the Universe, $\Gamma_\text{universe}$, that is simultaneously:\label{assumption:measure}
    \begin{enumerate}[\ref*{assumption:measure}1]
        \item mathematically precise, \label{assumption:rigor}
        \item empirically unambiguous, and \label{assumption:uniqueness}
        \item invariant under all gauge symmetries. \label{assumption:invariance}
    \end{enumerate}
    \item It is justifiable to interpret the weights given by the chosen measure in terms of the relative likelihood of the system being in a given region at a given time.\label{assumption:Boltzmann 2}
    \item There is an epistemologically meaningful and mathematically well-defined projection from the microscopic phase space of the Universe, $\Gamma_\text{universe}$, to a macroscopic phase space, $M_\text{universe}$.\label{assumption:epistomology}
    \item There exists a unique and exceptionally large state, defined to be the \emph{equilibrium state} $\Gamma_\text{eq}$, that is a typical macro-state on the phase space of the Universe at any given energy $E$; i.e.,
        $$\frac{\mu_\text{universe}[\Gamma_\text{E,universe}] - \mu_\text{universe}[\Gamma_\text{eq}]}{\mu_\text{universe}[\Gamma_\text{E,universe}]} \ll 1\,.$$\label{assumption:Boltzmann 1}
    \item Typicality arguments have explanatory power when applied to the Universe.\label{assumption:typicality}
    \item There is cosmological evidence for the PH being true.\label{assumption:observations} 
\end{enumerate}


\section{Objections to the PH} 
\label{sec:deconstructing_the_argument}

In this section, I will set the stage for the arguments motivating the considerations of \Sec\ref{sec:symmetries_and_measure_ambiguities}. I will recapitulate the three objections raised in \Sec\ref{sub:objections_to_particularism} and reformulate them in terms of the concepts and language I have developed in this chapter for describing the Boltzmannian programme.
\begin{enumerate}[I]
    \item \emph{Objections from mathematical and conceptual ambiguity:}. These objections question whether the formal quantities necessary for stating the PH can be given precise, unambiguous mathematical definitions.\label{worry: ambiguity}
    \item \emph{Objections from the breakdown of thermodynamic assumptions}. These objections grant \eqref{worry: ambiguity} but question whether the resulting formal quantities have the physical characteristics required for a Boltzmannian explanation  --- especially when gravitational interactions are taken into account.\label{worry:intuition}
    \item \emph{Objections from lack of explanatory force}. These objections grant both \eqref{worry: ambiguity} and \eqref{worry:intuition} but question the explanatory form and physical motivation of the typicality arguments used when applied to the Universe as a whole.\label{worry:typicality}
\end{enumerate}
Division of objections into the above categories emphasizes the reliance of the latter objections on being able to provide adequate responses to the former. If, for example, one cannot meet the standards of Category-\ref{worry: ambiguity}, then the framework must be rejected and the considerations of Categories \ref{worry:intuition} and \ref{worry:typicality} become irrelevant. We will see below that there are already significant worries raised at the level of Categories \ref{worry: ambiguity} and \ref{worry:intuition} even though a significant amount of philosophical literature is focused on evaluating objections falling into Category-\ref{worry:typicality}. I will now discuss several examples, taken as exemplars, of specific objections in order to illustrate each of the above categories. This analysis will help illustrate the importance of the distinct properties of the Liouville measure that provided the basis for the dilemma that we will present in \Sec\ref{sec:symmetries_and_measure_ambiguities}. It will also help to elaborate on the discussions of \Sec\ref{sub:objections_to_particularism}. Finally, this analysis will lead me to conclude that the time independence of the Liouville measure is the only good argument in favour of choosing it as a typicality measure.

\subsection{Objections from mathematical and conceptual ambiguity} 
\label{sub:case_c_measure_ambiguities}

In this section, I will primarily be concerned with issues arising from Conditions-\ref{assumption:measure} due to infinite phase spaces. Such phase spaces entail serious mathematical problems for measure-theoretic approaches to explanation. These problems stem from two distinct sources. The first arises because measures evaluated on an infinite interval can only be defined according to a limiting procedure that typically leads to physically significant regularization ambiguities. These problems are compounded in field theories because of a second source of ambiguity due to the phase space itself being infinite dimensional. As I mentioned in \Sec\ref{sec:preliminaries}, non-trivial translation-invariant measures do not exist for this case, and this often leads to ambiguities related to the truncation of the phase space. Ambiguities of these two kinds lead to a tension between mathematical precision (Condition-\ref{assumption:rigor}) and uniqueness (Condition-\ref{assumption:uniqueness}). To make matters worse, the purely mathematical problem of defining any measure on the phase space of general relativity invariant under all spacetime symmetries is far from being solved. This open technical problem is in fact one of the main formal obstructions to obtaining a canonical formulation of quantum gravity. With this in mind, it is advisable to explore various approximations to general relativity that render the computations of measures more tractable. But even in this simplified setting, one encounters immediate and troubling difficulties that are emblematic of the more general case.

Pioneering work in \cite{gibbons1987natural} that was elaborated on by several authors in both the physics \citep{Hawking:1987bi,Hollands:2002xi,Corichi:2010zp,Ashtekar:2011rm,Wald:2012zf} and philosophy literature \citep{earman2006past,frigg2009typicality,Curiel:2015oea} shows that the natural measure on homogeneous and isotropic cosmologies has infinite phase space volume. In the references listed, different schemes are provided for handling these divergences, and these schemes introduce ambiguities. A particular illustration of this will be outlined in detail in \Sec\ref{sub:dynamical_similarity_in_the_universe}. To resolve these mathematical ambiguities (of the first kind discussed above), new inputs, which are often physical in nature, must be introduced. It is thus paramount that the extra inputs needed to resolve these ambiguities neither conflict with other symmetry principles, in accordance with Condition-\ref{assumption:invariance}, nor implicitly assume what is trying to be explained: i.e., the time-asymmetry of local thermodynamic processes. Otherwise, the explanatory force of the PH is lost.

To illustrate the extent to which these ambiguities are problematic, consider the concrete results of different authors with different intuitions performing computations of the relative likelihood of cosmic inflation. Advocates for inflation \citep{Kofman:2002cj,Carroll:2010aj} proposed a measure according to which the probability of inflation was found to be infinitesimally close to 1. Inflation skeptics \citep{Turok:2006pa} proposed an alternative measure where the probability of inflation was found to be 1 part in $10^{85}$! This remarkably huge discrepancy reflects the extent to which individual beliefs can affect cosmologist's determinations of the appropriate physical principles used to justify their measure and the difficulties of resolving the tensions between Condition-\ref{assumption:rigor} and Condition-\ref{assumption:uniqueness}.\footnote{ The origin of this ambiguity lies in the insistence of treating the time-dependent measure of Equation~\eqref{eq:prob inflation} as a measure for counting \emph{solutions}. I will suggest an alternative way of thinking of such measures as giving weights for sets of states conditioned on knowledge about the current state. This interpretation will play a central role in my proposed explanation for the AoT.  } Any conclusions drawn on the basis of a typicality argument must be assessed in light of such remarkable disagreement between cosmologists.

Ambiguities of this kind are not improved when more realistic models including cosmological inhomogeneities are considered. Any preliminary hopes, such as those alluded to in \cite{callender2010past}, that adding an infinite number of degrees of freedom would help resolve these ambiguities can be seen to be in vain when explicit models are considered. This has been done, for example, in \cite{Wald:2012zf}. What was found there was that the additional degrees of freedom introduce corresponding regularization ambiguities of the second kind discussed above. It is therefore necessary to introduce new physical principles in order to resolve these ambiguities. Given the daunting nature of a full general relativistic treatment, these considerations raises serious doubts regarding the possibility of being able to attribute any meaningful notion of typicality to the Universe as a whole.


\subsection{Objections from the breakdown of thermodynamic assumptions} 
\label{sub:case_b_gravitational_considerations}

In this section, I will consider the unusual properties of gravitational dynamics that complicate our entropic intuitions for the Universe, assuming that a well-defined truncation of the phase space exists on which a Liouville measure can be defined. Consider the equilibrium state of a free gas. It is smooth, homogeneous and nothing like the current state of the Universe, which is characteristically clumpy and uneven. Those clumps comprise, among other things, star systems --- one of which supports the far-from-equilibrium biological system we find ourselves in. On the other hand, analysis of CMB temperature fluctuations reveals only a small $10^{-5}$ deviation from homogeneity. How can these observations be compatible with a low entropy past state? The standard response to this is that the gravitational contribution to the entropy should dominate at late times because of the unusual thermodynamic character of the gravitational interactions. This contribution is so great that it more than compensates for the decrease in entropy observed through the clumping of matter. Intuition for this comes from entropic considerations in Newtonian $N$-body self-gravitating systems, which have been used to model, for example, the dynamics of dust and stars in galaxies and galaxy clusters. But even in this simplified and well-tested setting there are difficulties that are emblematic of the considerations of \Sec\ref{sub:case_c_measure_ambiguities}.

Because Liouville volume is a volume on phase space, the inverse square potential due to gravity and the large momenta it can generate flip expectations for what constitutes a high and low entropy state. The steep gravitational potential well taps a large reservoir of entropy allowing for the kind of sizeable low entropy fluctuations we see in biological systems on Earth. These features as well as the difficulties they entail are reviewed nicely in \cite{Padmanabhan:2008overview,PADMANABHAN:1990book}, which gives detailed proofs of many of the results referenced below. This flipping of expectations is argued to occur not only for $N$-body systems, but also in a full-fledged general relativistic treatment of entropy. Thus, advocates of the PH (for example \cite{goldstein2004boltzmann,albert2009time}) emphasize the $N$-body intuition pump as providing an explanation for why the early homogeneous state of the CMB should be thought of as having low entropy and the current clumped state, which contains steadily accumulating stable records, as having high entropy. Moreover, this intuition was a primary motivation for early attempts at formulating an explicit PH such as Penrose's Weyl Curvature Hypothesis \citeyearpar{Penrose:1979WCH}.

The $N$-body intuition pump, however, also raises potential concerns. Firstly, if we follow the past state far enough into the early Universe, a full general relativistic treatment becomes unavoidable. But as we have already seen in \Sec\ref{sub:case_c_measure_ambiguities}, such a treatment suffers from troubling ambiguities, and it is not clear that the simple Newtonian intuition will remain valid. Another significant worry is the definition of equilibrium itself. The notion of equilibrium in gravitational systems is complicated by two sources of divergence (for details see \cite{Padmanabhan:2008overview}): i) the infinite forces particles exert upon each other when they collide, and ii) the infinite distances particles can obtain when ejected from a system. To cure these divergences, it is necessary to render the entropy finite by imposing additional constraints. This involves closing the system at some maximum size, so that particles are not allowed to escape, and forbidding two particles from being able to collide. This requires extra assumptions that must be grounded in physically acceptable principles. It is therefore paramount that these physical idealizations be well-motivated. But the fact that these idealizations break down under specified conditions implies difficulties in defining stable equilibrium for the system. Indeed, $N$-body systems are known to only have local --- but no global --- maxima \citep{Padmanabhan:2008overview}. Thus, gravitating systems do not have genuine equilibrium states, and Condition-\ref{assumption:Boltzmann 1} cannot be strictly satisfied. In absence of an equilibrium state, thermodynamic quantities such as macro-states and their entropy cannot be defined and Condition-\ref{assumption:epistomology} is strictly violated. While this is not problematic for local meta-stable systems like a galaxy, it can certainly be problematic for globally defined systems like the entire Universe. Moreover, even when local equilibria exist, there is still no guarantee that gravitational dynamics will actually steer the system towards these local equilibria in order to satisfy Condition-\ref{assumption:Boltzmann 2}. The crucial role of dynamics in the Boltzmannian argument has been emphasized in \cite{frigg2009typicality} and \cite{brown2001origins}.


\subsection{Objections from lack of explanatory force} 
\label{sub:case_a_typicality}

This section will firstly be concerned with the essential need to satisfy Condition-\ref{assumption:Boltzmann 2} by finding a valid justification for using Liouville volume as a typicality measure, assuming all concerns of Category~\ref{worry: ambiguity} and \ref{worry:intuition} have been resolved. In conventional statistical mechanical systems, this justification proceeds along two traditional routes. The first and oldest route relies on a theorem by \cite{birkhoff1931proof} that states that for ergodic systems the average time spent in a particular phase space region becomes roughly proportional to its Liouville volume if the timescales in question are much longer than the Poincar\'e recurrence time. Unfortunately, for almost all systems --- and certainly for the Universe --- the Poincar\'e recurrence time is significantly longer than the estimated time since the Big Bang. The second route, usually favoured for its practicality, is to argue that the system undergoes a process called \emph{mixing}. Roughly speaking, a system is mixed when the long-run evolution of the measure of a system becomes approximately homogeneous, and therefore Liouvillian. Many systems exhibit this property and the relevant mixing timescales can be computed explicitly. Unfortunately, \cite{Wald:2012zf} argue that the observed expansion of the Universe is too rapid to allow the large scale structures of the Universe to interact often enough for mixing to occur on these scales. This suggests that it is unreasonable to expect the Universe as a whole to undergo mixing. It would seem that in terms of conventional justification schemes for the Liouville measure Condition-\ref{assumption:Boltzmann 2} cannot be made compatible with the observational requirements of Condition-\ref{assumption:observations}.

It is possible to look for justification schemes satisfying Condition-\ref{assumption:Boltzmann 2} that do not originate from conventional statistical mechanical considerations. One proposal made by Penrose \citeyearpar{Penrose:1979WCH,penrose1994second} and later advocated  (either implicitly or explicitly) by \cite{goldstein2001boltzmann}, \cite{lebowitz1993boltzmann}, and \cite{albert2009time} is a version of the Principle of Insufficient Reason (PIR) as formalized by Laplace. In Penrose's version, a blind Creator must choose initial conditions for the Universe among the space of all possibilities. Being indifferent to which conditions to choose, the Creator assigns equal likelihood to each possibility according to the Liouville measure. Given the failure of standard justifications schemes, \cite{Wald:2012zf} point to Penrose's proposal as the only available alternative. Unfortunately, the PIR has a troubled history in the philosophy of science and suffers from several well-known difficulties. At least four prominent criticisms are identified in \cite{uffink1995entropyconsistency}. While some of these are addressed implicitly throughout this text, one line of criticism dating back to Bernoulli is noteworthy because it also directly puts into question the validity of Condition-\ref{assumption:epistomology}. In this line of criticism one derives paradoxes that originate in an incompatibility between the measures obtained when applying the PIR to different choices of partition for the microstates of a system. These paradoxes occur when the partitions correspond to \emph{disjunct coarse-grainings} or \emph{refinements} of each other \citep{norton2008ignorance}. There is nothing in the PIR that tells you which partitioning of the microstates is the ``correct'' one precisely because this would require some non-trivial knowledge about how these partitions may have been gerrymandered. Without direct knowledge of the ``correct'' partitioning of microstates, the PIR loses all explanatory power.

The only remaining justification for the Liouville measure is the uniqueness argument under time-symmetry. If one requires a time-independent measure, then the uniqueness of the Liouville measure under the requirement of being preserved by arbitrary Hamiltonian evolution does single it out. However, as we will see in \Sec\ref{sub:dynamical_similarity_in_the_universe}, an application of the PESA to cosmological systems puts into doubt any motivations for using the Liouville measure to establish a notion of typicality for models in the Universe.

Finally, let me mention a prominent ongoing dialectic between Price \citeyearpar{price2002boltzmann,price2004origins} and Callender \citeyearpar{callender2004measures,callender2004origins} on the explanatory power of the PH that questions the validity of Condition-\ref{assumption:typicality}. In this dialectic Price argues that the PH itself should require explanation in pain of applying a ``temporal double standard'' to a past state when an atypical future state would plainly require explanation. Callender responds by stating that contingencies rarely (or never) require explanation, and an initial condition such as a PH is a contingency of this kind. Indeed, I will revisit Price's temporal-double-standard objection in the conclusions (\Sec\ref{sec:conc:PH objections}) and show that my proposed solution in terms of a Janus--Attractor scenario provides a natural response.

\paragraph{Concluding remark} 
\label{sub:concluding_remark}

The analysis of this section has established that there are many concerns regarding the justification of the choice of typicality measure used to formulate a PH. In \Sec\ref{sub:case_b_gravitational_considerations} it was argued that self-gravitating systems have unusual thermodynamic properties and in \Sec\ref{sub:case_a_typicality} these arguments where combined with known facts about the Universe to suggest that conventional statistical mechanical justifications fail when applied to the Universe. Justifications that rely on indifference principles where also criticised on epistemological grounds. I conclude from this section that the only defensible justification for choosing the Liouville measure as a typicality measure is an argument based on its time independence. This time independence is essential to the Boltzmannian reasoning since, without it, the time dependence of the measure could itself be used to explain time asymmetry. And the Liouville measure really is the unique measure under this requirement.


\section{Dynamical similarity as a gauge symmetry of the Universe: counting what counts} 
\label{sub:dynamical_similarity_in_the_universe}

I will now argue, using the PESA, that dynamical similarity should be treated as a gauge symmetry in cosmology. This result will not only be essential to my reasons for rejecting the PH but will also form the basis of the new proposal that I will give in \chap\ref{ch:new_aot}. Additionally, I will show that, while the Liouville measure is, in general, not invariant under dynamical similarity, it is straightforward to find a measure for cosmology that is. Interestingly, as we will see, this measure is deemed to by physically plausible by cosmologist, in contrast to the time-independent measure build directly from the Liouville. An important property of the physically plausible measure is that it is highly time-\emph{dependent}. I will use this fact to raise serious doubts about whether a convenient measure can be found that can be both time-independent, as is assumed by the PH, and gauge-invariant, as is required by Condition-\ref{assumption:invariance}.

Recall from \Sec\ref{sub:generating_dynamical_similarity} that a dynamical similarity is a transformation generated by a vector field on phase space that rescales the action according to $S \to c S$ for some positive constant $c$. I will now show that modern theories of cosmology have dynamical similarities as (broad) symmetries. In the Standard Model of Cosmology, the spacetime geometry and matter content of the theory is separated into a background contribution and linear perturbations. The background is represented by a Friedmann--Lema\^itre--Robertson--Walker (FLRW) model, which is dynamically decoupled from the perturbations. Because of this decoupling, the symmetries of the background model must also be symmetries of the full theory. We can then restrict our attention to the FLRW model without loss of generality.\footnote{ This breaks down away from the regime where perturbation theory is empirically justified; i.e., when the Universe is not smooth. }

I will give a more mathematically complete description of the FLRW model in \Sec\ref{sec:cosmological_models}. For now, all we will need is to define the basic variables of the theory and state the equations of motion in terms of these variables. The geometry of spacetime in the FLRW model is represented by the volume, $v$, of a patch of space co-moving with respect to test particles travelling along geodesics. I will choose a time variable that is an affine parameter of these geodesics. The first time derivative of $v$ can be encoded in the Hubble parameter $H$ introduced in \chap\ref{ch:aot_prob}. The matter content of inflationary cosmology is represented by $n$ scalar fields $\phi^i$, the \emph{inflaton fields}, and their first time derivatives, $v_\phi^i \equiv \dot \phi^i$. See \Sec\ref{sec:cosmological_models} for a more complete definition of these variables. The equations of motion for the geometry and matter of the FLRW inflationary model are given by the \emph{Friedmann} and \emph{Klein--Gordon} equations respectively:
\begin{align}\label{eq:cosmo eoms}
    H^2 &= \frac{8\pi}3 \lf( \sum_i \frac {\lf(v_\phi^i\rt)^2}2 + V(\phi_i) \rt) & \dot v_\phi^i + 3 H v_\phi^i + \diby{V}{\phi^i} = 0\,,
\end{align}
where $V(\phi^i)$ is the inflationary potential, and we have set various constants (including Newton's constant and the scalar field masses) to one.\footnote{ I have also taken $k=0$ for simplicity. Generalisation to other values of $k$ is possible but does not add anything substantive to the analysis. }

In terms of the variables $\{ v, H, \phi^i, v_\phi^i \}$, it is straightforward to see that the theory defined by the equations of motion above is invariant under the transformation
\begin{equation}\label{eq:DS in cosmo}
    v \to c(t) v\,,
\end{equation}
where all other variables are held fixed. In \Sec\ref{sub:removing_dynamical_similarity_flrw}, I show that this is a dynamical similarity of the FLRW theory. This fact follows trivially from the observation that the equations of motion \eqref{eq:cosmo eoms} are completely independent of the variable $v$. Because of this, $c$ in \eqref{eq:DS in cosmo} can take arbitrary values at different times without affecting the dynamics of quantities in the set $A = \{ H, \phi^i, v_\phi^i \}$. This means that the equations of motion of the FLRW theory are already gauge-invariant with respect to dynamical similarity and autonomous in $A$. 

It is now straightforward to apply the PESA to investigate the role of dynamical similarity in inflationary cosmology. The equations \eqref{eq:cosmo eoms} define the complete dynamics of the background geometry in the FLRW inflationary model. They are first order and continuous, as long as $V(\phi^i)$ remains finite, and autonomous in $A$. I will investigate the continuity of these equations at the Big Bang in \Sec\ref{sub:light_cone_coordinates}. But for now, it is sufficient to restrict our attention to a range of values, far away from any potentially singular regions, for the variables in $A$ where the FLRW theory is empirically reliable. This means that the algebra $A$ satisfies the dynamical requirements of the PESA in the FLRW theory.

The algebra $A$ also satisfies the empirical requirements of the PESA. Removing any of the variables in $A$ would render the theory physically unviable since all of these variables appear in quantities that can be observed in cosmology.\footnote{ For example, the slow-roll parameters depend on these quantities. } This means that the necessity criteria is satisfied. But the sufficiency requirement is also satisfied --- at least in terms of determining the background structure. As we will see below, it is a well-established fact that the value of the \emph{scale-factor}, which is just the cubed root of $v$, is arbitrary in cosmology, and therefore does not correspond to any meaningful observable. This is reinforced by the fact that the dynamics of the scale factor are not fixed by the equations of motion. And since the background equations are decoupled from the perturbative equations, the same holds for the full inflationary theory. The PESA therefore tells us that a good interpretation of the FLRW inflationary theory is to identify the set $A$ as a generating set for the observable algebra of the theory. Since $A$ is invariant under dynamical similarity, we immediately find that \emph{dynamical similarities are \gauge symmetries of cosmology.}

I now turn my attention to the properties of the Liouville measure for the FLRW theory. Without even having to write down this measure we know from general considerations that it will not be invariant under dynamical similarity. This can be seen by noting that the symplectic $2$-form $\omega$ of a theory is required to rescale under a dynamical similarity because of the relation \eqref{eq:omega DS} so that the Liouville form, \eqref{eq:Liouville form}, is also rescaled. In the FLRW case, one can understand this because the phase space will contain $v$, which transforms non-trivially under dynamical similarity according to \eqref{eq:DS in cosmo}, and momentum variables conjugate to $\phi^i$, which must transform non-trivially so that $\omega$ has the correct weight under dynamical similarity.

To count solutions however, the Liouville measure must be restricted to a surface of constant clock time, which intersect all DPMs once, and the Hamiltonian constraint surface implied by the reparametrisation invariance of the theory. This is the general procedure for counting solutions described in \Sec\ref{ssub:counting_solutions_in_a_symplectic_theory}. A common way to implement this procedure\footnote{For a readable summary of how to do this, see \cite{Wald:2012zf}.} is to restrict to a single scalar field and choose constant-$H$ surfaces so that the Hubble parameter is treated as an internal clock for the system with the value $H = H^\star$. The Hamiltonian constraint can then be solved by eliminating the momentum conjugate to the scalar field. The restriction of the Liouville measure to these surfaces is then given by the Gibbons--Hawking--Stewart (GHS) measure \citep{gibbons1987natural}
\begin{equation}\label{eq:GHS}
    \mu_\text{GHS}(r) = \int_r \sqrt{ (H^\star)^2 - \tilde m^2 \phi^2 } \de v\, \de \phi\,,
\end{equation}
where $r$ is a region on the surface $H = H^\star$ that is compact in $\phi$ but not in $v$.

Importantly, the GHS measure is \emph{not} regarded as physically meaningful by cosmologists in part because of its non-compact domain in $v$, which causes a divergence, but also because of the non-physicality of $v$ itself. As is pointed out in \Sec IV of \cite{Wald:2012zf}, physical predictions do not depend on the initial value of the scale factor but only on $\phi$:
\begin{quote}
    Whether or not inflation occurs in a given universe only depends on the initial value of $\phi$. In particular, it does not depend on the initial value of $a$ --- as one would expect, since the value of $a$ does not affect the dynamics when the spatial curvature vanishes.
\end{quote}
This decoupling of $a$ is in-line with our own motivations for eliminating $a$. However, \cite{Turok:2006pa} go further stating:
\begin{quote}
    ... we identify the divergence as being due to the \emph{dilatation symmetry} of flat FLRW universes, and we show it is removed if one identifies solutions which cannot be observationally distinguished. [p. 11, emphasis added]
\end{quote}
Here, they explicitly discuss a ``dilatation symmetry'' as the origin of the non-physicality and use this to motivate the removal of the scale factor $a$, although they do not identify it directly with dynamical similarity.

Let us see how the removal of $a$ or, in our case, $v$ is performed. While the procedure was first introduced in \cite{Hawking:1987bi}, our treatment will follow most closely that of \cite{Wald:2012zf}. The idea is to eliminate the divergence due to the integral over $v$ by integrating over all of its possible values and normalising. The resulting measure
\begin{equation}\label{eq:prob inflation}
    \text{Prob}(r_\phi) = \lim_{v_\text{max} \to \infty} \frac{ \int_0^{v_\text{max}} \de v }{  \int_0^{v_\text{max}} \de v  } \frac{  \int_{r_\phi} \de \phi \sqrt{ (H^\star)^2 - \tilde m^2 \phi^2 }}{ \int_{r_{\phi_\text{max}}} \de \phi \sqrt{ (H^\star)^2 - \tilde m^2 \phi^2 }} \to \text{finite} 
\end{equation}
is finite due to the cancellation of the $v$-integrals in the numerator and denominator. The result depends only on the ratio of the integrals over the region $r_\phi$, which can be used to define inflation, and the finite region $r_{\phi_{\text{max}}}$, which is given in terms of the dynamical constraints of the theory.

From the perspective of the PESA, the integration over $v$ is motivated by requiring that the physical measure be invariant under what it unambiguously identifies as the \gauge symmetries of the theory. The integral over $v$ is an integration over the orbit of the dynamical similarity \eqref{eq:DS in cosmo}. The fact that the physicality of the measure \eqref{eq:prob inflation} and, conversely, the non-physicality of the measure \eqref{eq:GHS} has been independently acknowledged by practising cosmologists is evidence that the PESA is producing acceptable results.

The use of the gauge-invariant measure \eqref{eq:prob inflation} leads, however, to a serious problem for the PH. The measure \eqref{eq:prob inflation} is explicitly time-\emph{dependent} in the sense that it depends on the value $H^\star$ of the internal clock. This time dependence introduces \emph{significant} time-asymmetry into theory. As discussed in \cite{Wald:2012zf}, the values of the measure at different $H^\star$ span a range of $85$ orders of magnitude between the onset of inflation and the present.\footnote{ These are the same $85$ orders of magnitude resulting in the disagreement between inflation advocates and critics that was discussed in \Sec\ref{sub:case_c_measure_ambiguities}.  } This was identified as a kind of regularization ambiguity in the treatment of the divergence due to the integrals over $v$. But by identifying the $v$-integration as a reduction of dynamical similarity, which is a non-symplectic symmetry, we see instead that the time-dependence is due to the non-conservative flow of measures on contact spaces. Understanding how to adapt our reasoning to a time-dependent measure without introducing ambiguity will be the main focus of \chap\ref{ch:new_aot}, and will be central to our new understanding of the AoT. For now, note that the significant time-asymmetry introduced by this procedure poses a serious problem for PH accounts of the AoT.

\section{A Dilemma for the Past Hypothesis} 
\label{sec:symmetries_and_measure_ambiguities}

\subsection{The first horn: loss of explanatory power} 
\label{sub:the_origin_of_measure_ambiguities_in_cosmology}

As I have argued, the Liouville measure is indeed singled out as being the unique measure on phase space that is preserved by an arbitrary choice of dynamics. At first sight, this uniqueness appears to be particularly convenient for a particularist because a time-independent measure is very natural in the context of a PH. But time independence in the measure is more than a question of convenience in the context of a PH. In fact, as I will now argue, it is an essential ingredient for the PH.

Following \cite{price2002boltzmann}, the logic of the PH presented in \Sec\ref{sec:preliminaries} constitutes a contrastive explanation of the form: if A then B rather than C. The explanans A --- i.e., the PH itself --- is taken to explain the explanandum B --- i.e., the fact that typical processes are seen to overwhelmingly occur in a time-asymmetric way. The outcome C is then a typical member of a contrast class of outcomes that would be likely if not for A. The explanatory force of A comes from increasing the likelihood of B relative to C.

In the case of a PH, the contrast class is the set of worlds where typical processes overwhelmingly occur in a time-symmetric way. According to this logic, in order for the PH to be a good explanation of time-asymmetry, it must be the only significant source of time-asymmetry. Clearly this is consistent with the apparent FAPP time-symmetry of the form of the fundamental laws.

This consistency however is not sufficient. When a time-dependent measure is introduced into the formalism, the time-dependence of the measure could itself provide an explanation for the time-asymmetry of typical processes.\footnote{ We will see that time-dependence alone is not enough and that, to introduce time-asymmetry, we will have to introduce an extra structure on phase space called a \emph{Janus point}. } This is especially true if the time-dependence of the measure introduces a significant numerical temporal gradient as was shown in the previous section for the case of cosmological models. Moreover, the time-dependence of the measure introduces an ambiguity in terms of which moments in time should be used in order to obtain a unique measure on the space of models. Such an ambiguity can only be resolved by including some additional principle to the PH --- thus undermining much of its explanatory appeal. It is therefore essential to the logic of the PH that the measure employed be time-independent, and especially important that the measure not be badly time-dependent. Otherwise, we would have no reason to believe that processes would not occur in a time-dependent way even if the PH were not true. Note that these considerations hold regardless of any other justificatory considerations regarding the measure. This establishes the first horn of the dilemma.



\subsection{The second horn: violation of a gauge symmetry} 
\label{sub:symmetry_and_ambiguity}

At the end of \Sec\ref{sec:preliminaries}, I argued that a measure that is not gauge invariant will generically introduce a distinction without a difference. On top of badly violating epistemic expectations I also showed that this goes against standard practice in physics. I will now use this fact to state the second horn of our dilemma. As I argued in \Sec\ref{sub:dynamical_similarity_in_the_universe}, the PESA prescribes that dynamical similarity be a \gauge symmetry of inflationary cosmology. Thus, a measure that is not invariant under dynamical similarity will introduce a distinction without a difference. But as was shown in \Sec\ref{sub:dynamical_similarity_in_the_universe}, the Liouville measure is not dynamically similar. It follows that the use of the Liouville measure as a typicality measure in inflationary cosmology introduces a distinction without a difference, as reflected in the fact that cosmologists do \emph{not} regard the GHS measure to be physical. This is the second horn.

Now recall the first horn of the dilemma. Boltzmannian accounts of the PH must make use of the unique time-independent Liouville measure in order to retain their explanatory force. Thus, Boltzmannian accounts of the PH must face the following dilemma: either lose explanatory force or introduce a distinction without a difference.


\section{Final remarks: why we don't need the Past Hypothesis} 
\label{sec:prospectus_against_PH}

We have seen that Boltzmann-style explanations of time-asymmetry that make use of a PH depend upon a number of very restrictive conditions. The analysis of \Sec\ref{sec:deconstructing_the_argument} has uncovered several good reasons to question whether these conditions can ever be simultaneously satisfied. Broadly speaking we found that the nature of the phase space, dynamics and symmetries of general relativity provide reasons for pessimism regarding the prospects for providing and justifying a satisfactory notion of typicality for models of the Universe. A common response against critiques of this kind is to observe that strict insistence on mathematical rigour has often been unreasonable in the development of theoretical physics. Controversy over difficult technical problems such as defining a measure on the solution space of general relativity should not, it is argued, halt progress altogether. It should still be reasonable to advance conjectures regarding the plausible features of measures that may one day become available.

While such a strategy --- effective or not --- is available in response too much of the analysis of \Sec\ref{sec:deconstructing_the_argument}, it is no longer available in response to the dilemma of \Sec\ref{sec:symmetries_and_measure_ambiguities}. This is because the dilemma is the result of a simple symmetry argument applied to a very general way of formulating the laws of the Universe. To reject dynamical similarity as a gauge symmetry is to reject the arguments in favour of the PESA. To reject the Liouville measure is to abandon one of the most basic aspects of the Boltzmannian logic. To not require the gauge-invariance of the measure is to introduce a distinction without difference and to reject standard practice in particle physics and cosmology. None of these escape routes is particularly appealing. Even if one grants all the technical assumptions required by the PH, the dilemma persists.

On the other hand, a full-throated rejection of the PH as an explanation for time-asymmetry avoids the dilemma completely. But how, then, is one to explain the problem of the AoT? In the next section, I will use the significant time dependence of dynamically similar measures not as a potential obstruction but as the basis for a new and compelling explanation for the AoT independent of a PH.
\chapter{A new explanation for the arrow of time}
\label{ch:new_aot}

\begin{abstract}
    In this chapter, I present my proposal for explaining the Arrow of Time. The general scenario requires the presence of special features of the solution space; namely attractors and Janus points; and defines an Arrow of Time pointing from the Janus surface to an attractor for an observer that is in a state close to the attractor. This proposal requires neither a Past Hypothesis nor a time-asymmetric law, and therefore evades the impasse of \chap\ref{ch:aot_prob}. After defining the general scenario, I illustrate it in two concrete models: a non-expanding Newtonian $N$-body gravitational systems and homogeneous and isotropic cosmology. In the $N$-body model, I show that observers near an attractor typically see smooth states on the Janus surface. This solves the red-shift problem within this model. In the cosmological model, I show that observers near an attractor generically see a monotonic and divergent Hubble parameter in the past defined by my proposed Arrow of Time. This solves the red-shift problem within this model.
\end{abstract}

\ifchapcomp
    \tableofcontents
    \newpage
\else
    \cleardoublepage
\fi

\section{Introduction} 
\label{sec:introduction}

\subsection{A new hope} 
\label{sub:a_new_hope}


In this chapter, I will present a general scenario in which a significant AoT can arise in a theory with time-reversal invariant laws. I will show that this scenario is realised in models that capture the features of our world. In particular, I will give two physically well-motived models; namely, an $N$-body model and a cosmological model; that respectively solve the smoothness and red-shift problems introduced in \Sec\ref{sec:explanatory_target}. Remarkably, by removing dynamical similarity in an $N$-body system, I will show both that there are attractors and, more surprisingly, that observers near such attractors will \emph{typically see smooth states in their `past'}.\footnote{ Where I will be careful to define \emph{past} according to states on a history relative to those observers. } The combination of having a time-dependent measure and conditioning on being close to an attractor is what leads to a result so different from the usual expectations of the $N$-body theory. Similarly, removing dynamical similarity from the cosmological model presented below and imposing some reasonable physically-motivated constraints will result in attractors. Observers near such attractors will \emph{generically see states will large, monotonic Hubble parameters in their `past.'} These results, obtained for specific models, suggest a more general resolution of the problem of the AoT in terms of a unified model that combines the features of the ones presented here.

Before describing in more detail the concrete results of this chapter, let me first recall some key results from earlier chapters that will be necessary for constructing the argument. First, there is the PESA-based argument of \Sec\ref{sub:dynamical_similarity_in_the_universe} that concludes that dynamical similarity should be treated as a \gauge symmetry in FLRW cosmology. This argument says that the absolute size of the Universe, as encoded by the scale factor, is not an observable of the theory. I will show in this chapter that removing the scale factor from the observable algebra leads to an odd-dimensional contact system.

In \Sec\ref{sub:gauge_principle_for_DS}, I developed a general procedure for applying a generalization of the Gauge Principle to dynamical similarity. I showed that contact systems are formally similar to non-conservative, friction-like systems. The degree of non-conservation can be quantified in terms a quantity I called the \emph{drag} because of its formally similarity to the drag coefficient of a mechanical system.\footnote{ Specifically, the drag was defined as the Reeb flow of the contact Hamiltonian. See the discussion below Equation~\ref{eq:friction}. } I showed that the drag is the decay coefficient in time both of the Hamiltonian (see Equation~\ref{eq:Hc_eom}) and of the measure (see Equation~\ref{eq:mu evo gen}). Positive drag therefore signals non-conservation in terms of loss of energy and focusing of solutions.

In this chapter, I will focus on the behaviour of the drag in the $N$-body and cosmological models mentioned above. In both cases, I will find that the drag defines a significant temporal gradient away from a Janus point (which I will define in \Sec\ref{sub:janus_points}), where it is momentarily zero, under certain reasonable physical assumptions detailed below. On either side of the Janus point, the drag is monotonic along solutions, reaching attractors (which I will define in \Sec\ref{sub:dynamical_attractors}), asymptotically. This leads to a Janus--Attractor scenario, as described in \Sec\ref{sec:intro JA scenario} of the introduction and defined more rigorously in \Sec\ref{ssub:the_janus_attractor_scenario} below, where observers near an attractor will generically see a significant numerical gradient in the drag pointing from a Janus point to the attractor they are near to. This gradient then defines an AoT as guaranteed by our general considerations. In the Newtonian gravitational model, the gradient explains the growth of structure while, in the cosmological model, it explains the decrease of the Hubble parameter.

These AoTs arise from the generic behaviour of the drag, which is a scalar function on a contact space. Since contact systems arise from my general implementation of the Gauge Principle for dynamical similarity, JA-scenarios can occur when treating dynamical similarity a gauge symmetry. It is important to note, however, that the AoT itself is not a global feature of the theory. Rather, an AoT arises in the JA-scenario from directed gradients seen by a certain class of observers; namely those near attractors. Indeed, observers near attractors on different sides of a Janus point will see AoTs pointing in different directions (if such attractors exist). Moreover, observers near a Janus point will not be able to detect any significant gradient in the drag, and will therefore not be able to identify an obvious AoT of the kind I am proposing. But while AoTs in this picture arise for specific classes of observers, they are ultimately made possible by the existence of the Janus points and attractors themselves. \emph{These} structures, however, \emph{are} global features of the theory since they are sets determined by the contact dynamics. In the JA-scenario, an AoT thus emerges for certain observers because of global structures on the state space.

This understanding allows us to compare my proposal to existing explanations of the AoT. Importantly, the dynamical flow  --- and consequently the Janus points and attractors --- in my picture is invariant under a change of time orientation, and therefore does not privilege any particular time orientation. My mechanism for generating an AoT is, thus, not generalist because the laws themselves are time reversal invariant.

But while my picture is not generalist, it is also not particularist. Neither the existence of a Janus point nor of an attractor requires an hypothesis about a particular fact. Rather, these structures are \emph{generic} features of a dense set of solutions of the theory. Their existence can be proven rigorously using the mathematical properties of the dynamical laws. In particular, there is no need to postulate an \emph{atypical} early state.\footnote{ The behaviour on the Janus point, for example, is \emph{typical} in the $N$-body model and \emph{generic} in the cosmological model. } The explanation for the AoT in a JA-scenario is therefore neither particularist nor generalist. In this way, I evade the impasse of \Sec\ref{sec:the_dilemma} by showing, in \Sec\ref{sub:conc_gen-part_impasse}, how it is a false choice resulting from the assumption that the scale factor of the Universe is observable.

Let me wrap up these introductory remarks by summarising the main aspects of my proposal and how they can be seen to introduce an AoT. The basis of my proposal is the PESA-based argument for treating dynamical similarity as a gauge symmetry. This argument states that cosmological systems are best described as contact systems. In such systems, time-independent measures are no longer natural the way they are in symplectic systems. The problem of the AoT, which is ultimately a problem about how typical our highly time-asymmetric world is given the time reversal invariance of the laws, then takes on a different character. In particular, the drag appears as a dynamical quantity that naturally highlights certain highly time-asymmetric features of the world, making them seem more likely and, even, expected.

In \Sec\ref{sub:the_janus_attractor_scenario}, I will describe the JA-scenario in general and give mathematical definitions for the structures that give rise to an AoT. Then, I will carefully define the two models that realise a JA-scenario and show how they, respectively, solve the smoothness and red-shift problems. Before doing this, I will motivate these two models by describing them qualitatively and explaining how each provides an account of the phenomena I have set out to explain. But before moving on to a more complete treatment, I will address, in \Sec\ref{sub:time_laws_and_conventions}, an ambiguity in my procedure regrading the amount of drag in the system.

\subsection{The models} 
\label{sub:the_models}

\subsubsection{\(N\)-body model}\label{Nbody_motvation}

The first model I will consider is a Newtonian $N$-body system of self-gravitating point particles. This system idealises the large-scale structure of the Universe by representing galaxies as point particles interacting under their mutual gravity. In this model, the intension is to capture the growth of structure resulting from the gravitational collapse of over-dense regions of matter in the visible Universe. Red-shift is ignored for simplicity because we are interested periods of time after recombination where the effect of the red-shift is not significant in determining the qualitative features of local structure. Below, I will sketch how this model leads to one of the main results of this thesis; namely, that observers in states where lots of clumping has been seen to occur will typically see nearly smooth states in one time direction (i.e., their `past') and even clumpier states in the other time direction (i.e., their `future'). Details are given in \Sec\ref{sec:newtonian_gravitation_models}.

This model was introduced by Barbour, Koslowski and Mercati (BKM) in the three papers \cite{Barbour:2014bga}, \cite{barbour2013gravitational}, and \cite{barbour2015entropy} for the purposes of understanding the origins of the AoT. These papers formed the basis of the analysis appearing in \cite{barbour2020janus}. While my treatment in \Sec\ref{sec:newtonian_gravitation_models} will have significant overlap with those papers, there are important differences that are worth noting. First, there are some formal differences between our procedures. While they apply the Gauge Principle by performing an explicit reduction, I will retain full gauge invariance. Second, my analysis focuses on different aspects of the AoT. While I am primarily concerned with using this model as a way of addressing the smoothness problem, BKM were primarily concerned with understanding the emergence of local thermodynamic AoTs in terms of the accumulation of stable records. As a result, my approach will have different modelling assumptions and will focus on the behaviour of different mathematical quantities.\footnote{In particular, the main quantity that defines an AoT in my view is the drag, while in the BKM view, it is a function they refer to as the \emph{complexity}. I will call this function the \emph{$C$-function} (see Equation~\ref{eq:C-fun}) instead. In my work, the role of the $C$-function will be to establish that the AoT defined by the drag is consistent with a smooth initial state.} Finally, I will extend some of the BKM analysis to include a more direct quantitative measure of smoothness.\footnote{ See, for example, the analysis of \Sec\ref{ssub:C and uniformity}. } These differences notwithstanding, the analyses are not contradictory, and can be taken as serving complementary goals.

The idealisation I will use here is similar to, although less sophisticated than, the so-called \emph{Millennium Simulation} of \cite{springel2005simulations} that can reproduce the empirically adequate large-scale structure of the Universe by explicitly simulating an $N$-body system on a red-shifting background.\footnote{ While the Millennium Simulation is mainly an $N$-body simulation with initial conditions determined by the parameters of $\Lambda$CDM, it also involves some modelling of galaxy formation in addition to an expanding background.} Thus, with some improvements, the model used here could be upgraded to give an empirically adequate description of large-scale structure formation since recombination.

For now, my goal is not to achieve empirical adequacy but to understand the broad qualitative features of $N$-body systems that are relevant to the smoothness problem. In particular, I'd like to know what observers in an $N$-body Universe where lots of structure has formed will typically see in their `past,' where some definite procedure should be given to determine the `past' direction. Because I am treating the $N$-body system as a cosmological model, the PESA-based argument of \Sec\ref{sub:dynamical_similarity_in_the_universe} for the scale factor being unobservable should apply. This means that the overall size of the $N$-body system should be taken as gauge. In \Sec\ref{sub:ds_in_Nbody}, I will model changes of size of this kind as a dynamical similarity of the $N$-body system. I will thus apply the Gauge Principle for dynamical similarity developed in \Sec\ref{sub:gauge_principle_for_DS} to this system.

Given the reasonable physical constraints described below, I will show that the natural contact system obtained from this procedure contains a drag term whose value generically goes through zero so that there is a Janus point on generic solutions. The drag then grows monotonically away from the Janus point towards an attractor. The monotonicity of the drag results in a number of attractors that characterise the asymptotic behaviour of the $N$-body system.

What we can infer about these attractors is known from theorems derived in \cite{marchal1976final}. There, it is shown that the $N$-body problem for non-negative energy splits into at least two subsystems whose centre-of-mass motion asymptotes to that of a free particle. In general, there will be many of such subsystems when $N$ is large that expand in different directions so that the overall size of the system grows arbitrarily large compared to the maximum size of any subsystem. The attractors of the scale-invariant description thus resemble highly structured clumped states.

Contrastingly, the Janus points are characterised by arbitrarily more diffuse states of particles. These can then be interpreted as smooth `initial' states from the perspective of `late'-time observers near an attractor. Here, the distinction between `initial' and `late' can be given in terms of the value of the drag, which is zero at the initial state and maximum in the asymptotic future. Using this understanding, my results suggest that early states are necessarily much smoother than the states at late times.

Note that while the behaviour of the drag --- particularly the existence of Janus points and attractors --- can be proven generically from the dynamics using mathematical theorems, the link between the drag and the `clumpiness' of the state is less direct. I will investigate this link more carefully using the behaviour of a particular quantity that I will call the \emph{$C$-function} that is at the centre of BKM's analysis. I will argue that the value of $C$ is a good quantitative way to measure the amount of clustering, and takes a minimum when the state is smooth. I will then argue that a robust class of measures can be given that have the property that \emph{the initial state on a Janus surface is overwhelmingly likely to be near the minimum of $C$, and therefore nearly homogeneous.}\footnote{ I will be able to make this claim also in terms of more direct measures of homogeneity. } This claim, which I have emphasised earlier, forms the basis of my solution to the smoothness problem in the $N$-body model.

The physical conditions needed to derive these results are that the total energy of the original symplectic system, $E$, must be greater or equal to zero and that the total linear and angular momentum, $\vec P$ and $\vec J$, must vanish. The vanishing of $\vec P$ and $\vec J$ is required because non-zero values of such constants indicate the existence of a simple integral of motion whose conjugate variable could be used as a clock. As we saw at the end of \Sec\ref{ssub:counting_solutions_in_a_contact_theory}, when simple integrals of motion are available, it is straightforward to construct conservative descriptions of the contact system where the drag is exactly zero. Physically, this signals the presence of an external clock that is dynamically decoupled from the rest of the system. Such external structures could be argued to be unnatural in a relational theory of the Universe. Indeed, this is how such conditions were initially motivated in \cite{Barbour:2014bga}. In any case, the vanishing of $\vec P$ can always be achieved by going to centre-of-mass coordinates and the vanishing of $\vec J$ is consistent with cosmological observations in the sense that there does not seem to be a preferred rotational axis in the Universe.

The condition that $E \geq 0$ is required because negative total energy would allow for solutions where all the point particles could re-collapse under their mutual gravity. This would mean that the monotonicity of the drag --- and consequently the existence of the Janus points and attractors --- would not be guaranteed without extra constraints on the initial conditions. Treating the case of $E > 0$, however, requires a more general procedure for implementing the Gauge Principle for dynamical similarity when there are multiple dimensionful couplings. This procedure is described in \cite{sloan2021scale} and Section~III.C of \cite{bravetti2022scaling}. For simplicity, I will only consider the $E = 0$ case in this work.

Physically, the more general condition $E\ge 0$ can be motivated by noticing that this condition mimics a situation that is the well-known from general relativity, and which will be central to my considerations later, where a positive cosmological constant can be shown to prevent a re-collapsing Universe. Since I have ignored the ref-shift in this model, the simplest way to prevent the system from re-collapsing without a cosmological constant is to require $E\ge 0$. Ultimately, I would prefer a more realistic model with an $N$-body system expanding on an FLRW background, where only the requirement of a positive cosmological constant would be necessary. I will leave this to future work.

\subsubsection{FLRW cosmology}\label{sub:FLRW assumptions}

The second model I will consider is an FLRW cosmology coupled to a finite number of scalar fields. Below, I will sketch how reasonable physical assumptions in this model guarantee the existence of a global attractor such that observers near that attractor generically see states with large Hubble parameters in the time direction pointing away from the attractor (i.e., the natural `past' states of the observers). A detailed treatment will be given in \Sec\ref{sec:cosmological_models}.

We already encountered the relevant cosmological model in \Sec\ref{sub:dynamical_similarity_in_the_universe}. Models of this kind are used in inflationary cosmology to describe the evolution of the background geometry of the Universe. The effects of dynamical similarities in such models were first considered in \cite{Ashtekar:2011rm}. A more explicit treatment can be found in \cite{sloan2019scalar}, an action is given in \cite{sloan2021new}, and a geometric procedure on state space is developed in \cite{bravetti2022scaling}. Generalizations of these models to homogeneous but non-isotropic cosmologies were first given in \cite{KOSLOWSKI2018339} with an action given in \cite{sloan2023herglotz}. These results perform an explicit reduction of the theory by the action of dynamical similarity. None are concerned directly with the problem of the AoT. In this chapter, I will reproduce many of the key results from this previous work but using the general Gauge Principle developed in \Sec\ref{sub:gauge_principle_for_DS}. This gives the resources needed to easily switch between different representations of the system resulting from different gauge fixing conditions. In \Sec\ref{sub:light_cone_coordinates}, for example, I will use a particular representation that will allow me to easily understand the geometry of the solutions, including the behaviour at the Janus point.

By coupling this cosmological model to linear perturbations, it would be possible to produce an empirically adequate model of the power spectrum of the CMB using standard techniques in cosmology. But as in the $N$-body model, the goal for now is not to produce an empirical adequate model but, rather, to give a model that roughly captures the broad empirical behaviour of Hubble parameter in the Universe. Thus, I will not yet couple linear perturbations to the model and will assume homogeneity and isotropy throughout.\footnote{ The paper \cite{KOSLOWSKI2018339} relaxes these assumptions slightly by considering a finite number of anisotropies. } The purpose of this model is therefore to try to solve the red-shift problem in an idealised setting where it is assumed that the Universe is perfectly smooth. While the $N$-body model cannot address the red-shift problem because red-shift effects are explicitly ignored, the cosmological model given here cannot address the smoothness problem because inhomogeneities are explicitly ignored. Nevertheless, the hope is that these simple models, when taken together, make it plausible that a more realistic model could address both problems simultaneously.

In addition to homogeneity and anisotropy, I will make the following physical assumptions: i) a positive cosmological constant, ii) vanishing spatial curvature (i.e., $k =0$), and iii) the weak energy condition for the matter fields. As we will see in \Sec\ref{sub:a_janus_attractor_scenario_for_flrw}, these conditions are sufficient (but not necessary) to guarantee the monotonicity of the Hubble parameter, and consequently the existence of an attractor. More intuitively, these assumptions prevent the space-time from re-collapsing under its own gravity, preventing it from reaching a stable fixed point.

Physically, a positive cosmological constant is motivated by the fact that it is favoured by observations. The vanishing of spatial curvature is a simplifying assumption\footnote{ Lifting it would put mild constraints on the attractive basin of the attractor and require the more general procedure, alluded to above, for removing dynamical similarity in theories with more coupling constants of \cite{sloan2021scale} and Section~III.C of \cite{bravetti2022scaling}. } that is consistent with observations. And the weak energy condition, which formally requires that the matter density as seen by \emph{any} time-like observer be non-negative, is a relatively weak assumption consistent with all known local observations.

When these reasonable conditions are satisfied, I will show that the Gauge Principle for dynamical similarity developed in \Sec\ref{sub:gauge_principle_for_DS} leads, for this model, to a contact system where solutions generically have a unique Janus point, normally interpreted as the `Big Bang,' about which the drag increases monotonically towards a global attractor. The spacetime geometry of the global attractor is that of de~Sitter. Near the attractor, the Hubble parameter reaches a minimum and depends only on the cosmological constant. Because the attractor is global, we are led to the second key result of this chapter: observers in \emph{any} solution near the de~Sitter attractor will see an \emph{arbitrarily large numerical gradient in the drag pointing from the Janus point to the attractor.} This model thus provides an explanation of the large monotonic decrease of the red-shift factor, and solves the red-shift problem.

A rather remarkable consequence of applying the Gauge Principle for this model is that the contact system obtained after removing the dynamical similarity is Lipschitz continuous at the Janus point.\footnote{ See Section~\ref{sub:a_janus_attractor_scenario_for_flrw} and \ref{sub:light_cone_coordinates} for details. } This result, which was first obtained in a different representation in \cite{KOSLOWSKI2018339}, is notable because the original system is discontinuous at the Janus point, and this discontinuity is interpreted as the `Big Bang.' Thus, states that are singular in the original description are non-singular as contact systems. To be sure, such states are still remarkable points of the dynamics even in the contact system. But there is no obvious need to treat them as a fundamental breakdown of the dynamical equations. Thus, the dynamical-similarity-free description of the system has the potential to both solve the red-shift problem and remove the initial singularity.

\subsection{Time, laws and convention} 
\label{sub:time_laws_and_conventions}

Before giving a detailed treatment of the results sketched above, I will take a moment to address a potential issue that has been lurking in the background since \Sec\ref{ssub:counting_solutions_in_a_contact_theory}. This issue has to do with the conventionality of the drag in reparametrisation invariant contact systems. What we saw near the end of \Sec\ref{ssub:counting_solutions_in_a_contact_theory} was that the contact equations have a particular symmetry (given in Equation~\ref{eq:contact rep_sym}) for reparametrisation invariant theories that allows one to simultaneously rescale the contact form, redefine the Reeb vector, and change the time parametrisation of the system. This allows one to represent the contact dynamics using \emph{any} value for the drag.\footnote{The procedure for doing so is given after Equation~\ref{eq:measure ambiguity}.} When the contact system is obtained by applying the generalised Gauge Principle of \Sec\ref{sub:gauge_principle_for_DS} on a symplectic system that is dynamically similar, we saw that a conservative representation of the contact system (i.e., a system with zero drag) can always be given when an integral of motion is known in the original symplectic system.

On the one hand, this result might not be too surprising. An integral of motion, by definition, is a phase space function that is preserved under the evolution. Any variable that is canonically conjugate to such a phase space function will grow linearly in time by Hamilton's equations, and can therefore serve as a kind of external clock for the remaining degrees of freedom.\footnote{ An action-angle variable would be an example of such a clock. } It should not be surprising that one can then construct an equivalent drag-free contact system where this external clock is the parameter along the Reeb direction. Since reparametrisation invariant theories are invariant under different choices of parametrisation, this procedure will be gauge invariant. Indeed, from a physical perspective it is encouraging that one can always recover a symplectic subsystem with Hamiltonian dynamics whenever one can find a sufficiently isolated clock.

On the other hand, one may rightfully worry that the existence of a drag-free representation could undermine the entire programme of finding a solution to the problem of the AoT using a JA-scenario. If it is always possible to find a description of the system with a time-independent measure and no dissipative terms, then why not always use the resulting time-independent measure to count solutions in your theory? Doing so would reintroduce the problem of the AoT since worlds like ours, which have a significant amount of time-asymmetry, would undoubtedly have low weight according to this time-independent measure.

My response to this worry is to argue against the use of such measures in the cosmological setting because, as I will explain below, they are extremely cumbersome to construct in practice and lack the theoretical virtues normally expected of a good measure. In FLRW cosmologies and $N$-body systems (with zero linear and angular momentum), integrals of motion are very hard to identify because there are no genuinely isolated clocks --- at least not during the time periods that are relevant to explanations of the AoT. Mathematically, there are general procedures for finding integrals of motion (at least locally) provided the dynamical equations obey certain integrability conditions.\footnote{ For example, a proof of the \emph{Liouville--Arnold theorem} will give a prescription for obtaining a canonical transformation to action-angle variables, which are conjugate to a set of Poisson commuting integrals of motion. Identifying the generator of such a canonical transformation involves solving a differential equation similar to the Hamilton--Jacobi equation. For details about this procedure, see Chapter 10 of \cite{arnol2013mathematical}. } But even these procedures, which make strong assumptions about the regularity of the equations of motion, involve solving partial different equations so that the integrals of motion are expressible in terms of non-local operators on phase space. The presence of global symmetries can dramatically reduce the complexity of such a procedure, but such symmetries are notoriously absent from theories of the Universe, which are generally relativistic. Physically, this is because the universally attractive nature of gravity means that no degree of freedom can ever be completely isolated from any other.

For these reasons, it is well-established in the literature on general relativity that constants of motion, in general, are non-local operators on space-time. A classic formulation of this result was proved in \cite{torre1993gravitational}. As these results where being developed, Karel Kucha\v r called constants of motion in general relativity \emph{perennials} due to their forever unchanging nature, and described them thus:\footnote{ I include this quote in full because paraphrasing it simply could not do it justice. } \cite[p.24]{kuchar1993canonical}
\begin{quote}
    Perennials in canonical gravity may have the same ontological status as unicorns --- \emph{a priori}, these are possible animals, but \emph{a posteriori}, they are not roaming on the Earth. According to bestiaries, the unicorn is a beast of fabulous swiftness, strength, and beauty, but, alas, it can be captured only by a virgin. Corrupt as we are, we better stop hunting mythical beasts.
\end{quote}
In this spirit, I advise against seeking a measure that would require the capture of such an elusive prize.

Following Kucha\v r's analogy, the situation most often encountered in physics, where systems can be sufficiently isolated from their surroundings, is an Eden-like utopia where unicorns roam plentifully and humanity retains its innocence. The absence of absolute scale in cosmology, however, is the poison apple, condemning us to use measures that depend on time.

What is lost in recognising that dynamical similarity is a \gauge symmetry of the Universe  --- and it is indeed a great loss --- is the ability to find a measure that can rival the unicorn-like virtues of the Liouville measure of symplectic systems. It is worthwhile to list some of these now:
\begin{itemize}
    \item \emph{simplicity:} the Liouville measure takes a dramatically simple translation-invariant diagonal form when written in Darboux coordinates on phase space,
    \item \emph{universality:} it is preserved by \emph{any} choice of Hamiltonian,
    \item \emph{uniqueness:} it is the unique measure that is universal in the sense above,
    \item \emph{utility:} it has been hugely successful in the history of physics in terms of empirical adequacy, novel prediction, and explanatory power.
\end{itemize}
But such theoretical virtues lose their persuasiveness when they come into conflict with clear epistemic principles such as the PESA. And the PESA tell us that, in cosmology, our beloved Liouville measure double counts physically indistinguishable states. In other words: it counts as distinct states that differ only by an absolute value of the scale factor. Thus, in the cosmological setting, the symplectic Liouville measure is wholly inadequate. Perhaps suggestively, this is also the setting where the utility of the Liouville measure may also be questioned because it fails to place high weight on worlds like ours that have significant time asymmetry; i.e., it can't explain the AoT.

The challenge is to find a better measure that is invariant under dynamical similarity, but that has at least some of the rather formidable theoretical virtues of the Liouville measure of a symplectic system. However, we can only assess these virtues once we have seen explicit examples of the kinds of measures that arise from our generalised Gauge Principle. As I have outlined, the natural choice of measure is determined by the choice of the drag in the contact dynamics. What we will find in the model below is that, while there is no obvious unique drag, there are families of drags that are simple, universal and explanatory in ways that resemble the virtues of the Liouville measure in symplectic systems.\footnote{ Nevertheless, I regard it as a compelling open problem to find a choice of drag that will single out a unique measure that can rival the Liouville measure of a symplectic theory.}

We will see that the explanatory virtues we will find will arise when the behaviour of the drag is complimentary to the Janus and attractor structure of the theory. The JA-scenario is therefore central to how one can motivate a preferred representation of the dynamics. In order to be clear in stating my proposal, it is thus necessary to define the mathematical structures of the JA-scenario precisely. I will do this in the next section. 

\section{The Janus--Attractor scenario} 
\label{sub:the_janus_attractor_scenario}

In this section, I will describe the \emph{Janus--Attractor scenario} for obtaining an AoT in a theory with laws that are time-reversal invariant. To do this, I will first give the mathematical definitions of attractors and Janus points --- or, more specifically, the related notion of a \emph{Janus surface} --- and then show how an AoT can arise when both are present. I will then be able to illustrate more specifically why this scenario is neither generalist nor particularist, and why it allows me to evade the impasse discussed in \Sec{\ref{sec:the_dilemma}}. Later, I will apply this general procedure to the two models outlined above to illustrate independent solutions of the smoothness and red-shift problems.

I introduced the notion of time-reversal invariance in \Sec\ref{sub:discrete_symmetries}, where I referred to it as $T$-symmetry. There I defined $T$-symmetry rather simply as a broad symmetry (i.e., an isomorphism on the space of DPMs) induced by inverting the time parameter $t \to -t$ on the temporal domain of the DPMs. This definition can be generalised to quantum and general relativistic systems as long as the DPMs can be defined in terms of some sort of temporal orientation that can be flipped. In this case, I am considering systems where the DPMs are defined by the flows generated by a vector field on a state space.\footnote{ The state space in our case can be either a phase or a contact space. In \Sec\ref{sec:lagrangian_variational_principle} and \Sec\ref{sec:the_initial_value_problem} I illustrated just how general these considerations are. } Since these state spaces have an orientation,\footnote{ The existence of a Liouville volume-form on a phase or contact space is a sufficient condition for these spaces having an orientation. } $T$-symmetry says that swapping orientations of the vector field generating the flow will lead to integral curves that are also DPMs of the theory. But this is just another way of saying that the DPMs are \emph{undirected} integral curves on state space.

It is rather easy to see that the DPMs of the theories we are interested are indeed undirected curves on state space. Since any Hamiltonian system can be written as a Lagrangian system with an appropriate boundary term and since the value of the action is completely symmetric under the swapping of initial and final data, a change in the orientation of a curve must preserve the corresponding action, and therefore its solutions. Because no information about temporal orientation is affected by taking the quotient of a vector field, the contact system we obtain by quotienting by dynamical similarity should also preserve the action, and therefore $T$-symmetry should be a broad symmetry of the system. More straightforwardly, vector fields and differential forms must transform in a fixed way under a change of orientation for the consistency of the exterior algebra of oriented manifolds.\footnote{ For example, because vector fields pick up a minus sign under $T$-symmetry while scalar fields are invariant, one-forms must pick up a minus sign for the consistency of the interior product. } Thus, because Hamilton's equations \eqref{eq:Ham eqns} and the contact equations \eqref{eq:CVF def} are equations involving different forms, they must be invariant under a change of orientation. This can also be checked by examining the transformation properties of the equations of motion in Darboux coordinates.

This confirms that both symplectic and contact systems are indeed $T$-symmetric. However, as I have repeatedly emphasised, contact systems have a term proportional to the drag that mimics what happens in dissipative systems. I would now like to suggest that one way to quantify the effect of this drag is by showing that it can lead to attractors. Let me now define an attractor and illustrate how its existence is compatible with $T$-symmetry.

\subsection{Attractors} 
\label{sub:dynamical_attractors}

The concept of an \emph{attractor} is well-known in the study of dynamical systems. Loosely speaking, an attractor is a set $A$ that is invariant under the dynamical flow induced by a function, $f$, on a manifold, $M$. Attractors are particularly interesting when they are a proper subset of $M$ in which case there is a \emph{basin of attraction}, $B(A)$, that differs from $A$ by more than a set of measure zero. This basin of attraction is the set of points that flow into $A$ asymptotically. It is important to recognise that attractors can be proper subsets of attractors so that sometimes the word `attractor' is reserved for a \emph{minimal attractor}, which is an attractor with no proper subsets that are attractors.

There are many notions of attractor used in the literature that vary depending on their generality and intended application. I will adopt a standardly used definition given in \cite{milnor1985concept} that is well-suited to my purposes. These attractors are sometimes called \emph{measure attractors} because they are defined using an unspecified measure, $\mu$. However, the measure is only used to exclude sets of measure zero in the definition of $A$. My definition of an attractor will thus be independent of the particular measure chosen. Since we are working on phase or contact space, there will always be a preferred Liouville measure that allows us to explicitly implement this definition.

To make this definition more precise, I need to define a dynamical system with a bit more rigour. Consider a \emph{real dynamical system}, also called a \emph{flow}, defined by the tuple $(\mathcal T, M, f)$, where $\mathcal T$ in an interval on $\mathbbm R$, $M$ is a smooth manifold usually assumed to be compact,\footnote{ Compactness is useful for proving theorems in general dynamical systems. For example, when $M$ is compact, the $\omega$-limit set of an orbit is non-empty, compact and simply connected \citep{milnor1985concept}. But since I am interested in specific cases where attractors are already known to arise, I won't always need to assume compactness. If necessary, however, it is always possible to compactify $M$ by adding a point at infinity. } and $f$ is a continuously differentiable function $f: \mathcal T \times M \to M$. For such a system, $\mathcal T$ is the temporal domain parametrizing the flow, $M$ is the state space on which the flow is defined, and $f$ is the evolution function that defines the flow in time.

If $t \in \mathcal T$ is a time coordinate and $x\in M$ an instantaneous state, then it is customary to denote the evolution function as $f(t,x)$. This function then defines an \emph{orbit} $\gamma_x = \{ f(t,x): t \in \mathcal T\}$ passing through the state $x$. For our applications, the orbits are the integral curves of the vector field that satisfies either the Hamilton or contact equations and $M$ is either a symplectic or contact manifold. However, the concept of an attractor is more general so that this construction can be easily generalised. These definitions, thus, connect the formalism of this section to the language I have been using throughout.

At the base of the notion of an attractor is the notion of an \emph{$\omega$-limit set}. If there exists a sequence $(t_n)_{n\in \mathbbm N}$ in $\mathcal T$ such that
\begin{align}
    &\lim_{n\to \infty} t_n \to \infty \\
    &\lim_{n \to \infty}f(t_n,x) = y
 \end{align}
then $y$ is called an \emph{accumulation point} of the orbit of $\gamma_x$ through $x$. The $\omega$-limit set $\omega(x)$ of a point $x$ is then the set of all accumulation points of $x$. The set $\omega(x)$ thus acts like a kind of asymptotic limit of the flow passing through $x$.

The final ingredient I will need is a measure $\mu$ on $M$. In this case, one can always use the Liouville measure on a symplectic or contact manifold --- but any measure will do. Following \cite{milnor1985concept}, we can now define an \emph{attractor} as a closed subset $A\subset M$ satisfying the follow two conditions:
\begin{enumerate}
     \item The \emph{basin of attraction} $B(A)$, consisting of all points $x \in M$ for which $\omega(x) \subset A$, must have a strictly positive measure; i.e., $\mu(B) > 0$; and
     \item there is no strictly smaller closed set $A' \subset A$ such that $B(A')$ is equal to $B(A)$ up to a set of measure zero.
 \end{enumerate} 

It should be clear that this definition implements the intuitive notion I gave for an attractor at the beginning of this section. The first condition says that there exists a set of points; i.e., the basin of attraction; with non-zero measure that will asymptotically flow into $A$. This guarantees that the attractor is not a completely vacuous structure. The second condition ensures that all parts of $A$ play an essential role in its definition. The main role played by the measure is to exclude attractors that have basins of attraction that are sets of measure zero. These are rather formal considerations aimed at giving mathematically clean definitions of an attractor. We will not need most of these subtleties in our analysis. But it is useful to note, for our purposes, the conditions under which such rigorous definitions can be given. For example, while the notion of a measure played a kind of auxiliary role in my definition, the underlying state space still needs to be measurable for my definition to work. This could be an important issue for defining attractors in field theories with infinite dimensional state spaces.

The crucial aspect of this construction that will be important for my analysis is that an attractor is defined as the limiting set of some flow. One result of this is that an attractor is a fixed set of the flow; i.e., $A$ is such that $f(t,A) = A$ for all $t \in \mathcal T$. Moreover, the definition in terms of a measure guarantees that in the limit $t_n \to \infty$ the distance between $f(t,x)$ and $A$ goes to zero according to \emph{any} Riemannian metric on $M$. This allows us to talk more rigorously about a notion of ``closeness'' to an attractor: a point $x$ is getting close to an attractor as its distance to an attractor goes to zero. While the actual distance along an orbit to an attractor will depend on the metric chosen, its vanishing in the limit is metric-independent. More importantly, this gives us a way to identify the time orientation on an orbit for which the flow is `approaching' an attractor. We can say that the flow is approaching an attractor along the time orientation where increasing $t$ decreases the distance to $A$.

This construction is important because the definition of an $\omega$-limit set given above is not necessarily $T$-symmetric for any particular choice of $x$. In general, taking the limit $t_n \to +\infty$ of the flow from a point $x$ may lead to a different asymptotic structure than taking the limit to $-\infty$ from $x$. Ultimately, this is why I will be able to use the concept of an attractor to introduce an AoT for an observer at a certain point $x$.

This, however, doesn't mean that the orbit $\gamma_x$ passing through $x$ is not a DPM for either time orientation. In other words, the DPMs of a theory can be \emph{undirected} curves on state space even if those curves asymptote to different structures at either endpoint. All that $T$-symmetry implies is that the corresponding action is stationary for each choice of time direction along its DPMs. Hence, the $T$-\emph{asymmetry} of the $\omega$-limit set of a \emph{particular} choice of $x$ is not incompatible with the overall $T$-\emph{symmetry} of the theory.

We can see now that what creates the potential AoT is the singling out of special points, $x$, along an orbit --- namely those that are close to the attractor --- which I view as resulting from particular facts about special observers in the world. Using the notion of closeness to an attractor defined above, observers at those special points can indeed distinguish between time directions toward and away from the attractor along $\gamma_x$. That allows for the possibility for being able to identify an AoT relative to a particular measure, or family of measures, on the state space. But to fully realise this possibility, I will need to introduce a further notion: that of a Janus point.

\subsection{Janus points} 
\label{sub:janus_points}

The concept of an attractor allows us to distinguish between time directions that point towards and away from a particular attractor when starting from a point $x$ along a dynamical flow. This is a necessary ingredient to identify kind of AoT I want to establish for observers at $x$. But while an attractor can give a way to distinguish between time directions along an orbit, it can't give any good reason to align such directions with an AoT because an undirected orbit may start and end on two different attractors, and the symmetry between them gives no way to privilege any particular time direction.\footnote{ For example, there is no natural way to say that a point $x$ is closer to one attractor than it is to the other without introducing a preferred metric on $M$. } To give a full account of an AoT as I wish to formulate it, I will need to break the symmetry between the endpoints of a curve and show that there exists a significant monotonic gradient in some quantity towards one of the endpoints in question. To do this, I will introduce an extra structure along an orbit of the flow that I will call a \emph{Janus point}.

The idea of a Janus point was first introduced in \cite{Barbour:2014bga} and played a central role in a later book by Julian Barbour \citep{barbour2020janus}. In these works, the Janus point is understood at a state along an orbit out of which point oppositely directed AoTs --- hence the reference to the two-faced Roman god of beginnings and transitions. It is difficult, however, to find an explicit formal characterisation of a Janus point in terms of a general dynamical system. An early attempt to do so was given in \cite{Gryb:2021qix}, where a classification of Janus points was given in order to characterise some differences between the Janus points considered in cosmological and $N$-body systems. I will not need to make such a classification here and will attempt to provide a general definition that both implements the basic idea of a Janus point and works for all the different systems I want to consider. Note, however, that my presentation may differ in letter, though not in spirit, from other earlier constructions in the literature. Also, I should distinguish between a Janus point, which is a state space point, from Barbour's related concept of a point $\alpha$, which is a special, highly homogeneous point on configuration space.

The intuition underlying my construction is to understand a Janus point as a point $j$ along the orbit $\gamma_j$ of a real dynamical system where the flow is instantaneously preserving a measure $\mu$. Loosely speaking, the derivative of the measure, $\dot \mu$, has a zero at the Janus point (in a way that I will make rigorous below) that is unique in an open interval of $j$ on $\gamma_j$. This means that $\dot \mu$ must pass through zero at $j$ and that there exist two oppositely directed time orientations on $\gamma_j$, where $\dot\mu$ is definite on either side of $j$. The idea is that these different time orientations define candidate directions for an AoT about $j$.

To make my definition of a Janus point precise, it will be helpful to introduce the notion of a \emph{Janus surface} $\Janus$. Consider a flow $F= (\mathcal T, M, f)$ and measure $\mu$ on $M$. If we write $\mu(R) = \int_R \rho$ for some volume-form $\rho$ on $M$,\footnote{ This can always be done when $\mu$ is a measure on a smooth manifold that comes equipped with an exterior algebra. } then we can define the restricted measure $\mu_S(R)$ onto a codimension-1 submanifold $S \subset M$  as $\mu_S(R) = \int_R \rho\big |_S$.\footnote{ The restricted measure $\mu_S(R)$ defined in this way is equivalent to defining a measure for counting the solutions on the surface $R$ similarly to what we did for symplectic and contact systems in \Sec\ref{sub:measures_on_counting_solutions}. } A Janus surface $\Janus(F,\mu)$ is then a codimension-1 submanifold of $M$ transverse to the orbits of $f$ such that
\begin{equation}
    \frac{\de \mu_{f(t,\Janus)}}{\de t}\Big |_{t = 0} = 0\,.
\end{equation}
It should now be clear that the derivative $\frac{\de \mu_{f(t,\Janus)}}{\de t}$ is what I was referring to as $\dot\mu$ in my sketch above. In terms of $\rho$, this can be restated as a surface on which
\begin{equation}
    \Lie_X \rho \big |_\Janus = 0\,,
\end{equation}
where $X=\frac{\de}{\de t}$ is the generator of the flow. This latter definition is slightly more direct geometrically and convenient to use in practical calculations. A Janus point is then a point $j \in \Janus$.

This definition has several advantages for our purposes. First, it provides a natural scalar quantity: the focusing factor $-\frac {\dot\rho}{\rho}$ representing the rate at which the density of DPMs is being focused along the flow at a point in $M$. This quantity can serve as a candidate for a gradient to be used to define an AoT. As can be seen from these definitions, the zeros of this quantity define the Janus points and its sign determines the monotonicity of the measure. Second, if the measure $\mu$, used to define $j$, admits a reasonable physical interpretation as a counting of DPMs, then a potential AoT arising through this procedure would define a counting of DPMs that reflects a certain degree of time-asymmetry along the orbits. Thus, under the measure $\mu$, typical solutions would be time-asymmetric. In particular, one could compare the amount of time-asymmetry in the phenomena to the amount expected from the focusing of $\mu$. Finally, there is a natural way to implement this definition on a contact system in terms of the behaviour of the drag. Because the drag is just proportional to the focusing factor in a reparametrisation invariant contact system (as a result of Equation~\ref{eq:mu evo gen}) it is the natural structure to use to define an AoT. In such an implementation, the Janus point then becomes a point along the dynamics where the drag is zero so that the contact system is momentarily isomorphic to a symplectic system. 

A potential downside of my definition of a Janus point is its dependence on a particular measure $\mu$. Different choices of $\mu$ will in general have different Janus surfaces. Note also that the freedom to chose $\mu$ is the same freedom discussed in \Sec\ref{sub:time_laws_and_conventions} to choose the contact form $\eta$ and corresponding Reeb vector $R$ in a reparametrisation invariant contact system. Each $\eta$ defines a Liouville measure on the contact space via $\mu_c = \int_R \eta\wedge \de \eta^{(n-1)/2}$. Thus, the freedom to rescale $\eta$ by a positive function means that we can identify any $\mu$ with the Liouville measure of a particular choice of $\eta$. The Janus points of $\mu$ can then be found by finding the zeros of the drag in the representation of the flow where $\mu$ is Liouvillian.

We can thus think of Janus points and the AoTs that may potentially arise from them as features of a particular way of representing the dynamics of a contact system. Because this choice of representation and, correspondingly, of measure is not fixed by the formalism, it must be judged by the theoretical virtues that single it out from other choices. This introduces a certain degree of conventionality into the definition of a Janus point that does not exist when defining attractors alone because the definition of an attractor did not depend on any particular choice of measure. Thus, a certain degree of conventionality seems to be unavoidable in my proposal. I will defend the choices I make in my proposal more fully in \Sec\ref{ssub:jusitying_convension} of the conclusions.

\subsection{The Janus--Attractor scenario} 
\label{ssub:the_janus_attractor_scenario}

In this section, I will give a general scenario for establishing an AoT using the concepts of an attractor and a Janus point. The idea will be to consider an observer that attributes a state $x$ to the current state of the world, where $x$ is approaching an attractor $A$ that it is close to. For such an observer, an AoT will arise when the backward flow from $x$ reaches a Janus point $j$. This AoT points from $j$ to the set on the forward evolution of $j$, $\omega(j) \subset A$, on the attractor.

To see how this comes about, I will say that there is a \emph{Janus-Attractor(JA) scenario} in a real dynamical system $(\mathcal T, M, f)$ when there is a Janus surface $\Janus$ that has an intersection with the attractive basin $B(A)$ of an attractor $A \not\subset \Janus$ that has strictly positive measure on the Janus surface; i.e., $\mu_\Janus(B(A)) > 0$ for some measure $\mu_\Janus$ on $\Janus$. The condition that $A$ not be a proper subset of $\Janus$ ensures that there is some non-trivial flow of a set of positive measure away from some part of $\Janus$ onto $A$.

Let me now argue that an AoT will arise in a JA-scenario relative to a point $x \in B(A)$ that is near to $A$. Since $B(A)$ intersects $\Janus$ there will be a Janus point $j$ along $\gamma_x$ that is different from $x$ itself. This Janus point defines a time orientation from $j$ to $\omega(x)$ because it breaks the $T$-symmetry of $\gamma_x$ that would exist if $\gamma_x$ was bounded by two attractors.

Consider now the distance function $d(a,b)$, calculated with some Riemannian metric on $M$, between two points $a,b \in \gamma_x$. There is then a time orientation from $j$ to $\omega(x)$ that is such that increasing $t$ sends $d\lf(f(t,x),\omega(x)\rt)\to 0$ when $t\to\infty$. Closeness to the attractor can then be quantified by $d(x,\omega(x))$, which goes to zero when $x$ is `close' to $A$.\footnote{Clearly this notion is only metric-independent in the limit.} The inverse of this quantity, $1/d(x,\omega(x))$, is not only monotonic on the interval $(j,\omega(x))$ but grows unboundedly when $x$ is close to $A$. This quantity then defines \emph{a large temporally directed numerical gradient pointing from $j$ to $\omega(x)$ as needed for an AoT.}

What I need to argue for now is why a distance function $d$ should define a physically interesting AoT. One reason to believe that it should is that attractors are fixed sets of the dynamics. They therefore correspond to an equilibrium of the system. The stability of such equilibrium sets is guaranteed when one can find in the system a so-called \emph{Lyapunov} variable \citep{lyapunov1992general}, which is a function whose gradient along the flow is always negative (except at $\omega(x)$). A function bounded by the distance $d(x,\omega(x))$ is a special class of Lyapunov function (see Definition~5 of \cite{beretta1986theorem}). Whenever such a function $L(x)$ exists, one can always construct a function $S(x) = - L(x)$ that is monotonically \emph{increasing} along the flow and reaches its maximum near equilibrium. These functions have been identified with non-equilibrium \emph{entropy functions} in \cite{beretta1986theorem}. When there is a JA-scenario, a natural candidate for such an entropy function $S(t) = - 1/d(j,x)$ is the negative of the inverse of the distance $d(j, x)$ from the Janus point $j$ to the point $x = f(t,j)$. Closeness to an attractor is thus a reasonable notion for defining and AoT since such an AoT is aligned with the temporal direction in which the system is equilibrating such that a natural entropy function can be defined.

These considerations give us general reasons to believe that an AoT with certain acceptable physical characteristics will arise in a JA-scenario. However, there is no guarantee that for any particular physical system there will be a choice of entropy function that is simple and physically interesting. One possibility, which will be explicitly realised in the two models treated in this chapter, is that the drag behaves like $d(j,f(t,j))$; i.e., that the drag is monotonically increasing from the Janus surface.\footnote{Note that it is already defined to be zero there.} Measures with such properties are potentially interesting for a variety of reason. First, they highlight the geometric features of the state space in such a way that a Janus surface appears as a set with no focusing and attractors as sets that are maximally focused. This could be useful if the Janus points and attractors have interesting physical significance within the theory. Conversely, if the focusing itself can be understood in more physical terms, then that understanding could be used to explain the physical significance of the Janus points and attractors.

One possible way to understand the physical significance of the focusing is to interpret the measure as a way of quantifying the amount of dynamical variability in a region. This seems justified if the measure shrinks rapidly near attractors where the system is becoming dynamically fixed. In particular, for a region $R \subset B(A)$, the measure is a way of counting the number of distinct dynamical possibilities in $R$ that will converge onto the attractor $A$. This is indeed a way of quantifying the variability of the dynamical possibilities in the region $R$.

Under such an interpretation, the measure picks up epistemic significance: regions with smaller measure are more dynamically stable and therefore represent more reliable records of the state of the system. The Janus surface can then be thought of as a region where there is lots of dynamical variability in the orbits so that virtually no stable records can exist. In a JA-scenario, the orbits in a region become more dynamically stable over time with records getting more reliable as the system approaches the attractors. The AoT then points from the unstructured past states to the increasingly more structured future states, as might be expected.

In Sections~\ref{sec:newtonian_gravitation_models} and \ref{sec:cosmological_models}, I will detail the two models introduced in \Sec\ref{sub:the_models} and show that they implement a JA-scenario where the measure does indeed focus DPMs onto attractors. I will then apply the general arguments above and claim that the AoT resulting from each model does indeed match these epistemic expectations. I will also show that the $N$-body model exhibits the features required to solve the smoothness problem and that the cosmological model exhibits the features required to solve the red-shift problem.


\section{Solving the smoothness problem in Newtonian gravitation models} 
\label{sec:newtonian_gravitation_models}

In this section, I will define the Newtonian gravitational $N$-body model described in \Sec\ref{Nbody_motvation} and show that, when dynamical similarity is removed, the resulting system has attractors and a natural Janus surface when the reasonable physical assumptions of \Sec\ref{Nbody_motvation} are satisfied. I will then show that observers close to an attractor typically see smooth states in their past history when that history crosses the Janus surface. This result is based on an argument from the central limit theorem in the large $N$ limit, and is therefore robust against the choice of measure. This solves the smoothness problem in this model.

\subsection{Model definition} 
\label{sub:model_definition_Nbody}

The model I will consider here consists of $N$ self-gravitating point particles in $3$ spatial dimensions with masses $m_I$, where $I = 1,\hdots,N$. The positions of these point particles are represented by the configuration variables $q^i_I$, where $i=1,2,3$. The Lagrangian of the theory takes the form
\begin{equation}
    S_N = \int_{t_1}^{t_2} \lf( \frac 1 2 M^{IJ}_{ij} \dot q^i_I \dot q^j_J - V_N(q) \rt) \de t\,,
\end{equation}
where $V_N(q)$ is the Newtonian potential
\begin{equation}
    V_N = - G_N \sum_{I < J} \frac{ m_I m_J }{ r_{IJ} }\,
\end{equation}
$G_N$ is Newton's constant, $M^{IJ}_{ij}$ is the \emph{mass matrix}
\begin{equation}
    M^{IJ}_{ij} = m_I \delta^{IJ}\delta_{ij}\,,
\end{equation}
$r_{IJ}$ is the Euclidean distance between the $I^\text{th}$ and $J^\text{th}$ particle,
\begin{equation}
    r_{IJ} = \sqrt{ (q^i_I - q^i_J)(q^j_I - q^j_J)\delta_{ij} }\,,
\end{equation}
and $m_I$ is the mass of the $I^\text{th}$ particle. To simplify notation, I will revert to a single index $a$ for both particle and spatial indices so that $q^i_I \to q^a$. We then have $M^{IJ}_{ij} \to M_{ab}$ with suitable identifications. For most considerations, the only property I will need of the potential, $V_N(q)$, is its homogeneity of degree $-1$:
\begin{equation}
    q^a \diby{V_N}{q^a} = - V_N\,.
\end{equation}

Using these definitions, the Hamiltonian for the system can easily be seen to be
\begin{equation}\label{eq:ham n-body}
    H_N = \frac 1 2 M^{ab} p_a p_b + V_N(q)\,,
\end{equation}
where $M^{ab}$ is the inverse of $M_{ab}$ and $p_a = \diby{L}{\dot q^a} = M_{ab} \dot q^a$. Clearly, the Legendre transform is invertible when the mass matrix $M_{ab}$ is as well. The Hamiltonian equations of motion the take the form
\begin{equation}
    \dot f = \Lie_X f = \pb f {H_N}\,
\end{equation}
for all functions $f(q,p)$ on the phase space $\Gamma$ with symplectic form $\omega = \de p_a \wedge \de q^a$.

\subsection{Removing dynamical similarity} 
\label{sub:ds_in_Nbody}


I will now show that this model has a dynamical similarity using the results of \Sec\ref{subs:example_homogeneous_potentials}. The $N$-body problem is a particular case of the example considered in that section where we take $g_{ab}(q) = M_{ab}$ so that the quantities $m$ and $n$ of that section take the values $m = 0$ and $n = -1$. Using this, we find that $a = -2$ and
\begin{equation}
    D = 2 q^a \diby{}{q^a} - p_a \diby{}{p_a}\,.
\end{equation}
This gives us the following scaling properties under a dynamical similarity with parameter $c$:
\begin{align}
    t &\to c^3 t & q^a &\to c^2 q^a & p_a &\to c^{-1} p_a\,.\label{eq:ds Nby explicit}
\end{align}
This matches the dynamical similarities of the Kepler problem studied in \Sec\ref{ssub:the_kepler_problem}.\footnote{ To see this more explicitly, use $a = c^3$ and note that $r = r_{12}$, which scales like $q^i_I$. }

We can apply the Gauge Principle for dynamical similarity developed in \Sec\ref{sub:gauge_principle_for_DS} by defining the gauge-fixed surfaces $G_w$ on $\Gamma$ as the level surfaces of the function
\begin{equation}\label{eq:N-body gf}
    w = \tfrac 1 4 \log I\,,
\end{equation}
where we have defined the \emph{dilatational inertia} $I$ as
\begin{equation}
    I = M_{ab} q^a q^b\,.
\end{equation}
It is straightforward to verify that $D(w) = 1$. Given the above definitions, note that the gauge-fixed surfaces $G_w$ are also surfaces of constant $I$.

To find the invariant equations of motion, it is useful to compute the drag, which is the Reeb flow, $R(H_c)$ of the contact Hamiltonian on $G_w$. Using \eqref{eq:friction}, we find
\begin{equation}\label{eq:Nb focusing factor}
    R(H_c) = e^{(1-a)w} \pb{w}{H_N} = \frac{q^a p_a}{2 I^{1/4}}\,.
\end{equation}
The generator of the invariant contact equations is then
\begin{equation}
    \Xinv = {I^{3/4}} X - \frac{q^a p_a}{2 I^{1/4}} D\,.
\end{equation}
In terms of the Darboux coordinates on $\Gamma$, this gives the flow equations:
\begin{align}
    q^{a\prime} &= I^{3/4} M^{ab} p_b - \frac{q^b p_b}{I^{1/4}} q^a \label{eq:gf Nb H1}\\
    p'_{a} &= - I^{3/4} \diby{V_N}{q^a} + \frac{q^b p_b}{2 I^{1/4}} p_a\,. \label{eq:gf Nb H2}
\end{align}
It is straightforward to check that $I$, and therefore the gauge-fixing condition $w = w_0$, is preserved by this evolution.

This gives us a particular gauge-fixing of the equations of motion in $\Gamma$. It is, however, useful to use these gauge-fixed equations to construct a particular reduced representation of the dynamics on an invariant state space. This will also help us to compare to the formalism of \cite{Barbour:2014bga,barbour2013gravitational}.

Because $I$ is constant under this flow, it is possible to construct invariant quantities by noting the weight of the Darboux coordinates under $D$ and dividing by the appropriate power of $I$ to obtain quantities with weight zero. To find a suitable choice of Darboux coordinates $\{\hat q^a, \hat p_a, S \}$ on the reduced contact space, we can guess an expression for $\hat p_a$ from Equation~\ref{eq:gf Nb H1} and for $S$ from Equations~\ref{eq:gf Nb H2} and \ref{eq:Nb focusing factor}. This suggests the definitions
\begin{align}\label{eq:N-body Darboux}
    \hat q^a &= I^{-1/2} q^a & \hat p_a &= I^{1/4} \lf( p_a - \frac{q^b p_b}{I} M_{ac} q^c \rt) & S&= 2\frac{q^a p_a}{ I^{1/4}}\,,
\end{align}
which are constrained such that
\begin{align}\label{eq:Nb constraints}
    M_{ab} \hat q^a \hat q^b &= 1 & \hat q^a \hat p_a &= 0\,.
\end{align}
We can invert these definitions to obtain
\begin{align}\label{eq:inv hat var}
    q^a &= I^{1/2} \hat q^a & p_a &= I^{-1/4} \lf( \hat p_a + \tfrac 1 2 S M_{ab} \hat q^b \rt)\,.
\end{align}

We can find invariant flow equations by re-writing \eqref{eq:gf Nb H1} and \eqref{eq:gf Nb H2} using the definitions above to obtain
\begin{align}
    \hat q^{a\prime} &= \hat p^a \label{eq:GI Nb H1}\\
    \hat p'_a - (q^b p'_b) M_{ac} q^c &= - \diby{V_N(\hat q)}{\hat q^a} - \frac S 4 \hat p_a \label{eq:GI Nb H2} \\
    S' &= 2 M^{ab} \hat p_a \hat p_b + \frac{S^2}4 + 2 V_N(\hat q)\,. \label{eq:Nb S eom}
\end{align}
Note that only the trace-free part (where the trace is taken by contracting with $\hat q^a$) of $\hat p'_a$ is fixed by the equations above. This is because the trace of $\hat p'_a$ is fixed by \eqref{eq:Nb constraints} and \eqref{eq:GI Nb H1} to satisfy
\begin{equation}\label{eq:trace constraint}
    \hat q^a \hat p'_a + M^{ab} p_a p_b = 0\,.
\end{equation}
To obtain \eqref{eq:GI Nb H2}, we used the homogeneity of the potential to derive the expression
\begin{equation}
    I \diby {V_N(q)}{q^a} = \diby{V_N(\hat q)}{\hat q^a} - V_N(\hat q) \hat q^a\,.
\end{equation}
The equation for $S'$ can be computed using the contraction \eqref{eq:gf Nb H2} with $q^a$, the homogeneity of the potential, and \eqref{eq:trace constraint}.

We can compute the contact Hamiltonian for this system using the $N$-body Hamiltonian \eqref{eq:ham n-body} and the general definition \eqref{eq:contact H} of a contact Hamiltonian. Using the homogeneity of the potential we obtain:
\begin{equation}\label{eq:contact Ham N_body}
    H_c = \tfrac 1 2 M^{ab} \hat p_a \hat p_b + \tfrac 1 8 S^2 + V_N(\hat q)\,.
\end{equation}
Note that the equations of motion \eqref{eq:GI Nb H1} - \eqref{eq:Nb S eom} are simply the contact equations for this contact Hamiltonian subject to the constraints \eqref{eq:Nb constraints}. This results from the fact that, under such constraints (which imply $\hat q_a \de \hat q^a = 0$ and $\hat q^a \de \hat p_a = - \hat p_a \de \hat q^a$), the contact form can be written in $\{ \hat q^a, \hat p_a, S \}$ coordinates in Darboux form by taking exterior derivatives of \eqref{eq:inv hat var} as
\begin{equation}\label{eq:contact Newton}
    \eta = - e^{-w} \iota_D \omega_L =  I^{-1/4} \lf( 2 q^a \de p_a + p_a \de q^a \rt) = \de S - \hat p_a \de \hat q^a\,.
\end{equation}
We can now compute the drag directly in terms of the reduced variables and compare to \eqref{eq:Nb focusing factor}. Recall that the drag was defined as the Reeb flow the contact Hamiltonian. In the reduced coordinates, the Reeb direction can be read-off as $R = \diby{}{S}$. The drag is then $S/4$, which is the $S$-derivative of the contact Hamiltonian \ref{eq:contact Ham N_body}, in agreement with \eqref{eq:Nb focusing factor}. This gives us a fully reduced version of the $N$-body theory.

\subsection{A Janus-Attractor scenario for the \(N\)-body system} 
\label{sub:a_janus_attractor_scenario_of_the_smoothness_problem}

Let me now use the previous results to describe the generic behaviour of this system. I will first show that the natural contact form on the reduced space defines a Janus surface. Then I will describe the attractors for this system and, finally, show that there is a JA-scenario. This implies that there is a particular AoT pointing from the Janus surface towards the attractors of the theory. We will see that, in the representation we have provided, the drag behaves generically like the inverse of a distance function to the attractors so that the measure has all the nice epistemic features we hoped for at the end of \Sec\ref{ssub:the_janus_attractor_scenario}. Most importantly, the measure that we will find has the property that the states on the Janus point are typically smooth, homogeneous states when $N$ is large, offering a solution to the smoothness problem for this model.

Let us begin by finding the Janus surface of this theory and showing that it contains a generic set of solutions, where I'll understand `generic' to mean the entire solution space up to a set of measure zero. To do that, I will first prove an important property of the dynamics: the generic monotonicity of the drag, $S$. Let us require that the theory be reparametrisation-invariant so that we may interpret the evolution on any gauge fixed surface $G_w$ as physically equivalent. To achieve a reparametrisation-invariant description of this system, we know from \eqref{eq:Weierstrass} and the general considerations of \Sec\ref{sub:reparametrisation_invariance} that the Hamiltonian is a constraint of the form $H = N H_c$, where $N$ is a Lagrange multiplier that also indicates the time parametrisation used when undoing the Legendre transform. The net result of this is that the contact Hamiltonian is constrained to be equal to zero so that
\begin{equation}\label{eq:Ham constraint n-body}
    H_c = \tfrac 1 2 M^{ab} \hat p_a \hat p_b + \tfrac 1 8 S^2 + V_N(\hat q) = 0\,
\end{equation}
with all differentials of $\tau$ appearing in the contact equations replaced by $\de \tau \to N \de \tau$.

We can now use the vanishing of $H_c$ and \eqref{eq:Nb S eom} to prove the important relationship
\begin{equation}
    S' = M^{ab} \hat p_a \hat p_b\,.
\end{equation}
Because $M^{ab} \hat p_a \hat p_b \geq 0$, this tells us that $S$ is either monotonic or constant. Solutions for which $S$ is \emph{always} constant form lower-dimensional strata on the solution space, and are therefore sets of measure zero. We exclude them from our considerations here since we are interested in the generic behaviour of the system and because solutions with $\hat p_a = 0$ are not interesting dynamically. We then have that $S'>0$ so that $S$ has a zero at some instant during the evolution.

We can now use the general expression \eqref{eq:mu evo gen} to determine the time evolution of the privileged measure density $\rho = \eta\wedge(\de\eta)^{(n-1)/2}$ on contact space. Using the vanishing of the contact Hamiltonian \eqref{eq:Ham constraint n-body} and our expression of the drag in this system \eqref{eq:Nb focusing factor}, we find that
\begin{equation}\label{eq:Nb measure ev}
    \rho' = - \frac {3NS} 4 \rho\,.
\end{equation}
Our general definition of a Janus surface says that one exists when the measure density corresponding to this Janus surface has a time derivative equal to zero. Thus, the set of states where $S=0$ is a Janus surface. Since generic solutions contain exactly one zero of $S$, there is a unique Janus point on generic solutions.

Under the modelling constraints specified in \Sec\ref{Nbody_motvation}, the $N$-body problem is known to have an intricate asymptotic structure. This structure has been studied rigorously in \cite{marchal1976final}, and the results that are useful for our purposes have been conveniently summarised in Appendix~A.1 of \cite{barbour2013gravitational}. We will mainly make use of Theorem~1 of \cite{marchal1976final}, which states that as $t\to \infty$, excluding special cases of measure zero,\footnote{When $E = 0$, these cases are particle collisions and super-hyperbolic escape.}
\begin{equation}\label{eq:aysmptotic q}
    q^a \to A^a t + \mathcal O (t^{2/3})\,,
\end{equation}
where $A^a$ is a (possibly zero) constant and the $q^a$ must be expressed in centre-of-mass coordinates. The physical interpretation of this theorem is straightforward: for large times, the centre-of-mass coordinates of particles for which $A^a\neq 0$ grow linearly with time as the system splits into subsystems whose characteristic size grows at most with $\mathcal O (t^{2/3})$. Corollary 3 of \cite{marchal1976final} then states that, for the $E=0$ case we are treating, at least two particles must escape\footnote{ Excluding the measure zero set of solutions that exhibit superhyperbolic escape. } so that there are at least two subsystems for which $A^a\neq 0$.

The overall picture, that can also be reproduced in numerical simulations \citep{barbour2013gravitational}, is that of an $N$-body system gradually evaporating as it splits into progressively more tightly bound isolated subsystems whose centre-of-mass is moving with constant linear momentum. These subsystems also have emergent structures, studied in \cite{marchal1976final}, and behave like approximately isolated $p$-body systems, where $p < N$. On top of having constant linear momentum, they also have conserved non-positive energies and angular momenta \cite[Theorem~2]{marchal1976final}. In \cite{barbour2013gravitational,Barbour:2014bga}, these constants are proposed as candidate records of the full state of the system, becoming more reliable as $t\to \infty$. In this work however, I will mainly be concerned with the existence of attractors.

The asymptotic behaviour, \eqref{eq:aysmptotic q}, of the Cartesian centre-of-mass coordinates $q^a$ does not lead to attractors because the $q^a$ grow linearly with $t$. The insight here is to notice that attractors do emerge when dynamical similarity is treated as a gauge symmetry. When that is done, we find that the dilatational inertia $I$, whose square root is the characteristic size of the whole system, behaves like $I \to C t^2 + \mathcal O(t^{5/3})$, where $C = M_{ab} A^a A^b$, so that
\begin{equation}
    \hat q^a \to B^a + \mathcal{O}(t^{-1/3})\,,
\end{equation}
For some new constants $B^a$. This says that the configurations $\hat q^a$ accumulate at $\omega(\hat q^a) = B^a$. In general, the $B^a$ will depend on the choice of initial condition for $\hat q^a$.

To investigate attractors on the contact space, we need to know the behaviour of the remaining quantities, $S$ and $\hat p_a$, at late times. We are working with a reparametrisation invariant system, so we must ultimately parametrize the evolution using one of the state-space variables. Because $S$ is monotonic, this is the obvious choice. Since $S$ is a parameter of the flow, it must increase monotonically with $t$ rather than having a fixed point itself. Indeed, we can see from the fact that
\begin{equation}
    p^a = \dot q^a = A^a + \mathcal O(t^{-1/3})
\end{equation}
for large $t$ so that the definition of $S$ in \eqref{eq:N-body Darboux} tells us that
\begin{equation}
    S \propto t^{1/2}\,,
\end{equation}
which is monotonic as we showed more generally above.

Understanding attractors in a reparametrisation invariant theory is complicated by the fact that the standard definitions of attractors assume a fixed time parameter $t$ for the flow. An ambiguity occurs in defining the $\omega$-limit set for velocity variables because the asymptotic behaviour of a velocity can depend strongly on the parametrisation chosen.\footnote{ We choose to focus on velocity variables rather than momenta because the Legendre transform is non-invertible in a reparametrisation invariant theory, and therefore the momenta do not uniquely map to observations. On the other hand, velocities in terms of a particular choice of internal clock do have direct observational significance. } This ambiguity seems unavoidable. Fortunately, the variable $S$ is singled out as the unique choice of monotonic variable that can be easily constructed from the original physical quantities like the position and velocity of particles. Using it as a clock requires a choice of lapse of the form $N = 1/(M^{ab}\hat p_a \hat p_b)$, which guarantees that $S' = 1$. With this choice,
\begin{equation}
    \hat q^{a\prime} = N \hat p^a = \frac{ \hat p^a }{M^{bc}\hat p_b \hat p_c} = \mathcal O(t^{-1/6})
\end{equation}
since inserting the asymptotic expansions of $q^a$ and $p_a$ into the definition of $\hat p_a$ gives
\begin{equation}
    \hat p_a = \mathcal O(t^{1/6})\,.
\end{equation}
This means that, using $S$ as a time parameter, the velocities $\hat v^a \equiv \hat q^{a\prime}$ will accumulate at zero.

Finally, as long as there is one subsystem with at least two particles, then any initial conditions for that subsystem will converge to the same accumulation point. Thus, the set of points flowing into that accumulation point will have strictly positive measure.

It is now straightforward to show that the dynamically similar description of this system has attractors. For all initial conditions, $x_i = (\hat q^a_i, \hat v_i^a)$, leading to subsystems with at least two particles, there will be an attractor consisting of the $\omega$-limit sets, $\omega(x_i) = (B^a_i,0)$, of those initial conditions. Note that these attractors are guaranteed to be stable fixed points because of the existence of the monotonically decreasing Lyapunov function $1/S$, which goes to zero near the attractor. Similarly, there is a natural entropy function, $-1/S$, for this system that is monotonically increasing and reaches its maximum at equilibrium.

As a final step, we show that there is a JA-scenario in the model. Because there is a Janus surface at $S = 0$ and attractors at $S\to\infty$ for generic solutions, there is a time direction of growing $S$ along the orbits of the flow that points from a Janus point to an attractor (and an opposite time direction with similar properties for growing $-S$). An AoT will then arise for an observer close to one of those attractors pointing along that time direction.

\subsection{The smoothness problem} 
\label{sub:the_smoothness_problem_N_body}


\subsubsection{A measureless argument}

Our task now is to argue that the AoT defined by the JA-scenario described above can also explain early smoothness. On our way to doing that, let us first notice that the monotonic behaviour of the drag is helpful for explaining the existence of the attractors. Because $S$ is monotonic and also proportional to the drag, the system is dissipative along the AoT defined by the JA-scenario so that the kinetic energy and the phase space volume available to the states $(\hat q^a, \hat p_a)$ is slowly decreasing. This causes the system to seek out fixed points of the flow. Because the measure is focused on fixed points, we can interpret it as giving a measure of the dynamical variability in a set of solutions as discussed in \Sec\ref{ssub:the_janus_attractor_scenario}. The dissipative behaviour therefore provides an explanation for why the system is approaching an equilibrium state. We now need to argue that these equilibrium states are less smooth than the states near the Janus points.

To do this, it will be helpful to consider the properties of the (negative) of the scale-invariant potential $C = -V_N(\hat q)$. This quantity was initially called the \emph{complexity} in \cite{Barbour:2014bga} because highly clustered states were understood as being more `complex' than smooth states. But to avoid confusion with the term `complexity' that is standardly used to for studying chaos in dynamical systems, I will refer to this quantity simply as the \emph{$C$-function}.\footnote{ Conveniently, `C' can also refer to `clustering' or `clumpiness.' }  To see that $C$ is indeed sensitive to clustering, first recall that
\begin{equation}\label{eq:C-fun}
    C = - V_N(\hat q) = - I^{1/2} V_N(q)\,.
\end{equation}
The Newton potential, $V_N(q)$, is sensitive to the inverse of the two-point separation of particles in the system. When subsystems cluster, these separations shrink causing $V_N(q)$, and therefore $C$, to grow (i.e., to get more negative). Indeed, $-V_N(q)^{-1}$ can be used as a measure of the minimum separation between particles in a system \cite[Equation 1.11]{marchal1976final}. Similarly, the dilatational inertia $I$ gives a measure of the maximum separation in the system \cite[Equation 1.10]{marchal1976final}. Thus, when $I$ grows, bound subsystems can be interpreted as clustering in the sense that the size of any subsystem gets smaller compared to the overall size of the system. This combination means that $C$ is doubly sensitive to the amount of clustering. We thus expect states where the value of $C$ is large to be highly clustered.

Alternatively, it is known from direct numerical investigations, first performed in \cite{battye2003central}, that states with low values of $C$ are highly uniform. In particular, \cite{battye2003central} investigated the absolute minima of $C$ on an $N$-body configuration space.\footnote{ Note the discussion around Equation~(2.17) of \cite{battye2003central} to understand why their minimization problem is equivalent to a minimization of $C$. } These minima are called \emph{central configurations}. What was shown is that central configurations become very close to a uniform distribution as $N$ gets large. More specifically, numerical simulations in the large $N$ limit with equal masses show that the two-point separation distribution between particles in a central configuration approaches that of a perfectly uniform mass distribution. They do not quantify this in terms of a specific statistical criterion for closeness, but I will do this more carefully below. My analysis will confirm that \emph{minimum}-$C$ configurations are nearly uniform in the sense that they have high probability of having been sampled form a uniform distribution.

Further evidence that \emph{low}-$C$ configurations are also nearly uniform even when they are not at the absolute minimum can be found in \cite{Barbour:2014bga}. There, numerical simulations were performed by sampling from a uniform distribution on the unit-sphere in $\mathbbm R^{3N}$ (to remove overall scale dependence) after taking a quotient by the translational and rotational invariance of the system. What was seen (e.g., Figures~4 and 5) is that the probability density function (PDF) for $C$ peaks strongly close to the minimum of $C$ --- particularly for large $N$. In the next subsection, I will reproduce these simulations and give a way of estimating these distributions analytically. This analysis provides independent confirmation of those conclusions and gives a theoretical understanding of them. This tells us that states sampled from a uniform distribution are tightly peaked on low values of $C$.

From these considerations, I conclude that $C$ is a good (inverse) measure of smoothness because it is close to its minimum when states are nearly smooth and large when states are highly clustered.

Let us now come to the crucial point regarding the relevance of the AoT in the JA-scenario to the smoothness problem. Because $M_{ab} \hat p^a \hat p^b \geq 0$, the Hamiltonian constraint \eqref{eq:Ham constraint n-body} tells us that $C \geq S^2/8$. Since $S$ is monotonic and grows unboundedly, $C$ must also grow unboundedly as $t\to\infty$. This behaviour has been confirmed in numerical simulations (e.g., Figure~2 of \cite{Barbour:2014bga}). The unbounded growth of $C$ near an attractor tells us that the attractors are highly clustered states. Thus --- regardless of the initial state at the Janus point --- as one gets arbitrarily close to an attractor, the value of $C$ will be arbitrarily larger than it was near $j$. We then arrive at a general result: using the inverse of the $C$-function as measure of smoothness, \emph{observers near an attractor will generically see smoother past states.}

\subsubsection{The robustness of initial smoothness}\label{sub:robust smoothness}

The statement in the previous section explains early smoothness by showing that, as a system approaches an attractor, the early states appear more and more homogeneous compared to the current one. This statement is useful because it can be made without a measure. However, it makes no guarantee about the absolute smoothness of the states on a Janus surface or how close they might be to uniform distributions. One might then ask whether one can strengthen the statement by making use of a class of measures. After all, because smoothness comes in degrees, it doesn't make much sense to talk about initial smoothness in any precise way without the use of some measure. In particular, we'd like to known whether there is a natural class of measures under which low-$C$ states are typical on the Janus surface. If so, a measure in this class would tell us that typical past states are smooth.

A proposal for such a particular measure of this kind was given in \cite{barbour2015entropy}. There, a dynamically similar measure on the space of solutions was constructed using methods similar to those presented above. The dynamical similarity was gauged-fixed in that paper by fixing the magnitude of the momenta, $M^{ab} p_a p_a$, rather than the magnitude of the configurations, $M_{ab} q^a q^b$, as I have done here. It is thus difficult to compare directly with their results. Nevertheless, it is argued in Appendix~A of \cite{barbour2015entropy} that, in the $3$-body problem, the gauge-fixing they use leads to a measure on the space of DMPs that is both compact, so that the measure can be used to define a genuine probability distribution, and uniform over the configurations and momenta (see their Equation~67).

If similar properties continue to hold for larger $N$, then the PDF of any quantity measuring inhomogeneities will always be independent of the integration over momentum space. But since the measure on configuration space is uniform, the states sampled from that density are as likely to be uniform as they can be. For large $N$, we therefore expect the PDF of any quantity sensitive to inhomogeneities, such as $C$, to be peaked close to its minimum. This is indeed what was seen in Figures~4 and 5 of \cite{barbour2015entropy}.\footnote{ Note, however, that the measure used by BKM simply \emph{is} a uniform measure on two copies of the unit sphere. Thus, the samples taken from that measure \emph{define} the smooth states. Figures~4 and 5 can therefore not be used to illustrate that the states are smooth in the sense claimed by BKM. Rather, their results can be used to conclude that uniform states have low values of $C$. } I conclude that, according to the measure of \cite{barbour2015entropy}, states on the Janus surface are overwhelmingly likely to look like homogeneous states when $N$ is large. Since $N\sim10^{10}$ for $N$-body simulations of the Universe such as the Millennium Simulation (and each of those particles can represent many galaxies in the real Universe), this is clearly the limit relevant to our model.

The question I will turn to now is: what features of the measure used in \cite{barbour2015entropy} guarantee that smooth states are typical on the Janus surface? Answering this will help us to understand when the same conclusions will hold for a more general class of measures. Indeed, the argument above led to the desired result because we reasoned that functions of uniformity would generally be narrowly peaked for large $N$ when sampled from a uniform distribution.

Support for such reasoning can be leveraged upon a version of the Central Limit Theorem (CLT), which I will explain in detail below. If that reasoning were robust, then we would be able to define a large class of measures that will have the same properties as the uniform measure for large $N$. We will see that this is \emph{nearly} the case: the PDF of $C$ generally follows what one would expect from the CLT except that it has fat tails for large $N$. These fat tails restrict the class of measures that behave like the uniform measure at large $N$. Fortunately, these tails can be removed by introducing a modest cut-off on the short distance behaviour of the system. I will now explain how this can be done.

Consider an arbitrary measure density $\rho = f(\hat q, \hat p) \rho_0$ different from the uniform density $\rho_0$ because the function $f(\hat q, \hat p)$ is not equal to a constant. If the CLT theorem holds, then we'd expect the PDF, $g_X(x) = \int_{X = x}\rho_0$, generated by the measure $\rho_0$ of any quantity $X$ that is sensitive to inhomogeneities will be sharply peaked near its minimum for large $N$ (because smooth states have few inhomogeneities). That would mean that, in order for the PDF, $\tilde g_X(x) = \int_{X = x} \rho = f(x) g_X$, generated by any \emph{different} measure to \emph{not} be peaked near its minimum, $f$ will have to grow faster in $x$ away from the mean than a normal distribution decays. In other words, $f$ would have to grow faster than $\sim e^{x^2}$. This would put a strong constraint on the measures that behave differently from the uniform measure for large $N$. Moreover, it is not clear that there is any function naturally formed from the Hamiltonian that would have such growth. Either way, understanding the decay rates of the PDF of $C$ will help us understand how to define a class of measures where uniform states are typical.

Before we can draw any definite conclusions, however, we must first be more explicit about the limitations of the combinatorics arising in the proof of the CLT as applied to this case. We can do this by explicitly estimating various parameters of the PDF of the $C$-function. The $C$-function is the sum of $N(N-1)/2$ (normalised) inter-particle distances. A naive application of the CLT would then suggest that the standardized mean, $\bar \mu_C = \mu_C/\sigma_C$ of $C$, where $\mu_C$ and $\sigma_C$ are the mean and standard deviation of $C$, should scale like $\mathcal O(N)$. This tells us that the approximate width, $\sigma_C$, of the PDF of $C$ decreases like $\mathcal O(1/N)$ compared to the mean, giving us a way to quantify how tightly the distribution of $C$ is peaking as $N$ gets large.

This simple picture, however, tells a slightly incomplete story because the inter-particle separations are not completely independent random variables: they obey, for example, triangle inequalities and depend only on $3N$ particle positions. To get a more accurate estimate of the scaling of the moments of $C$ with $N$, we need to be more explicit. First, let us restrict the spatial extent of our sample of particles to a sphere of radius $R$. This introduces a new relative scale, $R/\sqrt{I}$, into the system. Physically, this can be interpreted as some large distance beyond which we have no epistemic access; e.g., the Hubble horizon.

We can then use known results regarding the two-point separation function, $p(r_{ij})$, of a uniform ball of radius $R$:
\begin{equation}\label{eq:Williamson}
    p(r_{ij}) = \frac{3r_{ij}^2}{R^3} - \frac{9r_{ij}^3}{4R^4} + \frac{3r_{ij}^5}{16R^6}\,.
\end{equation}
This distribution, called the \emph{Williamson} distribution, gives the probability of finding two particles separated by a distance $r_{ij}$ when sampled from a spatially uniform ball of radius $R$. A derivation of the Williamson distribution is given in \Sec 2.f of \cite{battye2003central}. Using this, it is relatively easy to find that the expectation value of any power $n>-3$ of $r_{ij}$ is (see also Equation~(2.30) of \cite{battye2003central})
\begin{equation}
    \mean{r_{ij}^n} = \frac{72 (2R)^n}{(n+3)(n+4)(n+6)}\,.
\end{equation}
For $n\leq -3$, the expectation values diverge --- a fact that will be important below.

\begin{figure}[H]
    \centering
    \includegraphics[width=0.95\textwidth]{\pdots 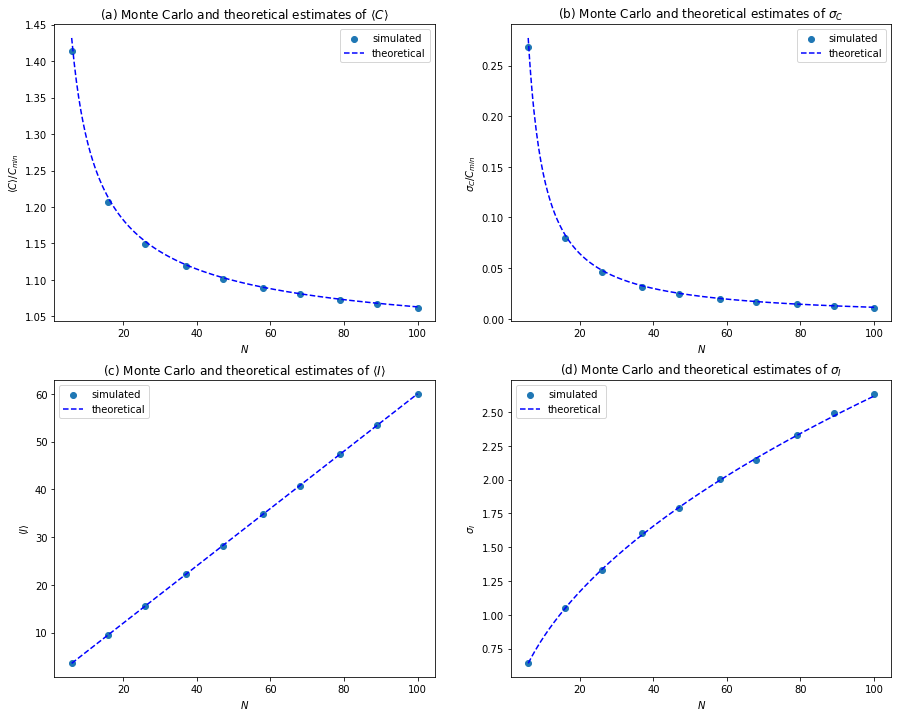}
    \caption{A comparison of two different estimates of: (a) $\mean{C}$, (b) $\sigma_C$, (c) $\mean{I}$ and (d) $\sigma_I$ using a direct Monte Carlo simulation (blue dots) and a theoretical estimate (blue dashed line) for different values of $N$. The number of samples is chosen to scale like $N^{-2}$ to keep the computation time constant for each data point with $10^6$ samples for $N=6$. All curves show excellent agreement between the directly simulated quantities and the theoretical estimates in accordance with expectations from the CLT. Unfortunately, similar trends do not hold for moments of order higher than $2$ as a consequence of the divergence of the integrals involved.
    }
    \label{fig:C_I_scaling}
\end{figure}

We can use the expression above to explicitly compute the $m^\text{th}$-order moments of $C$. This can be done because, while the inter-particle separations are not independent, their relationship to each other is symmetric. For the mean, this implies that each term in sum takes the same value. If we additionally use the approximation
\begin{equation}
    I \approx \mean{I} = \mean{\sum_{i=1}^N r_i^2} = N \mean{r^2} = N \frac{\int_0^R r^4 \de r}{ \int_0^R r^2 \de r} = \frac {3} 5 N R^2\,,
\end{equation}
where $r$ is the distance between a particle and the origin, which is valid for large $N$,\footnote{ The approximation $I \approx \mean{I}$ for large $N$ is justified because $I$ is a genuine sum of $N$ independent random variables; namely the distance of each particle to the origin. The CLT then tells us that $\mu_I/\sigma_I \sim \mathcal O(N^{1/2})$. Explicit calculation confirms this since, using the method explained in the text, $\sigma_I^2 = \frac {12}{175} N R^4$. Both of these results for $I$ agree with numerical simulations as can be seen from \fig\ref{fig:C_I_scaling}. This means that the PDF of $I$ is more tightly peaked than $C$ by $\mathcal O(N^{1/2})$. } we find:
\begin{equation}\label{eq:mean C}
    \mean{C} = \mean{ \sum_{i\neq j} \sqrt{I} r^{-1}_{ij} } \approx \frac {N(N-1)} 2 \sqrt{\mean{I}} \mean{r^{-1}_{ij}} = \lf( \frac {3N}{5} \rt)^{3/2} (N-1)\,.
\end{equation}
The higher order moments of $C$ can be computed similarly but involve binomial coefficients arising from raising the sum in $C$ to the power of $m$. Being careful to split this sum into diagonal terms involving $\mean{r_{ij}^{-2}}$ and off-diagonal terms involving $\mean{r_{ij}^{-1}}^2$, we find that the terms in the standard deviation of $C$ involving $N^2(N-1)^2$ cancel so that:\footnote{ This cancellation is the basis of the CLT. }
\begin{equation}\label{eq:std C}
    \sigma^2_C = \mean{\lf( C - \mu_c \rt)^2 } \approx \frac{N(N-1)}{2} \mean{I} \lf[ \mean{r_{ij}^{-2}} - \mean{r_{ij}^{-1}}^2 \rt]  = \frac {3^5}{10^3} N^2 (N-1)\,.
\end{equation}
This confirms the naive CLT-based argument that showed that $\sigma_C / \mu_C \sim \mathcal O(N^{-1})$.

Following the same reasoning, it is easy to see that the $m^\text{th}$ standard central moment, $\mu^m_C/(\sigma_C)^m = \frac{ \mean{ (C - \mean{C})^m } }{ \sigma_C^m }$, scales like $\mathcal O (N^{2-n})$ for $n>2$ due to cancellations similar to those seen above. Since the standardised moments tell us how the shape of a distribution deviates from normality, this scaling of the moments with $N$ suggests parity with the CLT. Where we get divergence from the CLT is that, for $m\geq 3$, the moments will involve terms of order at least as low as $\mean{r^{-m}}$. Because these term diverge, the higher order moments are not well-defined. So while the CLT gives us the correct scaling in terms of $N$ for the standardized moments of $C$, this scaling is irrelevant for the higher order moments of $C$, which simply do not exist. This tells us that the PDF of $C$ is likely fat-tailed.

To check the reliability of our idealisations, we can estimate the moments in a more direct way by performing a Monte Carlo simulation using $M\gg 1$ samples of $N$-body systems, compute the $C$-function for each of those samples, and use those results to estimate the statistics of $C$. \fig\ref{fig:C_I_scaling} shows the results of such a simulation for the mean and standard deviations of $C$ at different values of $N$.\footnote{ Units have been chosen so that the minimum value of $C$ is equal to one. A good estimate for large $N$ of the absolute minimum, $C_\text{min}$, of $C$ can be obtained from the lower bound derived in \Sec 2.d.ii of \cite{battye2003central}. Using their Equations (2.38), (2.13), (2.15) and (2.16), we find that $C_\text{min}= \lf( \frac 3 5 N (N^{2/3}-1) \rt)^{3/2}$. } Also plotted in \fig\ref{fig:C_I_scaling} are the Monte Carlo and theoretical estimates of $\mean{I}$ and $\sigma_I$ showing a sharp peaking behaviour in the PDF of $I$ for large $N$. The agreement between the two methods generally confirms that our theoretical idealisation is valid even for $N$ as small as $6$. To get an idea of what a typical distribution looks like, the Monte Carlo estimate of the PDF of $C$ for $N=20$ and $M=10^6$ is shown in blue in \fig\ref{fig:g_C simulated}.

For statistics depending on moments larger than $2$, such as the skewness and kurtosis, the estimates become unreliable even for large ($\sim 10^7$) values of $M$. This highlights the sensitivity of those statistics to outliers in the tail confirming our theoretical arguments about the divergence of moments larger than $2$. Together, the lines of evidence above provide a compelling argument that the PDF of $C$ is indeed fat-tailed.

The divergence of the moments of $C$ limits the space of functions $f(c)$ such that, using the notation defined above, the PDF of $C$, $\tilde g_C(c) = f(c) g_C$, generated by $\rho = f(c) \rho_0$ will behave like that of the PDF generated by the uniform distribution $\rho_0$. This is because $\int f(c) g_C\, \de c$ will diverge if $f(c)$ is a polynomial of order $3$ or higher so that the statistics of $C$ under $\tilde g_C$ are not well-defined. This will become problematic in the next section when we consider a particular natural choice of $\rho$ that will violate this condition.

To overcome this problem, I will impose a physically well-motived cut-off on the short distance behaviour of the system. In particular, it is not physically reasonable to assume that the particles in our model are indeed genuine point particles. If we assume that they have a typical size $r_c$, then $r_c$ sets the minimum size of the inter-particle separations of the theory. Because the divergences are sensitive to the most extreme outliers of inter-particle separation, a modest cut-off should cure the divergences without introducing unreasonable physics. Indeed, such cut-offs are commonly used in $N$-body statistical mechanics to cure the unphysical divergences in the $1/r_{ij}$ potential \cite{Padmanabhan:2008overview}.

We can find the mathematical requirements on such a cut-off by examining the source of the divergences in the moments. If we cut off each integral in $\mean{r^n_{ij}}$ at some minimum inter-particle separation $r_c$, then the definition $\epsilon = r_c/R$ tells us that, to leading order,
\begin{equation}
    \mean{r^m_{ij}} \sim \epsilon^{-m+3}
\end{equation}
for $m > 3$.\footnote{ For $m=3$, the divergence is $\sim \log \epsilon$. } These are also the leading divergences in the $m^\text{th}$ central moment $\mu^m_C$ of $C$. Using our previous result arising from the combinatorics of the CLT, we found that $\mu^m_C/\sigma^m \sim \mathcal O(N^{2-m})$ so that, for $m>2$, the $m^\text{th}$ \emph{standardised} moment $\tilde \mu^m_C = \mu^m_C/\sigma_C^m$ scales like
\begin{equation}
    \tilde \mu^m_C \sim \lf( N \epsilon \rt)^{2 - m} \epsilon\,.
\end{equation}
For small $\epsilon$, the standardised moments vanish provided $N \epsilon < 1$. However, if $m$ is large, then small differences in the numerical coefficients of each term could get amplified by raising the term multiplying $N \epsilon$ to a large power of $m$ if $N \epsilon \sim 1$. Thus, a safe cut-off should satisfy the following condition:
\begin{equation}\label{eq:cutoff cond}
    \frac 1 N \ll \epsilon \ll 1\,.
\end{equation}
A convenient choice that satisfies this condition is $\epsilon = N^{-p}$ for some choice of $p$ between $0$ and $1$. For large enough $N$, the simple choice $p = 1/2$ should lead to a reasonably small value for $r_c$ and be guaranteed to satisfy the condition \eqref{eq:cutoff cond}. However, the physical significance of this choice should be evaluated in the context of a specific model. Since I am not immediately concerned with questions of empirical adequacy, I will leave this determination to future considerations.

\subsubsection{A particular measure for early smoothness}\label{ssub:early_smoothness_measure}

Let me now consider a particular measure arising naturally from this analysis and see how it compares to the class of measures considered in the previous section. In my analysis, I have used the condition $I = I_0$ to fix the orbits of dynamical similarity because this condition led to very manageable contact equations. Interestingly, the natural measure density $\rho$ for this gauge fixing, when restricted to the space of DPMs on the Janus surface, satisfies the criteria specified above: that is, it behaves like a uniform distribution for large $N$ provided a cut-off satisfying the condition \eqref{eq:cutoff cond} on inter-particle separations is imposed.

To see this, note that using the contact form \eqref{eq:contact Newton}, the natural measure density for this theory is
\begin{equation}
    \rho = \de S \wedge (\de \hat p_a \wedge \de \hat q^a)^{3N}\big |_{\mathcal R}\,,
\end{equation}
where $\mathcal R$ is the surface defined by the constraints $\hat q = \sqrt {M_{ab}\hat q^a\hat q^b} = 1$ and $D = 2(\hat p_a \hat q^a)/\hat q  = 0$. To find the restriction of $\rho$ onto $\mathcal R$ we can use the Fadeev--Popov trick of inserting the magnitude of the Poisson bracket between the constraints and integrating over $\delta$-functions:
\begin{equation}
    \rho = \de S \wedge \int \lf|\pb{q}{D}\rt| \delta(q)\delta(D)\, (\de \hat p_a \wedge \de \hat q^a)^{3N}\,.
\end{equation}
Because these constraints are canonically conjugate, the Poisson bracket is $1$ and the integration gives
\begin{equation}
    \rho = \Omega^{3N} \de S  \prod_{a=1}^{3N-1} \de \theta^a_q \de p^\text{tf}_a\,,
\end{equation}
where $\Omega^{3N}$ is the determinant of the metric on the unit $3N$-sphere, $\theta^a_q$ are coordinates on that sphere corresponding to directions on configuration space, and $p^\text{tf}_a$ is a basis of vectors on $\mathbbm R^{3N-1}$ whose contraction is zero with a vector pointing in the directions specified by $\theta^i_q$.

We can now define a measure on the space of DPMs for this theory by restricting to the intersection of the Hamiltonian constraint surface $H = 0$ and the Janus surface $S = 0$. In this case, the Poisson bracket in the Fadeev--Popov determinant is replaced by $\Lie_X S = p_\text{tf}^2 = 2C$. If we choose spherical coordinates $(p_\text{tf}, \phi^m_p)$, where $m \in \{ 1,\hdots, 3N-2 \}$, on the trace-free momentum space, then we obtain an additional factor of $1/p_\text{tf}$ from integrating the $\delta$-function $\delta(H) \de p_\text{tf}$ over $p_\text{tf}$ and a factor of $\Omega^{3N-1}$ from the momentum-space spherical coordinates. The final result is then
\begin{equation}
    \rho_J = \lf( \sqrt{2C}\, \Omega^{3N}(\theta^i_q) \prod_{i=1}^{3N-1} \de \theta^i_q \rt) \lf( p_\text{tf}^{3N-2}\, \Omega^{3N-1}(\phi^m_p)  \prod_{m=1}^{3N-2} \de \phi_m^p \rt)\Big |_{p_\text{tf}^2 = 2C} = (2C)^{(3N-1)/2} \rho_0 \,,
\end{equation}
where $\rho_0$ is the \emph{uniform} measure on the tensor product of the spheres: $S^{3N}\times S^{3N-1}$.

The measure $\mu_J(R) = \int_R \rho_J$ has a number of interesting features. First, like the measure used by BKM, it is an integral over a compact space so that it is a probability measure. Second, the remaining momentum-space integral is simply one so that probability distributions over configuration space functions can be computed with the measure $\mu^q_J(R) = \int_R (2C)^{(3N-1)/2} \rho^q_0$, where $\rho^q_0$ is the uniform density on the unit $3N$-sphere. Finally, the measure density differs from the uniform density by the factor $(2C)^{(3N-1)/2}$, which is polynomial in $C$.

Using the definitions established in \Sec\ref{sub:robust smoothness}, let $g_C(c)$ be the PDF of $C$ generated by a uniform measure $\rho_0$ and $\tilde g_{C}(c)$ be the PDF of $C$ generated by $\rho_J$. Because the order of the polynomial of $c$ in $\tilde g_C$ grows with $N$, the fat-tailed nature of $g_C$ implies that $\tilde g_C$ will diverge for large values of $c\in C$ in an ever more severe way as $N$ gets large. To cure these divergences we therefore need to impose a cut-off as predicted in the previous section. Following the results of that section, the hard cut-off $r_c = R N^{-p}$ works very well when $p = 1/2$ for medium to large sized $N$. When implementing such a cut-off, we expect $\tilde g_C$ to converge to a normal distribution with mean $\mu_C$ and standard deviation $\sigma_C$ as given by Equations~\ref{eq:mean C} and \ref{eq:std C} when $N\gg 1$. Moreover, we will now see that these distributions will also be largely overlapping.

First note that for $N\gg 1$, $\text{Norm}(\mu, \sigma^2/N^2) \approx \text{Gamma}\lf(\frac{\mu^2 N^2}{\sigma^2}, \frac{\sigma^2}{\mu N^2}\rt)$, where Norm$(\mu,\bar \sigma)$ is the normal distribution with mean $\mu$ and variance $\bar \sigma^2$ and Gamma$(k,\theta)$ is the gamma distribution with shape $k$ and scale $\theta$. Let us now take $g_C(c/\mu_C)$ to be the PDF of $\text{Norm}(\mu,\sigma^2/N^2) \approx \text{Gamma}(k, \theta)$, where $\mu = 1$, $\sigma = \frac{3}{2^{3/2}}$ (to match Equations~\ref{eq:mean C} and \ref{eq:std C}) and we make the corresponding identifications for $k$ and $\theta$. Then we can use the fact that multiplying the PDF of a Gamma function by a polynomial of fixed degree simply shifts $k$ by a fixed amount. For our choice of parameters, this leads to $\tilde g_C(c)$ being approximately equal to the PDF of
\begin{equation}
    \text{Gamma}\lf(k+\frac {3N-1}2,\theta\rt) \approx \text{Norm}\lf( \mu + \frac{3\sigma^2}{\mu N}, \frac{\sigma^2}{N^2} \rt)\,,
\end{equation}
where approximations are taken to the lowest order in $1/N$. This means that, in units where $\mu_C = 1$, the mean is shifted by a constant factor of $\frac {\Delta \mu}{\sigma_C} = \frac {3\sigma}{\mu} \approx 1.591$ relative to $\sigma_C$. In other words, for large $N$ the distributions $g_C$ and $\tilde g_C$ are simply normal distributions with the same standard deviations, $\sigma_C$, and means shifted by a factor of $1.591 \sigma_C$. Such distributions have significant overlap with each other, which was what we set out to show.

\begin{figure}[H]
    \centering
    \begin{subfigure}[t]{0.5\textwidth}
        \includegraphics[width=\textwidth]{\pdots 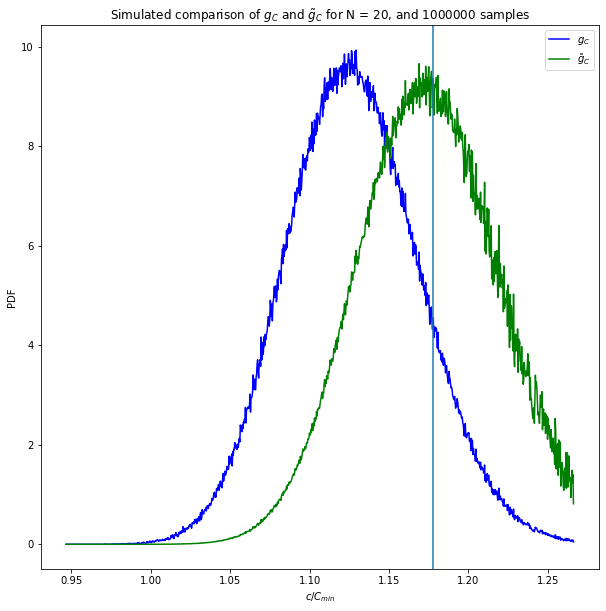}
        \caption{ Monte Carlo simulation with $M=10^6$ samples and $N=20$ particle showing $g_C$ in blue and $\tilde g_C$ in green. A cut-off of the form $r_c = R N^{-1/2}$ was used. }
        \label{fig:g_C simulated}
    \end{subfigure}%
    ~
    \begin{subfigure}[t]{0.5\textwidth}
        \includegraphics[width=\textwidth]{\pdots 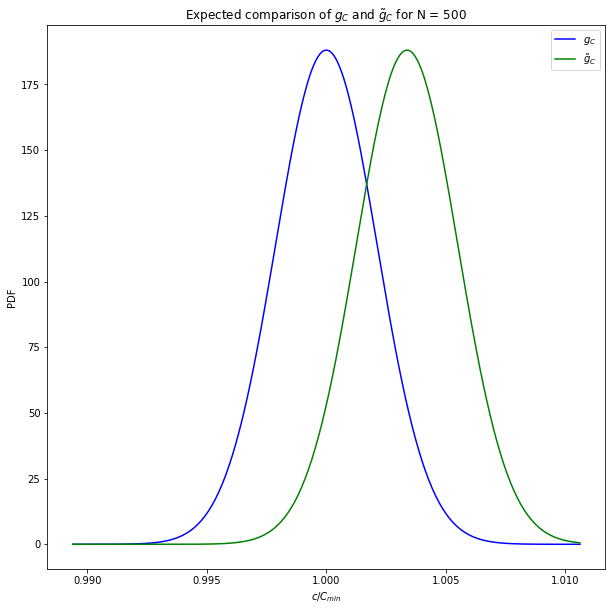}
        \caption{ Predicted behaviour of $g_C$ in blue and $\tilde g_C$ in green for large $N$ using the model discussed in the text. }
        \label{fig:g_C theory}
    \end{subfigure}

    \caption{ Comparison between $g_C$ and $\tilde g_C$ for different values of $N$. Good statistics for direct estimates are only computationally feasible for small $N$, where the effect of the cut-off on the tails is reduced. The results of a simulation for $N=20$ are show in \fig\ref{fig:g_C simulated}. This shows good overlap of $g_C$ and $\tilde g_C$ in line with theoretical considerations. A theoretical estimate of the overlap of the two distributions is given in \fig\ref{fig:g_C theory} using the approximations outlined in the text. This illustrates the expected overlap for large $N$ once a cut-off is imposed. }
    \label{fig:g_C comparison}
\end{figure}

In \fig\ref{fig:g_C comparison}, this behaviour is illustrated by plotting the results of a Monte Carlo simulation for $\tilde g_C$ (in green) compared to $g_C$ (in blue) for a cut-off of the form $r_c = R N^{-1/2}$. To implement the cut-off we have sampled the inter-particle separations used to compute $C$ directly from the Williamsion distribution modified to exclude tails smaller than $r_c$.\footnote{ We have also used the approximation $I = \mean{I}$, which is justified by our previous simulations. } This methodology is justified by our previous simulations that show a convergence of directly computed inter-particle separations to the Williamsion distribution for medium to large $N$.

Samples for $\tilde g_C$ have been obtained using a rejection method starting from a uniform distribution so that the statistics on the upper tail require significant computation time for large $N$. To get good statistics, we have restricted our plotted simulation to $N=20$, which is shown in \fig\ref{fig:g_C simulated}. The plot shows that the cut-off is successful at restricting the effects of the tail even for relatively small $N$. The shifted distribution $\tilde g_C$ already has a mean that closely matches the prediction made above, which is represented by the vertical blue line. \fig\ref{fig:g_C theory} shows the expected PDFs when $N=500$ based on the theoretical considerations above. Both plots show significant overlap of the distributions even for moderately low $N$.

I will conclude this subsection by noting that $\tilde g_C$ will only remain close to $g_C$ near the Janus surface. This is because away from the Janus surface $S$ grows monotonically, and the Hamiltonian constraint gives $p_\text{tf}^2 = 2C - S^2/4 $. The measure $\mu_J(R,S_0)$ projected onto an $S = S_0$ surface is then
\begin{equation}
    \mu_J(R,S_0) = \int_R (2C - S^2_0/4)^{(3N-1)/2} \rho_0\,.
\end{equation}
The extra factor depending on $S_0$ shifts the PDF of $C$ to be centred on $S^2_0/8$. For large $S_0$, this will be peaked on a value that is arbitrarily larger than the minimum of $C$. In other words, for timescales such that $S^2/8 \gg \text{min}(C)$; i.e., for ``late'' times when the system is approaching an attractor; likely states will be highly clustered.

This observation adds an important subtext to the usual arguments used in $N$-body systems suggesting that highly clustered states are typical because significant entropy is stored in the steep well of the Newtonian potential. Crucially, these arguments don't consider \emph{when} such states are likely to occur along solutions. We can see that by treating dynamical similarity as a gauge symmetry we find that highly clustered states, while numerous \emph{overall} in the state space, are overwhelmingly likely to be found near attractors, and not near Janus points. This insight completely changes our intuitions about the AoT. For interacting systems, the dynamics can be such that certain kinds of states may be more or less likely to occur at different times along a family of solutions. We will return to this point when discussing the potential implications for the Boltzmann Brain problem in \Sec\ref{sec:prospectus}. In the case treated here, this means that smooth states are exceedingly likely to be found near Janus points while clustered states are exceedingly likely to be found near attractors. The AoT defined by the JA-scenario for this model therefore points from smooth to clustered states. \emph{This solves the smoothness problem within this model.}

\subsubsection{The \(C\)-function and uniformity}\label{ssub:C and uniformity}

The validity of the argument given in the previous section relies on the $C$-function being a good proxy for inverse smoothness. In this section, I will try to further demonstrate that this is the case, first, by showing that small $C$-values do indeed correlate with more uniform distributions and, then, by showing that my results are consistent with a more direct measure of uniformity based on the distribution of 2-point separations of an $N$-particle system. I will make use of a well-known statistic, called the \emph{Kolmogorov–-Smirnov (KS)} statistic denoted by $D$ \citep{an1933sulla,smirnov1948table}, which is used to compare two different distributions by taking the maximum absolute difference of the CDFs of two distributions. This method is known to give a robust way to compute the distance between two probability distributions.

\begin{figure}[H]
    \centering
    \includegraphics[width=0.95\textwidth]{\pdots 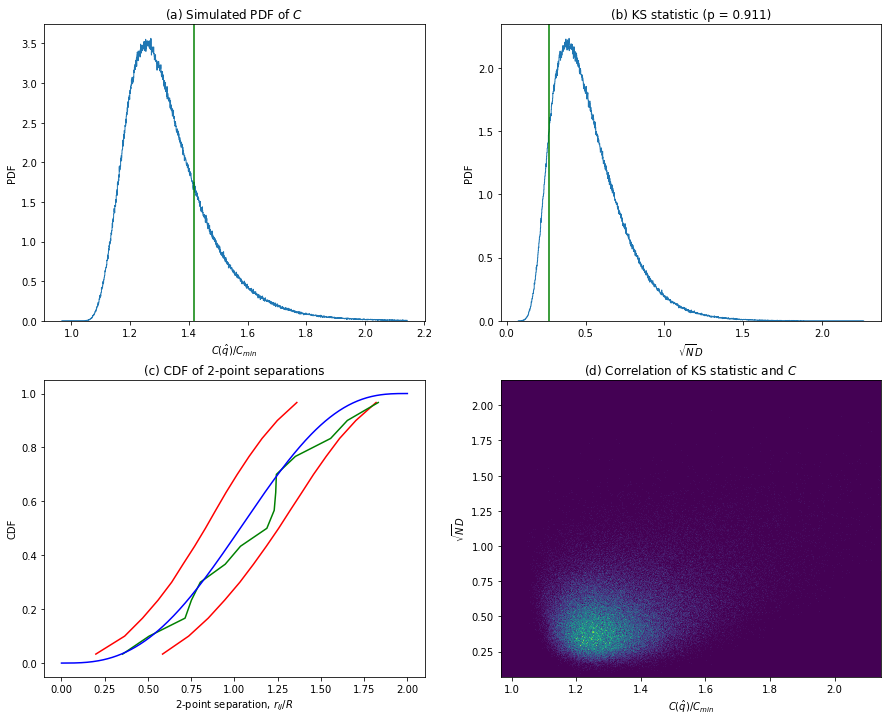}
    \caption{A Monte Carlo simulation of $6$-bodies with $3\times 10^6$ samples for initial conditions on the Janus surface illustrating the smoothness of likely states. (a) The simulated distribution $g_C$ in blue. A random state (green) sampled from $\tilde g_C$ lies in the thick part of the uniform distribution. (b) The PDF of the Kolmogorov--Smirnov (KS) statistic $\sqrt{N} D$ (blue) generated by comparing the two-point separation distribution of each $N=6$ sample drawn from the uniform measure $\rho_0$ with that of the Williamson distribution. A randomly chosen state (green) sampled from $\rho$ has small KS-statistic, and is therefore unlikely to have not been sampled from the uniform distribution $\rho_0$. (c) The simulated CDF of a random state (green) of $\rho$ compared with the theoretical CDF of a continuous uniform distribution (blue). Confidence intervals (red) of $80\%$ are computed from the Monte Carlo simulation sampled from $\rho_0$. (d) A $2$D histogram plotting the values of $C$ versus the KS statistic (bright colours mean greater frequency). Low values of $C$ are correlated with small KS statistics indicating that low-$C$ states are likely to have been sampled from a uniform distribution. }
    \label{fig:Nbody_sim}
\end{figure}

For the first demonstration, let us compare the simulated CDF of the 2-point separations of a sample of $N$ particles to the theoretical CDF of the 2 point separations of an ideal uniform distribution, which is given by the definite integral of the Williamson distribution \eqref{eq:Williamson}. Using the parameters of the Monte Carlo simulations of the previous sections, one can plot histograms for $C$ versus $\sqrt N D$\footnote{ This normalised value of $D$ is known to obey a particular distribution when the samples are independent. While the independence assumption is violated here (see below), I will nevertheless consider this normalised statistic in my analysis. } using many simulations of the $N$-body system. Such a plot is shown in \fig\ref{fig:Nbody_sim}(d) for $N =6$ and $M=3\times 10^6$ (the number of samples in the simulation). In these units, the minimum value of $C$ is chosen to be one while the maximum of $D$ is $\sqrt N$. This plot shows that low values of $C$ typically have low values of $\sqrt N D$, confirming that low values of the $C$ statistic are correlated with uniformity.

For the second demonstration, let us use the distribution of the KS-statistic, $D$, of $C$ to estimate the probability that a particular sample of $N$-particles has \emph{not} been drawn from a uniform distribution. Note that we are dealing with two distinct notions of uniformity: one is the Williamson distribution, which is the uniform distribution of a \emph{continuous} number of samples, and the uniform distribution, $\rho_0$, of a \emph{finite} number of samples $N$.

Since the inter-particle separations are not independent, we must estimate the distribution of $D$ using simulations. What I will do is compute the distribution of the value of $D$ obtained by comparing samples taken from $\rho_0$ with samples taken from the Williamson distribution. This gives a benchmark for what the $D$ statistic should look like for a finite sample of 2-point separations taken from a uniform distribution. The PDF obtained from such a simulation with the parameters used above is shown in blue in \fig\ref{fig:Nbody_sim}(b).

We can then compute the KS statistic for the 2-point separations of a set of $N$ particles sampled from the non-uniform distribution $\rho$ and use our benchmark to estimate how likely such a sample is to have not been sampled from a $\rho_0$ given its $D$-value. The vertical green line indicates the value of $\sqrt N D$ for a random sample taken from the distribution $\rho$. The $C$-value of this sample is shown in green in \fig\ref{fig:Nbody_sim}(a) along with the simulated distributions $g_C$ (blue). For this sample, $p = 0.911$ which means that there is only an 8.9\% chance that such a sample was \emph{not} drawn from a uniform distribution. There is, thus, not sufficiently strong evidence to exclude that the sample belongs to a uniform distribution. In order words, using the distribution of 2-point separations alone, we conclude that a typical sample taken from $\rho_0$ could have been sampled from the uniform distribution $\rho_0$.

We can gain more information about this particular sample of $N=6$ particles by comparing its CDF; i.e., the green line of \fig\ref{fig:Nbody_sim}(c); to the Williamson distribution (in blue).\footnote{ Units are chosen so that $R=1$. } Confidence bands of $80\%$ generated by the Monte Carlo simulation are shown in red. The green line clearly fits nicely to the uniform blue line within the expected error bars showing that a typical sample drawn from $\rho$ does not differ in any statically significant sense to a typical sample drawn from $\rho_0$.

The disadvantage of the direct approach, and the reason to favour the indirect one based on the $C$-function, is that the direct approach is much more computationally intensive\footnote{ This is the reason for treating relatively low values of $N$. } and lacks analytic control. Nevertheless, the results of this demonstration confirm, at least for small $N$, that the smoothness of states drawn from $\rho$ on the Janus surface can be confirmed \emph{directly} from the 2-point separations rather than having to infer smoothness \emph{indirectly} from the value of the $C$-function. 

\section{Solving the red-shift problem in cosmology} 
\label{sec:cosmological_models}

In this section, I will investigate the consequences of treating dynamical similarity as a gauge symmetry of an FLRW cosmology under the modelling constraints specified in \Sec\ref{sub:FLRW assumptions}. First, in \Sec\ref{sub:model_definition_flrw}, I will give a more complete description of the FLRW model itself that will compliment the brief remarks given in \Sec\ref{sub:dynamical_similarity_in_the_universe}. Then, in \Sec\ref{sub:removing_dynamical_similarity_flrw} I will identify the dynamical similarity of this model and apply the Gauge Principle, reproducing a known representation of the reduced system. In \Sec\ref{sub:a_janus_attractor_scenario_for_flrw}, I will show that this representation admits a JA-scenario that provides a solution to the red-shift problem in this model. Finally, in \Sec\ref{sub:light_cone_coordinates} I will investigate the nature of Big Bang singularities in the context of this model using a geometric description of configuration space. This will show how solutions can be extended in this framework by making particular clock choices and corresponding restrictions on the scalar field potential. These results are important both for illustrating how a JA-scenario can occur in a model with what is normally interpreted as a Big Bang singularity and for gaining a better understanding of various extensions of cosmological solutions motivated by the PESA.

\subsection{Model definition} 
\label{sub:model_definition_flrw}

The Friedmann–-Lema\^itre-–Robertson-–Walker (FLRW) model of general relativity is a cosmological model that assumes homogeneity and isotropy of the space-time metric over a spatial slice $\Sigma_t$. These space-times are globally hyperbolic, and the general space-time metric can be written in the form
\begin{equation}
    \de \tau^2 = -N^2(t) \de t^2 + a^2(t) \de \ell^2\,,
\end{equation}  
where $\de \ell^2$ is a metric on the homogeneous and isotropic surfaces $\Sigma_t$, $N(t)$ is the lapse function on $\Sigma_t$, and the speed of light $c = 1$. When $N=1$, the time variable $t$ gives a proper-time parametrisation along geodesics. The spatial hypersurfaces $\Sigma_t$ can have, at most, constant intrinsic scalar curvature that is normalised to $k = \{ 0, \pm 1 \}$. The variable $a(t)$, which is only a function of $t$, is called the \emph{scale factor}. As discussed in \Sec\ref{sub:FLRW assumptions}, I will take $k=0$ --- although my results can easily be generalised to other values of $k$. When $k=0$, the scale factor is best thought of as the relative size of a co-moving patch of space-time. I will take the matter content of the model to be a collection of $n$ scalar fields $\varphi^i$, with $i = \{ 1, \hdots, n\}$ with potential $\mathcal U(\varphi^i)$. Finally, as motivated in \Sec\ref{sub:FLRW assumptions}, I will assume a positive cosmological constant $\Lambda > 0$.

Under these assumptions, the ADM-decomposition of the Einstein action leads, after integrating over a co-moving patch and taking the $G_N = 8\pi$, to
\begin{equation}
    S_\text{FLRW} = \int \de t a^3 \lf[ - \frac 3 N \lf( \frac {\dot a}a \rt)^2 + \frac 1 {2N} \dot \varphi^2 - N( \Lambda + \mathcal U(\varphi^i) )  \rt]\,.
\end{equation}
These are the conventional variables used in cosmology, where the lapse, $N$, is usually taken to be one. For our purposes, it will be convenient to define the new variables
\begin{align}
    V &= \tfrac 2 3 a^3 & \phi^i &= \sqrt{\tfrac 3 2} \varphi^i
\end{align}
where $V$ is just a rescaling of the volume of a co-moving patch and $\phi^i$ is just a rescaling of $\varphi^i$. If we define the new potential $U(\phi^i) = \tfrac 3 2 \mathcal U(\varphi^i)$ then the action becomes
\begin{equation}
    S_\text{FLRW} = \int \de t V \lf[ - \frac 1 {2N} \lf( \frac {\dot V }V \rt)^2 + \frac 1 {2N} \dot \phi^2 - N\lf( \frac {3\Lambda}2 + U(\phi^i) \rt) \rt]\,.
\end{equation}
A straightforward Legendre transform leads to the Hamiltonian in the form $H_\text{FLRW} = N \Ham_\text{FLRW}$, where
\begin{equation}\label{eq:frlw ham}
    \Ham_\text{FLRW} = V \lf( - \frac {h^2}2 + \frac{\pi^2}{2V^2} + \frac{3\Lambda}2 + U(\phi^i) \rt) \approx 0
\end{equation}
is the Hamiltonian constraint of the FRLW theory resulting from reparametrisation invariance, $h = - 3 \frac{\dot a}a = -3 H $ is minus three times the \emph{Hubble parameter} $H$ and $\pi_i$ are the momenta conjugate to $\phi^i$. In these variables, the non-zero Poisson brackets of the theory are
\begin{align}
    \pb V h &= 1 & \pb{\phi^i}{\pi_j} &= \delta^i_j\,.
\end{align}
The lapse, $N$, is treated here as a Lagrange multiplier for the Hamiltonian constraint and not a canonical variable. Also, I have used a shorthand notation where the square of $\phi^i$ and $\pi_i$ are taken using some constant matrix $K_{ij}$, for $\phi^2 = K_{ij} \phi^i \phi^j$ and its inverse $K^{ij}$ for $\pi^2 = K^{ij} \pi_i \pi_j$.

In these variables, Hamilton's equations can readily be computed:
\begin{align}\label{eq:ADM FLRW}
    \dot V &= \pb V {H_\text{FLRW}} = - N V h & \dot \phi^i &= \pb {\phi^i}{H_\text{FLRW}} = N \frac{\pi^i}{V} \\
    \dot h &= \pb h {H_\text{FLRW}} = N \frac{\pi^2}{V^2} & \dot \pi_i &= \pb {\pi_i}{H_\text{FLRW}} = - NV \diby{U}{\phi^i}\,,
\end{align}
where I have used the shorthand notation $\pi^i = K^{ij} \pi_i$. Note that these equations of motion are subject to the constraint \eqref{eq:frlw ham}.

\subsection{Removing dynamical similarity} 
\label{sub:removing_dynamical_similarity_flrw}

This theory has a symmetry under dynamical similarity. According to the general analysis of \Sec\ref{sub:generating_dynamical_similarity}, we first need to find that there exists a function $\gamma$ such that \eqref{eq:H condition} is satisfied for the FLRW Hamiltonian (we use $\gamma$ instead of $\phi$ to avoid notational ambiguity with the scalar fields of this section). This leads to following condition:
\begin{equation}\label{eq:H cond frlw}
    \dot \phi^i \pi_i + \dot V h + \pb{\gamma}{H_\text{FLRW}} = a H_\text{FLRW}\,,
\end{equation}
for some constant $a$. The ansatz
\begin{equation}
    \gamma = A V h + B \phi^i \pi_i
\end{equation}
solves \eqref{eq:H cond frlw} for $A = -1$, $B = 0$ and $a = 1$. Using these values, \eqref{eq:D def2} tells us that the vector
\begin{equation}\label{eq:D gen FRLW}
    D = \pi_i \diby{}{\pi_i} + V \diby{}{V}\,
\end{equation}
generates a dynamical similarity that takes solutions to solutions.\footnote{ Note that our answer is independent of the form of $U$ since $D$ rescales the action in the appropriate way to be a dynamical similarity \emph{without} having to transform $\phi^i$. } We see from $D$ that the volume, $V$, and momenta, $\pi_i$, carry weight $+1$ under dynamical similarity while $h$ and $\phi^i$ are invariant. Exponentiating the action of $D$ using a gauge parameter $c$, we see that $V \to c V$ and $\pi_i \to c \pi_i$ so that $\dot \phi$ is invariant according to \eqref{eq:ADM FLRW}. This exactly reproduces the dynamical similarity of \eqref{eq:DS in cosmo} used in \Sec\ref{sub:dynamical_similarity_in_the_universe}. Note that we have required that the lapse $N$ be invariant under dynamical similarity. This implies no loss of generality because any possible transformation under dynamical similarity can be absorbed into the time parameter $t$.

We can use the conformal weight of $V$ to define a convenient gauge-fixing of $D$ using the surface defined by a level surface of
\begin{equation}\label{eq:cosmo gf V}
    w = \log V\,,
\end{equation}
which satisfies $\Lie_D w = 1$. The contact form on this gauge-fixed surface is
\begin{equation}
    \eta = - e^{-w} \iota_D \omega = \de h - \tfrac{\pi_a}{V} \de \phi^a\,.
\end{equation}
This suggests that the variable definition
\begin{equation}
    v^\phi_i = \frac {\pi_i}V\,,
\end{equation}
which satisfies $v^\phi_i = \dot \phi_i$ on-shell (where $\phi_i = K_{ij} \phi^j$) and can therefore be thought of as the velocity of $\phi^i$ on velocity phase space, puts $\eta$ into Darboux form with the invariant coordinates $\{\phi^i, v^\phi_i, h\}$.

In terms of these coordinates and using the value $a=1$ computed above, the contact Hamiltonian is
\begin{equation}
    H_c^\text{FRLW} = e^{-a w} H_\text{FLRW} = N \lf( -\frac{h^2}2 + \frac{v^2_\phi}2 + \frac {3\Lambda}2 + U(\phi^i) \rt) \,,
\end{equation}
which is constrained to be equal to zero. Since the Reeb vector field is $R = \diby{}{h}$ in these coordinates, we find that the drag is
\begin{equation}\label{eq:drag FLRW}
    R(H_c) = \diby{H_c}{h} = - N h\,.
\end{equation}
We can use this to compute either the gauge-fixed equations of motion for the original variables or the contact equations of motion for the contact variables. The gauge-fixed equations, generated by $\Xinv = e^{(1-a)w} X - R(H_c)D = X + N h D$ are:
\begin{align}
    \dot V &= 0 & \dot \phi^i &= N \frac {\pi^i}{V} \\
    \dot h &= N \frac{\pi^2}{V^2} & \dot \pi_i &= -N V \diby{U}{\phi^i} + N h \pi_i\,.
\end{align}
We note that because $a = 1$, the time parameter, which labels surfaces of constant proper-time, is already gauge-invariant. We can use the constancy of $V$ in this gauge to re-write these equations in terms of the invariant Darboux coordinates defined above. A very short calculation gives
\begin{align}
    \dot \phi^i &= N v^i_\phi \label{eq:contact eoms frlw phi} \\
    \dot v^\phi_i &= -N \lf( \diby{U}{\phi^i} - h v^\phi_i \rt) \label{eq:contact eoms flrw v}\\
    \dot h &= N v_\phi^2 \label{eq:contact eoms frlw h} \,.
\end{align}
It is straightforward to verify that these are identical to the contact equations generated by $H_c^\text{FLRW}$ in these coordinates.

\subsection{A Janus--Attractor scenario for FLRW} 
\label{sub:a_janus_attractor_scenario_for_flrw}

I will now show how a JA-scenario, and therefore an AoT that solves the red-shift problem, arises in this model. Let us apply the assumptions of \Sec\ref{sub:FLRW assumptions}. The condition $k=0$ has already been built in to the construction by restriction the form of the action. Given the energy moment tensor $T^{\mu \nu}$ of some matter field, the weak energy condition (WEC) requires that $T_{\mu\nu} x^\mu x^\nu > 0$ for all time-like vector fields in space-time. For homogeneous scalar fields in homogeneous slicing in a proper-time parametrisation (i.e., with $N=1$), this condition is satisfied when
\begin{equation}\label{eq:WEC}
    \frac{v^2_\phi}{2} + U(\phi_i) \geq 0\,.
\end{equation}
The WEC combined with the Hamiltonian constraint \eqref{eq:frlw ham} tells us that
\begin{equation}
    \frac {h^2}2 = \frac{v^2_\phi}{2} + U(\phi_i) + \frac {3\Lambda}2 > 0\,
\end{equation}
when $\Lambda > 0$. This means that $h$ can not have any zeros along the dynamical solutions. In the original GR picture where $\dot V/V = -h$, our assumptions then guarantee that the model universes do not re-collapse. Instead, they must either grow or shrink indefinitely depending on the choice of AoT.

To see an AoT arising from a JA-scenario let us note that \eqref{eq:contact eoms frlw h} ensures that $\dot h > 0$ so that $h$ is monotonic. Since $h$ cannot have a zero, this means that there must be two temporal limits that bound the evolution of the system. The first is the limit where the bound of the WEC \eqref{eq:WEC} is saturated. In this case, $h \to \pm\sqrt{3\Lambda}$ so that the geometric degrees of freedom of the system have a fixed point in this limit. Because this limit is reached for \emph{any} initial conditions on the scalar field, the basin of attraction of this fixed point has non-zero measure. Let us consider the branch where $h$ is \emph{negative} corresponding to a \emph{positive} a Hubble parameter, and note that the opposite branch can be obtained by flipping the sign of the coordinate $t$. As it is well-known, the spacetime geometry of a system with constant positive Hubble parameter is de~Sitter. Thus, de~Sitter spacetime is an attractor of this theory.

The opposite temporal bound of the solutions occurs when $h \to \pm\infty$. Taking again the negative branch, we see that this is a singular point in the dynamics since the equations \eqref{eq:contact eoms frlw phi} - \eqref{eq:contact eoms frlw h} are not Lipschitz continuous at this point when $N=1$. This reflects the non-extendibility of solutions in a proper-time parametrisation. We may thus interpret this second temporal bound as a Big-Bang.

But because a proper-time parametrisation loses its physical significance in the reduced system, we are free to consider alternative time parametrisations of the theory. The discontinuities of the equations of motion \eqref{eq:contact eoms frlw phi} - \eqref{eq:contact eoms frlw h} occur because of the divergence of $h$ in this limit. A simple remedy is to use a lapse function $N = |h^{-n}|$ for $n\geq 2$. The equations of motion are then Lipschitz continuous provided $|h^{-n}| \diby{U}{\phi^i} \to 0$ when $h\to\pm\infty$. This puts relatively mild constraints on the potential.

We will investigate these conditions in more detail in the next section. For now, let us specialise to the case $n=2$ and note that the drag \eqref{eq:drag FLRW} takes the form $R(H_c) = -1/h$. In the branch we're interested in, this means that the drag is zero at the Big Bang where $h\to-\infty$ so that it is a Janus point for \emph{all} solutions in this parametrisation. Moreover, because of the dynamical properties of $h$, the drag will grow monotonically and reach a maximum at the de~Sitter attractor. The drag is thus a natural non-thermodynamic entropy function for the system.

In the time variable selected by the lapse choice $N = h^{-2}$, the Janus point occurs at $t=0$ and the attractor at $t\to\infty$. The existence of both a Janus point and an attractor for \emph{every} solution of our model leads to a JA-scenario and defines an AoT in the direction of increasing $t$ for all solutions. Along this AoT, the Hubble parameter decreases monotonically from $+\infty$ at the Janus point to $\sqrt{3\Lambda}$ near the attractor. This realisation of the JA-scenario therefore provides a solution to the red-shift problem: for \emph{all} solutions of the model, \emph{an observer near an attractor will see an arbitrarily large and monotonic Hubble parameter in the temporal direction pointing away from the attractor}. This solution is stronger than the solution to the smoothness problem in the $N$-body model in the sense that it assumes only $k=0$, $\Lambda>0$ and the WEC, and doesn't require the use of a measure. 


\subsection{Extendibility through the Big Bang} 
\label{sub:light_cone_coordinates}

In the previous section, I gave a solution to the red-shift problem an FLRW cosmology that realises a JA-scenario. This solution has the feature that the Janus point is the classical Big Bang. Since the Big Bang is usually considered to be a singular point of the dynamics, one might worry about the $T$-invariance of the theory since solutions on either side of the Janus point don't seem to be smoothly connected. In this section, I will address this worry by showing that there are families of smooth continuations of this model, motivated by the PESA, that pass through the Janus point in such a way that the $T$-symmetry of the theory is manifest. The existence of such solutions shows that the features of the JA-scenario described in \Sec\ref{ssub:the_janus_attractor_scenario} hold for this model, and therefore that my explanation for the AoT applies.

One might, however, wish to pursue a more radical hypothesis that the smooth continuations below provide more empirically accessible parametrisations of the cosmological model than the usual proper-time parametrisations, which break down at the Janus point. This was the point of view advocated in \cite{KOSLOWSKI2018339}. To pursue such a view, it is necessary to give a set of conceptual and formal tools for understanding the empirical significance of these parametrisations. The geometric description I give below is an attempt to achieve that. Thus, the demonstrations below serve a double purpose: first, they show how $T$-symmetry is retained in the formalism through continuations of the solutions that are smooth through the Janus point and, second, they develop a geometric framework that could be used to try to better understand the empirical significance of these continuations.

To begin the construction, note that the kinetic term, $K_\text{FRLW}$, of $S_\text{FRLW}$ can be written as
\begin{equation}
    K_\text{FRLW} = \frac 1 {2N'} \eta_{\mu\nu} \dot q^\mu \dot q^\nu\,,
\end{equation}
where $\eta_{\mu\nu} = (-1, V^2 \mathbbm{1})$ is the Minkowski metric in the Rindler coordinates $q^\mu = (V, \phi^i)$ and $N' = NV$. Under this representation, the ``position'' of a Rindler observer is denoted by $V$ and the Rindler ``time'' is denoted by $\phi$. The Rindler horizon is located at $V=0$, where the curvature has a $\delta$-function singularity. Because of this, geodesics intersecting this horizon are incomplete. We will see in a moment why this fact about Rindler spacetime is relevant to initial singularities in this model.

Towards this end, let us rewrite the action for FRLW in geometric terms on configurations space so that we may understand its solutions geometrically. We can eliminate the Lagrange multiplier $N'$ from the action by inserting the equations of motion for $N'$ back into the action.\footnote{ It's not hard to show that such an elimination gives an equivalent theory without $N'$. } This brings the action into Jacobi form
\begin{equation}
    S_\text{FRLW} = \int \de t \sqrt{g_{\mu\nu} \dot q^\mu \dot q^\nu  }\,,
\end{equation}
where the Jacobi metric
\begin{equation}
    g_{\mu\nu} = \frac{ 2 \eta_{\mu\nu} }{- \frac{3\Lambda}2 - U(\phi^i)}\,,
\end{equation}
is conformal to the Rindler metric and equal to it (up to a constant factor) when $U(\phi^i) = 0$. With the action in this form, solutions can be seen to be geodesics of $g_{\mu\nu}$ following the general analysis of \Sec\ref{sub:jacobi_theory}. Since conformal transformation preserve null geodesics, the light-cones associated with $g_{\mu\nu}$ are those of Rindler space --- including the null geodesic defined by the $V=0$ hyper-surface.

We will now see how the Big Bang singularity arises in this picture. First, we note that Rindler space is well-known to be geodesically incomplete at $V=0$ because of the $\delta$-function singularity there. However, we are working in configuration space, and not spacetime, so that geodesic completeness in configuration space is not necessarily equivalent to geodesic completeness in spacetime. Moreover, for $U \neq 0$, the geometry of configuration space is \emph{not} purely Rindler. We will see below, however, that relatively mild conditions on $U$ guarantee that the configuration-space geometry approaches that of Rindler fast enough near the horizon so that the near-horizon solutions behave like the geodesics of Rindler.

Using this fact, it follows that because the time coordinate of our model is adapted to homogeneous temporal slices in a proper-time parametrisation, the vanishing of $V$ in this time parameter does represent a genuine degeneracy of the spacetime metric. Thus, the horizon at $V=0$ does represent geodesic incompleteness both in configuration space and in spacetime.\footnote{ For the latter you also need that the $V=0$ surface is reached in finite proper time. This is guaranteed because $N=1$ leads to a proper-time parametrisation of solutions in spacetime and the horizon is reached in finite time in this parametrisation. } We can therefore speak interchangeably of the horizon at $V=0$ as a singular surface in configuration space and a singular spacelike hyper-surface in spacetime.

\subsubsection{Extendibility of the free theory}

I will soon give a scheme for repairing the incompleteness on configuration space. But to get inspiration, I will first consider the case where $U(\phi^i) = 0$ so that the scalar fields are free. In this case, the configuration space is truly Rindler, and the solutions will look like free particles travelling along the straight lines of Minkowski space until they reach the Rindler horizon.

Given this interpretation, an obvious method for extending the solutions presents itself: extend the solutions through the horizon by passing to Minkowski variables, where the Rindler horizon appears only as a coordinate singularity. The $\delta$-function singularity at the Rindler horizon occurs, from this perspective, because the map from Minkowski to Rindler is non-bijective and, in particular, is degenerate on the horizon. I will show how this happens below. The completion of solutions then involves continuing geodesics from one region to another in Minkowski space. Note that while such an extension is straightforward in the free theory, when $U\neq 0$ the smoothness of this extension will depend on the properties of $U$ on the Rindler horizon in a way we will soon make explicit.

A particularly helpful set of coordinates that will be useful for our purposes because they are well adapted to describing the dynamics on the horizon are the so-called \emph{light-cone coordinates} of Minkowski space. These coordinates will be labelled $\{u^+, u^-, \alpha^I\}$, where $I = \{ 1,\dots,n-1 \}$, and take the form
\begin{align}\label{eq:u coord def}
    u^\pm &= V e^{\pm \Phi} & \varphi^i(\alpha^I) &= \frac {\phi^i} \Phi\,,
\end{align}
where $\Phi = \sqrt{ K_{ij} \phi^i \phi^j }$ and the $\varphi^i(\alpha^I)$ obey the constraint $K_{ij} \varphi^i \varphi^j = 1$ so that $\alpha^I$ are a choice of coordinates on the unit sphere. We will leave the choice of $\alpha^I$ unspecified so that the precise relationship to the constrained quantities $\varphi^i$ is left implicit.\footnote{ As an example, for $n =  3$ a natural set of coordinates would be the spherical coordinates $\alpha = (\theta, \phi)$ so that $\varphi = (\sin \phi \cos \theta, \sin \phi \sin \theta, \cos \phi)$. } Extending the space-time from Rindler to Minkowski can be done by allowing the $u^\pm$ to lie in the full range of the real line, $\mathbbm R$, rather than the positive real line, $\mathbbm R_+$. This map is not bijective because the Rindler coordinate $V$ is required to be strictly positive. Moreover, the Jacobian of this transformation in degenerate on the Rindler horizon when $u^\pm = 0$ and $V \to  0$ and $\Phi \to \pm \infty$ where the level surfaces of $V$ and $\Phi$ coincide.

In the light-cone coordinates, the $u^\pm$ label temporal and radial directions along the light-cones while the $\alpha^I$ label the compact directions of the light-cones. In these coordinates, the Minkowski metric becomes
\begin{equation}\label{eq:lightcone metric}
    \eta_{\mu\nu} = \begin{pmatrix}
                        0 & -2 & \mathbf{0} \\
                        -2 & 0 & \mathbf{0} \\
                        \mathbf{0} & \mathbf{0} & u^+ u^- \Phi^2 \Omega_{IJ}(\alpha^I)
                    \end{pmatrix}\,,
\end{equation}
where $\Omega_{IJ}(\alpha^I)$ is the metric on the unit sphere and $\Phi = \frac 1 2 \log\lf( u^+/u^- \rt)$ should be treated as a function of $u^\pm$. We will show below how to extend solutions through the horizon at $u^\pm = 0$. But before doing this, let us remind the reader that while these extensions are perfectly natural on configuration space, they are not well-motivated in general relativity because they pass through a region of configuration space where the geodesics \emph{of the resulting spacetime} are incomplete. As we have discussed above, in the homogeneous slices we have chosen, the condition $V = 0$ implies that the proper-time along a \emph{spacetime} geodesic is not well-defined. On the other hand, because $V$ scales under dynamical similarity, the value of $V$ is not observable in cosmological models, and therefore geodesic extendibility in spacetime is no longer a good physical criterion for a singularity.

\subsubsection{Interacting scalar fields}\label{sub:interacting scalar fields}

Let us now consider the extendibility of solution for $U \neq 0$. A potential worry arises because $\Phi \to \pm \infty$ at the horizon so that the conformal factor of $g_{\mu\nu}$, which contains $U(\Phi)$ in its denominator, could shrink to zero. This might then result in genuine inextendibility of the geodesics of $g_{\mu\nu}$ on the horizon. It is thus useful to explicitly check the properties of these solutions and, in particular, how these solutions are modified when we quotient by dynamical similarity.

To study the general solutions, we note that the momenta $p_\mu = \diby{S_\text{FRLW}}{q^\mu}$ can be written in terms of the inverse of $g_{\mu\nu}$ as
\begin{equation}
    p_{\mu} = g_{\mu\nu} \dot q^\nu\,.
\end{equation}
The Hamiltonian constraint $\Ham'$ in the Jacobi normalization then tells us that the momenta are unit vectors
\begin{equation}
    \Ham' = g^{\mu\nu} p_\mu p_\nu - 1 = 0\,.
\end{equation}
In terms of the Minkowski metric, this can be re-normalized to give the following representation of the Hamiltonian:\footnote{ Recall that $N' = N v$, where $N$ is the lapse normalization leading to a proper-time foliation in space-time. }
\begin{equation}
    H_\text{FRLW} = N' \lf( \tfrac 1 2 \eta^{\mu\nu} p_\mu p_\nu + \frac {3\Lambda}2 + U(q^i) \rt)\,.
\end{equation}
Recall from the derivations in \Sec\ref{sub:jacobi_theory} that Hamilton's equations for Jacobi theory, \eqref{eq:jac_eom}, are equivalent to the geodesic equations with metric $g_{\mu\nu}$ on configurations space in an arbitrary time parametrisation determined by a choice of lapse. But to study the properties of the equations of motion near the horizon, it is useful to explicitly introduce coordinates.

In that vein, since the kinetic term $\eta^{\mu\nu} p_\mu p_\nu$ is a scalar, we can write it in any coordinates we choose as long as we suitably transform $\eta^{\mu\nu}$. In light-cone coordinates, the metric is given by \eqref{eq:lightcone metric} and is diagonal. Inverting it leads to:
\begin{equation}
    H_\text{FRLW} = N' \lf( -2 p_+ p_- + \frac{\Omega^{IJ} p_I p_J}{2 u^+ u^- \Phi^2}  + \frac {3\Lambda}2 + U(\Phi, \alpha_I) \rt)\,,
\end{equation}
where $\{p_\pm, p_I\}$ are the momenta conjugate to $\{ u^\pm, \alpha^I \}$. Hamilton's equations then give
\begin{align}
    \dot u^\pm &= -2 N' p_\mp & \dot p_\pm &= \mp \frac {N'} {2u^\pm} \lf[ \diby U \Phi + \frac{\Omega^{IJ} p_I p_J}{u^+ u^- \Phi^2} \lf( \pm 1 - \Phi^{-1} \rt) \rt] \\
    \dot \alpha^I &= \frac {N'} {u^+ u^- \Phi^2} \Omega^{IJ} p_J & \dot p_I &= - N' \Phi \diby{\varphi^i}{\alpha^I} \diby{U}{\phi^i}\,.
\end{align}
The horizon in these coordinates is the joint union of the two surfaces defined by $u^\pm = 0$.

To get an idea of the significance of the different values of the parameters in this theory, it is helpful to explicitly solve the solutions when $U(\phi^i) = 0$, where all the momenta are constants of motion. Additionally, setting $p_I = 0$ implies $\dot\alpha^I = 0$ for all $t$. The remaining equations of motion can be easily integrated, and say that the light-cone coordinates grow linearly in time when $N' = 1$:
\begin{equation}\label{eq:free u eoms}
    u^\pm(t) = u^\pm_0 - 2 P_\mp t\,,
\end{equation}
where $P_\pm$ are the constants of motion associated with $p_\pm$ and $u^\pm_0$ are initial conditions. The Hamiltonian constraint then reduces to
\begin{equation}
    P_+ P_- = \frac {3\Lambda}4\,,
\end{equation}
which implies that $P_+$ and $P_-$ must have the same sign when $\Lambda>0$ (and neither can be equal to zero).

In Rindler space, both light-cone coordinates must be greater than zero: $u^\pm >0$. This means that, for $\Lambda > 0$, there are two branches of solution: one that starts at the horizon $u^+ = 0$ and one that starts at $u^- = 0$. For studying solutions coming \emph{out} of the horizon at $u^\pm = 0$ for $t>0$, the minus sign in \eqref{eq:free u eoms} indicates that we should consider parameter values for the momenta where $P_\pm < 0$. The two branches of solutions are symmetric under interchange of $+$ and $-$ so that there is no loss of generality in looking at only one. We can choose to describe the branch starting at $u^+ = 0$ by fixing $u^-$ as a clock.  Using the time translation invariance of the solutions, we can set $u^-_0 = 0$. Then solving for $t$ in terms of $u^-$ gives
\begin{equation}
    u^+(u^-) = u^+_0 + \frac {P_-}{P_+} u^-\,.
\end{equation}
The solutions are therefore straight lines in the $u^\pm$-plane with slope $\frac {P_-}{P_+}$ (or $\frac {P_+}{P_-}$ for those intersecting $u^- = 0$). The Hamiltonian constraint then implies that this slope must be greater than zero. This means that the solutions are time-like in configuration space (for $\Lambda > 0$ they are space-like and for $\Lambda = 0$ they are null). This restriction implies that no geodesic starting on $u^+$ will ever intersect $u^-$ (and vice versa).

In the large $u^-$ limit, the initial condition $u^+_0$ will eventually become irrelevant as the term growing linearly in $u^-$ dominates. In this case, solutions approach straight lines that intersect the origin at $(u^+, u^-) = (0,0)$. These are lines of constant $\Phi$. In other words, all solutions that differ only by the value of $u^+_0$ will converge to the point $\Phi = $const when the time variable $u^- \to \infty$. Since this space of such solutions has non-zero measure, the level surfaces of $\Phi$ are attractors of this model by our definition. On these attractors, the momenta, and therefore the velocities, of the scalar fields is zero. According to \eqref{eq:contact eoms frlw h}, this means that the Hubble parameter reaches a constant value and the resulting space-time is approximately de~Sitter. Our explicit construction of the solution space for the free theory in these variables has thus reproduced the general result of \Sec\ref{sub:a_janus_attractor_scenario_for_flrw} that the theory has a late-time de~Sitter attractor.

Let us now analyse the behaviour of the general solutions for $U \neq 0$ in light-cone variables. We will be particularly interested the extendibility of solution at the Janus point, which is the Rindler horizon in this construction. It is therefore useful to determine the near-horizon behaviour of the different variables appearing in the equations of motion. The variable $\Phi$ is logarithmic in $u^\pm$ and diverges according to $\Phi \to \mp \infty$ on the horizon; but the logarithmic nature of this divergence is such that $u^+ u^- \Phi^2 \to 0$. If we restrict to potentials that have a finite power series expansion in $\phi^i$, then the derivatives of $U$ are polynomials of positive powers of $\phi^i$. In general, these can thus be assumed to diverge logarithmically as $u^\pm \to 0$. Finally, the constrained variables $\varphi^i$ can be represented as first-order trigonometric functions of the compact variables $\alpha^I$ so that their derivatives can be assumed to remain finite on the horizon.

Putting together the considerations above, for the lapse function $N' = 1$, the equations of motion are non-integrable on the horizon. The most problematic terms are in $\dot p_\pm$, which have divergences that are linear-times-logarithmic in $u^\pm$ on the horizon. The $\dot \alpha^I$ terms also diverge linearly and the $\dot p_I$ terms logarithmically. Only the logarithmic divergences of $\dot p_I$ are eliminated by going to a proper-time parametrisation in space-time, which corresponds to $N' = \sqrt{u^+ u^-}$. This recovers our earlier results about the inextendibility of geodesics in spacetime.

We can, however, obtain integrable equations on configuration space by choosing a lapse that multiplies the time derivatives by an additional factor of $V^4 = (u^+ u^-)^2$. If we accept the claim that $V$ is not observable, then such a clock redefinition may be thought to relate physically equivalent descriptions, and the non-integrability of the equations of motion an artefact of a bad choice of gauge. We can understand this better by investigating the contact equations obtained by quotienting by dynamical similarity in light-cone variables.

\subsubsection{Dynamical similarity and extendibility}
\label{sub:DS and extendibility}

To find a natural gauge choice for dynamical similarity in light-cone variables, we first note that the function $\gamma$ and the parameter $a$, which were used to compute $D$ in \eqref{eq:D gen FRLW}, were found assuming that $N$ had no transformation properties under dynamical similarity. To compute the gauge-fixed equations we should then return to our original choice of lapse in which $N' = VN$. We then note that the canonical transformation to light-cone variables gives
\begin{equation}\label{eq:upm to h}
    h = p^+ \sqrt{\frac{u^+}{u^-}} + p^- \sqrt{ \frac{u^-}{u^+} } 
\end{equation}
so that $\gamma = - \sum_\pm u^\pm p^\pm$. This gives
\begin{equation}
    D = p_I \diby{}{p_I} + \sum_\pm u^\pm \diby{}{u^\pm}\,.
\end{equation}
A useful gauge-fixing condition for describing the solutions that intersect $u^+$ takes advantage of the fact that the light-cone coordinates that have weight one under a dynamical similarity. A simple choice is
\begin{equation}
    w = \log u^-
\end{equation}
so that $\Lie_D w = 1$ (with a corresponding choice of $w = \log u^+$ for the branch intersecting $u^-$).

For this choice of gauge, the contact form is
\begin{equation}
    \eta = - \de p_- - \frac{u^+}{u^-} \de p^+ + \frac{p_I}{u^-}\de \alpha^I \equiv - \de p_- + \sigma \de p^+ + \pi_I \de \alpha^I\,,
\end{equation}
where $\sigma = - u^+/u^-$ and is \emph{negative} in the Rindler wedge and $\pi_I = p_I/u^-$ is the momentum conjugate to $\alpha^I$ in the contact space. Recalling our choice of lapse where $N' = VN$, the contact Hamiltonian is
\begin{equation}
    H_c = N \sqrt{-\sigma} \lf[ -2 p^+ p^- + \frac{\Omega^{IJ} \pi_I \pi_J}{2 (-\sigma) \Phi^2} + \frac{3\Lambda}2 + U(\sigma, \alpha^I) \rt] = 0\,.
\end{equation}
Because the contact form is in Darboux form in these variables, the contact equations are
\begin{align}
    \dot p_+ &= \frac {N}{2 \sqrt{-\sigma}} \lf[ \diby{U}{\Phi} - \frac{\Omega^{IJ} \pi_I \pi_J}{ \sigma \Phi^2} \lf( 1 - \Phi^{-1} \rt) \rt] \\
    \dot \sigma &= 2N \sqrt{-\sigma} \lf( p_- + \sigma p_+ \rt) \label{eq:sigma eom}\\
    \dot \alpha^I &= \frac {N}{\sqrt{-\sigma}\Phi^2} \Omega^{IJ} \pi_J \\
    \dot \pi_I &= - N \sqrt{-\sigma} \lf( \Phi \diby{\varphi^i}{\alpha^I} \diby{U}{\phi_i} - 2 p^+ \pi_I \rt) \\
    \dot p_- &= - \frac{N \sqrt{-\sigma}}2 \lf[ \diby{U}{\Phi} + \frac{\Omega^{IJ} \pi_I \pi_J}{\sigma \Phi^2}\lf( 1 + \Phi^{-1} \rt) \rt]
\end{align}

Given the near-horizon behaviour discussed at the end of \Sec\ref{sub:interacting scalar fields}, $\sigma \to 0$ linearly as $u^+ \to 0$. Assuming, as we also did in that section, that $U$ has a finite power series expansion in $\phi^i$, then the system will be Lipschitz continuous at the horizon if we take $N = \sigma^2$ --- although any power greater than $(-\sigma)^{3/2}$ will do.\footnote{Note that, for any \emph{specific} choice of $U$, one can find a choice of $N$ that would be fine-tuned to that potential and lead to the Janus point arriving at finite clock time.\label{ftn:continuation}} For any clock of this kind, the drag
\begin{equation}\label{eq:cosmo measure focusing}
    R(H_c) = - 2 N \sqrt{-\sigma} p^+ = - 2 (-\sigma)^{5/2} p^+ \,,
\end{equation}
will go to zero provided $p^+$ is chosen to be finite on the horizon. Since $\dot p^+ \to 0$ at the horizon, this is a reasonable requirement. With this clock choice, the horizon at $u^+ = 0$ is a Janus surface. We therefore see that the extension procedure that worked so naturally in the free theory can be applied to the interacting theory as well --- provided one makes suitable restriction on the clock choice and potential. We leave the task of investigating the physical significance of such clock choices to future work.

We conclude from this that, given certain reasonable restrictions on the potential, there are suitable choices of parametrisation such that, after removing $V$, the dynamics of the system is Lipschitz continuous through the Janus point. Such solutions start and end on attractors, illustrating the $T$-invariance of the theory and realising the picture of the JA-scenario described in \Sec\ref{ssub:the_janus_attractor_scenario}. Finally, the geometric picture introduced here on configuration space gives a simple way to derive the general features of the solutions, such as the existence of attractors and the continuation of the solution in the $U=0$ case. These insights may be useful in understanding the empirical significance of these continuations in a more ambitious programmed aimed at questioning the empirical status of proper-time in general relativity.

\chapter{Conclusions}
\label{ch:conclusions}

\ifchapcomp
    \tableofcontents
    \newpage
\else
    \cleardoublepage
\fi

\section{Problems solved} 
\label{sec:conc problems_solved}

In this thesis, I studied gauge symmetry and the Arrow of Time. I posed three problems: Belot's problem, the smoothness problem and the red-shift problem. The first involves giving a definition of gauge symmetry while the second and third are empirical puzzles concerning the AoT. I offered solutions to these problems in \chap\ref{ch:pesa} and \chap\ref{ch:new_aot}. I will now summarise my findings and assess the extent to which I have addressed each problem.

\subsection{Belot's Problem}
\label{sub:conc_belot_problem}

In this section, I will evaluate my proposed solution to Belot's Problem and highlight its unique features. For convenience, let me restate Belot's Problem as stated in \Sec\ref{sub:belot_s_problem}:\medskip

\noindent {\bfseries Belot's Problem:}

\begin{quote}
    To find formal conditions on the symmetries of a theory that are, under good interpretive practice, necessary and sufficient conditions for a gauge symmetry.
\end{quote}
The proposed solution, developed in \chap\ref{ch:pesa}, is to identify a gauge symmetry with a transformation on the DMPs of a theory that preserves the smallest algebraic structure --- what I called the \emph{observable algebra} --- of the DPMs such that the theory is empirically adequate and the equations of motion are well-posed and autonomous. Here, `empirical adequacy' was understood in terms of the DPMs being able to provide a faithful representation of the intended target system in the sense of DEKI \citep{frigg2020modelling}, which I will recap below.

According to the definition above, a solution to Belot's Problem should inform good interpretive practice. Let me first focus on this aspect of the proposal before getting to the formal conditions. In \Sec\ref{sec:problems_solved}, I illustrated how the PESA could be used to distinguish good and bad modelling practices using concrete examples. Let me consider how this was achieved on general grounds.

The representational framework I used was the DEKI account described in \Sec\ref{sub:DEKI account}. Two elements of that account play an important role in my definition of gauge symmetry: the \emph{context}, $C$, needed for the representations to exemplify features of a target system, and the \emph{key}, $K$, needed to impute features of a target. Together, these structures specify what features of the representations are relevant to a theory and what procedures, idealizations, data manipulations, etc, are required to use a theory's models to impute actual features of the target.

What was seen is the examples of \Sec\ref{sec:problems_solved} is that the context and key of a theory have a strong effect on whether a symmetry of the theory is thought of as a gauge symmetry or not. One insight provided by my analysis is the recognition that the structures $C$ and $K$ are relevant to the definition of a gauge symmetry not because of some special aspects of gauge symmetries but because they are essential for giving good representations. The interpretative problems that arise in defining gauge symmetries are, therefore, no different from general problems that arise in building good models.

With this insight, it is possible to focus on the formal properties of gauge symmetries that distinguish them from other kinds of symmetries. What was found to be different about gauge symmetries was the difficulty of determining which constitutive structures of a model are essential (or not) for giving a faithful representation of the target. This was found to be particularly problematic given the fact that it is strongly advisable to do so. At a conceptual level, having a way to identify surplus structure, even implicitly, is necessary for epistemological clarity. And at a pragmatic level, we have seen that there are excellent reasons for having such clarity.

First, if the surplus structure of theory is causing mathematical difficulties --- such as divergences or discontinuities --- in building the theory's models, then one should find a way to remove such structure in order to get a mathematically workable theory. This is because the surplus structure is, by definition, not essential for describing the target, and therefore any problems it may cause should not be regarded as empirical. We saw an example of this in \Sec\ref{sub:light_cone_coordinates}, where the discontinuity of the cosmological equations of motion was found to be removable by applying the Gauge Principle to the theory's dynamical similarity. This example shows how identifying the appropriate surplus structure of a theory can dramatically alter the status of the conceptual problems of a theory.

Second, explanatory (or other) inferences within a theory may depend strongly on whether a particular representational structure is interpreted as being surplus or not. For example, a measure that counts as distinct states that are related by gauge transformations is giving extra weight to states that have a large gauge orbit under that measure. In \Sec\ref{sub:the_smoothness_problem_N_body}, we saw that this extra counting was the difference between expecting smooth states to be highly atypical (when dynamical similarity was treated as empirically relevant) and to be generic (when dynamical similarity was treated as a gauge symmetry).

These considerations highlight the formal conditions I used to define a gauge symmetry as well as the normative rules I proposed for implementing these conditions in concrete theories. My approach was inspired by that of \cite{dirac2001lectures}, where predictions about observable structure are required to be determinable from the laws and, conversely, predictions about surplus structure are required to be arbitrary. This means that equations of motion should be well-defined in terms of observable structure but underdetermined by surplus structure. The latter requirement is not logically necessary for an empirically adequate theory. Instead, it is a way to avoid the kind of faulty reasoning alluded to in the two preceding paragraphs. These formal conditions and norms combined with the interpretive framework used in this thesis thus provide a good solution to Belot's Problem and a motivation for the Gauge Principle.

\subsection{The generalist--particularist impasse}
\label{sub:conc_gen-part_impasse}

In \Sec\ref{sec:Price_taxonomy}, I presented a taxonomy due to Price that divided approaches to explaining the AoT into \emph{generalist}; i.e., approaches that postulate a general time-asymmetric law; and \emph{particularist}; i.e., approaches that postulate a particular fact --- normally a Past Hypothesis. I then proceeded, in \Sec\ref{sec:the_dilemma}, to give a series of worrying objections to both approaches. This led to a kind of impasse: how can we find an explanation for the AoT that is free of all such objections?

Let me now recall these objections and show that the JA-scenario, which I proposed as a general mechanism for explaining an AoT, can evade the impasse by evading all objections. Ultimately, the JA-scenario does so because it leads to explanations of the AoT that are neither generalist nor particularist, laying bare the limits of Price's taxonomy.

What we have seen is that a theory can have representational structures, such as measures, that are time-dependent even though the global models of the theory are time-reversal invariant. These time-dependent structures can introduce an AoT for certain classes of local observers, leading to a perceived AoT for those observers. In this way, a theory can have local time-asymmetry but still have global time-reversal invariance. In particular, for arguments relying on a measure, the possibility of time-dependence means that one must specify \emph{when} a state occurs in a history in order to assess its typicality, changing the overall explanatory structure of the theory. Let us take a moment now to see how this new insight plays out in each of the objections raised in \Sec\ref{sec:the_dilemma}.

\subsubsection{Objections to generalism}\medskip

\paragraph{Objection from redundancy} 

This objection, put forth in \cite{price2002boltzmann}, argued that particularist approaches are more economical than generalist approaches because generalist approaches must explain \emph{both} a time-asymmetric law \emph{and} the apparent low-entropy state of the early Universe. This is particularly difficult given the fact that the time-asymmetric law must reproduce, to a good approximation, our best microscopic laws of physics, which are time-symmetric. Here, the relevant notion of time-asymmetry is understood to be time-reversal invariance.

In the JA-scenario realised in the $N$-body model presented in \Sec\ref{sub:model_definition_Nbody}, a low-entropy (i.e., smooth) initial state was found to be \emph{typical} by a large class of natural measures because of an argument based on the Central Limit Theorem. This style of argument is available because the class of natural measures invariant under dynamical similarity is time-dependent. Importantly, we found that conditioning on a Janus point significantly changes the usual time-independent counting arguments by excluding the many clumped states found near the theory's attractors. This means that a JA-scenario can take advantage of the explanatory power provided by a time-dependent measure to explain an AoT without having to introduce a time-reversal invariant law. The explanation for the existence of attractors, Janus points and time-dependent measures is provided by the symmetry argument based on the PESA. Thus, invariance under dynamical similarity can provide an economical explanation of the smooth initial state, avoiding the objection from redundancy.

\paragraph{Objection from lack of independent motivation}

This objection is that the empirical success of our best fundamental theories belies any reasons, beyond explaining the AoT, for introducing a time-asymmetric law. While this objection was raised in response to the introduction of a non-time-reversal invariant law, and therefore does not strictly apply to our proposal, it could be restated as an objection to time-dependent measures.

To respond to this reframed objection, the argument presented in \Sec\ref{sub:dynamical_similarity_in_the_universe} is essential. There, it was shown that the scale factor of the Universe was not necessary for formulating an empirically adequate cosmological theory and, moreover, that the measure regarded by cosmologists as physical was both time-dependent and invariant under dynamical similarity. Thus, the empirical success of our best theories is not in conflict with treating dynamical similarity as a gauge symmetry. Furthermore, the PESA has given us independent motivation for treating dynamical similarity as a gauge symmetry in cosmology. Doing so led, in \Sec\ref{sec:cosmological_models}, to a theory of cosmology in which an AoT appears generically in the behaviour of the Hubble parameter. Thus, my explanation for the AoT evades the objection form lack of independent motivation by introducing a PESA-based symmetry argument consistent with best practices in cosmology.

\paragraph{Objection from historical progress}

This objection considers the historical evolution of our thinking about directions in space and argues for thinking analogously about directions in time. Our fundamental theories give no a priori privilege to any particular direction in space. Rather, privileged directions emerge approximately from contingencies such as the presence of a large massive body like the Earth. Similarly, goes the reasoning, something analogous should be hold of directions in time, favouring particularist approaches over generalist ones.

Nothing, however, in this reasoning singles out a PH as the analogous condition for a time direction: attractors and Janus points can play an equivalent role in such an argument. Indeed, because the AoT seen by an observer in a JA-scenario is local and contingent, it is arguably more analogous to the spatial case than a PH, which is a global condition on all DPMs of the theory. The JA-scenario thus evades the objection from historical progress by treating time directions analogously to spatial directions.

\subsubsection{Objections to particularism}\medskip \label{sec:conc:PH objections}

\paragraph{Objections from mathematical and conceptual ambiguity}

These objections stem from difficulties is attributing a meaningful and mathematically precise notion of typicality to the states of the Universe. Without such a notion, it is impossible to run the Boltzmann-style reasoning underlying particularist approaches.

In a JA-scenario, the AoT arises from the presence of attractors and Janus surfaces, and it not fundamentally linked to Boltzmann entropy. Attractors can be given precise mathematical definitions in terms of the $\omega$-limit sets of the dynamical flow as outlined in \Sec\ref{sub:dynamical_attractors}. Similarly, Janus surfaces can be defined in terms of the level surfaces of the drag, which is a smooth function on contact space, in $N$-body and cosmological systems as demonstrated in \Sec\ref{sub:a_janus_attractor_scenario_of_the_smoothness_problem} and \Sec\ref{sub:a_janus_attractor_scenario_for_flrw}. The notion of entropy we do recover is a non-equilibrium entropy \emph{function} that is smooth and well-defined. We thus evade the mathematical difficulties, discussed in \Sec\ref{sub:case_c_measure_ambiguities}, that are normally encountered in those models. Shifting the focus from entropy to attractors and Janus points thus provides an improvement in the level of mathematical rigour with which one can define an AoT. This doesn't mean, however, that it is not possible to recover the usual notions of entropy when appropriate, as I explain below.

Further mathematical and conceptual advantages can be gained by removing the global scale variable by constructing a gauge theory of dynamical similarity. In the $N$-body model, for instance, the lack of a strict equilibrium state arises because the overall scale of the system grows monotonically. Removing this scale and parametrizing the dynamical flow by its momentum leads to a true equilibrium, in the form of an attractor, as shown in \Sec\ref{sub:a_janus_attractor_scenario_of_the_smoothness_problem}. This is in part because the scale contains the only non-compact part of the state space.

I should note, however, that explanations relying on particular choices of measure, such as those concerning initial smoothness, do involve certain amount of convention. This will be addressed in \Sec\ref{sub:conc_smooth_redshit}.

\paragraph{Objections from the breakdown of thermodynamic assumptions}

These objections question whether the thermodynamic assumptions underpinning Boltzmannian explanations of time-asymmetry in isolated systems of free-gases apply to self-gravitating systems in the Universe.

In this regard, it is important to note that the explicit models I used to solve the smoothness and red-shift problems model equilibration towards attractors using monotonic dynamical quantities within those systems. It is precisely the process of equilibration that is usually in question when applying Boltzmannian reasoning in the Universe. But unlike the Boltzmann entropy, the quantities in our models can be proven to be monotonic by rigorous theorems. As I have emphasised, the resulting functions have the behaviour one would expect of an entropy in a non-equilibrium thermodynamic system.\footnote{ See \Sec\ref{ssub:the_janus_attractor_scenario} for details. } This raises the question of whether the JA-scenario can be seen as a more general model for equilibration in thermodynamic systems.

The recovery of local (in space) thermodynamic AoTs from these global features of the theory is a result, in my approach, of the less controversial application of thermodynamic concepts to stars and other stellar systems such as galaxies and galaxy clusters. For example, using my explanation of the smoothness of early states, one can assume the approximate homogeneity of early matter and justify the use of periodic boundary conditions when defining the entropy of matter in a large co-moving patch of the early Universe. Thus, my approach evades the objections from the breakdown of thermodynamic assumptions by replacing the notion of Boltzmann entropy with more rigorously defined functions precisely when use of the Boltzmann entropy is questionable.

\paragraph{Objections from lack of explanatory force}

Objections of this kind question whether a PH itself can really be explanatory unless it can be given an independent motivation. Since no PH is given in a JA-scenario, the criticism does not strictly apply. Instead, one could question the explanatory force of a JA-scenario along a similar vein by demanding an explanation for the presence of attractors and Janus points. The most important factor leading to a JA-scenario in our models is the gauge status of dynamical similarity. The strength of the explanation therefore rests, primarily, on the strength of the argument for treating dynamical similarity as a gauge symmetry. The justification for this can be found in the PESA.

Further assumptions, however, are required to obtain a JA-scenario in the two models I have considered. Consider first the cosmological models. There, attractors are guaranteed to exist by the various modelling assumptions put forth in \Sec\ref{sub:FLRW assumptions}. These include the presence of a positive cosmological constant and matter obeying the Weak Energy Condition. As I argued in \Sec\ref{sub:FLRW assumptions}, these assumptions are well-motivated physically. Moreover, they have no obvious dependence the existence of an AoT. The claim:
\begin{quote}
    dynamical similarity, a positive cosmological constant and the Weak Energy Condition \emph{explain} the AoT apparent from the behaviour of the red-shift,
\end{quote}
therefore has considerable explanatory force.

For the $N$-body model, I proved in \Sec\ref{sec:newtonian_gravitation_models} that smooth distributions on the Janus surface and highly clumped attractors are guaranteed to exist for a generic set of solutions given the energy of the system is non-negative. Since the non-negativity of the energy is meant to mimic the behaviour of a positive cosmological constant, the claim:
\begin{quote}
    dynamical similarity and a non-negative energy explains the smooth early state of many self-gravitating particles
\end{quote}
also has considerable explanatory force. Thus, unlike a PH, the modelling assumptions leading to the explanation of the relevant aspects of the AOT in these models can be given independent motivation.

Finally, the JA-scenario is well-equipped to address Price's `temporal-double-standard' objection against the PH. In this objection, Price is questioning whether it is legitimate to hypothesise a special condition in the past when it would be unacceptable to hypothesise a similar special condition in the future. In the JA-scenario, the $T$-symmetry of the solutions about the Janus point ensures that no such double standard applies. DPMs can be bounded by attractors at both ends and the AoT is relative to a Janus point and a particular observer near one of the attractor. Thus, for every observer that is some distance from an attractor and experiencing a particular temporal arrow, there is a corresponding observer the same distance from the other attractor experiencing the opposite arrow. `Past' and `future' are, therefore, relative concepts in this picture. Ultimately, there is no temporal double standard because $T$-symmetry is retained throughout at the level of the definition of the DPMs.

\paragraph{The dilemma for the Past Hypothesis}

It should be clear the JA-scenario avoids the dilemma afflicting particularist explanations of the AoT that was presented in \Sec\ref{sec:symmetries_and_measure_ambiguities}. The second horn; i.e., introducing a distinction without a difference by using a measure that is not invariant under dynamical similarity; is explicitly avoided by applying the Gauge Principle to dynamical similarity. This should not be surprising as it was the second horn that partly motivated the proposal in the first place. While evasion of the second horn is sufficient to escape the dilemma, we must also check that a slight reformation of the first horn; i.e., the worry that the freedom to choose a measure on contact space could lead to a loss of explanatory power; does not present a problem for the new proposal. I will address that worry in \Sec\ref{ssub:jusitying_convension}.

\subsection{The smoothness and red-shift problems}
\label{sub:conc_smooth_redshit}

I have already given a justification for the modelling assumptions introduced in \Sec\ref{sub:the_models} and shown how these assumptions lead to a powerful explanation of the AoT. I will now assess the extent to which I have been successful in modelling the particular empirical phenomena I set out to explain. This means assessing my solutions to the smoothness and red-shift problems.

\subsubsection{Smoothness problem}\medskip

Recall from \Sec\ref{sec:explanatory_target} that the smoothness problem was the problem of explaining the relative smoothness of the early state of the Universe. It thus might seem strange to seek a solution in the form of an $N$-body model on a fixed Newtonian spacetime. It's important, however, to note that the intuitions that lead to the smoothness problem originate from standard considerations in the thermodynamics of gravitational $N$-body systems. In particular, as was described in \Sec\ref{sub:case_b_gravitational_considerations},\footnote{ See that section for references. } it is well-known that smooth states are very low entropy in $N$-body systems because of the negative heat capacity of self-gravitating Newtonian systems. Moreover, clumped states can have arbitrarily high entropy because of the divergences in the $1/r$ potential. These observations are the primary motivation for the claim that the early smooth state is highly atypical.

It is thus a significant, and rather unexpected, result of \Sec\ref{ssub:early_smoothness_measure} that smooth states on the Janus surface were found to be \emph{typical} when using the natural, time-dependent measures on the contact space. Note that this result required a short distance cut-off on the $1/r$ potentials, which otherwise lead to fat-tailed distributions in the $C$-function. Note also that the states on the Janus surface are not perfectly smooth in the sense that the distribution of the $C$-function only approaches that of a homogeneous distribution as $N$ gets large. However, applying a cut-off on the $1/r$ potential is physically well-motivated and common practice in $N$-body mechanics because realistic celestial bodies and galaxies are not point particles, and this leads to deviations from the $1/r$ potential in practice. Moreover, the distribution of matter in the early Universe is \emph{not} completely homogeneous --- a fact that forms the basis of modern observational cosmology. Thus, deviations from homogeneity could be interesting empirically in more realistic models of the Universe.

I cannot, however, say that I have given a fully satisfactory solution to the smoothness problem. For that, I would need to show that the smoothness result continues to hold for early states in more realistic models of the Universe. This will involve placing an $N$-body system on an expanding background and, eventually, treating a full general relativistic system.

Nevertheless, what my model shows is that early smooth states can, indeed, be typical provided one uses a natural time-dependent measure for assessing typicality rather than the time-independent Liouville measure used in standard treatments. The move to a time-dependent measure means that one must condition on the state in order to make inferences about how typical they are. Thus, it is a valid inference to say that observers near an attractor will find smooth states to be typical near a Janus point. Conversely, it is not valid to say that smooth states are typical in general since there are many clumped states near attractors. The key insight is that, while clumped states still dominate the state space overall, these are exceeding likely to be found near `late'-time attractors and not `early'-time Janus points.


\subsubsection{Red-shift problem}\label{ssub:conc_red_shift}\medskip

The red-shift problem, as stated in \Sec\ref{sec:explanatory_target}, is the problem of explaining why the Hubble parameter, which measures the relative rate of red-shifting, was so large and monotonically decreasing in the past. The solution presented in \Sec\ref{sub:a_janus_attractor_scenario_for_flrw} shows that, for homogeneous and isotropic cosmologies with an arbitrary number of scalar fields, the Hubble parameter is divergently large at the Big Bang and shrinks monotonically to a de~Sitter attractor provided $k=0$, $\Lambda>0$ and the Weak Energy Condition is satisfied. Gauge-fixing dynamical similarity in this system reveals the attractor structure and suggests a family of time-dependent measures consistent with those used by cosmologists. This solution therefore explains the behaviour of the red-shift under the given assumptions.

Let us now evaluate the restrictiveness of these assumptions and the prospects of generalising them to more realistic models. First, it is important to note that we only have reliable evidence of the behaviour of the Hubble parameter until the average temperature of the Universe was not much higher than the energy levels probed in particle physics experiments. Thus, it is not strictly necessary to solve the red-shift problem in all the epochs leading up to the Big Bang. Perhaps a safe strategy would be to consider only the epochs up to the onset of inflation.\footnote{ Unless we want to consider alternatives to inflation. } Moreover, we must also be able to solve the smoothness problem up to the appropriate epoch in order to justify the assumption of homogeneity and isotropy.

I will therefore assume that we have at hand a good solution to the smoothness problem and lift the requirement of strict monotonicity of the Hubble parameter before the onset of inflation. In doing this, I can consequently lift the requirement of $k=0$ and the Weak Energy Condition (although the former is perhaps preferable to the latter) at the cost of introducing mild constraints on the initial momenta of the scalar fields.\footnote{ I will leave the explicit computation of these constraints to future work. } Note, however, that our knowledge of the empirical behaviour of the Hubble parameter near inflation is limited such that any such constraints do not pose serious problems for the empirical problem I have set out to solve.

While I have emphasised the role of removing the scale factor in defining a smooth evolution through the Big Bang as an independent motivation for treating dynamical similarity as a gauge symmetry, not much about my solution to the red-shift problem required identifying the Janus point with the Big Bang itself. As long as my assumptions guarantee monotonicity of the Hubble parameter up to the empirically accessible epochs, my solution to the red-shift problem remains essentially intact.

A more important feature that a general model must have for describing empirical observations is a de~Sitter attractor. While there is some freedom in specifying what one means by the early state of the Universe, the current state is very well-known to be approaching that of de~Sitter. Fortunately, there exists a growing literature on the so-called \emph{Cosmic No-Hair Conjecture} suggesting that de~Sitter attractors arise generically in solutions to the Einstein equations with only very mild restrictions on the matter content.\footnote{ For an early formulation of the conjecture for homogeneous cosmologies, see \cite{wald1983asymptotic}. For a modern statement under general conditions, see \cite{andreasson2016proof}. For an introduction aimed at philosophers of physics, see \cite[Chap VII, \S 7]{belot2023accelerating}. For a more detailed analysis of the philosophical points, see \cite{doboszewski2019interpreting}. } This suggests that the primary assumption underpinning a more general solution to the red-shift problem is simply the assumption of a positive cosmological constant. This then leads to a relatively robust explanation of the cosmological AoT in terms of a JA-scenario.

\subsubsection{Justifying convention}\medskip
\label{ssub:jusitying_convension}

Any explanation that makes use of a JA-scenario must justify the conventions used to represent the dynamics on the contact space. This is because, as we saw in \Sec\ref{ssub:counting_solutions_in_a_contact_theory}, the contact flow equations are invariant under a transformation (\eqn\ref{eq:contact rep_sym}) that simultaneously rescales the contact form and Hamiltonian. Because the contact form fixes the Reeb vector (by finding the unique solution to \eqn\ref{eq:CVF def}), we can understand the freedom due to this transformation as the freedom to redefine the drag, which was defined to be proportional to the Reeb flow of the contact Hamiltonian.

Let us focus now on the behaviour of the drag and see how the choice of contact form affects this. This will allow us to assess the theoretical virtues of different choices and compare them with those of the Liouville measure of symplectic theories, where the drag is zero. Let us recall these virtues from \Sec\ref{sub:time_laws_and_conventions}:
\begin{itemize}
    \item \emph{simplicity:} the Liouville measure takes a dramatically simple translation-invariant diagonal form when written in Darboux coordinates on phase space,
    \item \emph{universality:} it is preserved by \emph{any} choice of Hamiltonian,
    \item \emph{uniqueness:} it is the unique measure that is universal in the sense above,
    \item \emph{utility:} it has been hugely successful in the history of physics in terms of empirical adequacy, novel prediction, and explanatory power.
\end{itemize}

Let me first rule out a naive choice of contact form, analogous to the Liouville measure in symplectic systems, for which the drag is zero. To write down such a contact form, one needs an integral of motion.\footnote{ This was proved using \eqn\ref{eq:measure ambiguity}. } In the cosmological case, however, integrals of motion are usually complicated non-local functions of the state space that depend explicitly on the Hamiltonian for the reasons discussed in \Sec\ref{sub:time_laws_and_conventions}. This is essentially because there are no interesting isolated degrees of freedom. Thus, in the cosmological case, the contact form is neither simple --- because it is a complicated non-local function on contact space --- universal --- because it depends on the Hamiltonian --- nor unique --- because it is not clear what (complicated) integral of motion should be used. Moreover, contact forms of this kind do not appear to be very useful as, to my knowledge, they do not appear in broader cosmological applications. I thus conclude that contact forms for which the drag is zero do not have \emph{any} of the theoretical virtues possessed by their symplectic analogue.

\paragraph{\(N\)-body models}

I now turn my attention to the contact forms used in the models considered in this thesis. In general, the choice of drag depends on the choice of gauge fixing for the dynamical similarity as a consequence of \eqn\ref{eq:measure ambiguity}. In the $N$-body case, the dynamical similarity was fixed by fixing the value of the dilatational momentum, $I$, as in \eqn\ref{eq:N-body gf}. This sets the spatial size of the $N$-body system and is a simple scalar one can form from the configuration coordinates and metric. This gauge choice leads to simple equations of motion that can easily reproduce the equations of the reduced system used in \cite{Barbour:2014bga}. Thus, aside from being conceptually simple, in that the dynamical similarity is parametrized entirely by the spatial scale, it is also a choice that leads to a simple mathematical description.

On top of being simple in the senses given above, my chosen gauge-fixing is a member of a large universality class of gauge-fixings that lead to measures with smooth distributions of matter at the Janus point. The existence of this universality class is guaranteed by a general argument, given in \Sec\ref{sub:robust smoothness}, that is based on the Central Limit Theorem. The combinatorics of the Central Limit Theorem guarantee that the measure would have to grow exponentially with the $C$-function (defined by \eqn\ref{eq:C-fun}) in order to not be smooth at the Janus point. While this behaviour is not as universal as the conservation of the Liouville measure, it still ensures a good degree of robustness in the smoothness of distributions near a Janus point.

One potential drawback of working with reparametrisation invariant contact systems is that the choice of gauge, and therefore the choice of drag, is not unique. It might be that some new principle could be found that would lead to a unique choice of drag, but I have not yet found a principle worthy of investigation here. Note, however, that the lack of uniqueness does not impact my arguments about the AoT: the features of the phenomena that I am interested in explaining can be modelled by taking asymptotic limits in, for example, the number of particles (for smoothness at early states) and time (for the existence of attractors), and the behaviour of the system in these limits only depends on the choice of universality class for the drag.

Regarding the utility of the chosen gauge fixing, one simple observation is that my choice lead to a monotonic drag, and therefore introduces an AoT. Thus, my gauge-fixing is useful because it provides a resolution of the smoothness problem and an explanation of the AoT. This is not helpful in justifying the explanation of the AoT itself but can be used as part of a more general argument for justifying the choice of drag.

I conclude that, in the $N$-body model, there exists a universality class for the choice of drag where the drag shares many of the theoretical virtues of the Liouville measure in a symplectic theory. Most notably, we lose the ability to single out unique, natural choice of drag, as it was possible in the symplectic theory. Instead, we gain the ability to explain initial smoothness and the AoT.

\paragraph{Cosmological models}

As a last consideration, let me assess the theoretical virtues of the contact form used in the cosmological models. To be precise, consider gauge-fixing the dynamical similarity by fixing the value of spatial volume in a homogeneous slice as was done throughout \Sec\ref{sec:cosmological_models}. I showed in \Sec\ref{sub:removing_dynamical_similarity_flrw} that this gauge fixing leads to a remarkably simple parametrisation of the dynamics that immediately reproduces the Friedman and Klein--Gordon equations in their standard form. The chosen gauge fixing is also conceptually simple in that it attributes the action of dynamical similarity directly with changes of spatial size --- as was done in the gauge fixing used for the $N$-body model.

In addition to being simple, my choice of drag is also universal in the sense that any smooth monotonic function of the drag will lead to the same attractor structure. Unlike the smoothness problem, which can only be stated in terms of a measure, the red-shift problem concerns the monotonicity of the Hubble parameter, and this can be achieved with an entire functional family of drags. The one aspect of the red-shift problem that is more quantitative is the largeness of the Hubble parameter in the past. But as was discussed in \Sec\ref{ssub:conc_red_shift}, the relative insensitivity of my solution to the nature of the Janus point illustrates the general robustness of the approach. Thus, the lack of empirical knowledge about early-universe physics leaves considerable freedom in the choice of drag. I hope that future investigations will enable a fixing of this freedom.

As with the gauge fixings used in the $N$-body problem, the gauge-fixings used in cosmology are not unique. While this does not affect the conclusions about the AoT, it does leave open whether some new principle should be imposed to guide future investigations.

Finally, the gauge fixing used in cosmology is useful. First, as was shown in \Sec\ref{sub:dynamical_similarity_in_the_universe}, the measure proposed by \cite{Hawking:1987bi}, and the one used extensively by cosmologists, for counting solutions in cosmology is invariant under dynamical similarity and was derived under the same gauge fixing condition that I used for the spatial volume. Second, if we further fix the lapse function to be proportional to an inverse power of the Hubble parameter, as was done in \Sec\ref{sub:a_janus_attractor_scenario_for_flrw}, then the dynamics become Lipschitz continuous at the Big Bang. This gives a solution to what many consider to be a serious problem for general relativity as applied to our Universe: the breakdown of the Einstein equations at the initial singularity.

I conclude from this that, as in the $N$-body model, the choice of drag shares many of the theoretical virtues of the Liouville measure. In the cosmological case, the universality is slightly stronger than in the $N$-body case owing to the fact that the red-shift problem, in contrast to the smoothness problem, is not obviously a counting problem. Finally, there is even the potential to address worries related to the nature of the Big Bang.

\section{Prospectus}
\label{sec:prospectus}

In this section, I will discuss some of the many open questions raised in this thesis that could be pursued in future work. This list is non-exhaustive and, by its nature, speculative. I hope, however, that it can give an indication of the generality of my analysis and how it might be usefully applied.

\subsection{Applications of the PESA}

In \chap\ref{ch:pesa}, I defined the PESA and, in \Sec\ref{sec:problems_solved}, applied it to some simple examples. The examples were chosen to reflect simple situations that illustrate puzzles associated with symmetry studied in the literature. Much more work remains, however, to consider more extensive and elaborate examples. One such example involves applying the PESA to systems that exhibit the Aharonov--Bohm effect discussed in \Sec\ref{ssub:reduction}. While we've restricted attention to classical systems in this thesis, the PESA applies to very general dynamical systems, including those with quantum evolution. The Aharonov--Bohm effect would therefore consist of a useful quantum mechanical example. In general, it would be interesting to apply the PESA to cases where the observable algebra might be expected to include global degrees of freedom associated with the fibre-bundle structure of a gauge theory. This would allow for a comparison between reduction and retention of explicit gauge invariance, often called `sophistication' in the recent literature \citep{dewar2019sophistication,read_martins_sophistry}. The PESA could be used to distinguish contexts where either reduction or sophistication is preferable.

The PESA could also be applied to gauge theories that impose spatial boundary conditions on the gauge degrees of freedom. This would include, for instance, the examples involving asymptotically flat general relativity discussed in \Sec\ref{ssec:intro symmetry and its problems} of the introduction. In particular, a proposal made by \cite{Hawking:2016sgy} suggests that classical black holes can be characterised by an infinite head of supertranslation `hair,' which can be distinguished using BMS charges at infinity. These extra charges, might represent an extension of the observable algebra according to the PESA, which may be relevant to the black hole information paradox.

\subsection{ Equilibration in thermodynamic systems}

The process of equilibration in thermodynamic systems is, in general, a difficult problem to model. Several approaches exist that are successful at describing different kinds of systems, but there is no universally accepted paradigm for addressing all possible situations.\footnote{ While the Boltzmannian approach described in \Sec\ref{sec:preliminaries} works well for ideal gasses, it is limited by many of the criticisms discussed in \Sec\ref{sec:deconstructing_the_argument}. Other approaches to equilibration involve variants of the approach developed in \cite{ottinger2005beyond}, which are widely used in polymer physics, or the geometric approach using contact dynamics that was developed in \cite{bravetti2019contact}. } It is interesting to note, however, that thermodynamic systems, as described by a statistical mechanical partition function, are invariant under dynamical similarity. This is because the Boltzmann constant, $\beta$, sets the units for volume on phase space and can therefore be used to parametrize the flow of a dynamical similarity. However, as it is well-known, the Boltzmann constant drops out of any thermodynamic quantities. It would, thus, be interesting to use the PESA to study the gauge status of dynamical similarity in thermodynamic systems that are out of equilibrium.

If, indeed, an analysis in terms of the PESA would suggest applying the Gauge Principle to the dynamical similarity of such systems, then the same general considerations that led to an AoT in \chap\ref{ch:new_aot} might be able to be used to model equilibration in general thermodynamic systems. A symmetry argument of this kind might also be useful in motivating existing approaches to equilibration such as those developed in \cite{ottinger2005beyond} and \cite{bravetti2019contact}.

\subsection{Recovering other arrows of time}

As was mentioned in \Sec\ref{sec:aot_prob intro}, there have been several attempts to explain many (or all) relevant arrows of time once one has in hand a particular solution to the smoothness or red-shift problems (e.g., \cite{Penrose:1979WCH,Penrose:NewMin}, \cite{albert2009time} or \cite{Rovelli:2018vvy}). There are, however, good reasons to doubt that current proposals are sufficient.\footnote{ See \cite{sep-time-thermo} for a list of reasons. } One remaining question is to explain the emergence of an epistemic arrow of time; namely, to find an explanation for why we seem to have more reliable records of the past than of the future. The JA-scenario provides a potential mechanism for answering this question.

In \Sec\ref{ssub:the_janus_attractor_scenario}, I suggested a potential way to interpret the time-dependence of the measure in terms of the dynamical variability of solutions. As solutions converge in time, the state-space volume of a region transverse to the dynamical flow gets smaller and smaller, shrinking to zero as the solutions approach an attractor. The dynamical stability of the attractor therefore suggests that the attractor represents a kind of stable record of the past state. In the $N$-body model, for instance, there are many attractors on the state space so that approaching an attractor allows an observer to make inferences about the initial conditions on the Janus surface corresponding to the trajectory the observer is on. Unpacking this process might lead to insights about how an epistemic arrow of time could emerge in a JA-scenario.

\subsection{ Initial conditions and the CMB power spectrum } 

In the $N$-body model considered in \Sec\ref{sec:newtonian_gravitation_models}, `early' states were found to be \emph{nearly}, but not \emph{completely}, homogeneous in qualitative agreement with cosmological observations. This surprising fact raises the question of whether a more realistic cosmological model, combining a dynamical background with local scalar-field perturbations, could come close to reproducing the more detailed observations of the CMB power spectrum. In particular, one could develop a way to describe a dynamically similar theory of cosmological perturbations and identify a natural choice of measure for explaining the density fluctuations of the CMB. To be clear, this would be a highly speculative project, and one that may require a quantum theory of fluctuations on a curved background. However, the payoff is considerable given that the project would be addressing the very nature of inflation.

\subsection{ Boltzmann Brains }

The problem of Boltzmann Brains has re-emerged as a problem of interest in modern cosmology.\footnote{ For a philosophical introduction to the problem with applications to modern cosmology, see \chap IX of \cite{belot2023accelerating}. } The problem arises from a simple combinatorial argument: there are many more states in the state space of the Universe that contain isolated brains then there are states containing brains in bodies on a planet like the Earth. Using standard Boltzmannian arguments (such as the \emph{mixing} arguments criticised in \Sec\ref{sub:case_a_typicality}), the Liouville measure can be used to estimate the chance that, at any given time, the Universe can be found in a particular range of states. This suggests that it is exceptionally more likely for our brains to spontaneously appear out of a fluctuating soup of matter than it is for them to appear in human bodies as the result of some evolutionary process on the Earth.

If the argument of \Sec\ref{sub:dynamical_similarity_in_the_universe} is correct, however, and dynamical similarity should be taken seriously as a gauge symmetry of the Universe, then there is no longer a good reason to think that a time-independent measure, like the Liouville measure, is the appropriate measure to use for assessing the typicality of states. As a result, a simple combinatorial argument, which has no input from the dynamics, can no longer deliver the explanatory punch needed to motivate the Boltzmann brain problem.

In the dynamically similar case, it is not correct to ask whether brain configurations are typical at \emph{any given} time but whether they are typical at \emph{particular} times. In this regard, the class of measures considered in \Sec\ref{sub:the_smoothness_problem_N_body} for solving the smoothness problem tells a very different story. What we found was that smooth states were typically found near Janus points while clumped states were typically found near attractors. This suggests that states with brains, which are themselves relatively clumpy, are more likely to be found near attractors, where gravitational collapse has triggered the formation of star systems and planets. In order words, the dependence of the measure on the gravitational Hamiltonian changes the counting argument about the likelihood of states with brains in a way that suggests that they should be much more likely to be found in star systems than in a fluctuation about a smooth state in the early Universe.

While this argument seems promising as it stands, a more detailed analysis and a direct comparison to the problems arising in the recent literature could be revealing.


\subsection{ Quantum theory }

One immediate worry about dynamical similarity is that a quantum version of its Gauge Principle will lead to a radically different quantum theory. This is because $\hbar$, the parameter that sets the scale of quantum effects, carries units of angular momentum and should rescale under dynamical similarity. In particular, the rescaling of the symplectic $2$-form under dynamical similarity means that the Poisson bracket, which is the classical analogue of the quantum mechanical commutator, is also not invariant. It seems doubtful that it could be possible to construct a dynamically similar quantum theory that is empirically equivalent to standard quantum theory without preserving the standard commutation relations between position and momentum.

While these considerations present a serious challenge to extending the Gauge Principle for dynamical similarity developed here beyond the classical limit, they also present an opportunity to generalise the quantum formalism for cosmology. Because it is possible to recover a symplectic theory from a dynamically similar one whenever there is an isolated clock system,\footnote{ So that a simple integral of motion can be identified, and the drag can be set to zero using the formalism developed in \Sec\ref{ssub:counting_solutions_in_a_contact_theory}. } it should always be possible to recover the standard quantum formalism for most quantum mechanical applications. But when no obvious dynamically isolated clock is available, such as in the early Universe, a more general quantum formalism may be necessary. This could be particularly relevant for constructing a possible alternative to inflation, where the quantum formalism has been applied in the cosmological setting to explain empirical phenomena.

Note that there are good epistemic reasons to think that dynamical similarity \emph{should} be enforced as gauge theory in cosmology --- independently of the PESA. While $\hbar$ sets the scale of quantum effects in the Universe, it does so only through ratios; e.g., the ratio of the Bohr radius to the radius of the Hubble horizon. Only changes of such ratios are observable through red-shift. This means that the scale of quantum effects should, in principle, be invariant under dynamical similarity. Such an argument, when combined with the PESA, could serve as a valuable foundational principle for the development of a generalised formalism for quantum gravity that may lead to an empirically adequate theory of cosmology.

\section{ Summary and final remarks }
\label{sec:final_remarks}

In this thesis, I accomplished two main goals. The first, which was achieved in Part~\ref{part:foundations}, was to give a universal definition of gauge symmetry containing a minimal set of dynamic and epistemic ingredients. The second, which was achieved in Part~\ref{part:aot}, was to give a new explanation of the Arrow of Time (AoT) that evades the standard objections plaguing existing proposals. Let me now recap how each of these goals was achieved.

\subsection{Part I: gauge symmetry}

My proposed definition, articulated by a principle I called the \emph{Principle of Essential and Sufficient Autonomy (PESA)}, used an existing account of representation (i.e., the DEKI account of \cite{frigg2020modelling}) to make a clean separation between puzzles resulting from standard difficulties in model-building and those resulting directly from gauge symmetry. The proposal combined the elements of Dirac's sufficient criteria for gauge symmetry, according to which gauge structures are those that are underdetermined by the equations of motion (see \cite{dirac2001lectures} and \Sec\ref{ssub:the_dirac_algorithm}), with Caulton's suggestion to increase gauge structure until empirical adequacy cannot be maintained \citep{CAULTON2015153}.\footnote{ Strictly speaking, Caulton's suggestion was to increase what he called \emph{analytic symmetries}, which I take to correspond to \emph{gauge symmetries} in the language I have introduced. } I formulated my proposal as an attempt to solve what I called \emph{Belot's Problem}, which I defined in \Sec\ref{sub:belot_s_problem}.

The output of this analysis, presented in \Sec\ref{sec:statement_of_the_pesa}, was a set of definitions and norms for determining whether and how a gauge symmetry should be implemented for a given theoretical context. When applied to several well-studied examples, such as Galileo's Ship and the Kepler $2$-body problem (see \Sec\ref{sec:problems_solved}), my proposal was able to distinguish different theoretical contexts and suggest corresponding representations of the relevant symmetries that matched expectations. When applied to global reparametrisation invariant systems, which are not as well-studied in the philosophical literature, my proposal led to the conclusion that time evolution should not be treated as a gauge transformation, undermining the motivations that lead to the \emph{frozen formalism problem} in quantum gravity. On the way to achieving these results, I developed, throughout \chap\ref{ch:rep_sym}, a compact way of synthesising Noether and Dirac's treatments of gauge symmetry showing how the various constraints and theorems obtained in these two formalism result directly from degeneracies in the variational procedure from which a theory's equations of motion are derived.

Finally, as a bridge to the second main goal of the thesis, I combined, in \Sec\ref{sec:dynamical_similarity}, the PESA and the general variational formalism that I developed to give an implementation of the Gauge Principle for dynamical similarities. This showed that the gauge-fixing of a Hamiltonian theory results in a contact system characterised by a function I called the \emph{drag}, which appears in the equations of motion in a form reminiscent of the drag coefficient of a damped harmonic oscillator. Unlike in standard Hamiltonian systems, the drag can be non-zero and corresponds to the decay coefficient of the Hamiltonian and the analogue of the Liouville volume-form, which can both change in time. Thus, when gauge-fixing dynamical similarity, energy is not necessarily conserved and solutions may focus or diverge along the dynamical flow.

\subsection{Part II: The arrow of time}

In Part~\ref{part:aot} of the thesis, I proposed a new mechanism for explaining the AoT and illustrated how this mechanism works in two highly relevant models: a self-gravitating Newtonian $N$-body system and a particular FLRW cosmology. I called the new proposal the \emph{Janus-Attractor (JA) scenario}, whereby an AoT is seen by an observer near an \emph{attractor} of a theory that points from a \emph{Janus point}, which the observer interprets as the `past,' towards the nearby attractor, which the observer interprets as the `future.' This scenario is realised in the models above when dynamical similarity is treated as a gauge symmetry. In the gauge-fixed models, the drag was shown to grow monotonically from the Janus point, where it is zero, to the attractor when certain physically motivated assumptions are satisfied.\footnote{ In the $N$-body model, these assumptions were that the system have non-negative energy and did not undergo super-hyperbolic escape (see \Sec\ref{Nbody_motvation}). In the cosmological model, these assumptions were that the cosmological constant be positive, the spatial curvature vanish, and the Weak Energy Condition be satisfied (see \Sec\ref{sub:FLRW assumptions}). }

I began my analysis in \Sec\ref{sec:explanatory_target} by describing two empirical problems that, I argued, are important manifestations of the AoT in our Universe. The first problem, which I called the \emph{smoothness problem}, involves explaining the relative smoothness of the early state of the Universe. The second problem, which I called the \emph{red-shift problem}, involved explaining the wildly out-of-equilibrium behaviour of the red-shift, manifest in the large monotonic values of the Hubble parameter, in the past states of the Universe. In \Sec\ref{sec:newtonian_gravitation_models}, I then showed that, in the $N$-body model, an observer near an attractor typically sees smooth states on the Janus surface using a natural measure in the gauge-fixed theory. This solves the smoothness problem in that model.

For the cosmological model, I showed in \Sec\ref{sec:cosmological_models} that the Hubble parameter was monotonic and divergent towards the Janus point, which is normally interpreted an initial singularity, and that simple parametrisations of the gauge-fixed theory extend the cosmological solutions \emph{smoothly} through the Janus point. This solves the red-shift problem in that model. In \Sec\ref{sub:conc_gen-part_impasse}, I then compared my solutions to existing strategies for explaining the AoT; namely the generalist and particularist strategies; and found that my solutions are not vulnerable to their most prominent objections.

\subsection{Final remarks}

Central to my proposal is the claim that dynamical similarity is a gauge symmetry of modern cosmology. This claim is justified by the PESA because, as was shown in \Sec\ref{sub:dynamical_similarity_in_the_universe}, dynamical similarity transformations relate empirically indistinguishable states in cosmology. This observation is essential because it is the gauge-fixing of dynamical similarity that results in the behaviour of the drag, and the drag, in turn, is responsible for the AoT in the JA-scenario. It also raises serious questions about the validity of a Past Hypothesis, whose motivations, as we saw in \chap\ref{ch:against_PH}, rely heavily on the assumption of a time-independent measure.

The drag is the decay coefficient of the natural density on the gauge-fixed state space. Thus, any empirical or conceptual problems that require a counting procedure that makes use of a measure must be reconsidered in light of the time-dependence of that measure when the drag is non-zero. The difference between the measures that arise in my proposal and the time independent Liouville measure is that the Liouville measure counts as distinct any states that are global rescalings of the Universe. But if one only counts the states that truly matter for representing the phenomena, then one should use a state-space measure that is time-\emph{dependent}.

What I have shown is that a shift to a time-dependent measure on state space can have radical implications for the explanatory structure of a theory. No longer can one simply count entire \emph{solutions} by projecting the \emph{state space} measure onto a surface of constant time $t$. Such a procedure will not work with a time-dependent measure because a count of this kind will depend on a particular value of $t$, which is not a structure that is naturally available in the space of solutions. The old luxury of passing from state-space counts to solution counts, which forms the basis of so much intuition in statistical mechanics,\footnote{ Since a counting of solutions can be obtained from taking a time-average of the states on those solutions, this luxury is related to the luxury of assuming that the Liouville measure gives you time averages. } is no longer available in cosmology. Instead, one can only ask questions that condition on particular moments in time. These are questions like: ``given an observer is close to an attractor, how likely is it that states along the current dynamical history are smooth at the Janus point.'' Fortunately, it is exactly these kinds of questions that happen to be relevant to the AoT.

My proposed solution to the problem of the AoT therefore suggests a shift in conceptualization: to count what counts involves asking the right kinds of questions of the right kinds of things. And when this is done in the way I have proposed in this thesis, there is no remaining puzzle about the AoT. An AoT is exactly what observers like us should expect to see.


\backmatter

\thumbfalse

\bibliographystyle{apacite}
\bibliography{dissertation}

\end{document}